\documentstyle[12pt,aasms4]{article}
\def\kms{km\thinspace s$^{-1}$\ }

\def\vsini{$v${\thinspace}sin{\thinspace}$i$}
\def\deg{$^\circ$}

\received{16 April 1997}
\accepted{ }
\journalid{337}{15 January 2000}
\articleid{11}{14}

\slugcomment{Submitted to: Astrophysical Journal Supplement}

\def\deg{$^{\circ}$\ }
\begin{document}

\title{Doppler Imagery of the Spotted RS CVn Star \\
	HR 1099 (= V711 Tau) from 1981 - 1992\footnote{Based on observations
collected at UCO/Lick Observatory, McDonald Observatory, and the European 
Southern Observatory.}}

\author{Steven S. Vogt, Artie P. Hatzes\altaffilmark{2}, Anthony A. Misch}
\affil{UCO/Lick Observatory, Board of Studies in Astronomy and Astrophysics \\
    University of California, Santa Cruz, CA 95064}
\and
\author{M. K\"urster}
\affil{Institut f\"ur Astronomie, Universit\"at Wien, T\"urkenschanzstr. 17,
    A-1180 Wien, Austria}


\altaffiltext{2}{now at McDonald Observatory, The University of Texas
at Austin, Austin, TX 78712}


\begin{abstract}

Author One, Author Two, 
   Authors ad infinitum (affiliations if desired belong in parentheses, 
   but don't include full snail mail addresses) 

We present a set of 23 Doppler images of the spotted RS CVn star HR
1099 ( = V711 Tau = HD 22468) obtained from 1981 to 1992. HR 1099
shows a large cool polar spot which has persisted for the 11 years of
this study and other low-latitude spots which come and go on
relatively short (less than 1-year) timescales, and which can emerge
anywhere on the star. The polar spot has variable protuberances which
look very similar to the time-variable vertical extensions of the
Sun's polar coronal hole. The area of the polar spot and its
extensions shows marginal evidence of being periodically variable in
time with a period of about 3 yrs. and an amplitude of about 1\%,
perhaps indicative of a weak cycle, but not yet conclusive.

Comparison of our Doppler images with previously published `few-spot'-
model fits to the light curves shows that such simple spot model
solutions are often misleading and nonunique, particularly when the
light curve amplitude is small. Moreover, these spot-model fits do not
recover the existence of the polar spot. The Doppler images show quite
good agreement among multiple images at a given epoch, and between
different Doppler imaging research groups using completely independent
data sets and imaging software.

Our (cool spots only) Doppler imaging solutions generally do
well-reproduce the published light curves, however in one instance the
difficulty of fitting light curves suggests that at least one hot spot
was present on HR 1099 during one observing season.  Variations in the
mean brightness of the system, at the observed 0.05 magnitude level,
seem to correlate with spot area, particularly the polar spot,
indicating that the mean light level is a pretty good proxy of spot
area on HR 1099.

While the polar spot with variable extensions was always present,
isolated spots also frequently appeared at both mid and low
latitudes. On several occasions, isolated prominent spots emerged and
then disappeared on or near the equator.

The `migrating photometric wave' on HR 1099 is due not to a simple
longitudinal migration of spots on a differentially rotating star, but
rather to changes in the spatial distribution of a few spots (some of
which move but most of which are fixed in longitude) that emerge and
then disappear. So, at least for HR 1099, the phase drift of this
migrating photometric wave minimum contains very little unique
information about differential rotation or spot migration.

The tracks of two long-lived spots suggest that some spots which
emerge at low or intermediate latitudes migrate up to the pole in a
clockwise spiral (slower than the orbit), and then apparently merge
with the polar spot. If these dark spots trace magnetic flux, then
some of the magnetic flux which emerges at lower latitudes migrates
pole-ward and merges with the polar spot flux. It is not yet clear
whether this flux is of the same or opposite polarity to the polar
spot, and thus whether these northward-migrating low-latitude spots
reinforce or cancel the polar spot field. One of the high-latitude
spots also appeared to get stretched in longitude as it approached the
polar spot, and its overall track is quite reminiscent of the annulus
of toroidal field found by \cite{don92b} encircling the polar spot of
HR 1099 in 1990.9.

In general, the spots are very tightly locked to the orbital frame of
the system, and most disappear before they have had a chance to
migrate significantly. Like solar coronal holes, they show very little
evidence for shear due to differential rotation. A few selected
long-lived features gave longitudinal migration rates of 1 part in 300
to 1 part in 3600 of the rotation period, in the sense that
intermediate and low latitudes rotate slightly slower than the orbital
angular velocity, while the pole and highest latitudes appear to be
synchronized to the orbit. The implied differential rotation is thus
of opposite sign and about a factor of 50 less than for the Sun. The
rotation rate versus latitude bahavior can be well fit with a variety
of formulae including the Maunder formula. One of the best fits is
provided by a rotation period vs. latitude that is proportional to the
surface strength of a centered axisymmetric magnetic dipole field,
with the pole synchronized to the orbit, and lower latitudes rotating
more slowly. We believe that these starspots are not measuring
photospheric differential rotation. Instead, like solar coronal holes,
their lack of shearing and nearly solid-body rotation may be enforced
by a multi-kilogauss, axisymmetric, nearly current-free
quasi-potential global magnetic field. Our Doppler images also agree
very closely with the Zeeman Doppler imagery of \cite{don92b} and
support their finding that regions around the edge of the polar spot
and within bright spots show largely monopolar fields of at least 300
- 700 G strength.

The large permanent cool polar spots, the very low observable
differential rotation of starspots, and the evidence of strong,
essentially unipolar magnetic fields associated with them leads us to
believe that HR 1099 and other rapidly rotating RS CVn stars harbor
quite strong (multi-kilogauss) axisymmetric global magnetic dipole
fields. These fields have historically been largely hidden from view
by their high degree of rotational symmetry, by being concentrated in
the low surface brightness dark spots, and by these stars' high degree
of rotational line broadening. We propose that the starspots on HR
1099 and other rapidly rotating RS CVn stars are, by analogy with
solar coronal holes, large unipolar magnetic regions tightly frozen
into multi-kilogauss axisymmetric dipole fields in these stars. Since
the large cool polar spots, the signature of these dipoles, are not
present on more slowly rotating RS CVn stars, we believe that they
must be dynamo-induced fields rather than remnant fossil fields.

\end{abstract}


%
%

\section{Introduction}

HR 1099 (= V711 Tau = HD 22468) is one of the brightest and
best-studied of the spotted RS CVn stars. It was first discovered by
\cite{bop76}. Many of its fundamental parameters were published by
\cite{fek83}. It is a binary system consisting of K1IV + G5 V
stars. The K1IV star is the more active of the pair, and features
prominent spot activity. The G5V companion is also active, showing
H$\alpha$ in emission. Because it is one of the brightest members of
the RS CVn class, HR 1099 has received considerable attention at all
regions of the spectrum over the past 20 years, and the literature on
this star is now vast. For brevity, we will refer to many of these
references in later sections of the paper rather than trying to
summarize them all here. Of greatest interest for the present paper is
the history of spot activity on this star, which has been extensively
monitored for about three decades by many researchers. These many
studies point to a picture of an RS CVn binary system with an
extremely active and heavily spotted early-K subgiant component.

The evolution of the large cool spots on the K subgiant give rise to
variations in the shape, amplitude, and phase of the light curve of
the system. In particular, the phase migration and small period
variations of the light curves - quantities easy to measure quite
accurately from the photometry - have often been interpreted as
evidence of longitudinal migration of spots on a differentially
rotating star. By analogy to the solar sunspot `butterfly diagram',
some researchers have been led to the conclusion that spots were
forming above and below some intermediate latitude of co-rotation. A
`butterfly diagram' pattern of spot latitude formation gives rise to
light curves which could drift slowly in either phase direction as the
spots occupied latitudes which were differentially rotating slightly
faster or slower than the co-rotation latitude. Simple models using
two or three circular spots have often been used to successfully model
the light curves. However, the inherent non-uniqueness of such
solutions has always been a worrisome weakness.

The many years of broadband photometric monitoring has been extremely
useful for verifying the presence of evolving spots on the surface of
HR 1099 and deducing basic spot parameters such as size, temperature,
longitude, latitude, etc.. Photometric monitoring has perhaps even
detected evidence of stellar differential rotation and longterm spot
cycles on some RS CVn stars. But the inferences and conclusions drawn
from such studies are always indirect since only the integrated flux
from the star is measured, and the integrated flux contains little
unambiguous spatial information on the actual spot distribution. For
example, the observed migration of the light curve minimum with phase
(`the migrating photometric wave') could as easily arise from subtle
changes in the spot distribution as from actual longitudinal motion of
a given spot feature, as is often assumed.

A spatially resolved image of the spot distribution on the star is
needed to resolve these ambiguities. Such maps can be obtained using
the Doppler imaging technique for stars with sufficient rotation and
suitable inclination. Doppler images show unambiguously where spots
first emerge, and how they then migrate with time. A many-year set of
resolved images of the star can then be examined directly to follow
the longitudinal and latitudinal motions of individual spots, and
thereby reveal patterns of differential rotation on the star. Such
longterm image sets will help in interpreting the large and very
valuable body of photometric monitoring data on the class of active
late-type spotted stars, many of which are not amenable to Doppler
imaging. HR 1099 is well-suited for Doppler imaging by virtue of its
intermediate inclination, its intermediate rotation rate (which
produces pronounced but not excessive line broadening), and its
apparent brightness. It was thus selected as one of our first
candidates to follow in detail for many years by Doppler imaging.

The first Doppler image of HR 1099 was presented by \cite{vog83} and
by \cite{vop83}. It showed two large spots on the star in 1981, one
roughly circular spot very near the pole with a narrow attached lane
descending to intermediate latitude, and a second, slightly elongated
spot, near the equator. An improved version of the 1981 Doppler image
along with images for 1984 and 1985 were presented by \cite{vga87} and
by \cite{vgb87}. In each case, the Doppler image of HR 1099 was
dominated by a large cool spot straddling the pole with an attached
lane descending to lower latitudes and accompanied by one or two
smaller spots near the equator. \cite{vop83} were struck by the strong
resemblance of this polar spot and attached lane to the X-ray image of
CH1 (Coronal Hole No. 1) on the Sun, presented by
\cite{tim75}. \cite{vop83} proposed that starspots are also large
unipolar magnetic structures, analogous to solar coronal holes, the
only difference being the strength of their magnetic fields. On the
Sun, coronal holes are unipolar magnetic regions of only 10-20 Gauss
field strength, too low to affect energy transport mechanisms in the
photosphere. Thus they show up only very weakly in photospheric images
of the Sun. But their fields do significantly effect the low density
solar corona, leading to a reduction of both temperature and density
in the solar wind outflow above the holes, and producing dramatic
`dark' regions (holes) in X-ray images of the hot corona. However,
\cite{vop83} speculated that the fields in the spots of RS CVn stars
are several orders of magnitude stronger, reaching
sunspot-umbrae-level strengths of several kilogauss over most of the
area of the unipolar magnetic region. Such a field strength is
sufficient to dominate convective heat transport mechanisms in the
photosphere (as in sunspot umbrae), lowering the surface temperature
and thereby giving rise to prominently visible cool photospheric
spots. Since the large polar spot appeared to be a permanent feature
of HR 1099, \cite{vop83} also speculated that the low-latitude
isolated spots may migrate poleward to join the polar spot, thereby
maintaining its persistence.

Direct observational support for the idea that starspots might be
large unipolar magnetic regions of kilogauss field strength came from
the spectacular Zeeman Doppler imagery work of \cite{don90},
\cite{don92a}, and \cite{don92b}. They first reported the detection of
a largely monopolar magnetic region of about 1 kilogauss near
quadrature (phase 0.85) on HR 1099 in August, 1989. This magnetic
region covered about 18\% of the total stellar surface and may have
been coincident with a photospheric hot spot. They then found that the
magnetic signature from the K1 subgiant component varied with
rotational phase, and also that both the intensity and magnetic
signatures (Stokes I and V parameters) varied significantly from 1989
to 1990. Their results suggest that large time-variable monopolar
regions are present on HR 1099.

The first actual Zeeman-Doppler images (longitudinal field component)
of HR 1099 were presented by \cite{don92b} for epochs 1988.9 and
1990.9. There were large polar spots on the star in both years. The
1988.9 image also showed two warm regions (300K hotter than the
photosphere) slightly above the equator, one of which seemed to be
associated with a region of enhanced magnetic activity from a
(incomplete) 1989.6 epoch Zeeman-Doppler image taken 8 months
later. The authors suggested that the warm regions of 1988.9 were
areas of emerging radial or poloidal magnetic field. Correspondence of
the magnetic field structure with the hot and cool spots was
dramatically demonstrated in their 1990.9 epoch magnetic and
brightness images. Here, their brightness image showed a prominent,
roughly circular, cool polar spot with attached proturbence, and a
single smaller cool spot near the equator. Their simultaneous
Zeeman-Doppler image revealed that the polar spot was surrounded by a
(clockwise) ring of -300 Gauss magnetic field, and the smaller
equatorial spot was spatially associated with a (counterclockwise)
monopolar region of +700 Gauss field. The authors concluded that the
fields within the dark spots were not being directly detected due to
lack of photons from the dark regions, but were strongly implied from
the close association of magnetic signatures with the detailed shape
and boundaries of the spots. They also speculated that the toroidal
field structure surrounding the polar spot was due to winding of a
portion of the dipole field line distribution by differential
rotation. We believe that this Zeeman Doppler imaging work provides
strong support for the idea that starspots are large unipolar magnetic
regions, similar in structure to solar coronal holes, but posessing
magnetic field strengths several orders of magnitude larger.

Over the years, the idea of large cool polar spots on RS CVn stars has
had its share of skeptics, no doubt because the spots were not
detected by broadband photometric spot modelling, and also perhaps
because they had no obvious analog among sunspots. Many mechanisms
were proposed - limb brightening, differential rotation, gravity
darkening, problems with the line flux profile radiative transfer
calculations, chromospheric line-filling, out-of-focus spectrographs,
etc. - which might give rise to spurious polar spot features. But,
over the past decade, many other Doppler imaging researchers have
independently found large polar spots in their images of rapidly
rotating RS CVn stars. Furthermore, some more slowly-rotating spotted
RS CVn stars have also now been found by \cite{hat93} which do {\it
not} show these polar spots. \cite{hat96} also recently completed a
detailed study of the reality of polar spots using the observed
inclination dependence of the mean shape of the line flux profiles
from spotted stars.  Their modelling simulations argue strongly
against gravity darkening, differential rotation, limb brightening,
equatorial bright bands, chromospheric line-filling, and unknown
effects in line radiative transfer physics as possible causes of
spurious polar spots. Rather, the observed inclination dependence of
the flattening of the line flux profiles in RS CVn stars is quite
naturally and simply explained by the presence of large, cool polar
spots.

We have spent the past decade monitoring the spot distributions on
several of these stars. In the present paper, we present a set of 23
Doppler images of HR 1099 obtained between 1981 and 1993. In
hindsight, HR 1099 has been a rather difficult star to study for
Doppler imaging since the spots change on timescales of less than one
year, requiring more than one Doppler image per season to obtain a
fair sampling of all the changes in their distribution. Where
possible, more than one image per year has been obtained.

We compare theoretically generated light curves from our images with
observed light curves and also attempt to tie in many other
observations of activity on HR 1099 with our spot maps. It is hoped
that researchers studying this star will be able to make use of these
images to better interpret their wide variety of optical, radio,
ultraviolet, and x-ray observations taken over the same time interval.

\section{Observations}

Spectral observations for the Doppler imaging represents a combination
of data sets taken at three different observatories: Lick, McDonald,
and The European Southern Observatory (ESO). During the course of
collecting the observations five different spectrographs were used for
data acquisition. Tables 1-23 list the journal of observations which
include rotation phase, the Julian Day of mid-exposure, the exposure
length, the signal-to-noise per pixel, and the observing station.
(Lick 80"= 80'' coude spectrograph camera, McDonald coude= 2.1m coude
spectrograph; ESO = 1.4m CAT + CES; Lick HS is the Hamilton; McDonald
SE=2.1m + Sandiford echelle). In all observations where strong
telluric lines were present the observation was divided by an
appropriately scaled spectrum of a rapidly rotating hot star.

\subsection{Lick observations}

The 1981 data were obtained at Lick Observatory using a double-pass
echelle and Intensified Dissector Scanner (IDS) system. From 1982
through 1988 observations of the Ca I 6439 {\AA} line were obtained
using the coud{\'e} spectrograph of the Shane 3-m telescope at Lick
Observatory and a Texas Instruments 800 $\times$ 800 CCD detector. A
modified Bowen-Walraven image slicer was used to reformat the light
from a 3 arc-second hole to a 0.67 arc-second slit. The resulting
resolving power was 48,000 and typical signal-to-noise was about
200-300 per 15 $\mu$m pixel. A resolution element consisted of 3.5
pixels. Since the beginning of 1990, all observations of HR 1099 have
been made with the Hamilton Echelle spectrograph (\cite{vogt87}) at
the coude focus of the Shane 3-m telescope. The Hamilton yielded a
resolving power and signal-to-noise roughly equivalent to the
conventional coude spectrograph, but with improved throughput and much
greater spectral coverage.

\subsection{McDonald Observatory Observations}

The McDonald data set was obtained using the 2.1-m telescope at
McDonald Observatory. From 1988 until early 1992 data was taken using
the the coud{\'e} focus of McDonald Observatory's 2.1-m telescope. A
1200 gr mm$^{-1}$ grating in second order and a Tektronics 512$\times$
512 CCD were used resulting in a dispersion of 0.038 {\AA}
pixel$^{-1}$ and a resolving power of about 65,000. The wavelength
coverage was 23~{\AA} centered on 6430 {\AA}.

Beginning in late 1992, data at McDonald Observatory were acquired
using the Sandiford Cassegrain Echelle Spectrograph, again at the
2.1-m telescope. This instrument is a prism cross-dispersed echelle
mounted at the Cassegrain focus of the 2.1-m telescope
\cite{msbb93}. It is used with a 1200$\times$400 Reticon CCD and
provides a dispersion of about 0.05 {\AA} pixel$^{-1}$ at 6500
{\AA}. The instrumental resolution for the observations was 50,000.
The spectrograph setup was chosen such that a wavelength coverage of
about 5700--6900 {\AA} was obtained.

\subsection{ESO Observations}

Spectroscopic observations for four of the images were taken at ESO
using the 1.4\,m--CAT telescope equipped with the CES spectrograph,
the short camera, and an RCA CCD. All spectra were recorded at a
dispersion of $3.47~{\rm \AA ~mm}^{-1}$ and centered at $6444$~{\AA
}. Since the CCD has 1030 pixels of size $15\mu $ in the dispersion
direction, the resulting length of the measured spectrum was $\approx
52.5$~{\AA }. A resolving power of $\lambda /\Delta \lambda =50,000$,
equal to a spectral resolution of $0.129$~{\AA } was used.  Data
reduction for this data set was performed with the MIDAS software
developed by ESO and involved the usual steps of bias subtraction,
flatfield division, removal of cosmic ray hits, spectrum extraction
(i.e.~summation of relevant CCD columns), wavelength calibration, and
continuum normalization.

\section{Imaging the Photospheric Features}

\subsection{Stellar Parameters}

In order to image a stellar photosphere one needs the stellar
inclination, projected rotational velocity ({\vsini}), rotation
period, and the spectral line profile at each location on the stellar
surface. For the stellar inclination the value of the orbital
inclination, $i$ = 33 $\pm$ 1{\deg} determined by Fekel (1983) was
adopted. The {\vsini} was determined by fitting the mean Ca I 6439
{\AA} profile with a synthetic profile generated using model
atmospheres and a disk integration scheme. The mean profile was
calculated by summing individual observations spanning a rotation
period.  This minimizes the distortions due to spot features and gives
a slightly better approximation of the immaculate (i.e. unspotted)
line shape than does a single observation. The value we obtained was
{\vsini} = 40 $\pm$ 1 {\kms}, consistent with the one found by Donati
et al. (1992). Synthetic spectral lines were generated using model
atmospheres of \cite{bell76}. Profiles were generated at 20 limb
angles to account for linb darkening. A macroturbulent velocity
$\xi_{RT}$ = 4 {\kms} was also used.

We used the HR 1099 orbital ephemeris of \cite{fek83} for calculating
all phases. This formula expresses the heliocentric Julian Day of
conjunction (active KIV spotted star in front) as: $$HJD = 2442766.080
+ 2.83774{\pm} 0.00001 E$$ Thus Phase 0.0 corresponds to conjunction
with the spotted star in front and, on all our Doppler images, the
secondary sits fixed at Phase = 0.5. When looking for phase drifts in
a highly synchronized orbital system, one must also consider the
effects of period errors which accumulate phase errors over
time. \cite{fek83} also found the orbital period to be constant over
60 years. Using three times his formal error of the period estimate,
we find no more than 0.015 phases (5.4\deg) of longitude error (with
respect to the orbit) over the entire 11.1 year span of the datasets
analyzed in this paper. Thus the adoption of this constant period is
more than adequate to compare longitudes of all features on HR 1099
over those 11.1 years.

\subsection{Reconstruction Technique}

The spot distribution was derived from a time series of spectral line
profiles using the maximum entropy method (MEM), the details of which
can be found in Vogt, Penrod, \& Hatzes (1987). Because the problem of
deriving a surface distribution from spectral line profiles is
non-invertible (i.e. there are a number of possible solutions), other
constraints must be imposed on the image. The maximum entropy method
uses the criterion that the image be the smoothest solution (i.e the
one with the least amount of spatial information) that still fits the
observed data to the level of the noise.

In the Doppler images presented here, the additional constraint was
imposed that only cool spots (i.e. temperatures below the photosphere)
be allowed. This was done for two reasons. First, it is
well-established that most photometric varations of RS CVn stars can
be explained by cool spots. In fact, virtually all spot temperatures
that have been derived for RS CVn-type stars have been cooler than the
photosphere. Second, and more importantly, simulations indicate that
an unconstrained MEM (one that can find hot and cool spots) will find
both, even from a data set that was constructed using only cool
spots. Thus, unless one has very accurate {\it simultaneous}
photometry to further constrain the temperature of the spots, it is
risky to allow the method to find hot spots. Since we generally did
not have this simultaneous photometric data, we chose to constrain the
method to find only cool spots rather than try to interpret hot
features that may well be artifacts.

\subsection{Thresholding the Images}

\subsubsection{Using Spectral Lines}
	
The maximum entropy digital image reconstruction method produces a
maximally-smoothed image with a continuous transition from spotted to
photospheric regions. This is the case even if the original
distribution consistes of only two temperatures.  This makes it
difficult to discern spot boundaries (edges) when comparing different
images. To facilitate comparisons betwen images, as well as to display
the more prominent features of the spot distribution, the `raw' MEM
images were converted into a two-temperature image in the following
manner.  First a temperature threshold was selected, below which any
image pixel was regarded as being `spot', and above which a pixel was
regarded as being unspotted `photosphere.' All `spot' pixels were
replaced by a temperature of 1200 K below the photospheric temperature
of 4750 K, and all `non-spot' pixels by the nominal photospheric
temperature. This ``binary'' distribution was used to calculate a set
of predicted spectral line profiles that were then compared to the
observed profiles. The threshold level was adjusted until a
satisfactory match to the observed line profiles was
obtained. (Another artifact of MEM is that low-latitude spot features
tend to be warmer than high latitude features due to their small
projected areas. Consequently the threshold level for these
low-latitude features was about 100 K warmer than for high latitude
features.)

Although this approach disallows any possibility of imaging penumbrae,
such details are probably not warranted by the dataset anyway. The
process does capture the essence of the fact that most sunspots, when
viewed at low spatial resolution, are probably umbral-dominated
structures. These thresholded images will often be referred to as
``spectral'' images (since they were derived purely from spectral
data) and will be shown as a combined two-temperature and grayscale
version. The black areas of the Doppler images are regions where spots
are highly certain and the presence of darkening is relatively
insensitive to the details of thresholding. These regions thus
represent the two-temperature distrubution that best fits the observed
spectral line profiles. Regions of the image that did not survive the
thresholding process are shown in their original grayscale level. By
showing both the `raw' and the thresholded versions of the Doppler
image solution as gray and black respectively, we have sought to focus
the viewer's attention on the most certain features (black), while
still preserving other more subtle features (gray), some of which may
also be real.

\subsubsection{Using Photometry}

None of our spectral data were accompanied by simultaneous
photometry. Published light curves, when available, are still useful
even if not contemporaneous with our spectral data. If the photometry
is contemporaneous, then it provides additional constraints to the
Doppler image, primarily in fixing spot areas. If the light curve is
not contemporaneous with our Doppler image then it may provide us with
useful knowledge about spot evolution. This is particularly important
since the time sampling of our images can often be quite poor, so a
spot distribution derived from the light curve can fill in gaps
between images.

The reconstruction technique, in its current implementation, does not
incorporate light curve fitting. However, the derived Doppler image
combined with light curve information can be used to produce a
photometric spot model in the following manner. The spectral image was
used as a starting point for fitting the observed light curve. First,
the best fit to the light curve was obtained by increasing or
decreasing spot areas (essentially setting a new threshold level)
until the predicted light curve best matched the observed. If the fit
to the light curve was still poor then individual spot features were
removed or added. Additions were generally only made to those regions
of the MEM image that indicated the presence of a feature (even if it
was at a low level). This way the photometry could bring out low-level
features which were not certain in the Doppler image and did not
survive the thresholding process, but for which the photometry
indicated a feature must be present.

In some instances, the photometric data were acquired significantly
far from the time of our spectral data so that spot migrations may be
important. In this case, not only were spot areas from the original
MEM image allowed to change, but individual spot locations were also
allowed to move to obtain a better fit to the light curve. Finally, if
the predicted light curve was still a poor fit, additional spots were
added at arbitrary locations. The final spot distributions which
provided the best fit to the observed light curve will often be
referred to as the ``photometric'' image since the final spot areas
and locations were driven more by the light curve than by the spectral
data.

\subsection{Limitiations of Poor Phase Coverage}

In trying to discern spot evolution and migration patterns we were
often forced to derive images from spectral data with sparse phase
coverage. It is worth mentioning here the resulting limitations. For
high latitude features sparse phase coverage has minimal effects. The
inclination of HR 1099 means that all regions of the star above a
latitude of 57{\deg} are always in view. Features above this latitude
(e.g. polar spot) can be accurately recovered with only a few
observations.

The largest effect of sparse phase coverage is on low-latitude
features. If a low-latitude spot were to lie in the middle of a large
phase gap it would be missed outright by the reconstruction
process. If this feature is seen only at one phase, then it may appear
in the raw MEM image, but be lost by the thresholding process. For
example if, at a given observation, such a spot is near the limb of
the star, it will only effect the spectral line profile in the wings
where the distortions are more subtle than for features at line
center. The reconstruction process would thus put only a weak,
low-level feature near the location of the spot. This feature would
not survive the thresholding process since there are no additional
observations to help us judge whether such a low-level feature is
indeed real. On the other hand, if the low-latitude spot is seen at
line center in only one observation, it will produce a strong
distortion in the line profile, but its latitude will remain
completely ambiguous. Thus in the thresholding process a high-latitude
feature (e.g. a polar appendage) may be favored at the expense of a
low-latitude feature at the same longitude.
	
Mirroring (i.e. the appearance of ghost images at low-latitudes from
high-latitude features) is another subtle effect which may be present
at some level even with reasonably good phase coverage. If the star
has an inclination of 90{\deg} then there is an equal probability that
a feature lies in either hemisphere. The spot distribution below the
equator will thus be a mirror image of the opposite hemisphere. As the
inclination of the star is decreased this ambiguity is removed and MEM
tends to favor placing the spot at the positive latitudes. However,
even for low-inclination stars there may still be some residual spot
features visible at negative latitudes.

The appearance of ghost or mirror images is also a concern when the
phase coverage is sparse. To fit a distortion in a line profile at a
given phase, MEM places spots along a constant radial velocity chord.
In order for MEM to ``chose'' whether the feature actually lies at low
or high latitude it must have additional information at other phases;
the temporal behavior of distortions from low-latitude spots is
different from that from high-latitude spots. If the phase coverage is
poor, then blurring of features in latitude may cause ghosts to appear
at low latitudes.

\subsection{Pre-1981 Gleanings}

Our first Doppler image of HR 1099 was obtained in 1981. Prior to
that, our knowledge of spots was limited to relatively simple (1-spot
or 2-spot) models fit to broadband light curves.  We begin our
discussion of the present 11-year image-set by reviewing the published
perception of starspots on HR 1099 from data and modelling done before
1981.

\cite{wei78} presented coordinated ultraviolet, optical, and radio
observations of HR 1099 taken in Sept. 1976. They observed V-band
variations as well as variable H$\alpha$, L$\alpha$, and Mg II
emission, and strong Ca II H and K emission. Their observations
provided strong evidence that significant chromospheric activity was
present in this binary system and was associated with the K0IV star.

\cite{dor82} modelled a set of light curves collected between 1963 and
1981. As is typical of most spotted RS CVn stars, HR 1099 exhibited
large light variations each observing season, with dramatic changes in
the light curve's shape, amplitude, and phase from season to
season. They were able to obtain good fits to the rather complicated
changes in the light curves throughout the entire 12-year period with
a simple `2-spot model' with cool spots on the K0IV star. The spots
were found at latitudes as high as 65$^{\circ}$. The complex changes
in the light curve were explained by fairly simple systematic changes
in the sizes, latitudes, and longitudes of the two spots. They also
noted a maximum in spot area which occured in late 1978, at which time
some 20\% of the K0IV star was covered with cool spots. They were also
able to place a lower limit of 5 years on the length of any spot
cycle. They noted a steady increase in the phase of photometric
minimum (photometric wave advances with phase), with a migration
period of about 9.5 years. Finally, they also noted that observed
radio flaring seemed to occur at a time when the spots were changing
rapidly in size and position. \cite{kan89} presented a similar `2-spot
model' analysis of essentially the same (1963-1982) light curve
set. Their fits to these light curves yielded cool spots on the K0IV
star which also produced a photometric `migration wave' which advances
with phase with a migration period of about 7.6 years, and came from 2
dominant spots fixed at 45$^{\circ}$ latitude. The advancing phase of
the migrating wave then implied, with their model, that the star is
rotating more slowly than the orbital period at 45$^{\circ}$ latitude.

A large radio outburst observed on February 20, 1978 by \cite{fel78}
drew further attention to this star and set off a large number of
other observations (many of these are described in the series of
papers introduced by \cite{hal78}). \cite{fel78} interpreted the
observed radio emission as nonthermal gyrosynchrotron radiation from a
volume whose characteristic dimension was several times larger than
the stellar diameters but comparable with the binary star separation.

\cite{bar83} presented light curves from 1979-81 and interpreted the
remarkable changes in the light curve shapes as due predominantly to
latitudinal and longitudinal migration of two distinct and
constant-area spots on a differentially-rotating star whose
co-rotating latitude was somewhat away from the equator. They compiled
data on the phase of photometric minimum from 1976 to 1981 and
concluded that the wave migration rate was neither constant, nor in a
single direction. From 1977 to 1979, the migration rate was constant
at about 5 years/cycle towards increasing phase, but, as shown by
\cite{par81}, from 1976 to 1977 the migration rate was much slower and
moved toward decreasing phases. This was a key observation since, up
to that time, the migration rate had appeared quite uniform and in the
same direction for many years. From 1979 to 1980, the migration rate
was also towards advancing phases, but slowed considerably from that
of 1976 to 1979.

\cite{dor81} presented photometry of HR 1099 from 1977-78 and 1979 and
derived basic spot parameters from simple `2-spot' model fits to the
photometry using cool spots on the chromospherically active K0IV
star. They determined the spot temperatures to be about 1800 K cooler
than the photosphere, and (circular) spot radii to be about
26$^{\circ}$ to 32$^{\circ}$. Their solutions placed spots as high as
48$^{\circ}$ in latitude, with spot areas 10-14\% of the stellar
surface. They noted a significant change in the amplitude, shape, and
phase of the light curve from 1978 to 1979. They also reported their
strong suspicion that a spot cycle may be in progress on HR 1099, with
spot locations migrating from high to low latitudes over a few-year
interval. \cite{lod88} presented a two-component circular spot model
which used both penumbral and umbral regions for each spot. They found
spots at latitudes as high as 60$^{\circ}$. They reached the same
basic conclusion as previous modellers of the 1977 to 1981 light
curves, that the changes were predominantly due to movement of
constant-area spots in longitude and latitude.

\cite{ayr82} reported emission features of such high-temperature
species as C II and C IV from HR 1099 in 1980.5. These emission
features were clearly associated with the K star and were interpreted
as a patchy brightness distribution on and above the K star surface,
spread out in velocity by the star's rapid rotation.

\subsection{The 1981 Doppler Image}

So, prior to obtaining an actual image of a spotted star, the
published perception of the spots on HR 1099 was of a small number of
roughly circular cool spots, covering a total combined area of as much
as 20\% of the stellar surface. These spots were believed to sit at
mostly intermediate latitudes, with some spots reaching as high as
65\deg latitude. These spots were thought to maintain roughly constant
area from year to year, and the dramatic changes in the light curves
from were ascribed mostly to migration of constant area spots in both
longitude and latitude.  The `migrating photometric wave' was
interpreted as due to motion of spots on a differentially-rotating
star whose co-rotating latitude was somewhat away from the equator,
though the details were not clear since the migration rate was
variable and the direction could also change.

Our first Doppler image (\cite{vop83}) of HR 1099 was for the year
1981. It was a rather preliminary and crude version because, at that
time, we were approaching the inversion problem by `trial and error'
iteration. Nevertheless, it showed something quite unexpected from the
previous photometric spot models and from solar analogy; a large spot
straddling the pole. This polar spot was always in view in this
low-inclination system, and thus contributed little modulation to the
light curve. The polar spot had an attached lane which descended to
lower latitudes. There was also a second, slightly elongated spot near
the equator. Since that preliminary image was made, we have vastly
improved the Doppler imaging technique, incorporating principles of
maximum entropy digital image reconstruction (\cite{vph87}).

\begin{figure}
\plotone{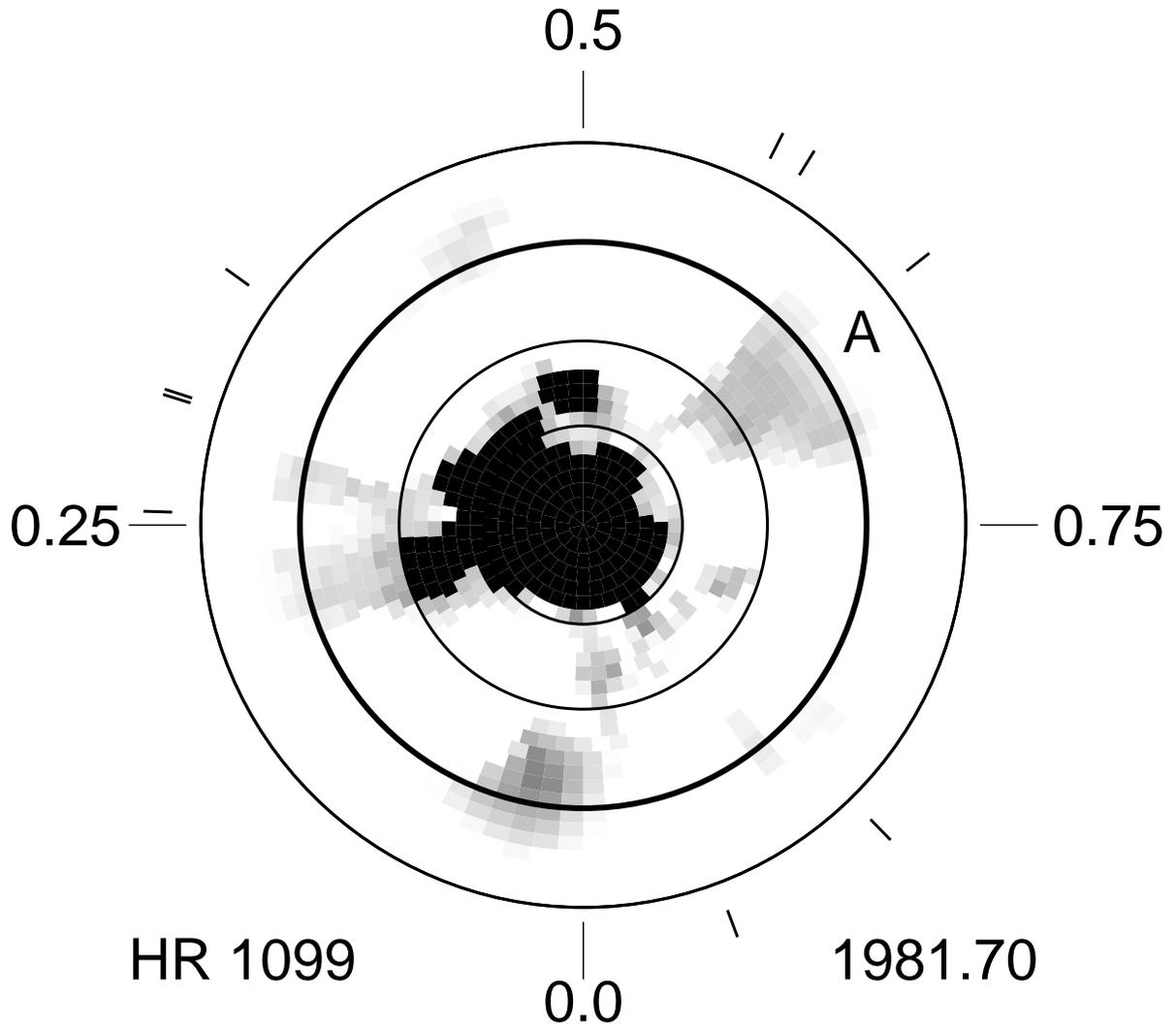}
\caption{HR 1099 raw Doppler image for 1981.7}
\label{fig:1981.7_raw}
\end{figure} 

\begin{figure}
\plotone{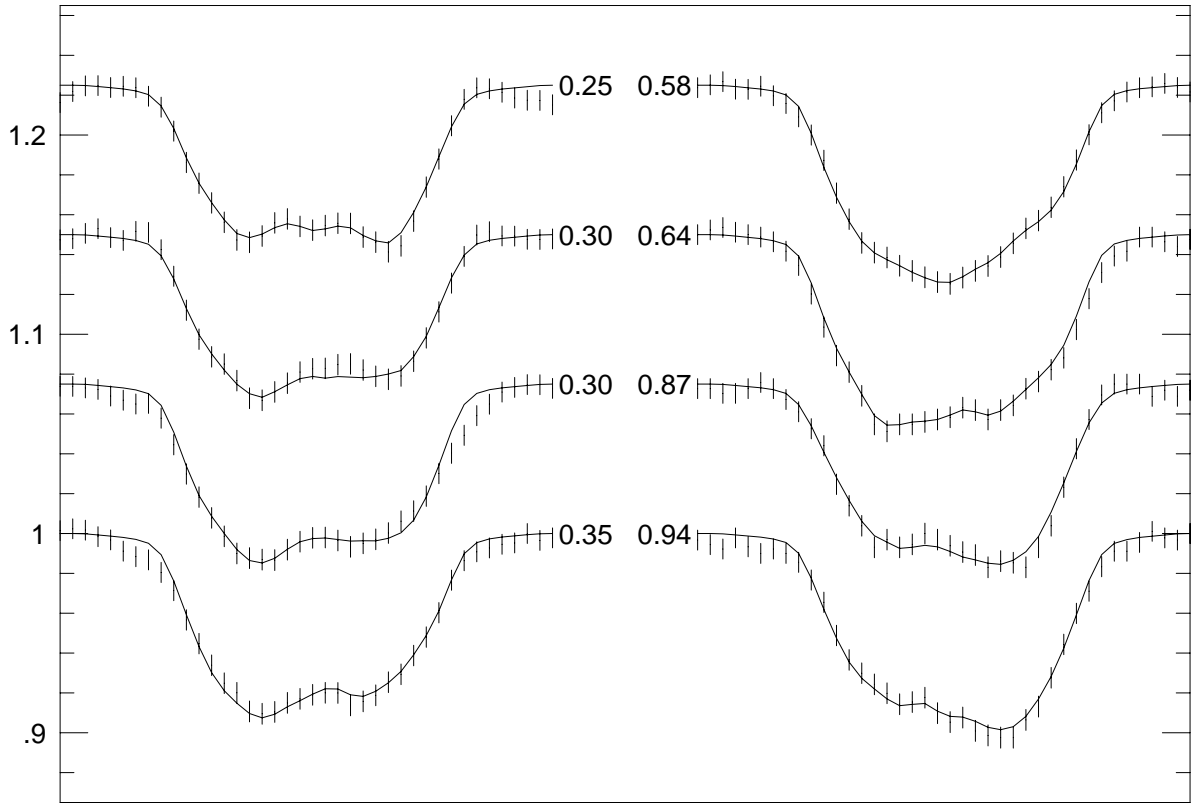}
\caption{Observed spectral line data (vertical bars) and fits (solid lines) 
for the 1981.7 image} 
\label{fig:81fits}
\end{figure} 

The most recent (raw) version is shown in
Figure~\ref{fig:1981.7_raw}. HR 1099 is actually viewed from an
inclination angle of about 33\deg and, also being spherical, makes
representation of the Doppler image of the star difficult, requiring
either highly distorted mercator-like projections, or 3-d spherical
renderings (globes) at multiple phases. To make intercomparison of
images as simple as possible, we have rendered all the Doppler images
of this paper as `flattened polar' projections, such that only one
minimally-distorted image is required for each epoch, and all areas of
the star visible from Earth are shown simultaneously in a single
image, including latitudes below the equator, down to the limit of
visibility of about -30\deg. The bold latitude line in
Figure~\ref{fig:1981.7_raw} represents the stellar equator, with finer
lines for latitudes of +30\deg and +60\deg also provided. The pole of
the star is thus at the center of the figure and the outside edge of
the star represents a latitude of about -30\deg. Phases at which the
actual line flux profiles were obtained are denoted as tick marks
arranged radially around the star. The spectral line profiles used in
the modeling are shown as vertical bars in
Figure~\ref{fig:81fits}. The length of each bar represents the error
in the flux measurment. The MEM fit to the profiles are shown as solid
lines.

Figure~\ref{fig:1981.7_raw} is called a `raw' image in the sense that
it was derived only from the line profile information, with no attempt
to fit the broadband lightcurve. It shows a large polar spot of about
55\deg diameter straddling the pole with an attached protuberance at
about phase 0.2 - 0.3. In our original `trial and error iteration'
image (\cite{vop83}) there was also another large isolated spot at
latitude 21\deg and phase 0.65.  However, in our more recent and
powerful MEM imagery, this feature essentially disappeared, leaving
only a low-level remnant in Figure~\ref{fig:1981.7_raw} near latitude
20\deg and phase 0.65. While neither the line profiles nor the
photometry require a feature here, evidence of a similar feature has
been reported by others, so in the following discussion we will refer
to it hereafter as Feature A.

\begin{figure}
\plotone{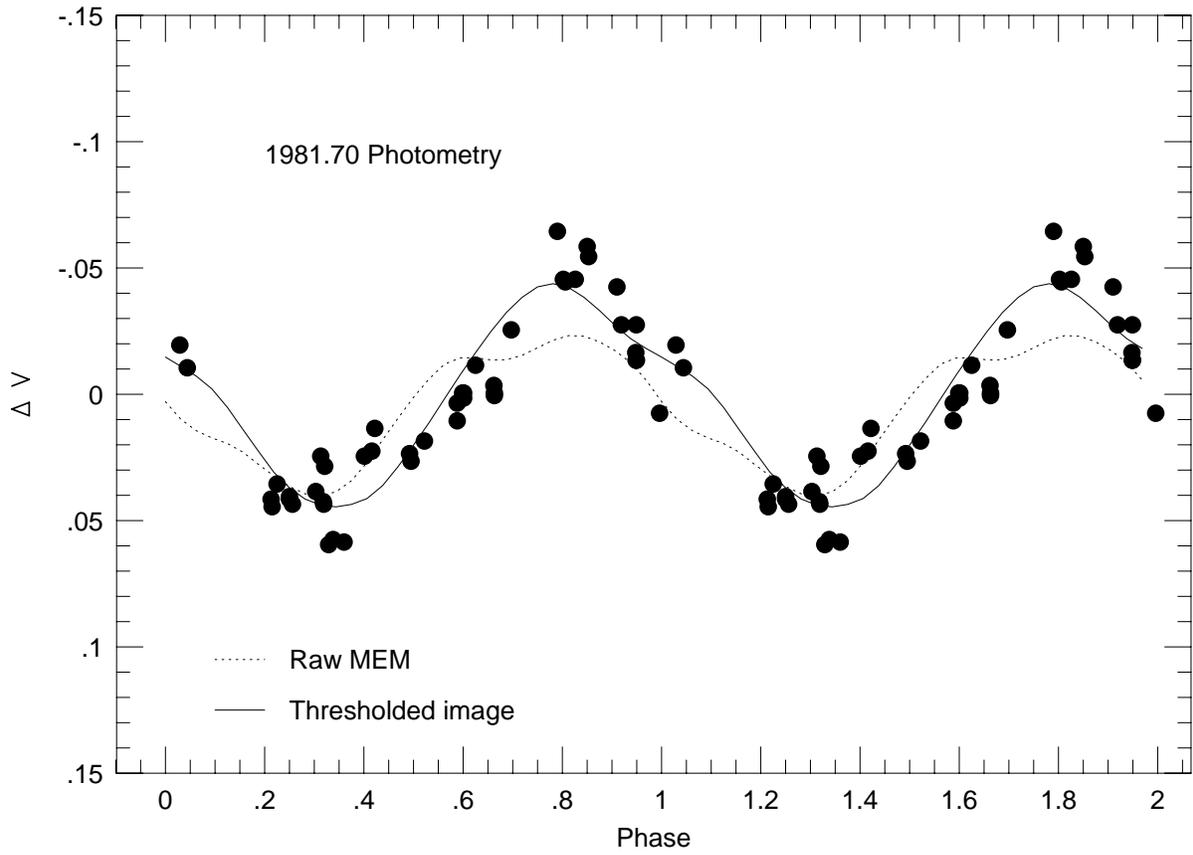}
\caption{HR 1099 light curve for 1981.70 - 1981.80}
\label{fig:1981.7_light}
\end{figure}

\cite{rod86} presented photometry from the 1981-2 observing season (in
conjunction with the 1980 and 1981 IUE observing campaigns, and worked
out `two-spot' model fits to the photometry. Their 1981.7-8 light
curve data are presented as the points in
Figure~\ref{fig:1981.7_light}. The theoretical light curve predicted
from our raw Doppler image of Figure~\ref{fig:1981.7_raw} is shown as
the dotted line.  Clearly its amplitude is too low, and is not an
acceptable fit to the observations. However, the fit from the
thresholded image (2-temperature distribution) does provide an
adequate fit to the data.

\begin{figure}
\plotone{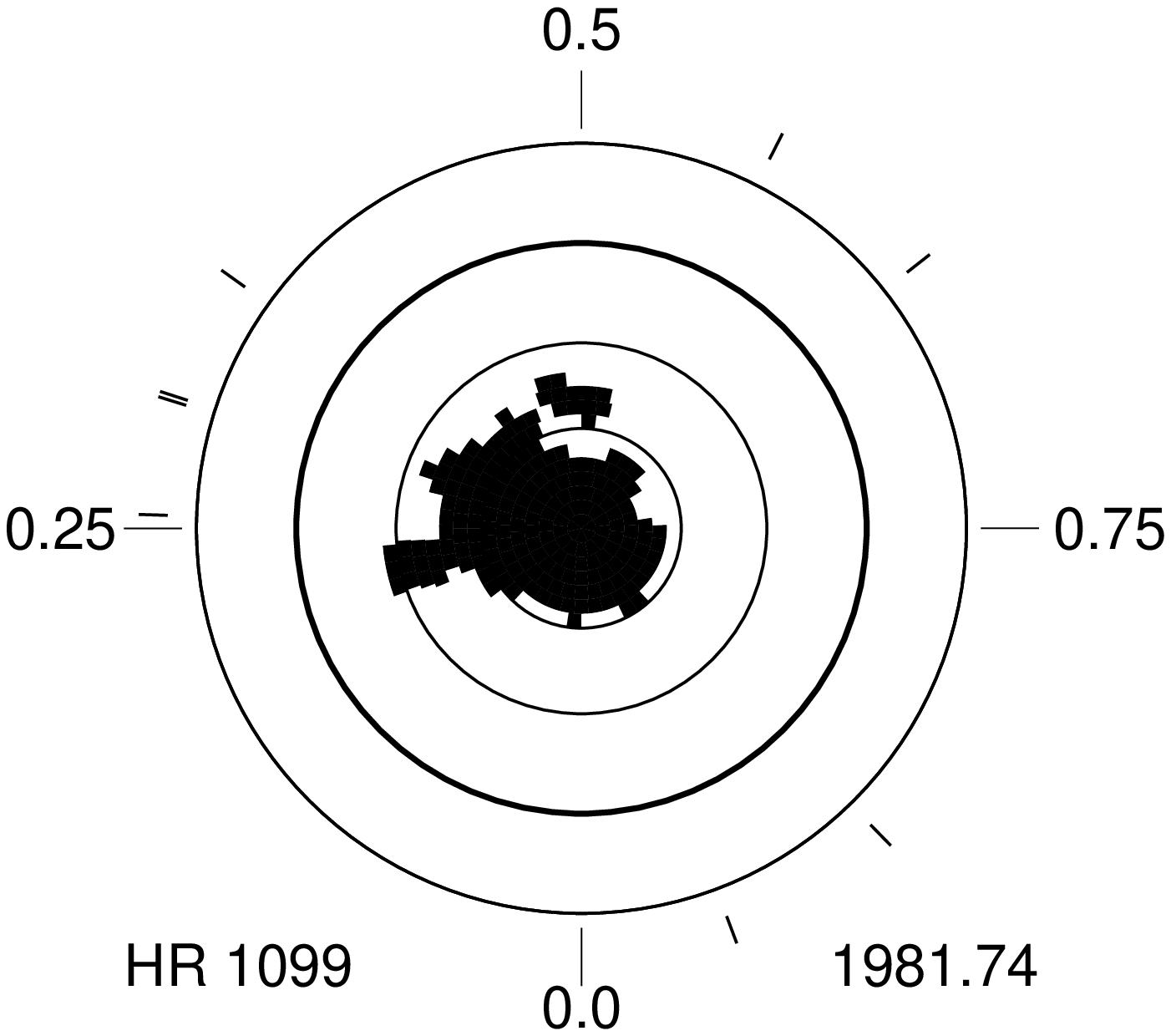}
\caption{HR 1099 thresholded Doppler image for 1981.7}
\label{fig:1981.7_image}
\end{figure}

The best fit to the photometry with a thresholded image was obtained
{\it without} the low latitude phase 0.65 spot (Feature A). This spot
is actually at a fairly low level in the MEM image and the line
profile fits of the thresholded image are totally consistent with not
having a spot there at all. Our best and final Doppler image, obtained
after further threshholding to improve the light curve fit and still
fit all the line profiles is shown in Figure~\ref{fig:1981.7_image}.
Its predicted light curve is shown as the solid line in
Figure~\ref{fig:1981.7_light}. The polar spot was pretty much
unchanged by the additional constraints of light curve fitting, but
Feature A completely disappeared.

\begin{figure}
\plotone{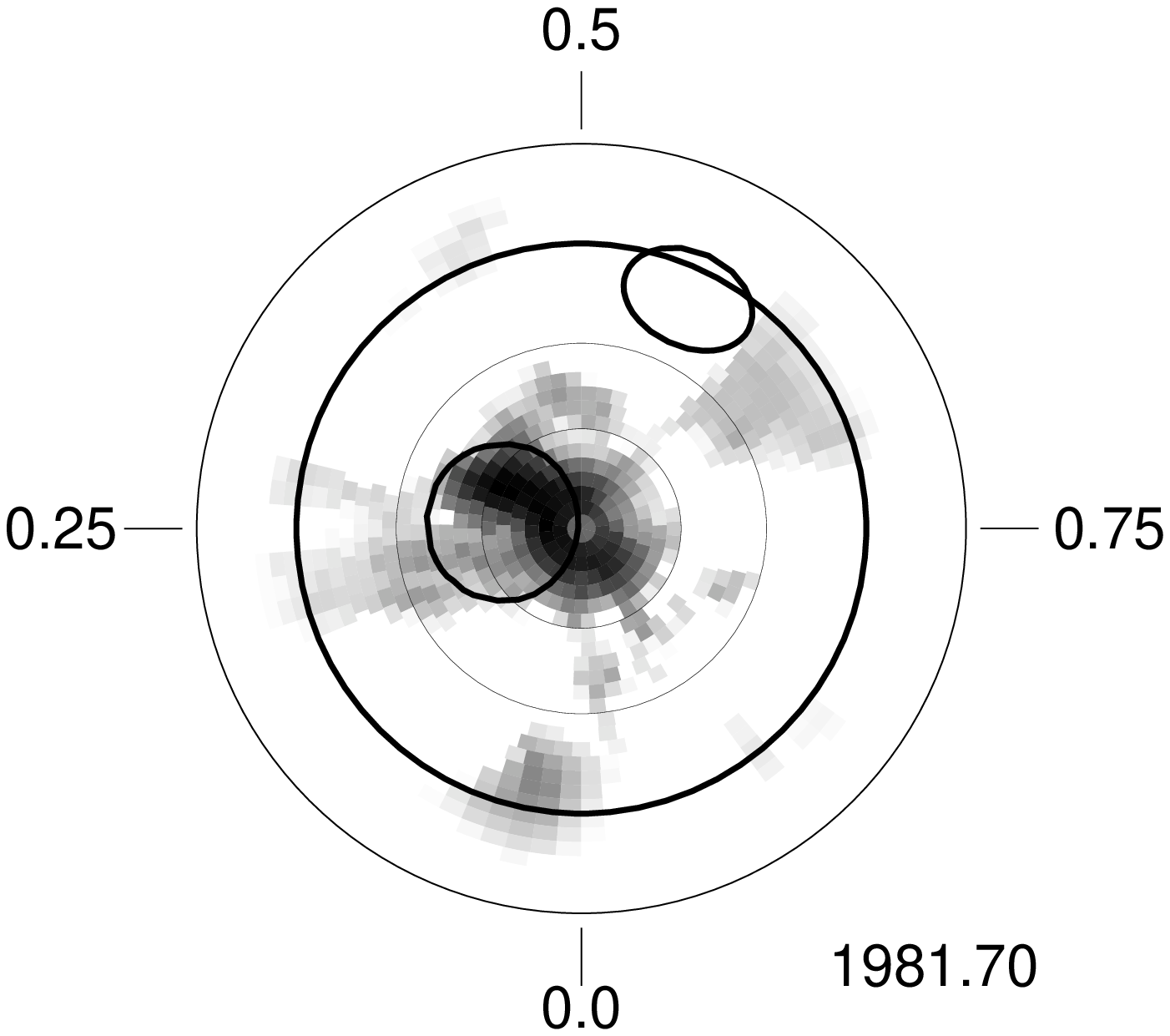}
\caption{The 2-circular-spot photometric model of Rodono et al. superimposed
on the raw MEM Doppler image}
\label{fig:solution1}
\end{figure} 

Figure~\ref{fig:solution1} shows the \cite{rod86} spot model
superimposed on a grayscale version of the MEM image (without
thresholding) from 1981.  Clearly their spot at high latitude
corresponds to our polar spot protuberance near phase 0.3 in
Figure~\ref{fig:1981.7_image}. This feature has smaller diameter and
does not straddle the pole as does the corresponding feature in the
Doppler image primarily because the photometry is not sensitive to
spot regions very near and symmetrical about the rotation pole. Their
spot 2 was of diameter 30\deg in 1981.7 and situated at +10\deg
latitude and at their phase 0.57. This phase corresponds to phase 0.59
in our image and is somewhat near Feature A in our original and raw
Doppler images. Their Spot 2 was also growing in diameter and shifting
toward earlier phases, reaching a diameter of about 48\deg by the time
it reached phase 0.49 at epoch 1981.9. It would be natural to then use
such a result, longitude migration of a single spot, to derive the
differential rotation of the star. However doing so in this case would
have been quite misleading since Feature A probably does not exist.

\cite{rod86} may have been guided by our original published HR 1099
image from \cite{vop83} (which {\it did} show a prominent Feature A)
in their choice of a spot solution which fit their photometry. Thus
the initial agreement between their 2-spot solution and our original
Doppler image looked promising. However, since Feature A did not
survive our more modern MEM imaging technique and our further
constraints of threshholding to fit the photometry, its existence is
quite suspect.  We cannot, however, absolutely rule out the existence
of Feature A, and could tolerate some spot at that general location on
the star and still be consistent with all the line profile and
photometric data.  Unfortunately, the 0.02 magnitude scatter in the
photometry does not allow us to be more definitive since the reality
of features such as Feature A hinge crucially on fitting inflections
in the light curve at these levels.  But, to within the S/N of the
data, Feature A is certainly {\it not} required by either the line
profile fits or the photometry. Our Doppler imaging method, by design,
yields the simplest image (i.e the one with the least amount of
information content) consistent with all the data. It is thus an
`Occam's razor' approach to a highly non-unique inversion
problem. Thus there may be image structure present on the star beyond
what is revealed in the maximum entropy image, but the data cannot
unambiguously support such structure. By the same token, the minimal
information content nature of our Doppler images assures that any
structure present in the image is fully demanded by the dataset. Since
the much simpler final image we present, without any Feature A, is
fully consistent with both the photometry and all the line profiles,
we must conclude that Feature A probably does not exist. The end
result is that, overall, the agreement between our final threshholded
Doppler image, and the 2-circular-spot solution of \cite{rod86} is not
very good.

\cite{lod88} also presented a `2-spot' model solution to the 1981
light curve data of \cite{rod86}. Their solution provided a basically
adequte fit to a single light curve, but their fit was clearly not as
good as the fits to multiple light curves by \cite{rod86}. Not
surprisingly, \cite{lod88} derived a quite different characterization
of the spots, further indicating how far off the mark the `2-spot'
approach can stray. They found two spots of diameter 58\deg and 42\deg
respectively, separated by 110\deg in longitude, and both situated at
23\deg latitude. They did not include a phase zero point in their
solution, so their only longitude information was the relative
longitude separation of the spots.

\begin{figure}
\plotone{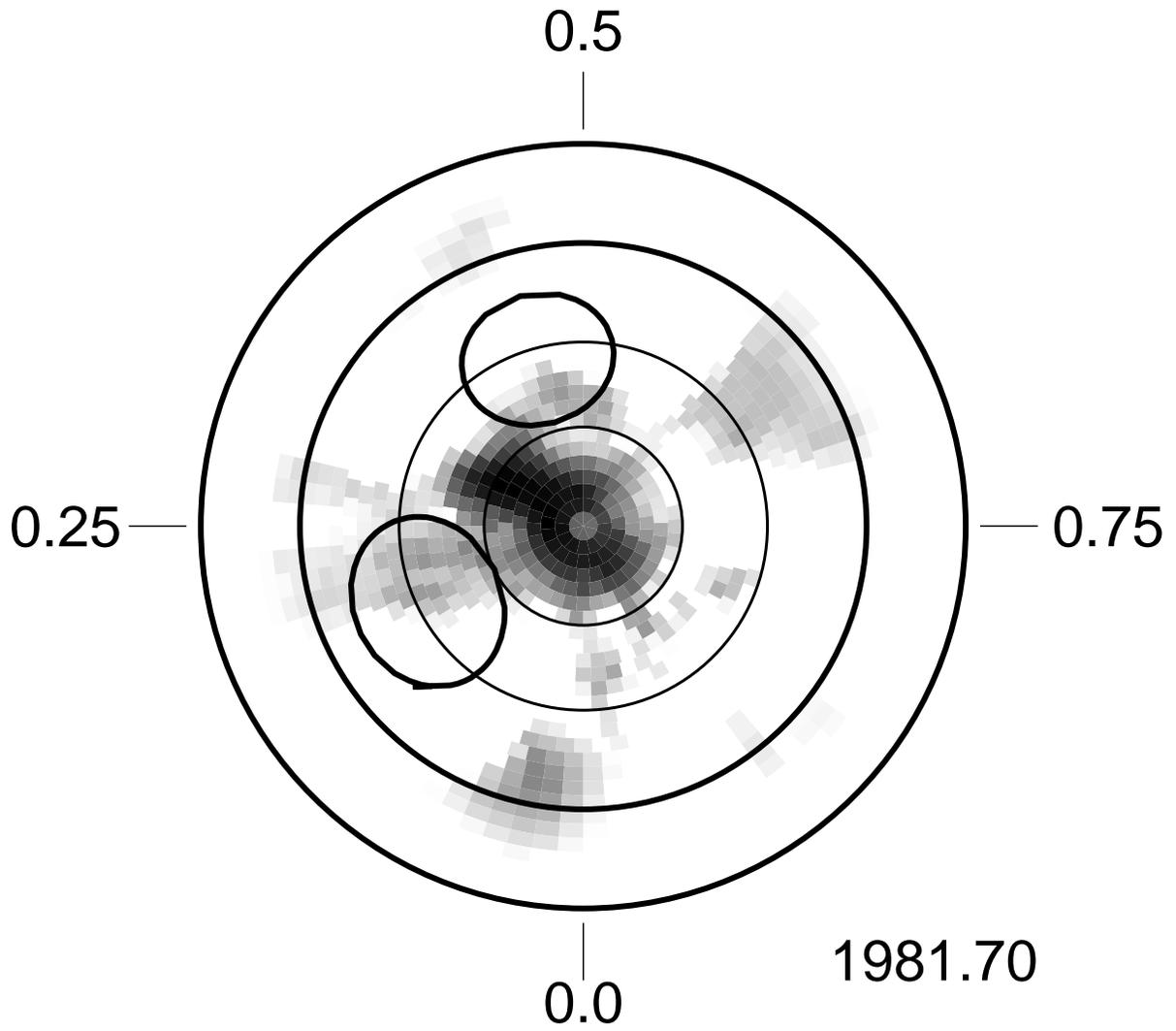}
\caption{The 2-spot 1981.9 photometric model of Kang \& Wilson superimposed
on the raw MEM Doppler image of 1981.7}
\label{fig:solution2}
\end{figure} 

Another 2-spot model fit to the light curve for this epoch is provided
by the results of \cite{kan89}. Their solution is shown as the circles
in Figure~\ref{fig:solution2}. Again however, they fit only a single
epoch curve for 1981.9; this curve had only 8 observed points.  Their
`2-spot' fit to the 1981.9 light curve of \cite{ruc83} yielded two
spots, both at 35\deg latitude.  Their first spot was at their
longitude 345\deg (our phase 0.458) and was 42\deg in diameter, while
their second spot was at their longitude 244\deg (our phase 0.178) and
had a diameter of 47\deg.  Again, while they did achieve a respectable
fit to the light curve, that light curve did not contain enough
constraints for a meaningful spot solution (6 spot parameters were
derived from a light curve with only 8 observed points). Their 2-spot
solution is a clearly not an accurate representation of the spot
distribution. Their conclusions about spot longitude migration and
spot cycles drawn from these solutions must be judged accordingly.

So we see three different attempts at deriving the spot distribution
at epoch 1981.7 via `2-circular-spot' model fits to light curves. All
three yielded respectable light curve fits but gave quite different
spot solutions. None of the three agreed well with our Doppler
image. Our Doppler image is certainly not to be regarded as an
absolute standard against which other solutions should be
judged. Doppler images also have some degree of nonuniqueness, but the
images are much more highly constrained (by a large set of intensity
measurements at many different velocities) than are the photometric
solutions. Also, the `maximum entropy' criterion of the images
guarantees that every feature shown in the image is actually required
by the data, and that there is no more image information than is
minimally necessary to reproduce all the data. We thus believe that
the Doppler images are likely to be a much more highly constrained and
accurate representation of the spot distribution.

The lesson to be learned here is that `2-spot' modeling of light
curves is, at best, a non-unique process, largely because it uses so
few constraints for the image solution, that - as demonstrated above -
multiple solutions can be found, each of which fits the light curve
equally well. At the very least, the technique requires very accurate
fitting of the subtle inflections in high quality photometry. A
relaxed light curve fit and/or noise in the data can easily mask the
true details of the spot distribution, yielding highly non-unique
solutions. A reduced light curve amplitude also can contribute to
non-uniqueness. 2-spot and 3-spot solutions must always be viewed in
that light. In particular, drawing conclusions about spot locations,
spot migration, or differential rotation from such modelling can be
quite misleading.

There were some other observations of HR 1099 reported during this
epoch. \cite{rod87} presented IUE emission line flux observations
obtained during October, 1981. They found stellar line fluxes to be
hundreds of times larger than solar values and found a close spatial
correlation between the cool spots and the plage-like features giving
rise to the line emission. This was interpreted as an indication of
large spot areas down in the photosphere with overlying magnetic
loops, which then give rise to plages higher in the
atmosphere. \cite{byr87} combined ground-based optical and IUE
satellite-ultraviolet observations of HR 1099 in 1981 to show that
solar-like densities are present in the outer atmosphere of the star,
and that total radiative losses were at least two orders of magnitude
larger than on the Sun.  \cite{lin89} published a detailed study of a
well-observed flare of Oct. 3, 1981 on HR 1099. They concluded that
they had observed a region of turbulent infall on the K1IV star near
phase 0.2. That might place it over the protuberance of the polar spot
in Figure~\ref{fig:1981.7_image}. \cite{nat86} reported some H-alpha
line emission observations taken in the late 1981 which showed that
86\% of the emission was coming from the K1IV star.

\subsection{The 1982 Doppler Image}

Our next Doppler image was obtained almost exactly one year later at
epoch 1982.74. Unfortunately, we did not have a contemporaneous light
curve for this exact epoch. \cite{and88} presented light curve data
for the interval 1982.78 through 1983.17, but most of these data are
an uncomfortable 2-3 months later than our image epoch, and we were
concerned about the light curve varying over this long interval.

\begin{figure}
\plotone{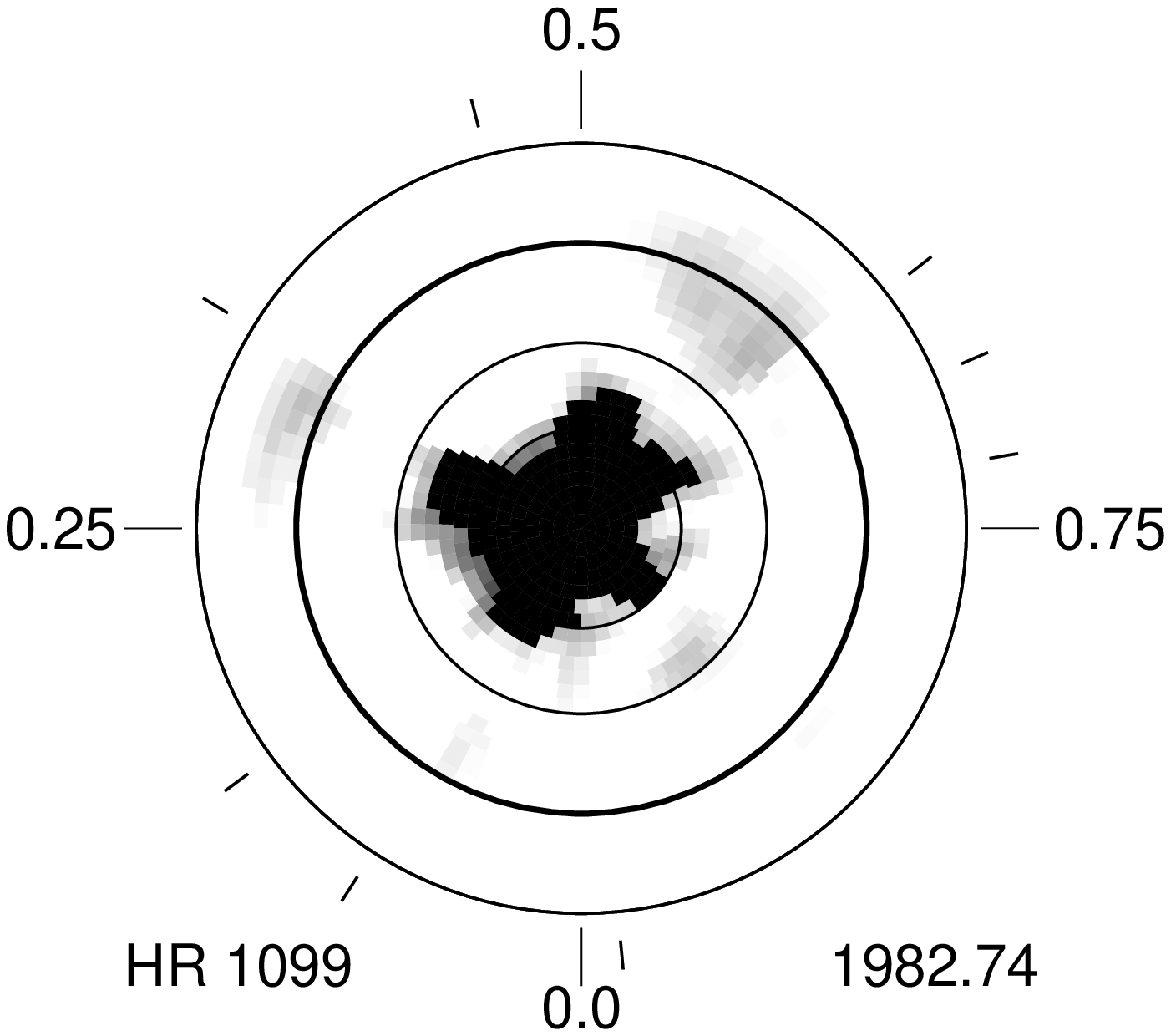}
\caption{HR 1099 raw (unthreshholded) Doppler image for 1982.74}
\label{fig:1982.74_raw}
\end{figure} 

\begin{figure}
\plotone{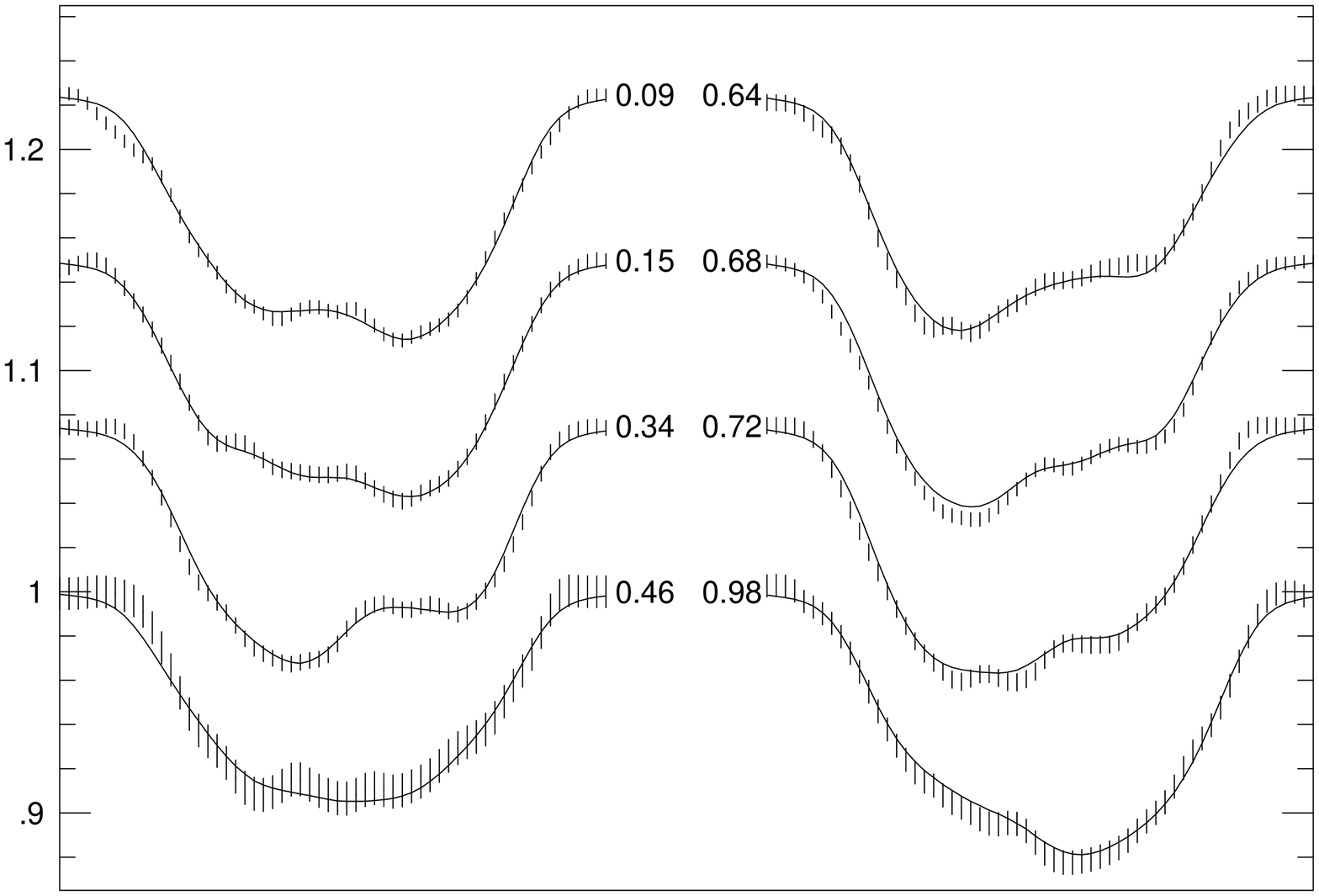}
\caption{The spectral line profiles and line fits for the 1982.74 image}
\label{fig:82fits}
\end{figure} 

As usual, we first derived a Doppler image from the line profile
information alone, and this is shown in
Figure~\ref{fig:1982.74_raw}. Figure~\ref{fig:82fits} shows the
spectral line profiles and fits. In the image one sees the same basic
polar spot straddling the pole, but now substantially larger in area,
and with more protuberances. It is not clear that any of these
protuberances can be identified with the one from the previous year
(and thereby used to measure the relative rotation rate of the polar
spot) since the polar spot now looks quite different.

\begin{figure}
\plotone{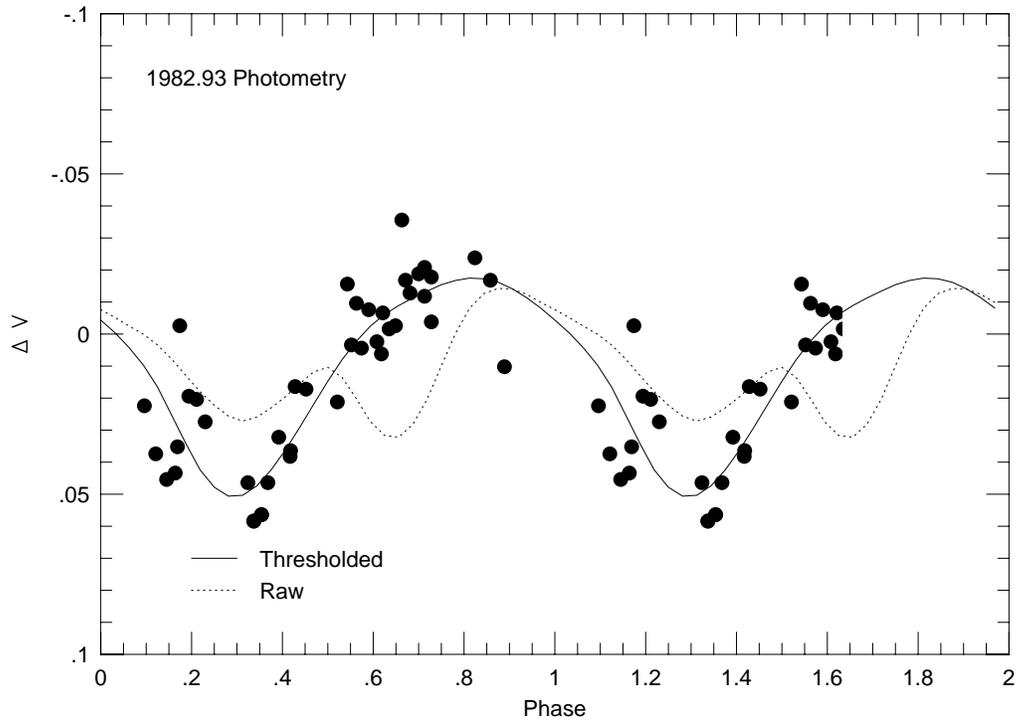}
\caption{1982.93 light curve of HR 1099 used for the 1982.74 image}
\label{fig:1982.74_light}
\end{figure}

\begin{figure}
\plotone{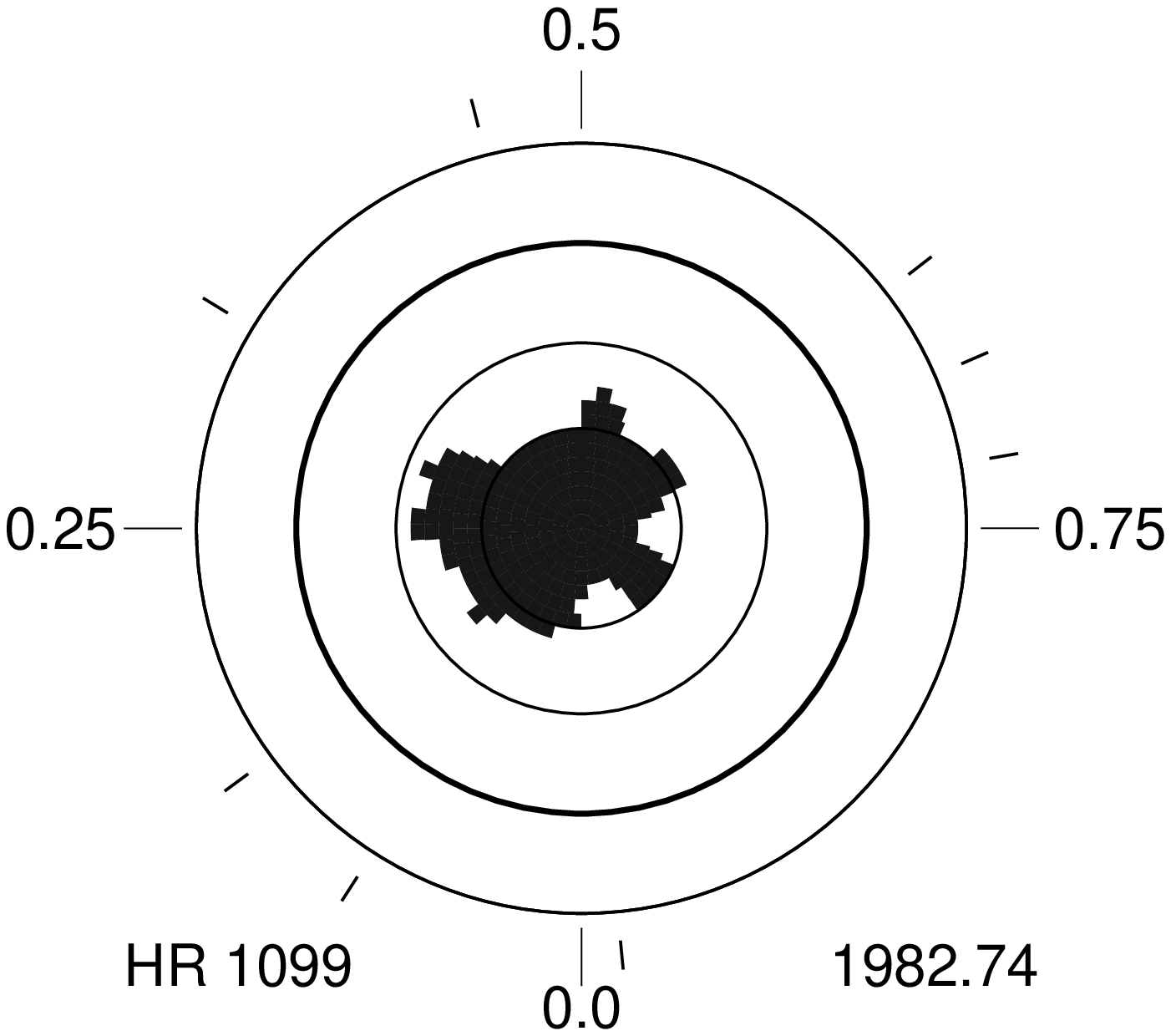}
\caption{HR 1099 thresholded Doppler image for 1982.74}
\label{fig:1982.74_image}
\end{figure}

The 1982.93 light curve data of \cite{and88} is shown in
Figure~\ref{fig:1982.74_light}. Our raw image produced a light curve
(dotted line) which fit these data marginally well, even with the
light curve taken nearly 2-3 months after the image solution
centroid. Only a slight bit of threshholding was needed to produce the
final predicted light curve (solid line) which fit quite well. The
resultant final threshholded image is shown in
Figure~\ref{fig:1982.74_image} and is essentially identical to the
original unthreshholded image. Again, the low-latitude spot near last
year's Feature A, as well as all the other low-latitude spots, did not
survive the threshholding constraint.

\begin{figure}
\plotone{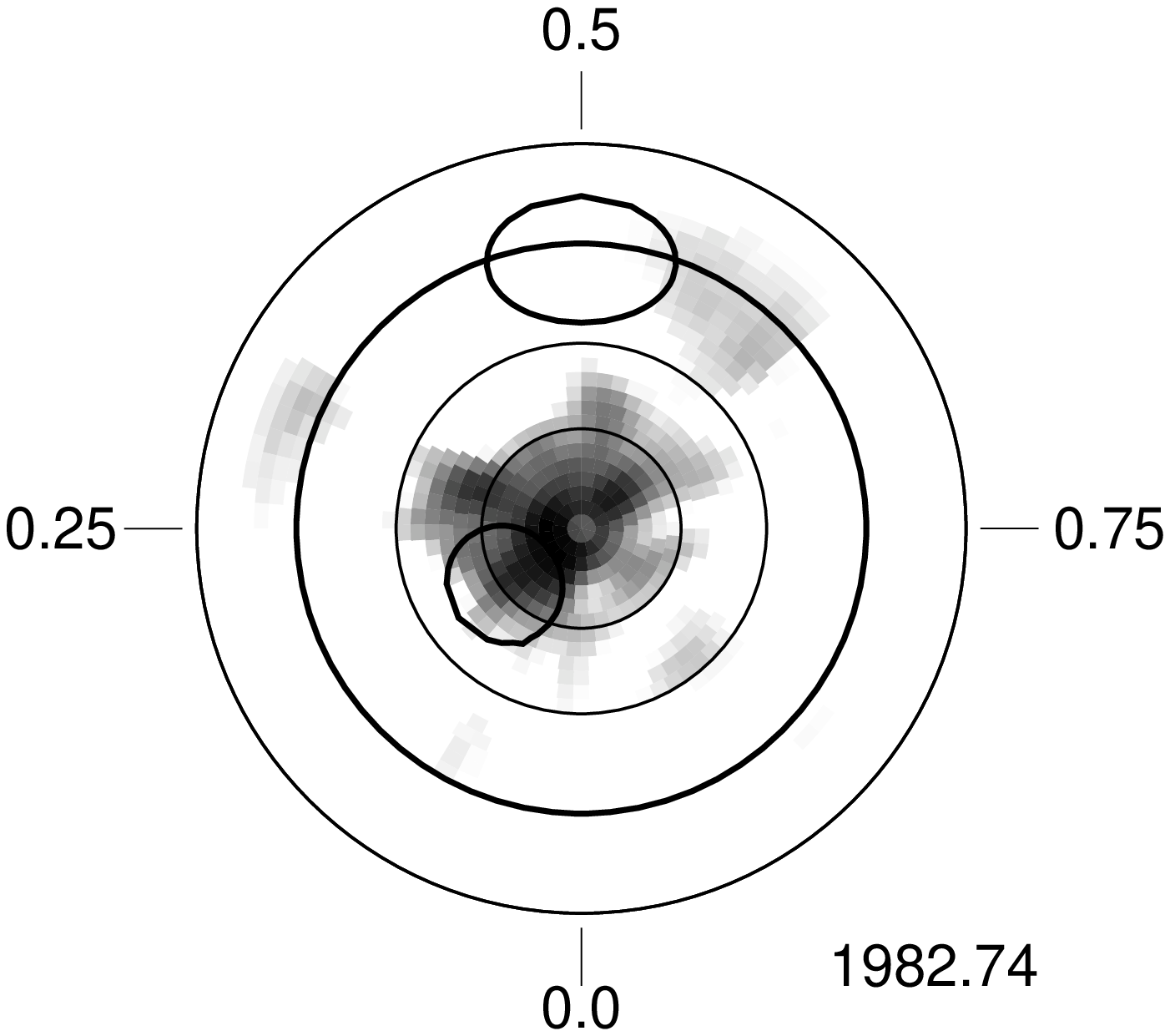}
\caption{The 2-spot photometric model of Andrews et.al (1988) superimposed
on the raw 1982.74 Doppler image}
\label{fig:solution3}
\end{figure} 

\cite{and88} also presented a 2-spot solution from fits to their
photometry. This is shown in Figure~\ref{fig:solution3} along with a
grayscale version of the raw Doppler image. Their Spot 1 was of
diameter 36\deg at latitude 60\deg and phase 0.15. Their Spot 2 was of
diameter 40\deg at latitude 5\deg and phase 0.5. While their Spot 1
does reflect some sense of the protuberance on our polar spot near
phase 0.2, we see no evidence of their Spot 2. Overall, their 2-spot
solution is not a good representation of the spot distribution from
our Doppler imagery.

\cite{les84}, using a five station VLBI, detected a sub-milliarcsecond
radio component from HR 1099 at epoch 1983.2, during a strong radio
outburst at orbital phase 0.226, when the polar spot's protuberance
was facing the earth. Their data suggested gyrosynchrotron emission
from a power-law energy distribution of electrons in a magnetic field
of strength about 30 gauss and from an area 75\% of the diameter of
the K-subgiant.

\subsection{The 1984 Doppler Image}

\begin{figure}
\plotone{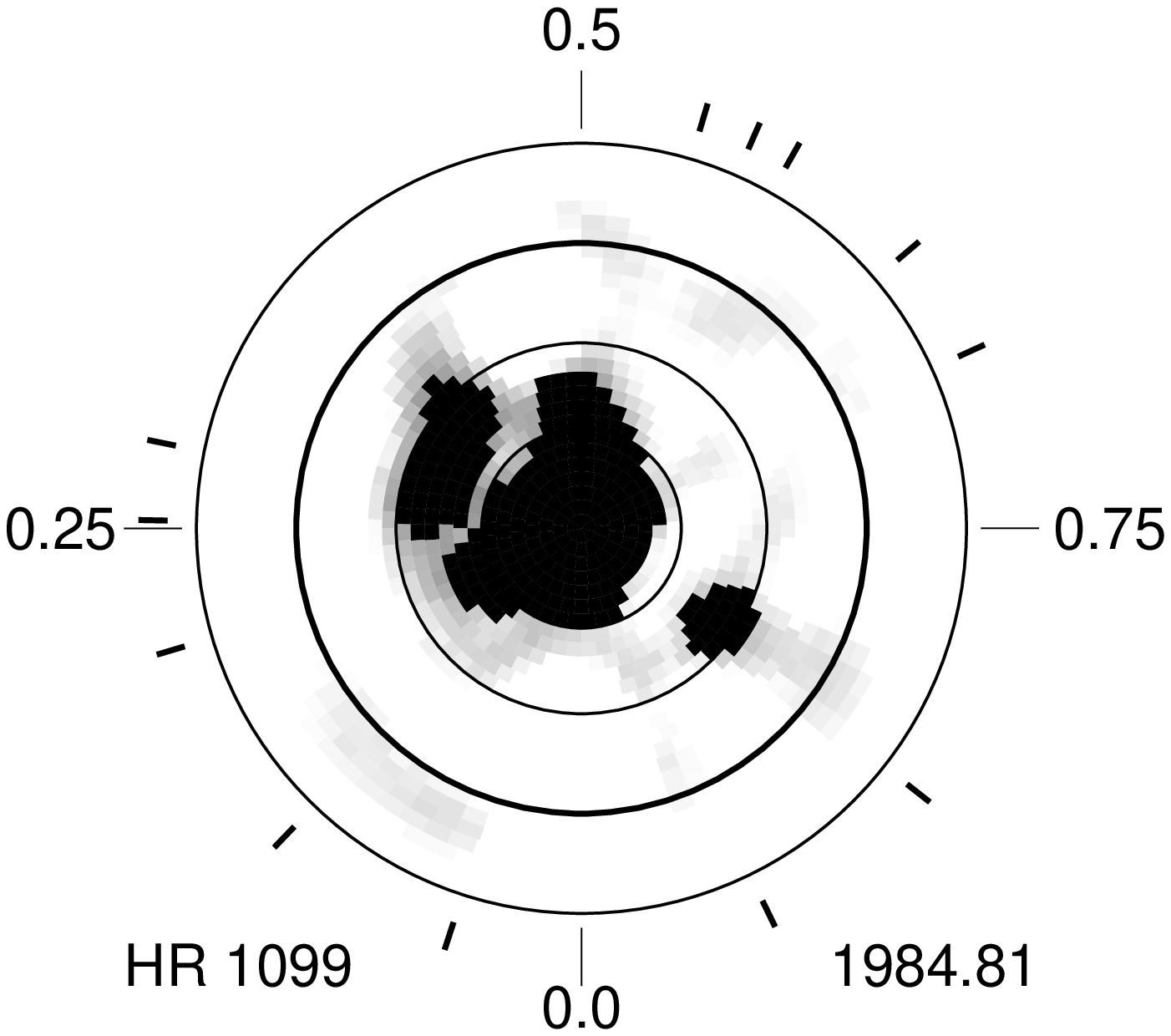}
\caption{HR 1099 raw (unthreshholded) Doppler image for 1984.81}
\label{fig:1984.81_raw}
\end{figure}

\begin{figure}
\plotone{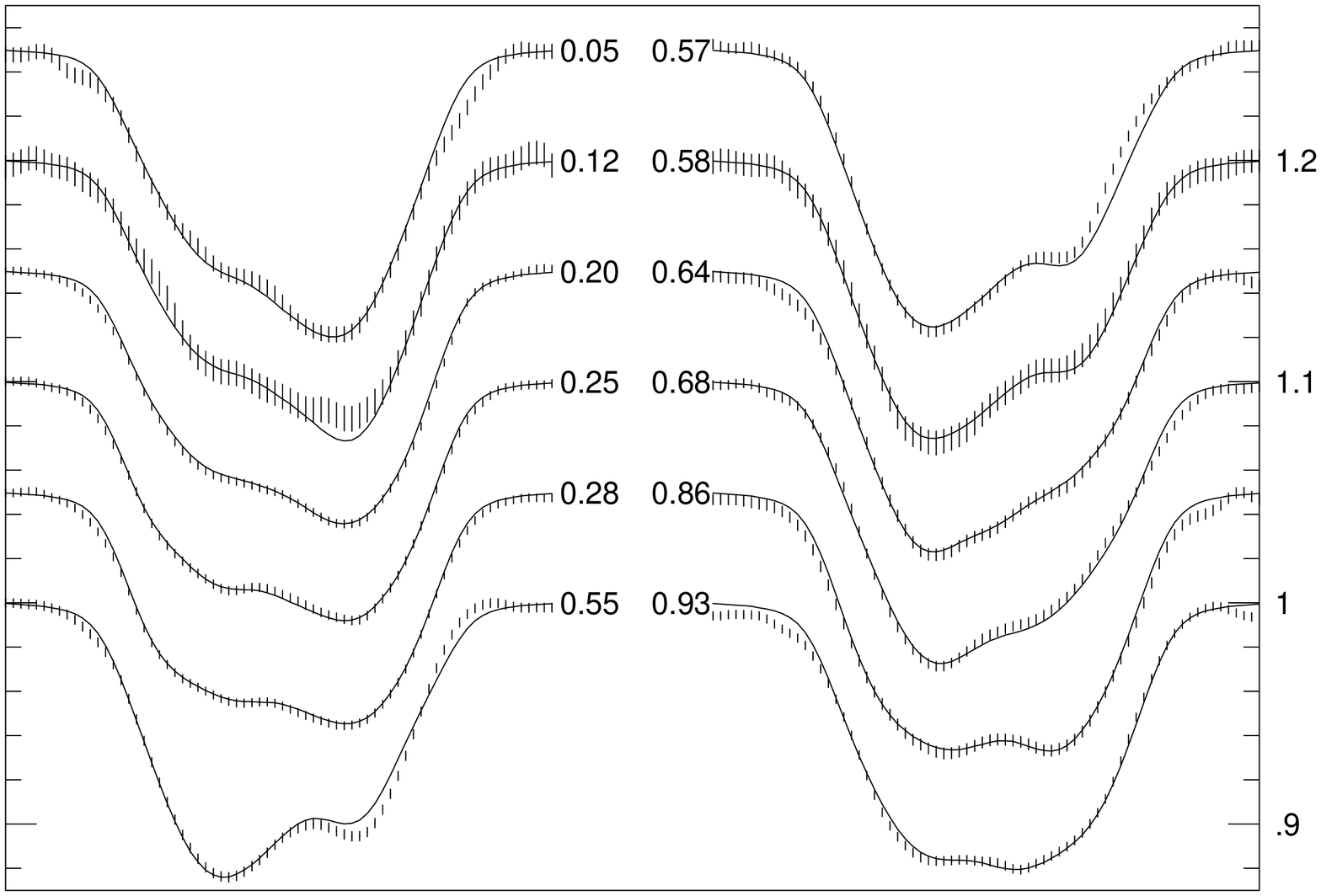}
\caption{The spectral line profiles and fits for the 1984.81 image}
\label{fig:84fits}
\end{figure}

\begin{figure}
\plotone{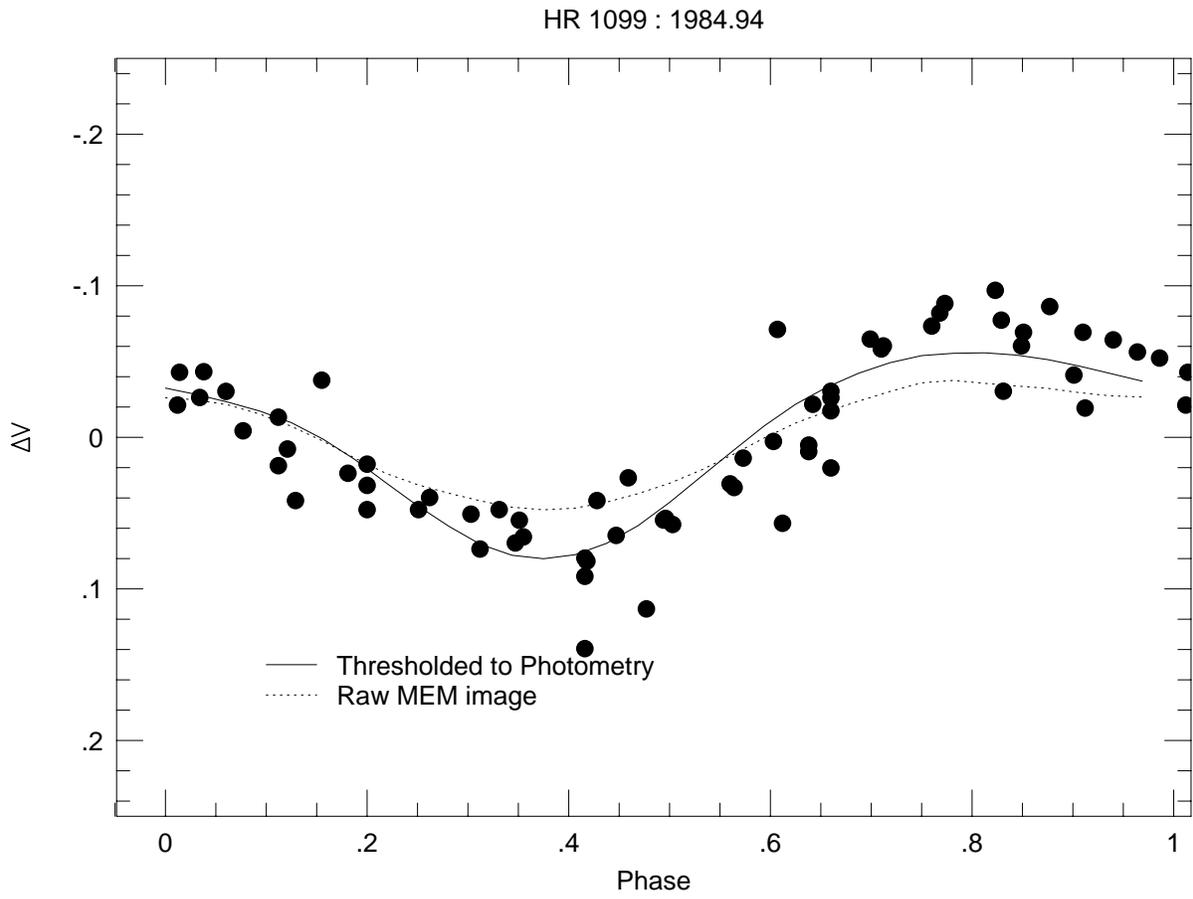}
\caption{1984.94 light curve used for the 1984.81 HR 1099 image}
\label{fig:1984.81_light}
\end{figure}

\begin{figure}
\plotone{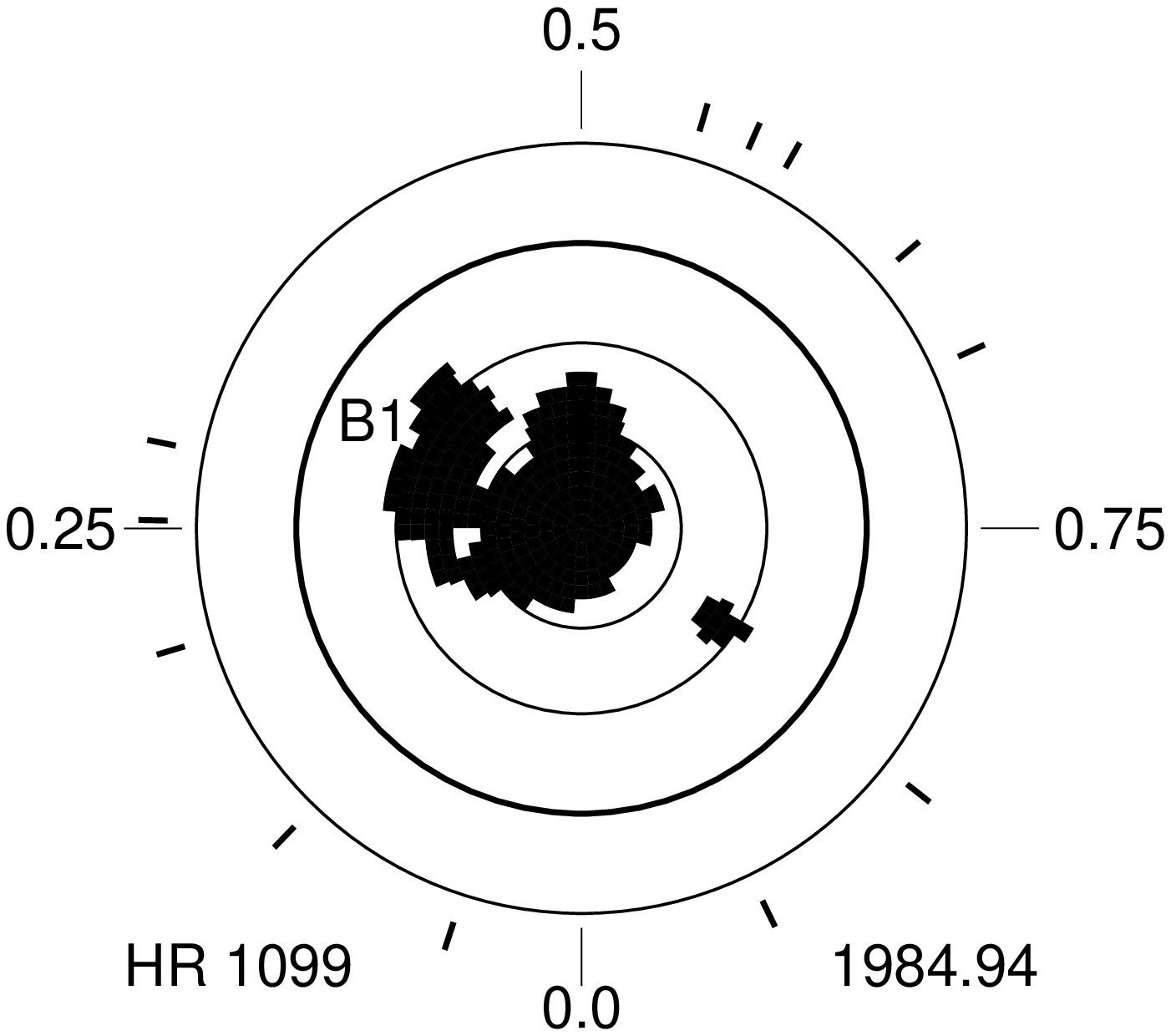}
\caption{HR 1099 thresholded Doppler image for 1984.94}
\label{fig:1984.81_image}
\end{figure}

Unfortunately, due to bad weather, we were unable to obtain a Doppler
image for the 1983 season. The image set resumes with a Doppler image
for 1984.81. Our raw version, unthreshholded by any light curve
constraints, is shown in Figure~\ref{fig:1984.81_raw}. The spectral
line profiles and fits are shown in Figure~\ref{fig:84fits}. Light
curves for the 1984-5 observing season were published by \cite{str89}
and by \cite{moh93}. The amplitude was about 0.12 magnitudes and the
shape rather similar to that of the previous and following years. We
were unable to find any spot model fits to these light curves for this
epoch. Our fits to the 1984.94 light curve of \cite{str89} are
presented in Figure~\ref{fig:1984.81_light}. The predicted light curve
(dotted line) from the raw MEM image again fit the measured light
curve reasonably well but was slightly too low in amplitude. A slight
threshholding adjustment yielded the final image of
Figure~\ref{fig:1984.81_image}, and brought the light curve amplitude
(solid line in Figure~\ref{fig:1984.81_light}) up to the observed
level. As can be seen, there is no significant difference (other than
spot area scaling) between the raw and the
photometrically-threshholded images. In this case, the light curve was
taken some 1-2 months later, but the close agreement between image
solutions, with and without the light curve constraints, indicates
that the spot distribution was quite stable over this season's entire
observing interval.

The polar spot is still present and still straddling the pole, but
with ever-changing shape as protuberances come and go, or perhaps
shift in phase.  The polar spot changes shape to such an extent that
again we cannot find a simple rotation that uniquely matches previous
polar spots and that would allow us to derive relative rotation (if,
indeed, the true evolution is really that simple). The true situation
may be a mixture of both spot rotation and evolving protuberances.
The polar spot has a prominent, nearby, almost-attached spot at phase
0.32 and latitude 41\deg. We shall refer to this spot as `Feature
B1'. This feature is truly resolved from the polar spot. There is also
now a small and isolated spot near phase 0.85 and latitude
37{\deg}. Again, it is not known whether this isolated spot is at all
related to any of the features in the 1982.74 image. It also seems
unlikely to be a phase ghost since, at a latitude of 37{\deg} it is
seen in the line profiles at about 5 phases.

\subsection{The 1985 Doppler Image}

\begin{figure}
\plotone{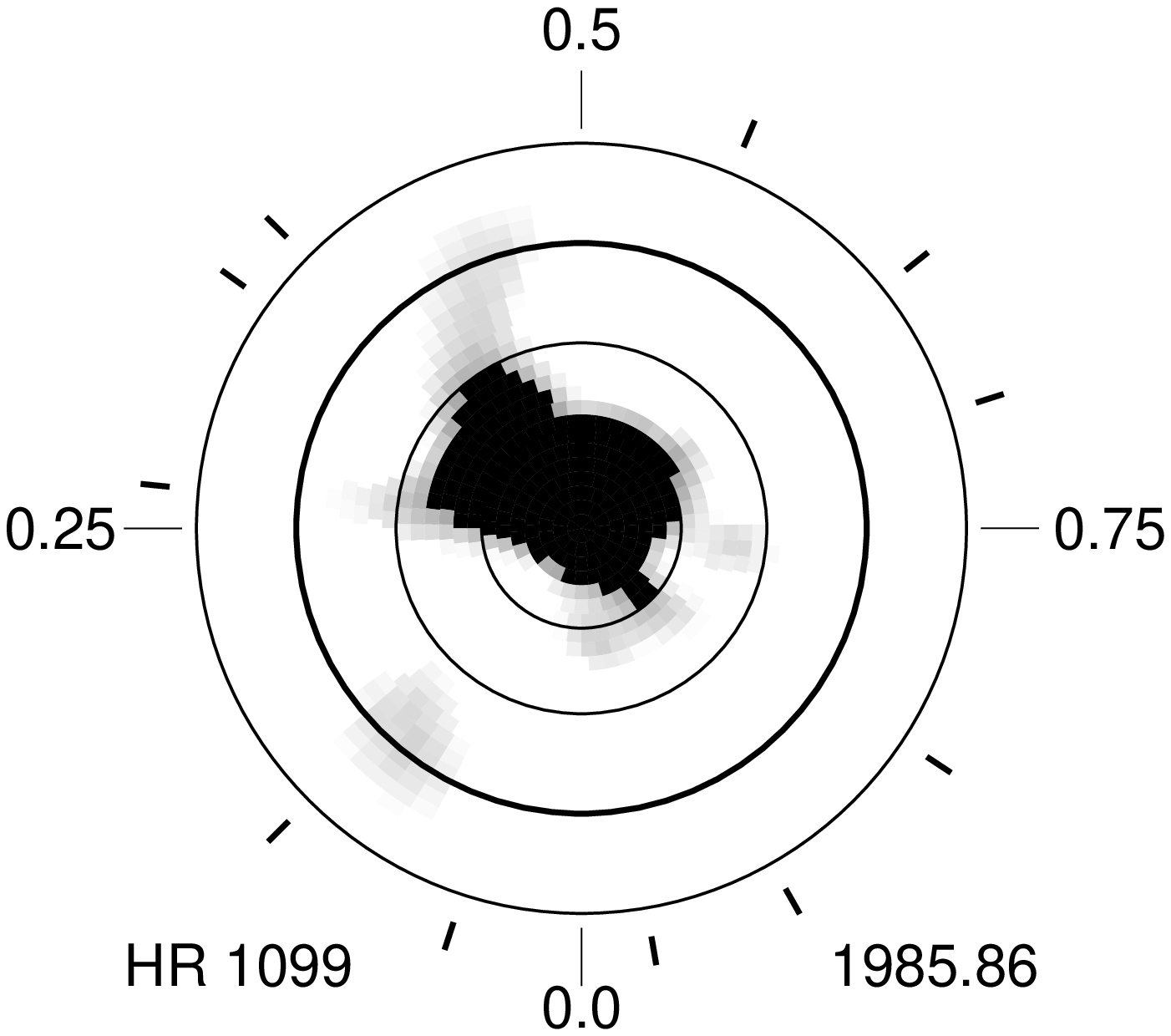}
\caption{HR 1099 raw (unthreshholded) Doppler image for 1985.86}
\label{fig:1985.86_raw}
\end{figure}

\begin{figure}
\plotone{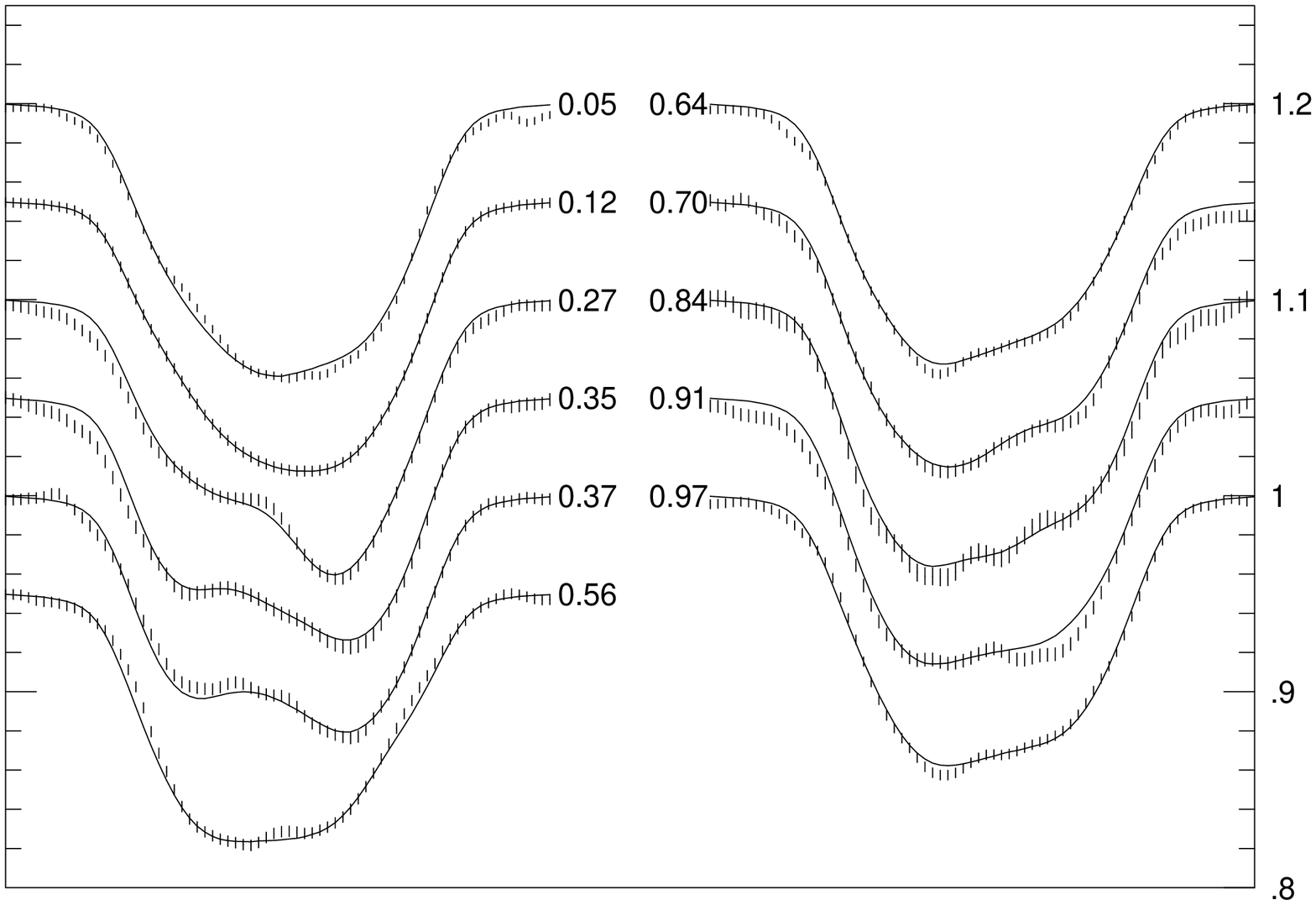}
\caption{The spectral line profiles and fits for the 1985.86 image}
\label{fig:85fits}
\end{figure}

\begin{figure}
\plotone{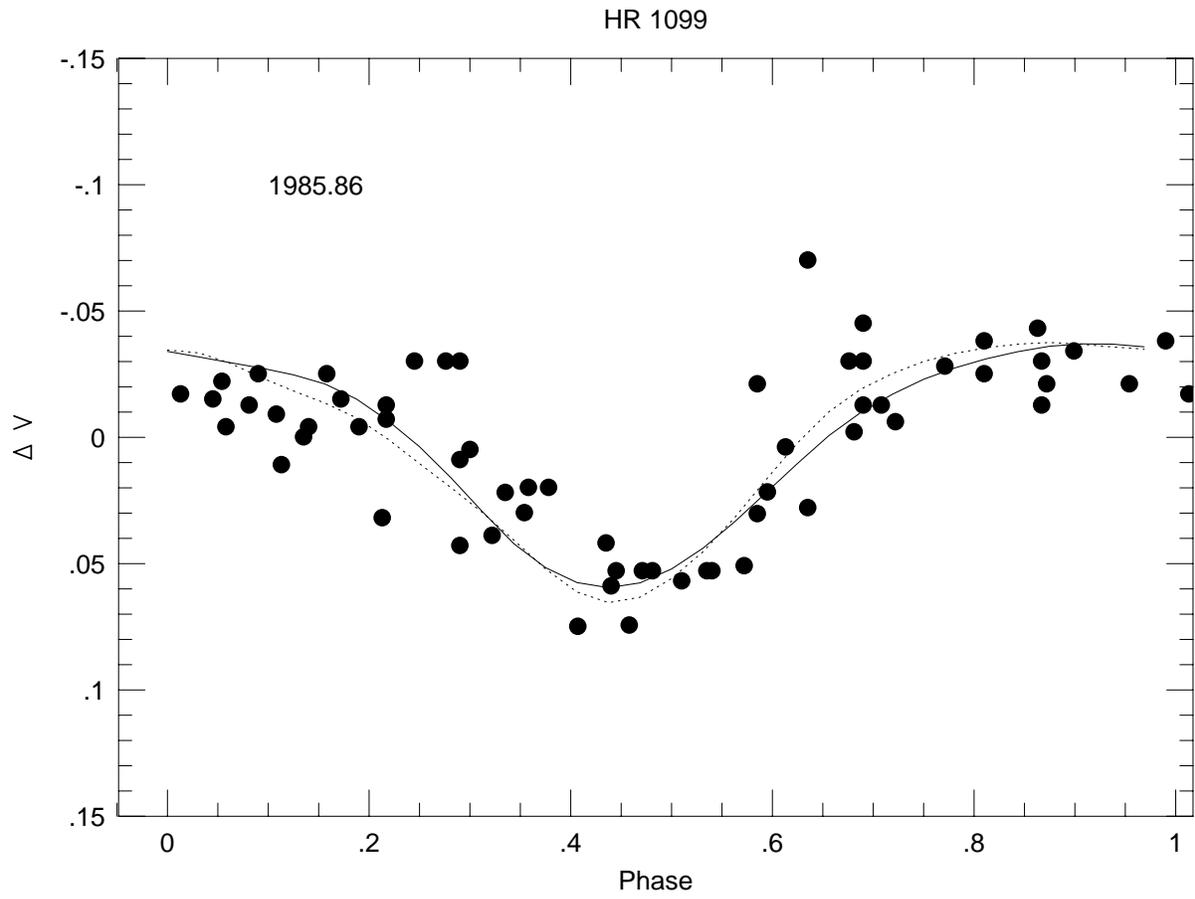}
\caption{HR 1099 1985.75 light curve used for the 1985.86 image}
\label{fig:1985.86_light}
\end{figure}

\begin{figure}
\plotone{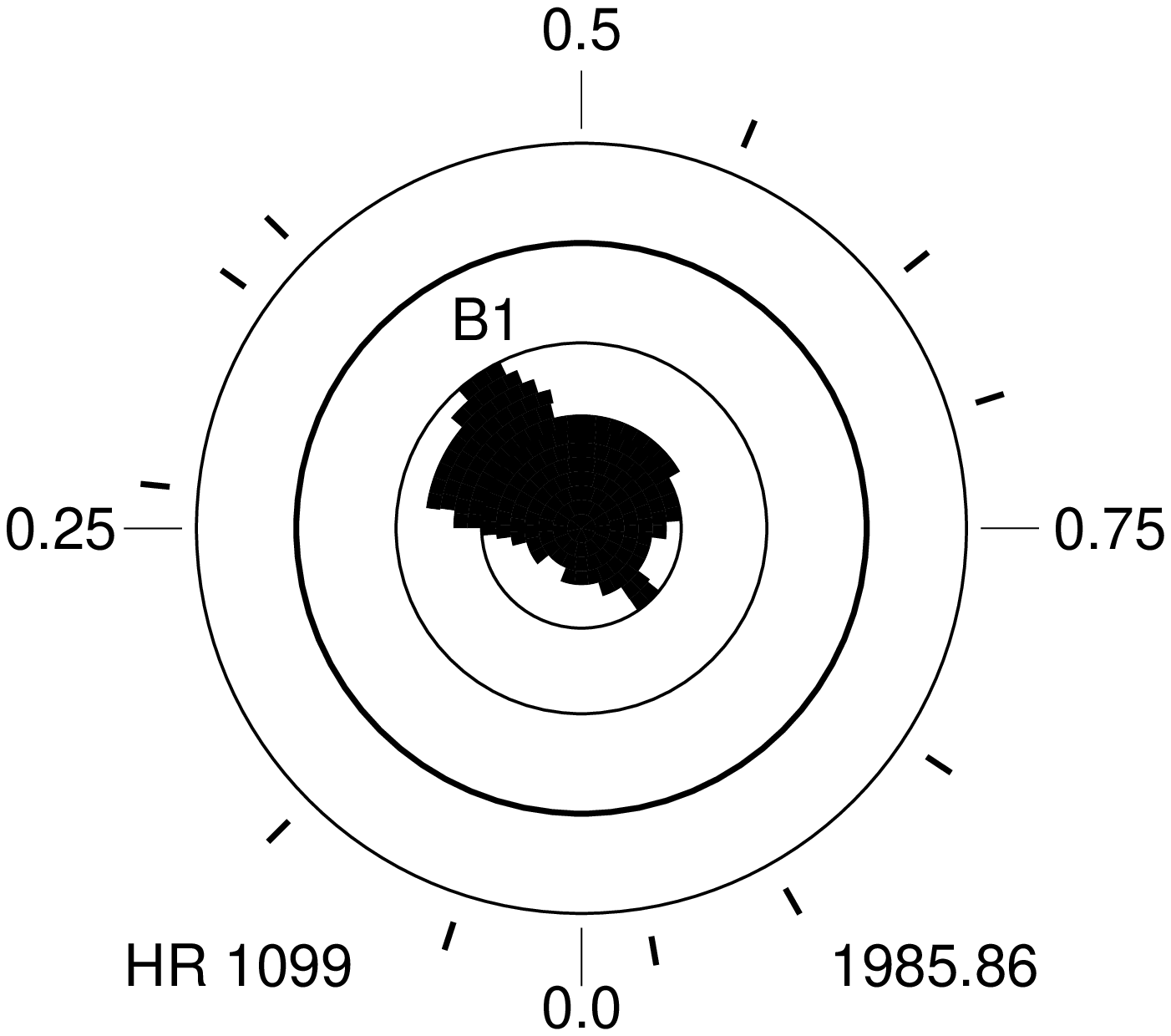}
\caption{HR 1099 thresholded Doppler image for 1985.86}
\label{fig:1985.86_image}
\end{figure}

The raw Doppler image for 1985.86 is shown in
Figure~\ref{fig:1985.86_raw} and the spectral line profiles and fits
in Figure~\ref{fig:85fits}. Light curves for this season were
published by \cite{str89}, by \cite{moh93}, and by
\cite{cut90}. Figure~\ref{fig:1985.86_light} shows the 1985.75
photometry of \cite{str89} (points) along with our predicted light
curves from the unthreshholded (dotted line) and threshholded (solid
line) Doppler image solutions. Again the predicted light curve from
the raw MEM image fits the observed photometry quite
well. Thresholding the image improved the fit only slightly, and again
produced no significant change in the image. The final threshholded
image is shown in Figure~\ref{fig:1985.86_image}.

At this epoch, there were no isolated low-latitude spots, only the
omnipresent polar spot with a large protuberance again near phase
0.35.  We note a similarity in both shape and area of this
protuberance to Feature B1 from 1984.81. It looks as though Feature
B1, first seen at latitude 41\deg and phase 0.32 in 1984.81, simply
moved poleward to 50\deg latitude and clockwise to phase 0.35 in
1985.86, with little change in either its shape or area. Feature B1
thus seems to be slowly circling the pole in a clockwise direction,
and may even be merging with the polar spot as it approaches from
lower latitudes. If correct, the migration rate implied from Feature
B1 (with respect to the co-rotating frame of the orbit) is about 9\deg
-- 17\deg per year at latitudes of roughly 37\deg - 50\deg. This
corresponds to about 1/3600 of the orbital period and in the sense
that high latitudes are rotating slightly more slowly than the
orbit. Note that a 3-sigma error on the orbital period determined by
\cite{fek83} amounts to a longitude error due to period uncertainty of
only 0.5\deg. Of course, with a time sequence of only two images, many
other interpretations are possible. In particular, it is not clear
what if any role the other time variable protuberances on the polar
spot may have played. But the similarity of the shape of the phase 0.3
protuberance in 1985.86 with Feature B1 of 1984.81 leads us to suspect
that we are seeing Feature B1 circle clockwise around the pole.

\subsection{The 1986 Season Doppler Images}

\begin{figure}
\plotone{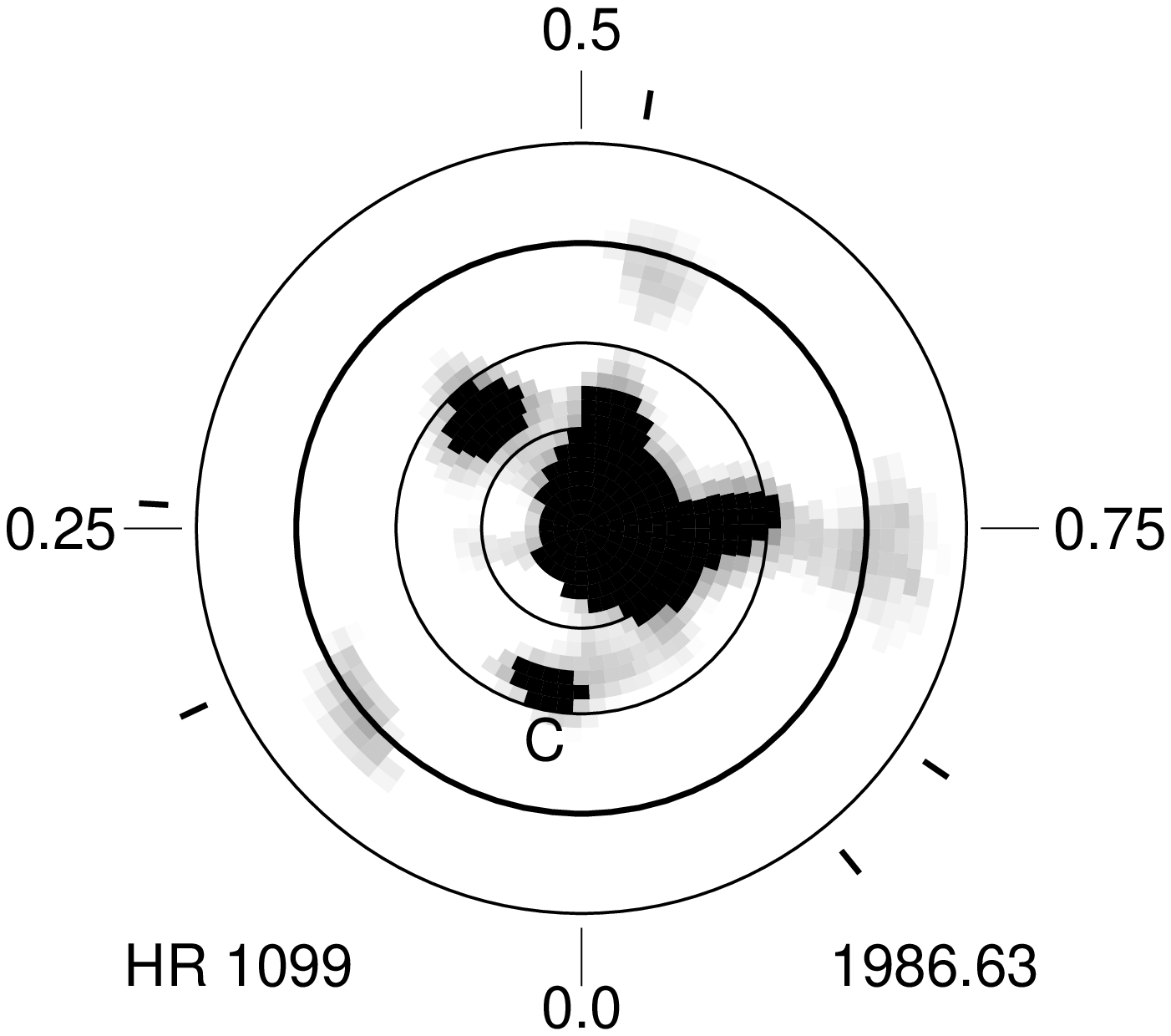}
\caption{HR 1099 raw (unthreshholded) Doppler image for 1986.63}
\label{fig:1986.63_raw}
\end{figure}

\begin{figure}
\plotone{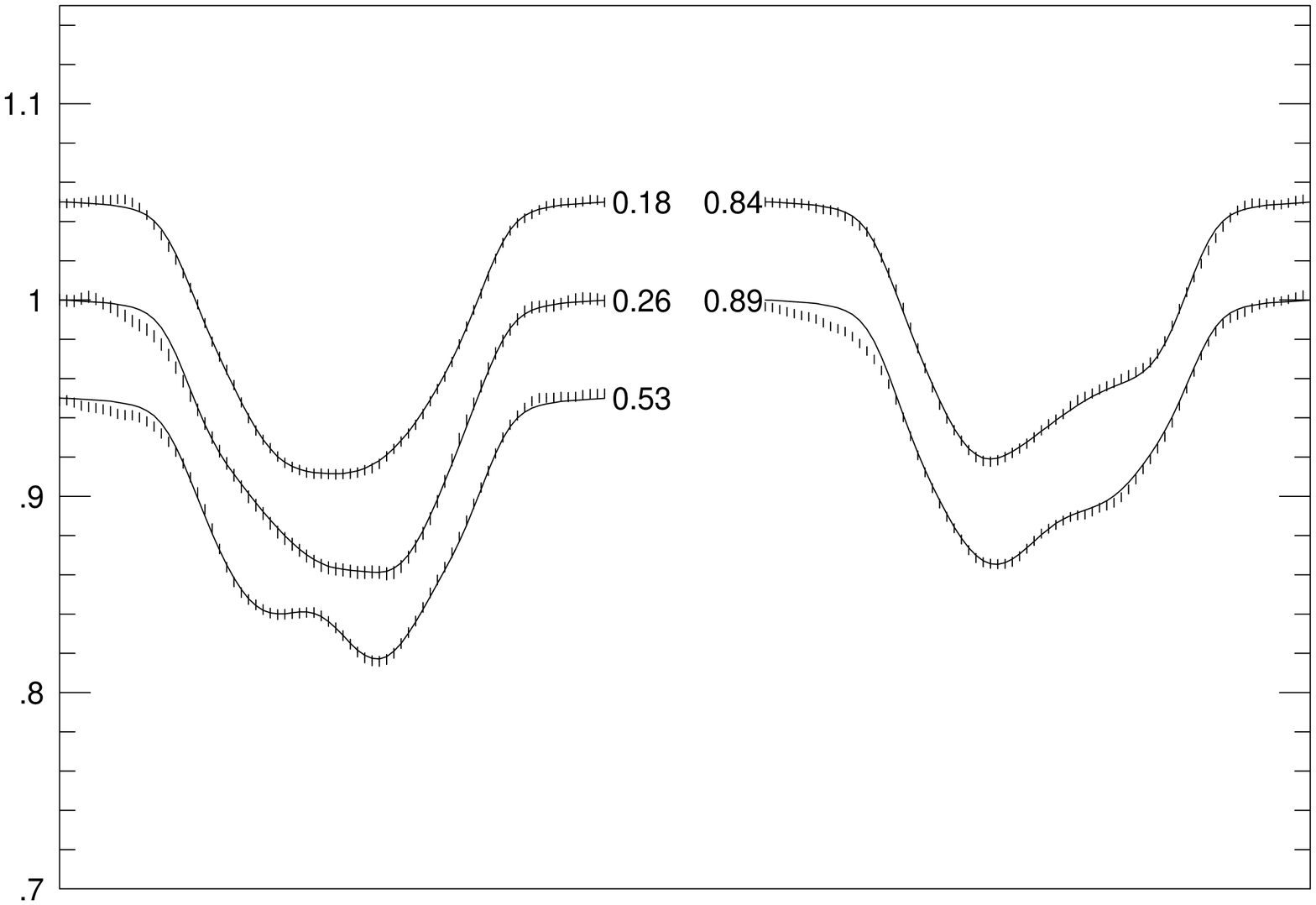}
\caption{The spectral line profiles and fits for the 1986.63 image}
\label{fig:1986.63fits}
\end{figure}

\begin{figure}
\plotone{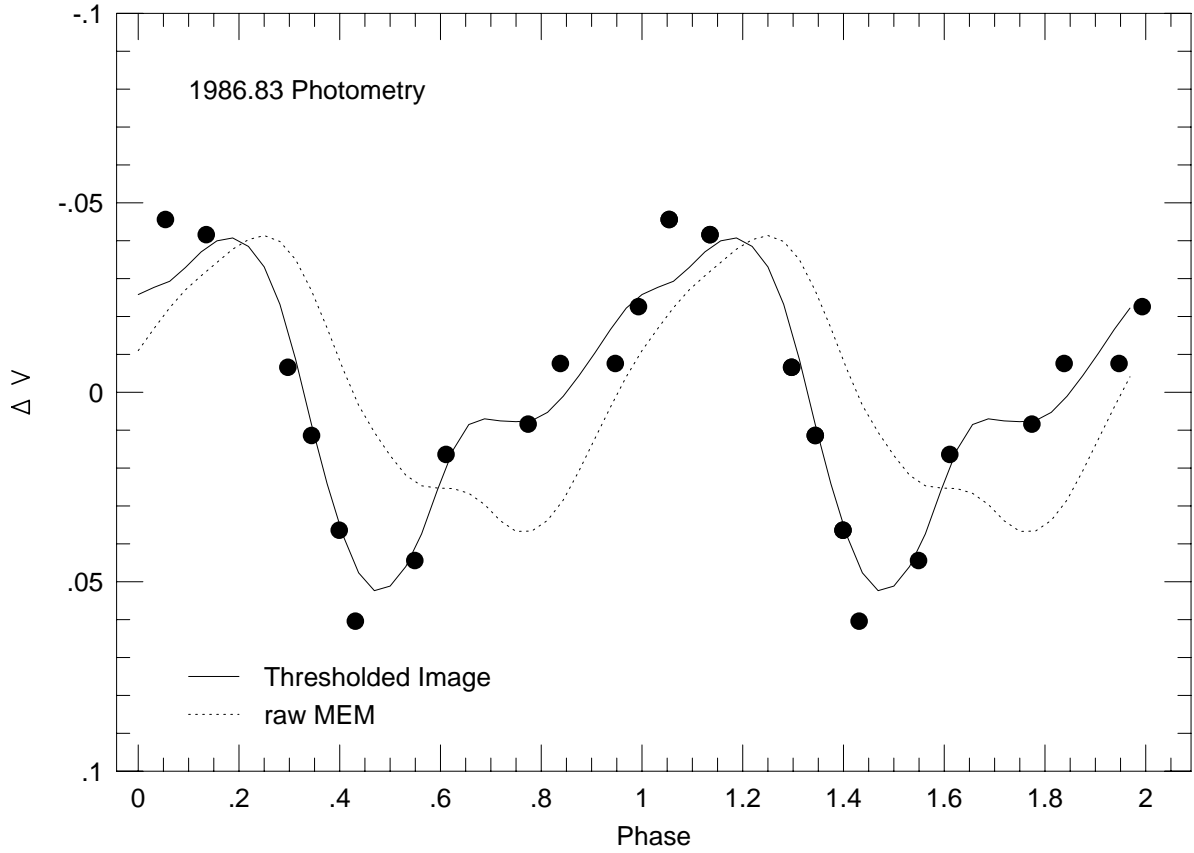}
\caption{1986.83 light curve used for the 1986.63 HR 1099 image}
\label{fig:1986.63_light}
\end{figure}

The 1986 season was the first for which we managed to obtain two
images, separated by only 5 months rather than the usual 12-month
gap. Our unthreshholded raw image for 1986.63 is shown in
Figure~\ref{fig:1986.63_raw} and the spectral data and fits in
Figure~\ref{fig:1986.63fits}. Light curves for 1986-7 were published
by \cite{str89}, \cite{cut90}, \cite{moh93}, and by \cite{mek87}.
Figure~\ref{fig:1986.63_light} shows the 1986.83 light curve of
\cite{mek87} (points) along with our predicted light curve (dotted
line) from the raw Doppler image. The light curve amplitude is modest
because the low latitude spots and polar appendages are roughly evenly
distributed in longitude. In fact, there is probably little hope of
recovering much about the complex spot geometry of this epoch from
2-spot or 3-spot fits to this low amplitude light curve, and, indeed,
no spot models were found in the literature for comparison with this
season's Doppler images.

\begin{figure}
\plotone{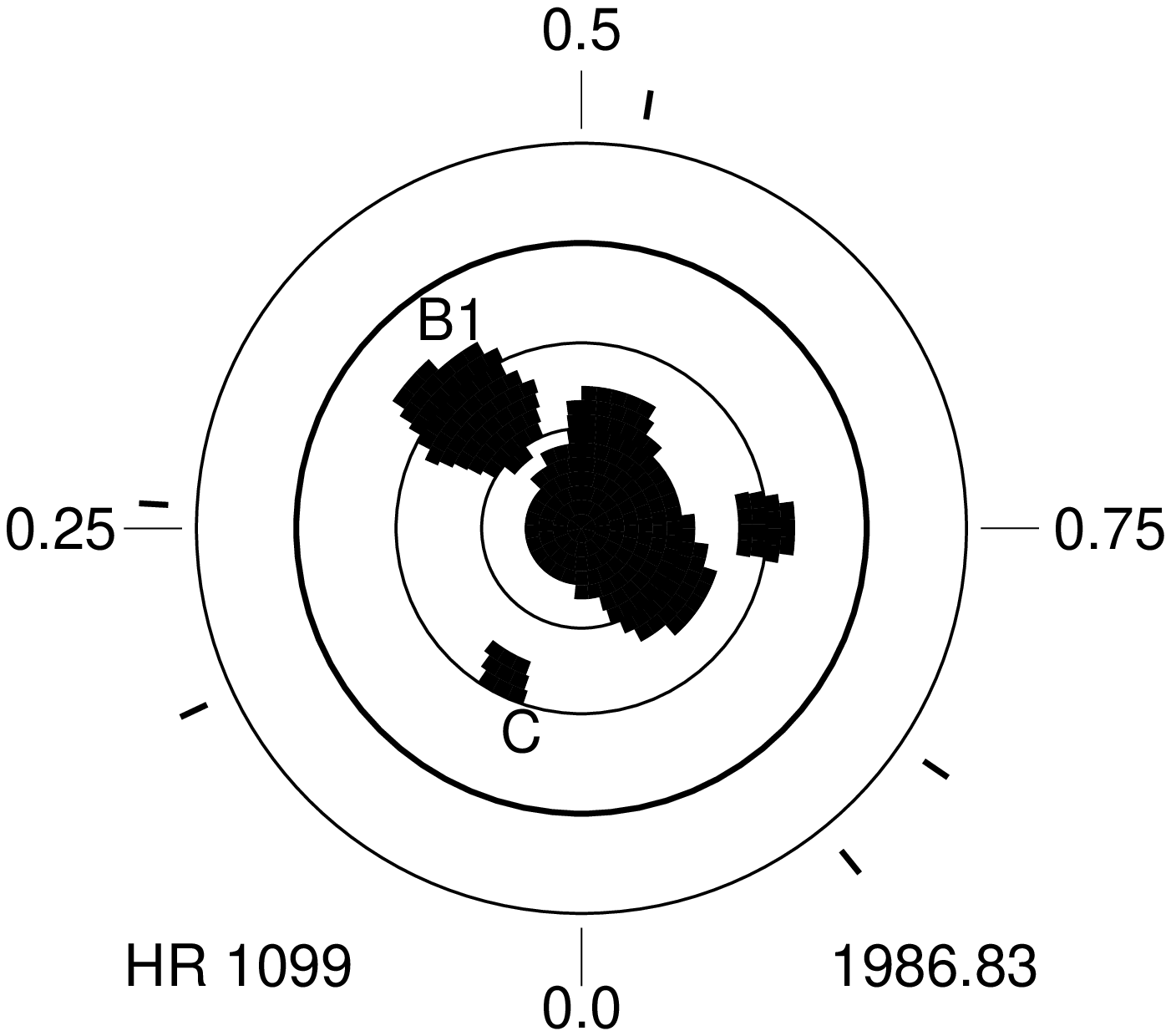}
\caption{HR 1099 thresholded Doppler image for 1986.83}
\label{fig:1986.63_image}
\end{figure} 

The predicted light curve from our 1986.63 raw MEM image clearly did
not fit the 1986.83 photometry adequately. (This is not entirely
surprising since the light curve was taken some 2.4 months later, and
furthermore, large phase gaps in our data may have caused the imaging
process to miss some low-latitude spots). We then attempted to fit the
light curve with a slightly modified version of the Doppler image. The
image that gave the best fit to the photometry is shown in
Figure~\ref{fig:1986.63_image}; the resulting photometric fit is shown
as the solid line in Figure~\ref{fig:1986.63_light}. The
photometrically revised image (Figure~\ref{fig:1986.63_light}) has a
slightly larger area for the mid-latitude spot at phase 0.38 (this
spot is in the middle of our phase gap and therefore not
well-constrained by the line profiles) as well as a slightly larger
polar appendage at phase 0.88. This latter change may be due to
changes in the polar spot between the time of the photometry and
Doppler imagery, but we cannot say for sure. Also the narrow
protuberance on the polar spot near phase 0.75 in the original Doppler
images became a detached spot in the photometrically-constrained
image.

There is an isolated mid-latitude spot (hereafter referred to as
Feature C) seen near phase 0.03 and latitude 38\deg in the 1986.63 raw
image, and at phase 0.11 and latitude 45\deg in the 1987.05 image (see
below). The best fit to the 1986.83 light curve required moving
Feature C from phase 0.03 in the 1986.63 raw image to slightly higher
phases (about 0.08) in the photometric image. We did this by assuming
that the feature was migrating at a constant rate in longitude, and
simply used the migration rate of this spot derived from the 1986.63
and 1987.05 images to interpolate its position for the 1986.83
photometrically-constrained image. This yielded a predicted light
curve (solid line in Figure~\ref{fig:1986.63_light}) which then agreed
quite well with the 1986.83 photometry. There is a slight indication
that Feature C was also moving northward since its latitude in the
1987.05 image is about 45\deg, whereas it was 38\deg in
1986.63. Perhaps, like Feature B1/B2, it is following a northward
clockwise spiral path toward eventual merger with the polar spot.

\begin{figure}
\plotone{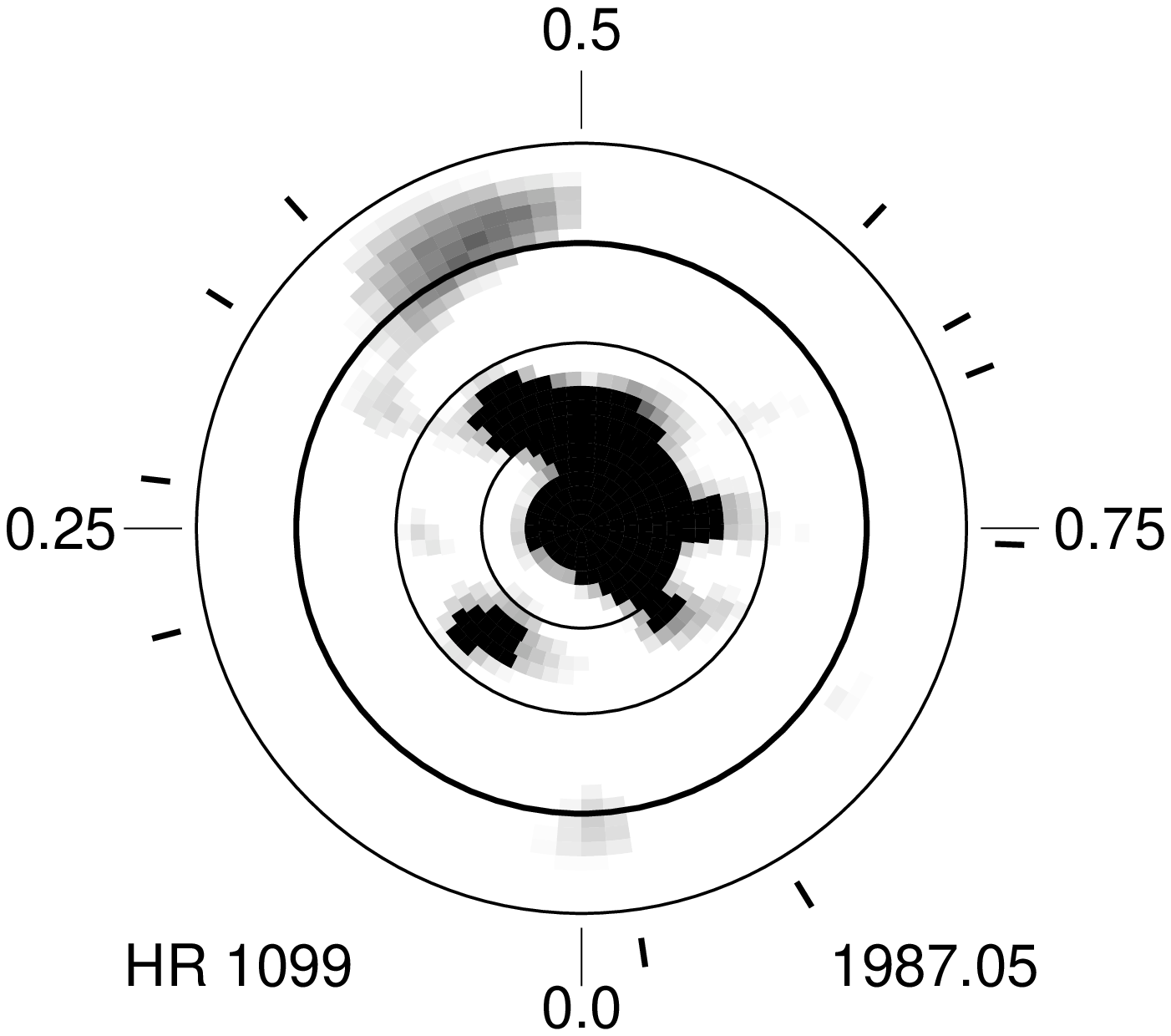}
\caption{HR 1099 raw (unthreshholded) Doppler image for 1987.05}
\label{fig:1987.05_raw}
\end{figure} 

\begin{figure}
\plotone{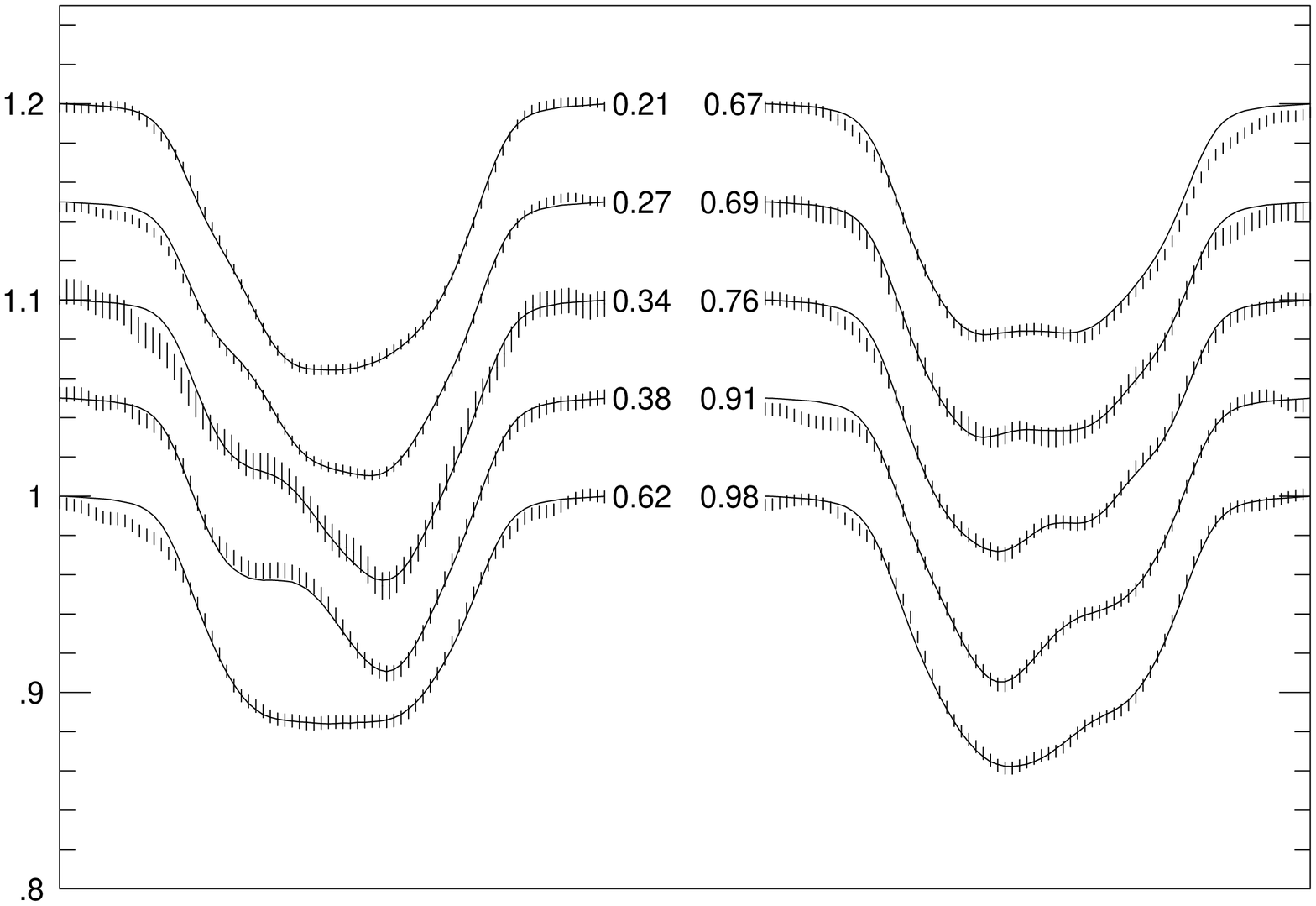}
\caption{The spectral line profiles and fits for the 1987.05 image}
\label{fig:1987.05fits}
\end{figure} 

\begin{figure}
\plotone{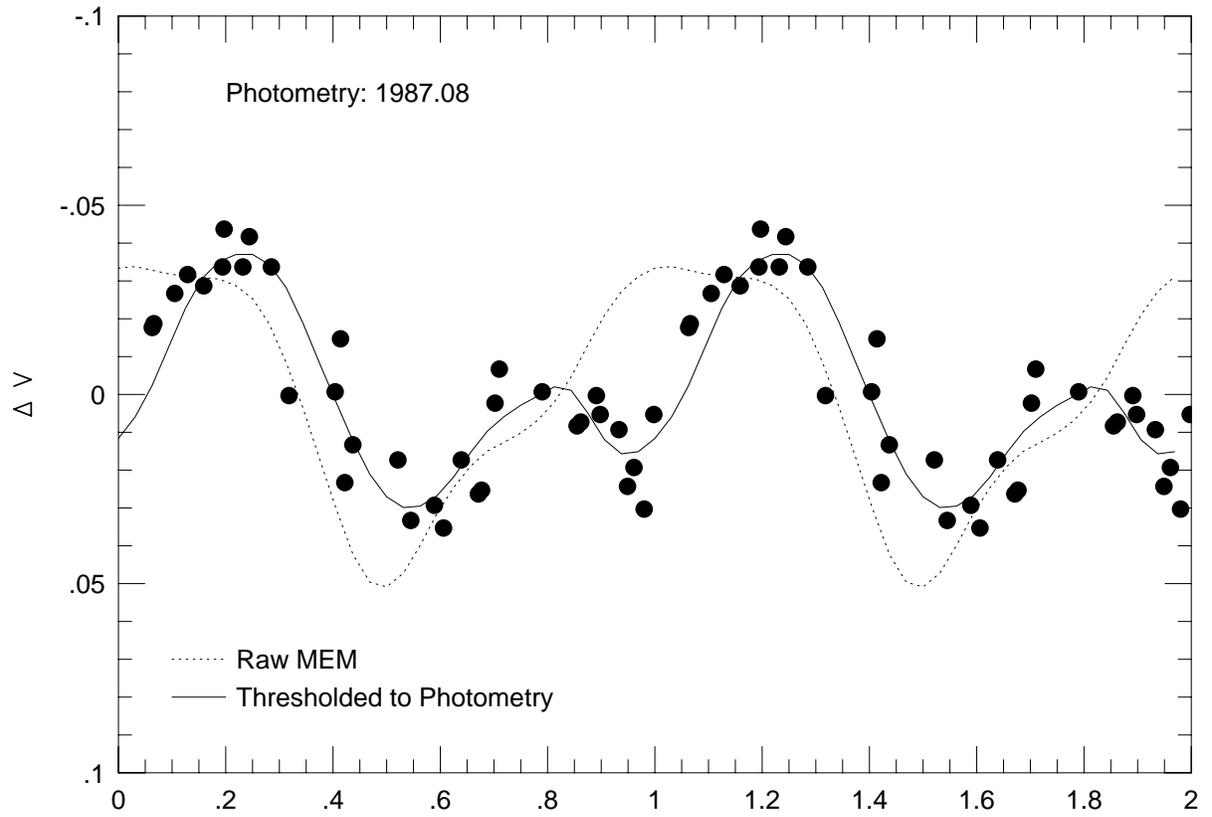}
\caption{1987.08 light curve used for the 1987.05 HR 1099 image}
\label{fig:1987.05_light}
\end{figure} 

\begin{figure}
\plotone{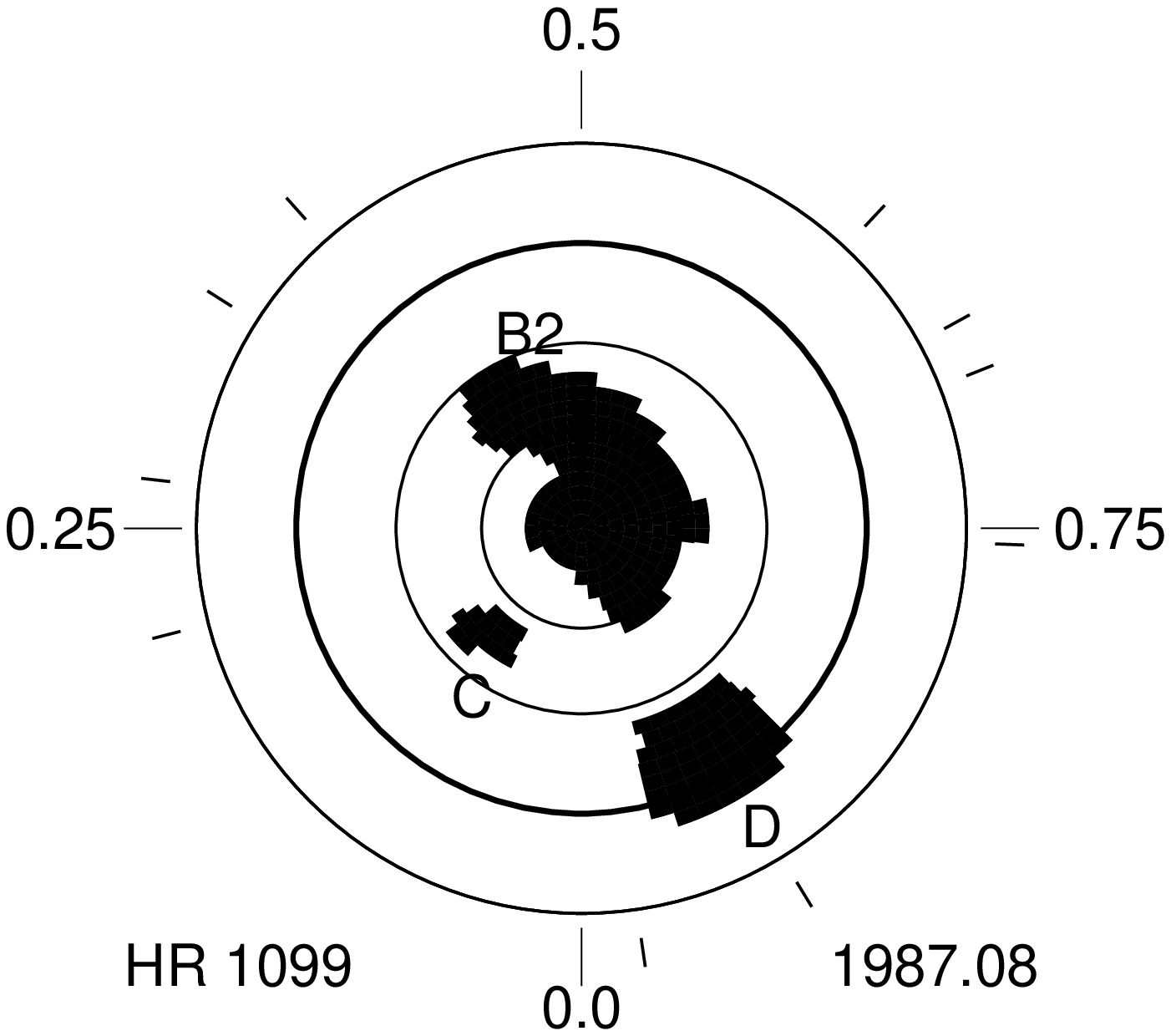}
\caption{HR 1099 thresholded Doppler image for 1987.08}
\label{fig:1987.05_image}
\end{figure} 

For the 1987.05 epoch, the raw spectral image is shown in
Figure~\ref{fig:1987.05_raw}. Note the spurious low-latitude
`mirroring' of the high-latitude spot appendage at phase 0.38. The
spectral line profiles in Figure~\ref{fig:1987.05fits}. The predicted
light curve from this image as fit to the 1987.08 photometry of
\cite{moh93} is shown as the dotted line in
Figure~\ref{fig:1987.05_light}. The fit of the predicted light curve
for the 1987.05 raw MEM image was again clearly inadequate.

Figure~\ref{fig:1987.05_image} shows the
photometrically-constrained image that best fits the observed light
curve. It is identical to the original Doppler image derived from the
spectral line profiles except for the large equatorial spot at phase
0.915 that seems to be required by the photometry. We hereafter refer
to this spot as Feature D. The predicted light curve for this image is
shown as the solid line in Figure~\ref{fig:1987.05_light}. Clearly,
the 1986.83 light curve shows significant differences to the one in
1987.08. The latter light curve developed a pronounced dip near phase
0.9 signifying the rapid appearance of Feature D sometime after
January 1987. The \cite{moh93} light curve was made between 5 Jan -- 8
Mar 1987, with the bulk of the measurements taken after 21 Jan. This
is somewhat after our Doppler image (which was derived from data
covering 12-19 Jan 1987). Feature D was therefore probably missed in
the Doppler image due to limited phase coverage and lack of time
resolution.

Here, we were able to use photometry taken at a slightly different
time to fill in the time gaps in our images and glean information
about spot evolution. The combination of the photometry, together with
the line profiles, tells us that there is rapid spot evolution
occuring that is not being adequately reproduced by our infrequent
images. We have tried here to weave the photometry and line profile
constraints together as best we can, but it is clear that, in this
regard, we are working near the sampling limits of our data set.

We believe, from combining both the imagery and light curve fitting,
that Feature C is real and was migrating clockwise as discussed
above. It was also well-isolated from other spots, and had similar
size, shape, and latitude in all images. Furthermore, the time span of
these images being only 5 months, this assumption has a reasonable
chance of being correct. If we are indeed tracking the same spot, then
Feature C, like Feature B1, is also migrating clockwise. A least
squares fit to the phases for Feature C of 0.03, 0.08, and 0.11 in
1986.63, 1986.83, and 1987.05 respectively gives a longitudinal
migration rate of 64\deg $\pm$ 13{\deg} yr$^{-1}$ at latitude
40{\deg}, or about 1 part in 723 of the orbital period. Again, the
implication is that the spot longitude migration rate (differential
rotation?) is much smaller (2-3 orders of magnitude) than the Sun and
again in the sense that intermediate latitudes of HR 1099 are rotating
more slowly than the orbital rate.

Assuming a constant longitudinal migration rate for Feature C of
64{\deg} yr$^{-1}$, places Feature C at about phase 0.89 in the
1985.86 image (Figure~\ref{fig:1985.86_image}), interestingly close
to, but perhaps just coincidentally at the phase of the small
projection and low-level feature at phase 0.88 on the polar spot in
1985.86. Extrapolation of Feature C to 1984.81 puts it at phase 0.71,
well away from the low-latitude spot seen there at phase 0.85, and not
near any other obvious mid-latitude feature. Extrapolation of Feature
C to the 1987.75 image puts it at phase 0.23 in
Figure~\ref{fig:1987.75_image}, slightly past but quite near the
narrow projection on the polar spot at phase 0.21 and latitude 45\deg
- 70\deg in the 1987.75 image, as will be discussed in the next
section.

The polar spot is again present in both images this observing season,
with several large protuberances and intermediate-latitude spots. A
study of TiO absorption detected in January 1987 by \cite{hue87}
indicated a polar spot area of 10\%. Our Doppler image also yields an
area of 10\% for the polar spots in both of the 1986-7 images, in
excellent agreement with \cite{hue87}.

In both the 1986.83 and 1987.08 images, a persistent feature remains
at or near the location of Feature B1 from 1984.81 and 1985.86. Some
spot activity has thus been present on this particular area of the
star (phase 0.3-0.4 and latitude 30\deg to 60{\deg}) for at least 2
years. It is difficult to say with certainty what is happening with
Feature B1. It seemed to have merged with the polar spot by 1985.86
but perhaps it was only that our phase coverage was not sufficient to
have resolved it from the polar spot. Clearly it looks detached again
in 1986.83, and then a bridge forms and it becomes reconnected in
1987.08. At this point, it seems to have changed its basic shape, so
we now change its name to Feature B2 to reflect the fact that it may
still be associated with the original Feature B1, but is now being
tracked as a possibly different feature. Whatever the case, Features
B1/B2 seem to show little or no longitudinal migration with respect to
the orbit. This is yet another indication that the high latitude spots
of HR 1099 seem to be rotating precisely at, or only slightly slower
than the orbital angular velocity. Feature B2 may even be discernable
for several more years in the image set as it follows a slow clockwise
drift around the polar spot.

\subsection{The 1987 Doppler Image}

\begin{figure}
\plotone{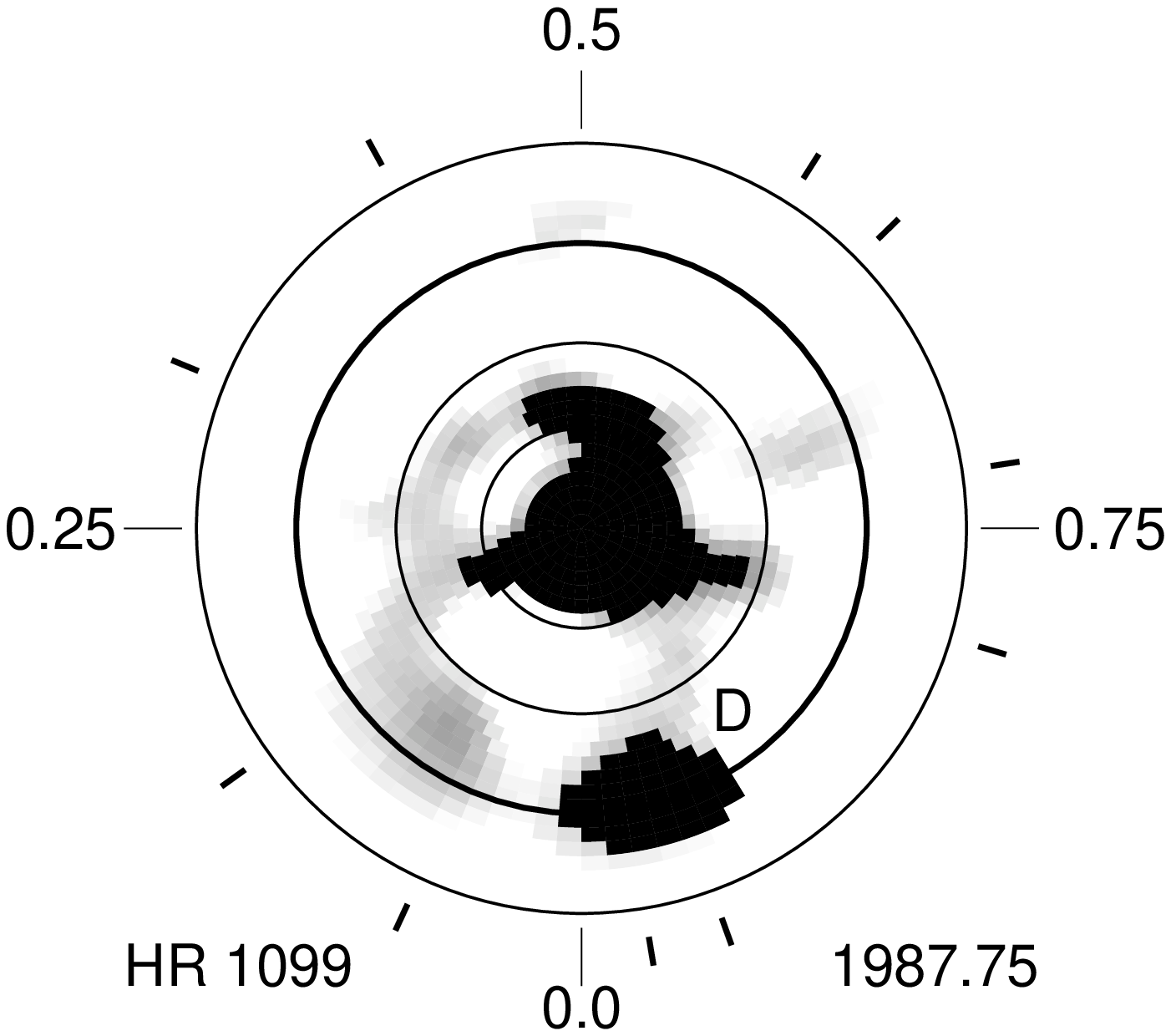}
\caption{HR 1099 raw (unthreshholded) Doppler image for 1987.75}
\label{fig:1987.75_raw}
\end{figure}

\begin{figure}
\plotone{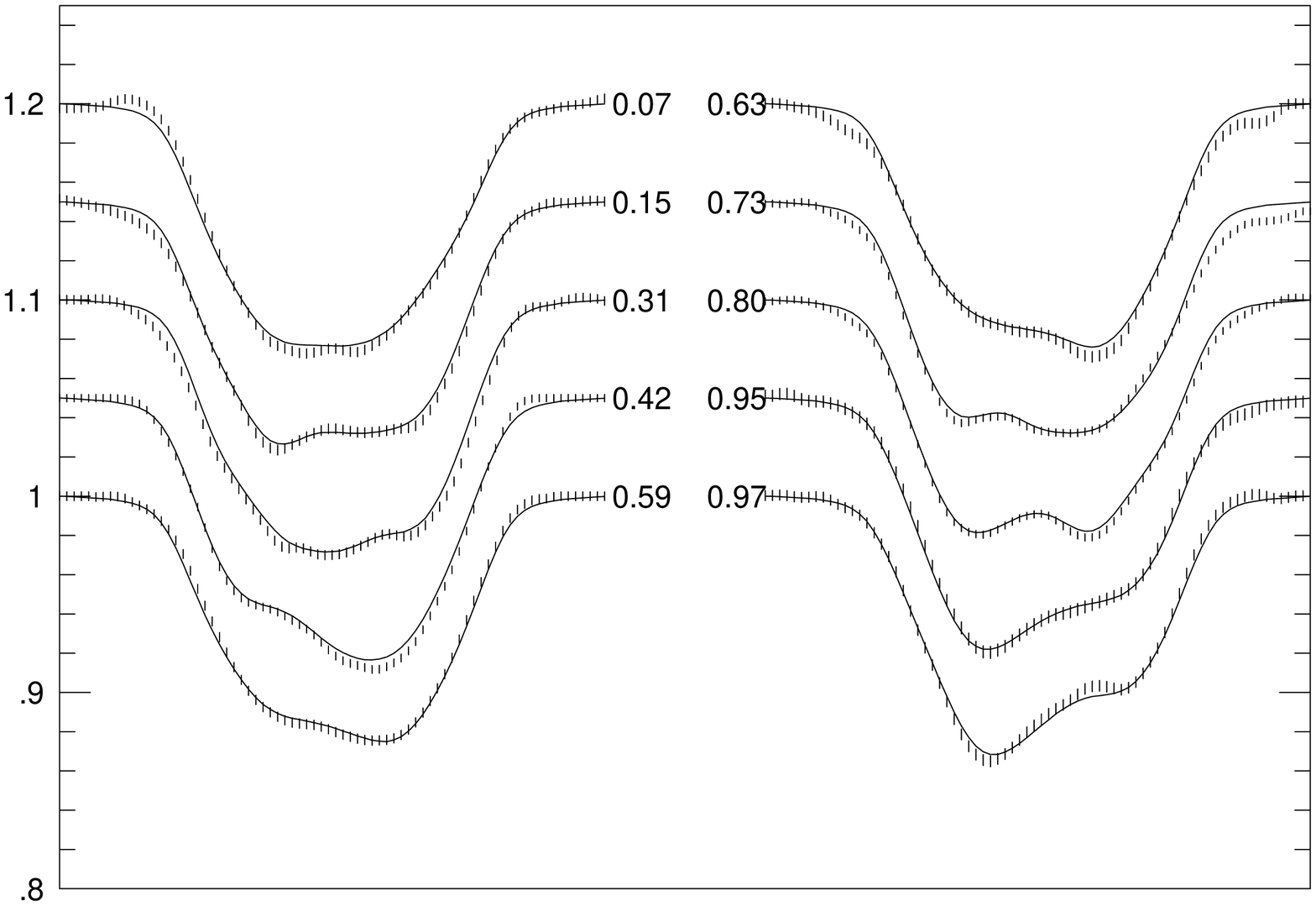}
\caption{The spectral line profiles and fits for the 1987.75 image}
\label{fig:1987.75fits}
\end{figure}

We obtained only a single Doppler image of HR 1099 in the 1987-88
observing season at epoch 1987.75. The raw unthreshholded image
solution is shown in Figure~\ref{fig:1987.75_raw}. The spectral line
profiles and fits for the Doppler image are shown in
Figure~\ref{fig:1987.75fits}. Light curves for epochs 1987.17 and
1988.07 were presented by \cite{moh93}, and a curve for 1988.16 was
presented by \cite{rod92}. The light curve amplitude was only about
0.05 magnitudes at this time and would have resulted in fairly
ambiguous 2-spot or 3-spot model solutions. In any case, no spot model
solutions were found in the literature to compare with our Doppler
image of this year.

\begin{figure}
\plotone{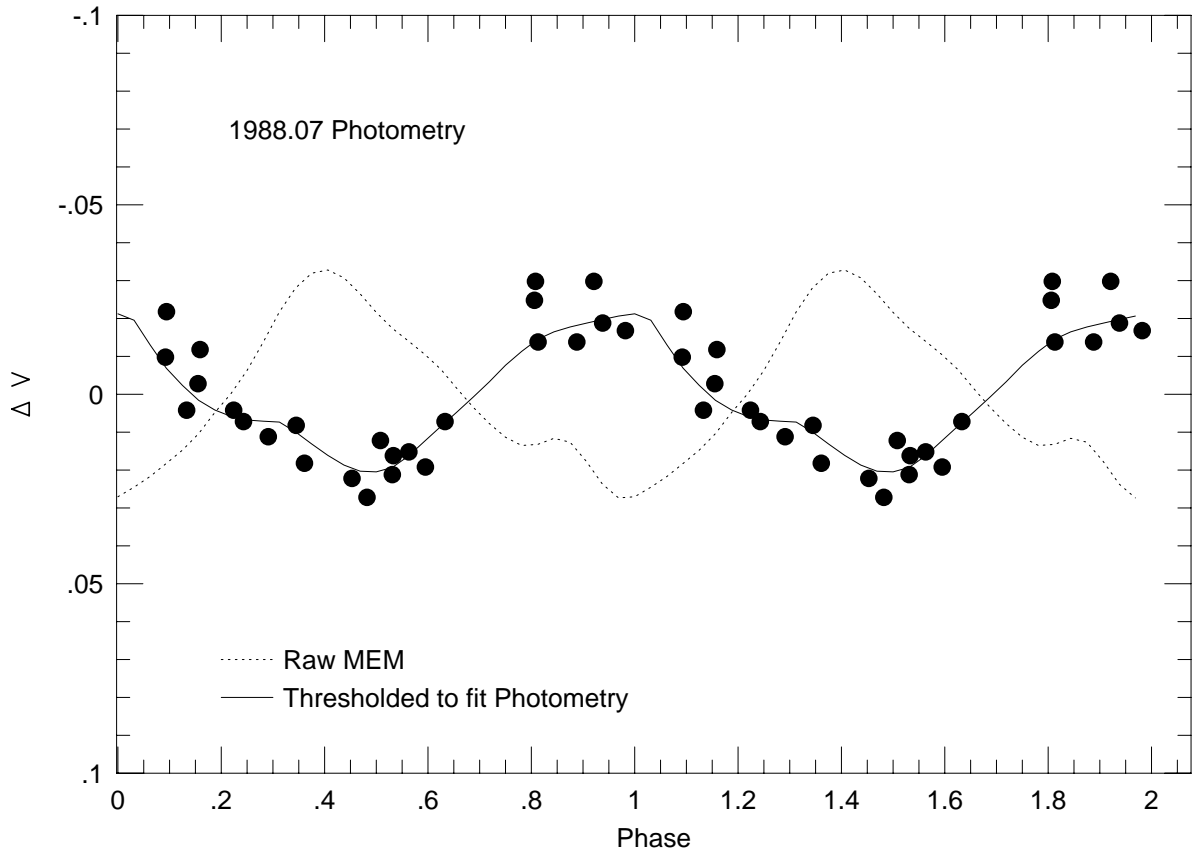}
\caption{1988.07 light curve used for the 1987.75 HR 1099 image}
\label{fig:1987.75_light}
\end{figure} 

\begin{figure}
\plotone{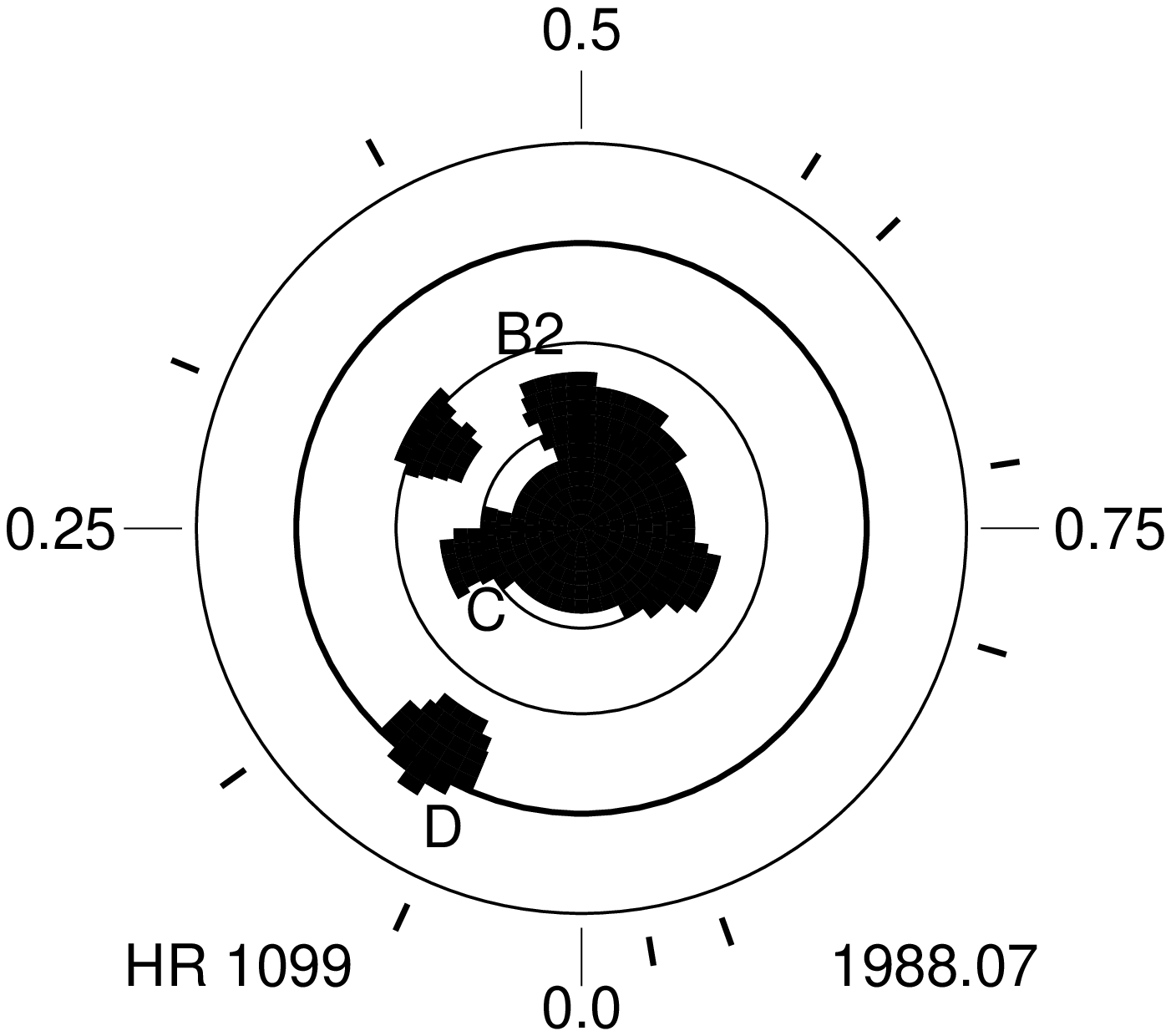}
\caption{1987.75 Doppler image of HR 1099 thresholded by the 1988.07 photometry}
\label{fig:1987.75_image}
\end{figure} 

Our raw image shows a polar spot with a large protuberance near phase
0.5 which looks quite similar to the polar spot protuberance named
Feature B2 of the 1987.08 image, as though this is still Feature B on
its slow clockwise migration around the pole. There is also a
pronounced equatorial spot at phase 0.96 which probably corresponds to
Feature D from the previous (1987.08) image but advanced in phase by
about 0.045.

Figure~\ref{fig:1987.75_light} shows the light curve for epoch 1988.07
(points) from \cite{moh93} along with our predicted light curve
(dotted line) from the unthreshholded 1987.75 Doppler image of
Figure~\ref{fig:1987.75_raw}. Here, the predicted light curve from the
raw MEM image was a quite poor fit to the photometry. In fact, the
predicted photometry is 180$^\circ$ out of phase with the actual
photometry! Again this is no great surprise since the Doppler image
and photometry were taken an uncomfortable 4 months apart.

Once again an attempt was made to fit the photometry using a modified
version of the 1987.75 raw Doppler image with the results shown in
Figure~\ref{fig:1987.75_image}. The epoch of this image is labeled as
1988.07 to reflect the mid-time of the photometric observations. The
light curve from this distribution is shown as the solid line in
Figure~\ref{fig:1987.75_light}. The polar spot of the original Doppler
image retained its shape. Only the sizes of some of the appendages
increased slightly. An additional mid-latitude spot was added at phase
0.3 at the location of a low-level spot in the raw Doppler image. The
main discrepancy between the raw (Figure~\ref{fig:1987.75_raw}) and
photometrically-constrained (Figure~\ref{fig:1987.75_image}) images is
the requirement that the low-latitude spot (Feature D) seen at phase
0.96 in the former, be shrunk somewhat and moved to phase 0.095 in the
latter.

There are a number of hypotheses which can account for the discrepancy
in location of this low-latitude feature between the two images. Could
Feature D from the photometrically-constrained image actually have
been present at the time the spectral data were acquired? There is a
low-level spot in the spectral image that coincides with this location
in the photometric image, but it is not as strong as the low-latitude
feature at phase 0.96. The phase coverage of the spectral data is such
that any information about low-latitude spots centered on phase 0.09
must come from the two observations centered on phase 0.06 and 0.15,
respectively.  Possibly the spot feature in the spectral image appears
so weak because it had recently emerged and was present at the time of
the observation at phase 0.15, but not at phase 0.06. Because spectral
distortions resulting from this spot appear only at one phase, MEM, in
trying to fit both profiles puts weak spot features at this
location. If true then this implies a rapid emergence time for
magnetic flux since the time difference for the two observations is
only one day! There is some evidence to support this. Placing a very
cool ($\Delta$T = 1200 K) spot at the location required by the
photometric image results in an excellent fit to the observed spectral
line profile at phase 0.15, but a rather poor one at phase 0.06,
suggesting the presence of a spot only at the time of the later phase.

Another explanation, and the one we favor on grounds of pure
simplicity, is that the low-level equatorial feature at phase 0.095 in
the raw image is not actually a real spot. Rather, the appearance of a
spot at this location in the photometrically-constrained image is
simply due to the fact that 4 months later - at the time the light
curve was observed - Feature D seen at phase 0.96 in the 1987.75 raw
Doppler image had migrated to phase 0.095.

Finally, it may well be that the low-latitude spot that appears in the
photometric image represents a different spot feature entirely.
Unfortunately, with so few Doppler images to guide our interpretation
we can never be sure which scenario is correct.

Assuming we are tracking the migration of Feature D, we can derive a
longitudinal migration rate for this isolated spot. It moved from
phase 0.915 in 1987.05 to phase 0.96 in the 1987.75 raw image, and
then on to phase 0.095 in the 1988.07 photometrically-constrained
image. These three positions are not consistent with a strictly
constant drift rate, but a least squares solution gives 57{\deg}
yr$^{-1}$ clockwise. As we shall see though, it is likely that the
spot sat fixed for a while before moving off. In this case, taking
only the last two positions would give a better estimate of the
terminal migration rate and yields 152{\deg} yr$^{-1}$ at latitude
7{\deg} for Feature D. We also found that this maximum migration rate
was consistent with the longitude extents of all the Feature D
spots. All Feature D spots were at least as large as the expected
phase smearing due to such migration in each image.

Feature B2, and/or some remnant thereof, seems to persist near phase
0.3 and latitude 37{\deg}, and a piece which may have calved off seems
to continue its slow clockwise arc around the pole, showing up as a
large protuberance projecting from the polar spot between phases
0.4-0.7. The detailed shape of this phase 0.4-0.7 protuberance
(Feature B2) on the polar spot in the 1988.07 photometric image is
very similar to the same Feature B2 protuberance in the 1987.08
spectral image, but advanced by about 16{\deg} $\pm$ 4{\deg} in
longitude between the 1987.08 image and the 1987.75 image (whose epoch
by the light curve constraints is actually 1988.07), leading us to
suspect that we are seeing slow clockwise migration of this edge of
the polar spot. Again, if we assume that this is the same feature,
making a slow clockwise circle around the pole, we get a migration
rate of about 16\deg $\pm$ 3{\deg} yr$^{-1}$ at this latitude, quite
similar to what we previously derived for Feature B1 (which may in
fact be the same feature). It is difficult to say what latitude this
feature corresponds to as it also appears to be moving northward as it
proceeds clockwise. Also, above 70\deg it merges with, and becomes
indistinguishable from, the permanent polar spot, so much of its area
may be northward of 70{\deg}. We'll use its mean latitude of about
55\deg from the 1988.07 epoch with the caveat that it appears to be a
rather rigid structure and may be part of a single, totally rigid
polar spot, in which case its effective latitude is probaly much
higher. There is a definite sense between the 1987.08 and 1987.75
images that Feature B2 is spiraling northward and clockwise towards a
merger with the polar spot.

This same sense of a northward, clockwise spiraling migration towards
the polar spot is suggested by Feature C. As mentioned in the previous
section, extrapolation of the (assumed constant) 64{\deg} yr$^{-1}$
clockwise longitudinal migration of Feature C from the 1986.63 image
to the 1987.75 image puts it at phase 0.23 in
Figure~\ref{fig:1987.75_image}, slightly past but quite near the
narrow projection on the polar spot at phase 0.21 and latitude 45\deg
- 70\deg in the 1987.75 image. Now, between the 1986.63 and 1987.05
images, Feature C also moved 7.5\deg northward in 0.42 years or about
18{\deg} yr$^{-1}$. This northward rate would then place it near
latitude 58\deg in our 1987.75 image, right at the latitude of the
phase 0.21 protuberance on the polar spot in the 1987.75
image. Furthermore, though the assumed constant migration rate of
64{\deg} yr$^{-1}$ places it 0.02 in phase beyond this protuberance,
it is likely that the migration rate decreases with increasing
latitude, and we should probably be taking a slightly smaller
migration rate appropriately averaged over latitude. In any event, the
correspondence between the expected position of Feature C
(extrapolated from its migration trajectory) and the phase 0.21
protuberance on the polar spot in the 1987.75 image is quite good, and
may be further indication that there is a systematic clockwise
northward spiraling flow of some intermediate-latitude spots up to
eventual merger with the polar spot.

\subsection{The 1988 Season Doppler Images}

We obtained two images this observing season at epochs 1988.79 and
1989.11. Light curves for 1988.80, 1988.87, and 1988.97 were presented
by \cite{rod92}, though no accompanying spot models were
given. \cite{moh93} presented a light curve for 1989.11 and
\cite{cut92} also presented a light curve for the 1989.09 epoch. The
light curve was quite complex, but relatively low in amplitude. For
most all of our other images which had contemporaneous or
near-contemporaneous light curves, we were able to find satisfactory
fits to both the line profiles and the light curve simply by properly
threshholding the Doppler image, or by slight modifications of the
spot distribution to reflect spot evolution. However, for both images
this season, we were quite unable to reproduce the light curve from
any reasonably adjusted version of the raw MEM image.

\begin{figure}
\plotone{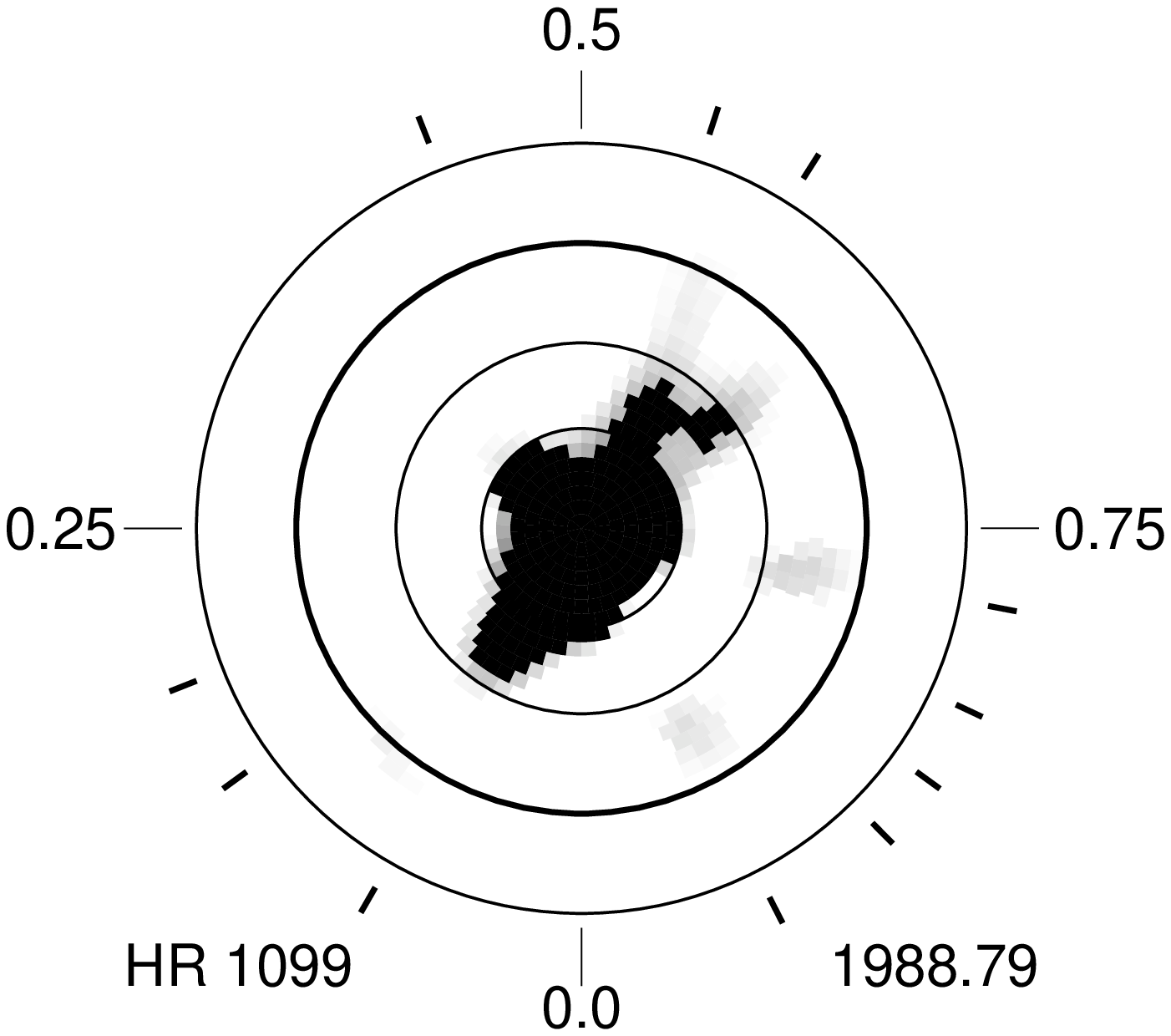}
\caption{HR 1099 raw (unthreshholded) Doppler image for 1988.79}
\label{fig:1988.79_raw}
\end{figure} 

\begin{figure}
\plotone{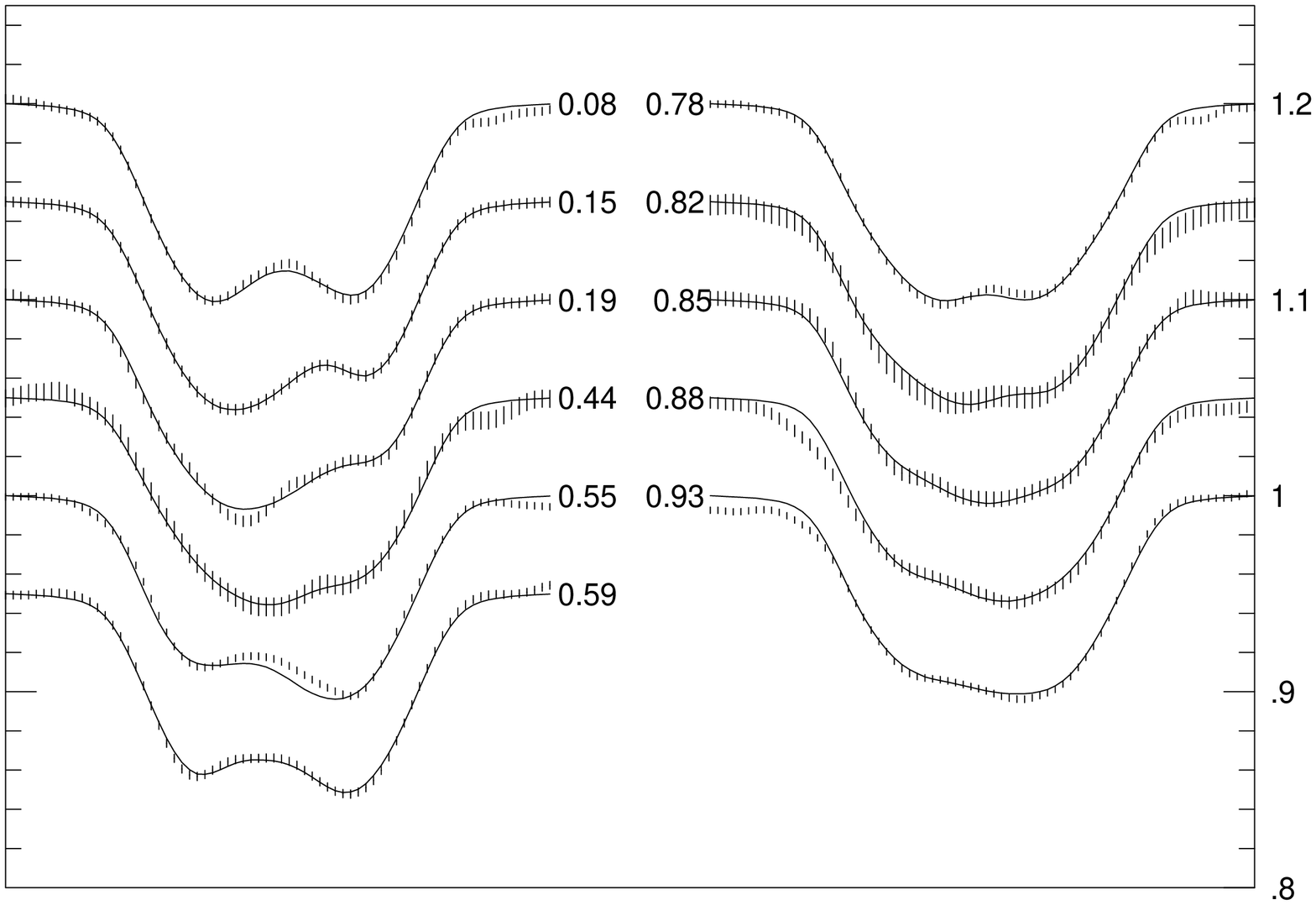}
\caption{The spectral line profiles and fits for the 1988.79 image}
\label{fig:1988.79fits}
\end{figure} 

\begin{figure}
\plotone{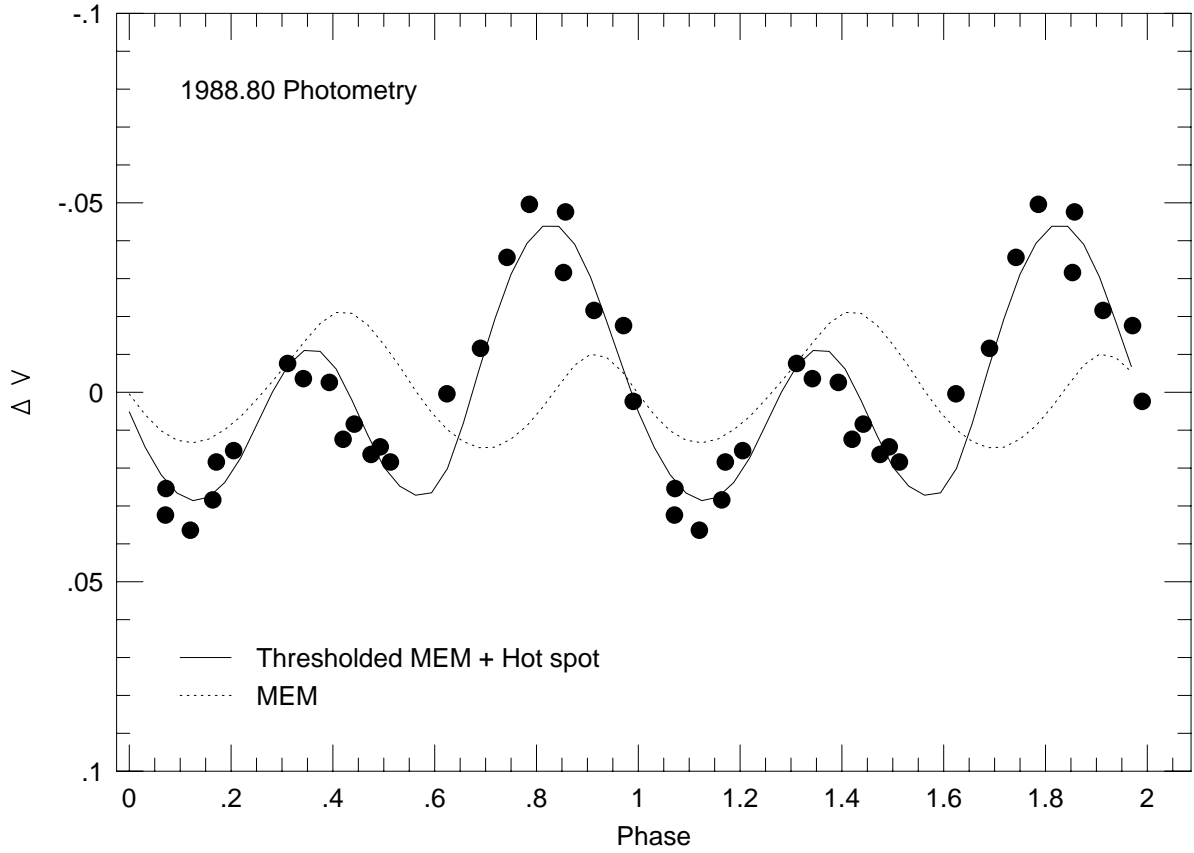}
\caption{HR 1099 light curve for 1988.80}
\label{fig:1988.79_light}
\end{figure} 

\begin{figure}
\plotone{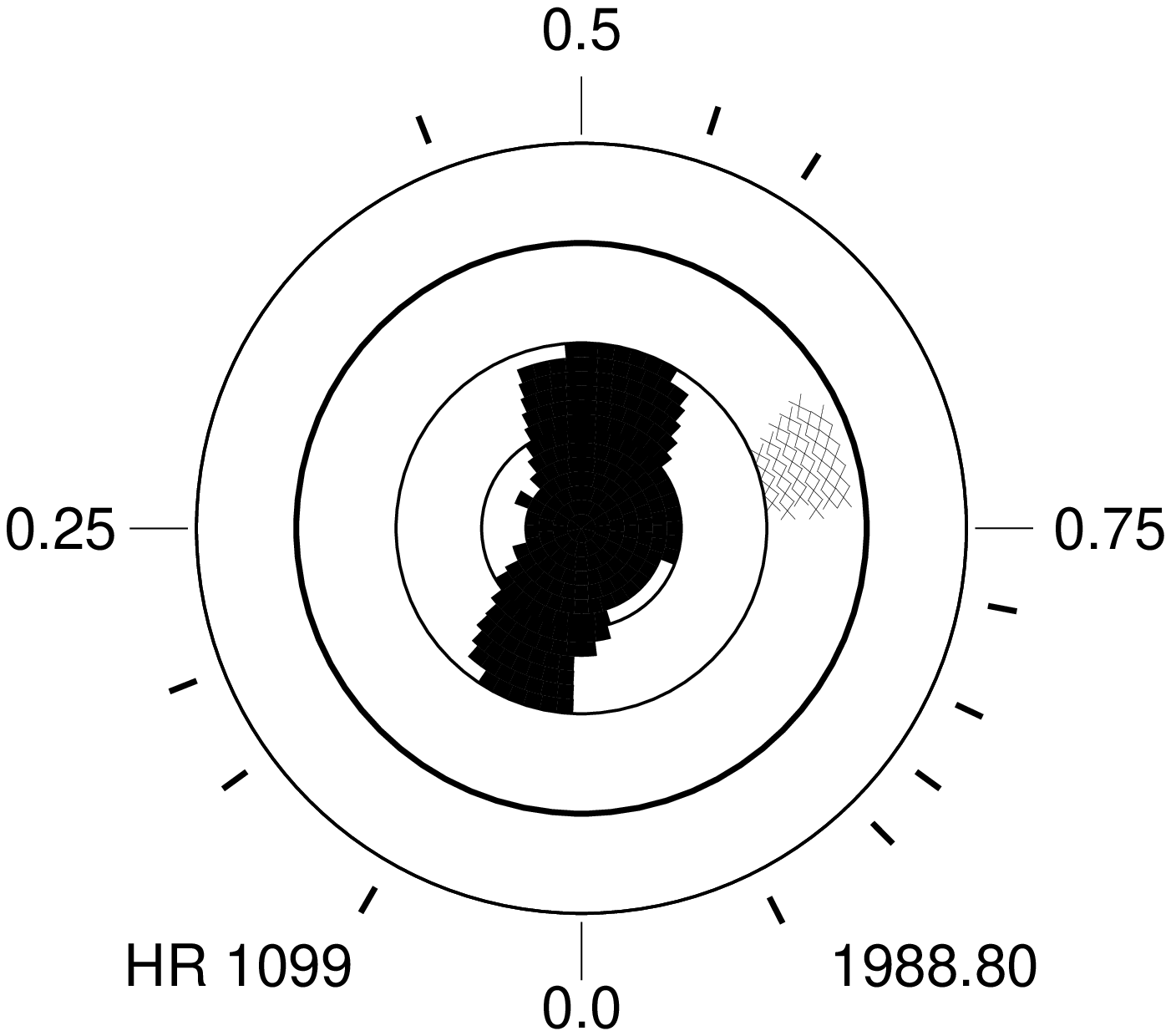}
\caption{HR 1099 photometrically-constrained Doppler image for 1988.80. The
cross-hatched region represents a hot spot 400 K warmer than the photosphere.}
\label{fig:1988.79_image}
\end{figure} 

The spectral image for 1988.79 is shown in
Figure~\ref{fig:1988.79_raw} and the spectral line fits in
Figure~\ref{fig:1988.79fits}. The predicted light curve from this
image is shown in Figure~\ref{fig:1988.79_light} as the dotted line
along with the 1988.80 light curve (points) of \cite{rod92}. As can be
seen, the predicted curve is much too low in amplitude and does not
fit well.

It quickly became apparent that thresholding and/or adding cool spots
alone could not fit the observed light curve; the Doppler image would
have to be modifed to such an extent that it would no longer fit the
observed spectral line profiles. Clearly, some feature was required to
significantly raise the brightness level near phase 0.80. This was
accomplished by placing a hot spot with a temperature 400 K above the
photosphere at the equator at phase 0.71. All other cool spot features
remained essentially the same as in the raw Doppler image. This
photometrically-constrained 1988.80 image is shown in
Figure~\ref{fig:1988.79_image}, and the light curve from this image is
shown as the solid line in Figure~\ref{fig:1988.79_light}.

\begin{figure}
\plotone{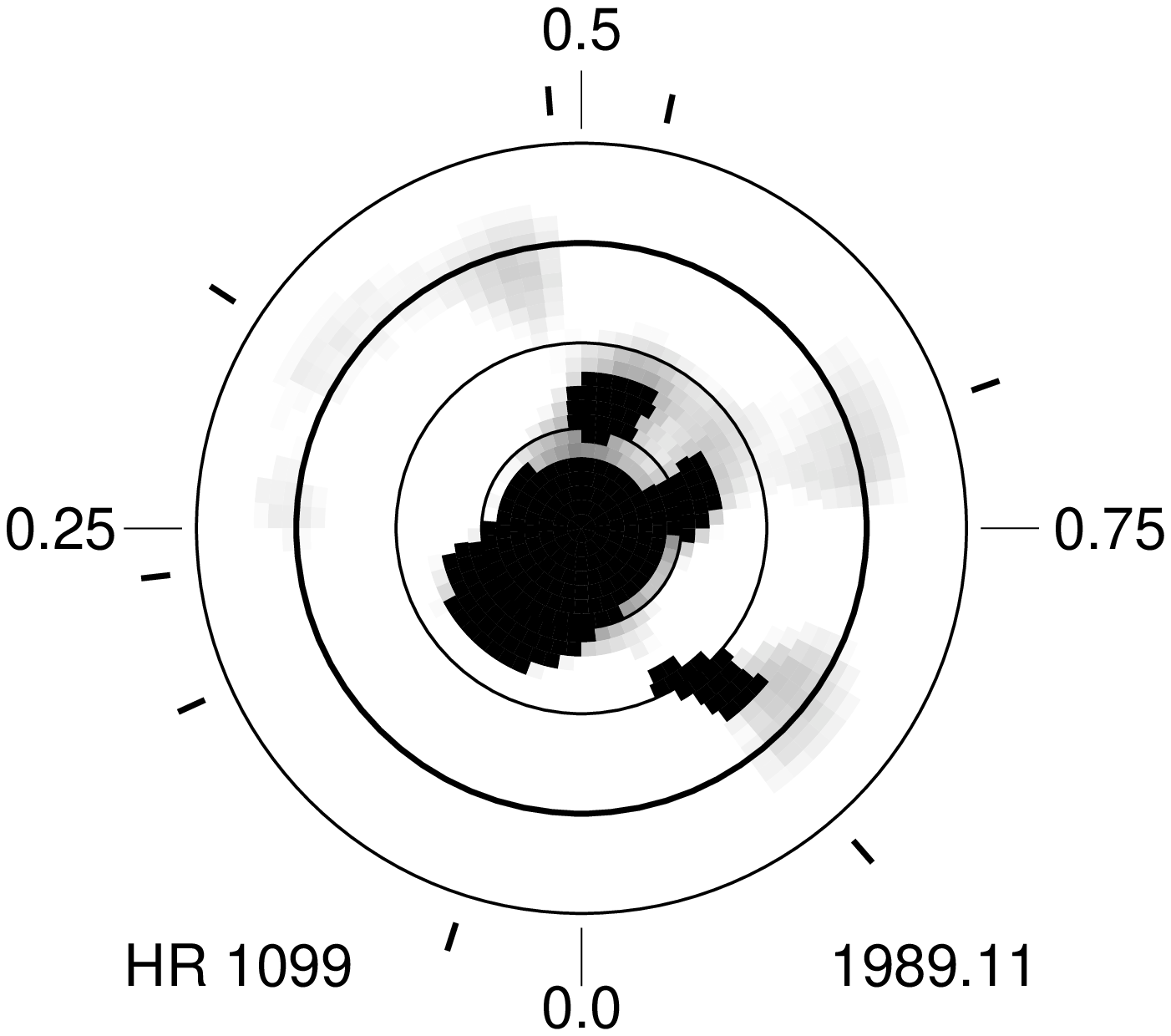}
\caption{HR 1099 raw (unthreshholded) Doppler image for 1989.11}
\label{fig:1989.11_raw}
\end{figure}

\begin{figure}
\plotone{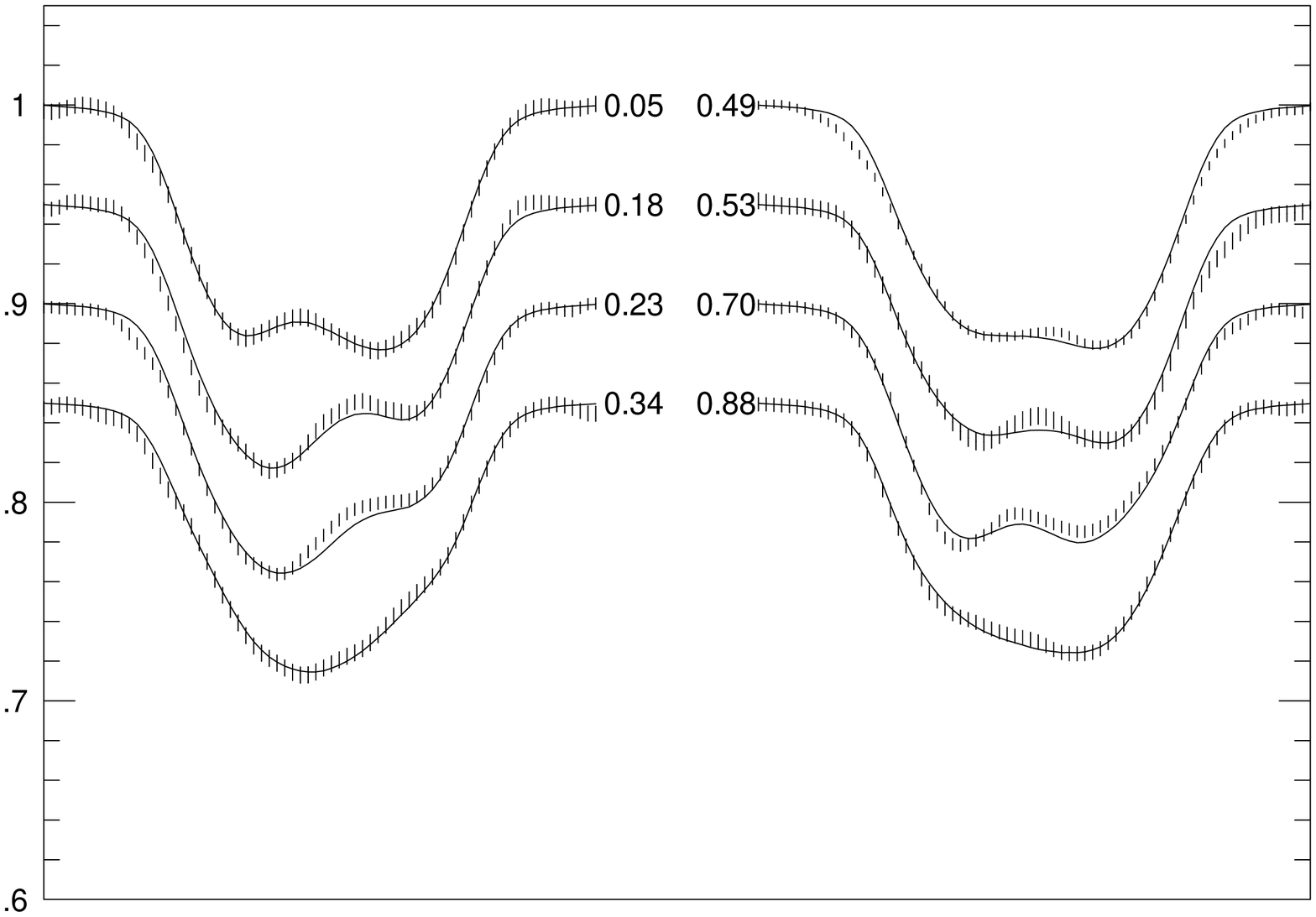}
\caption{The spectral line profiles and fits for the 1989.11 image}
\label{fig:1989.11fits}
\end{figure}

\begin{figure}
\plotone{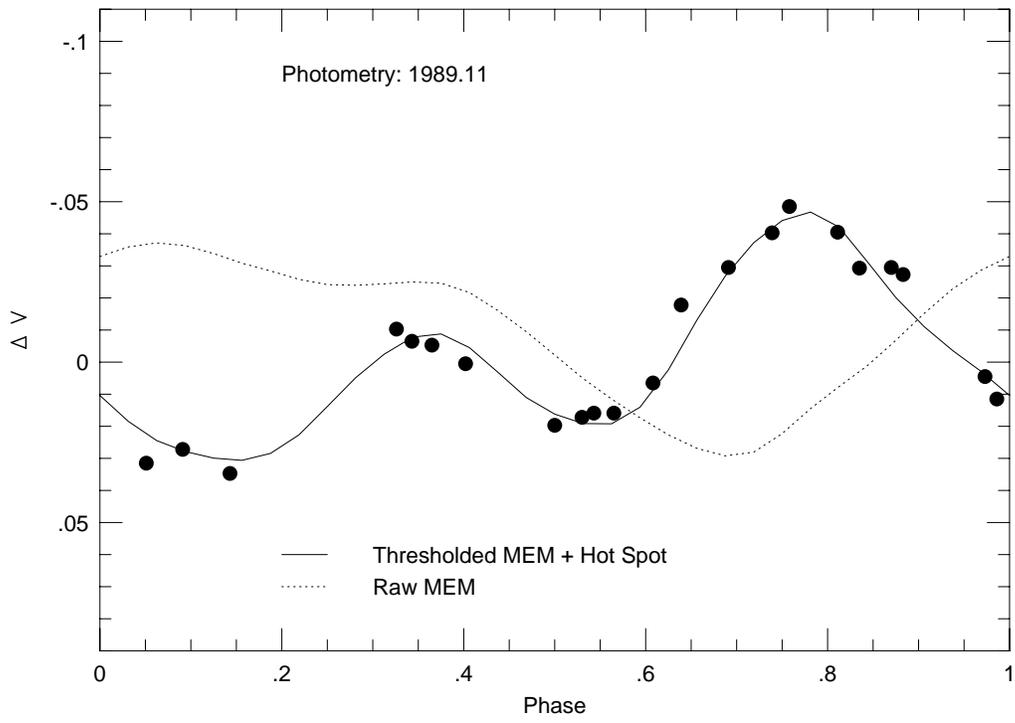}
\caption{HR 1099 light curve for 1989.11}
\label{fig:1989.11_light}
\end{figure}

We encountered similar difficulty in deriving the 1989.11 image
solution in that no simple combination of dark spots was found which
could adequately fit both the line profiles and photometry
simultaneously. Figure~\ref{fig:1989.11_raw} is our raw unthreshholded
MEM image for 1989.11. The spectral line profile and fits are shown in
Figure~\ref{fig:1989.11fits}. The predicted light curve from the raw
image is shown by the dotted line in Figure~\ref{fig:1989.11_light}
along with the 1989.11 photometry (points) of \cite{rod92} which have
been re-phased to the \cite{fek83} ephemeris.

Again, this raw MEM image gave a very poor fit to the photometry. We
were, however, able to get a decent fit by including a hot spot with a
temperature of 400-500 K above the photosphere placed near the equator
at phase 0.74 (a spot at a similar location was also required to fit
the photometry accompanying the 1988.79 image) and by increasing the
area of one of the polar appendages.

\begin{figure}
\plotone{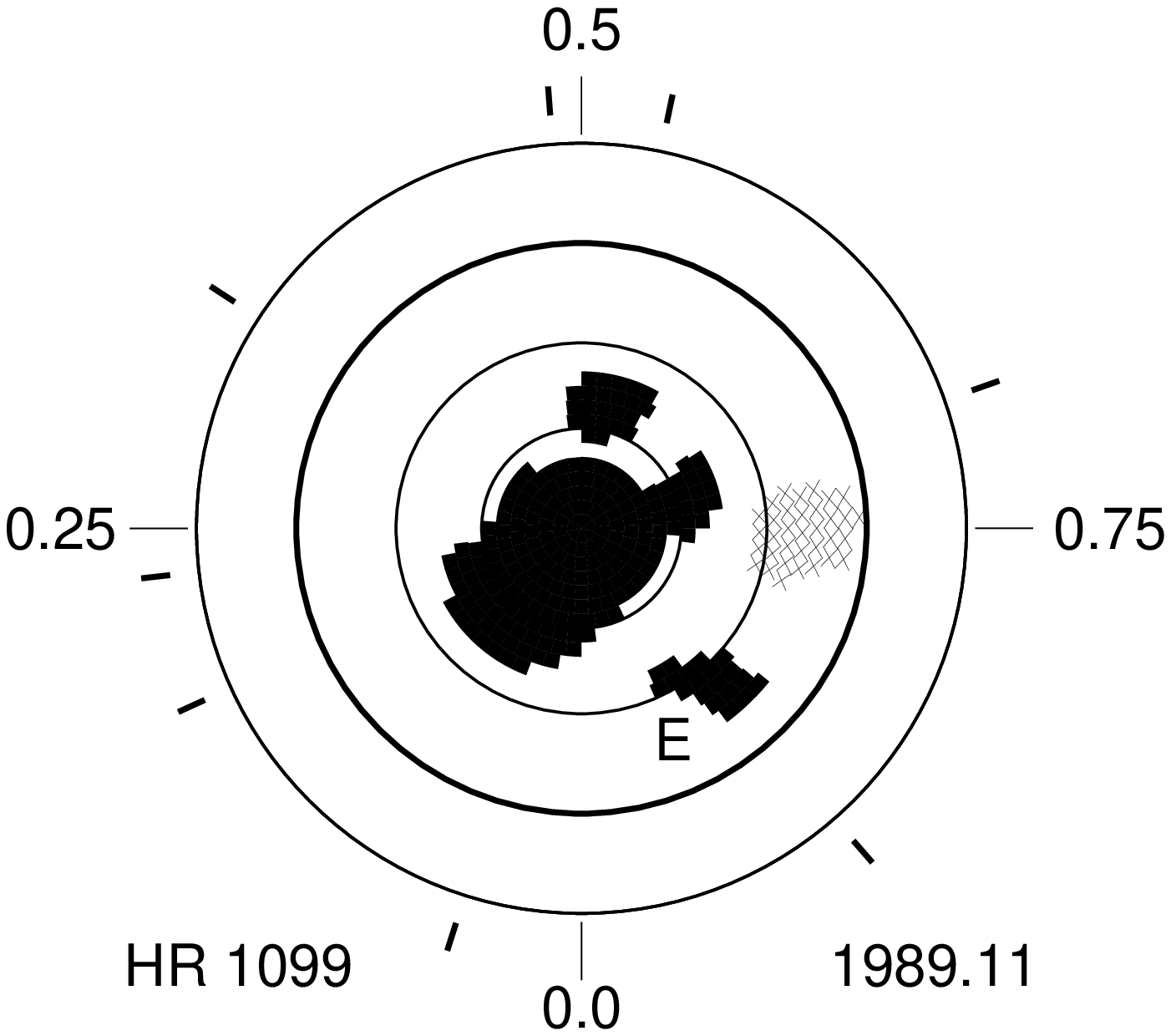}
\caption{HR 1099 photometrically-constrained Doppler image for 1989.11. The
cross-hatched region is a hot spot at a temperature of 400-500 K above the 
photosphere.}
\label{fig:1989.11_image}
\end{figure} 

The photometrically-constrained image, which fits adequately both the
line profiles and the light curve is shown in
Figure~\ref{fig:1989.11_image} and its predicted light curve is shown
as the solid line in Figure~\ref{fig:1989.11_light}. The basic shape
of the polar spot was largely unchanged in the process, although its
area increased somewhat. The most striking change again was the
apparent need to have a low-latitude hot spot near phase 0.74.

We were quite uneasy with including hot spots in our solutions since
we deliberately constrained the MEM method to only image features {\it
cooler} than some quiescent photospheric temperature. In most other
cases, this has been adequate and has produced good fits to both line
profiles and, with a little threshholding, to light curves as well,
even when the light curves were added after the fact. But it was quite
clear in this year's images that we simply could {\it not} achieve a
satisfactory fit with only dark spots. Our result certainly does not
confirm the presence of optically-visible hot spots on RS CVn stars,
but it is perhaps corroborated by \cite{don92b} who also found
evidence for a hot spot in their 1988.9 image at the same latitude and
near phase 0.78, as will be further discussed below.

If prominent, optically-visible hot spots do exist on HR 1099, then
they must not be very common. In our 11 years of Doppler imaging with
23 independent images, there was evidence for hot spots in only two
images and they were both from this 1988-89 observing season. While
one may well criticize our constraint of imaging only cool spots, in
most cases our Doppler imaging shows that it is a good assumption, and
that indications of prominent hot spots are quite apparent from the
solution fitting. Furthermore, adding these hot spots does not seem to
alter the gross features of the dark spot component of the image. It
is clear that it would be preferable to implement a version of Doppler
imaging which allows both hot and cool spots, but in the absence of
very accurate simulataneous photometry, one runs the considerable risk
of introducing spurious hot spots. It would be crucial to have
excellent simultaneous multi-color photometry to be sure of hot spots.

The prominent feature near the polar spot at about phase 0.53 in the
final photometrically-constrained 1989.11 image
(Figure~\ref{fig:1989.11_image}) may be the persistent signature of
Feature B2, though it's difficult to calculate a migration rate since
the structure of that feature has now changed considerably. There is
now no sign of Feature D but there is a very large protuberance on the
polar spot at phase 0.09 whose origin is difficult to connect with any
previous feature due to gross changes in the spots. Its longitude is
consistent with the phase 0.095 of Feature D in the 1987.75 image, but
is not consistent with Feature D's position if it simply continued its
152{\deg} yr$^{-1}$ migration rate observed previously. So, if Feature
D were related to this major protuberance, it would have had to have
come to a sudden stop in longitude migration between 1987.75 and
1989.11.

In other work, \cite{buz91} reported a dramatic reduction of
`extraphotospheric' emission on the primary star at about epoch
1988.71 which they ascribed to the disappearance of a prominence on
the K star. They also reported seeing evidence for mass transfer in
this binary system. A `Doppler snapshot' of HR 1099 computed by
\cite{dem92} for the Oct. 1988 - Sept. 1989 interval suggested the
presence of a large polar spot (from the flattening in the cores of
the spectral line profiles), but no actual image was presented.

\cite{don92b} presented a detailed Doppler image of HR 1099 at epoch
1988.9, very close in time to our image of 1988.80 in
Figure~\ref{fig:1988.79_image}. Comparing our Doppler images with
their image (Figure 4 of their paper) shows quite good agreement
between these two completely independent Doppler imaging solutions,
both for the shape and size of the polar spot, even for two low-level
features near latitude 15\deg and phases 0.78 and 0.92 in our raw
image (Figure~\ref{fig:1988.79_raw}). While most of the \cite{don92b}
data were obtained over a 3-month interval from Oct. 11 through
Dec. 22, one of their crucial phases (phase 0.855) for Stokes V
observations was obtained 4 months later, on March 24, 1989. At that
time, they detected quite a large monopolar circular polarization
signal near phase 0.855, indicative of a 1 kG monopolar region near
the equator. They also found indications that the region was bright,
and, though dark regions may also have been associated with that
region, the polarization signature would have been dominated by any
associated bright region. They also remarked that their largest
fitting residuals occurred for the phase 0.855 line profile core.

It seems likely that this region corresponds with the emergence of
Feature E in our 1989.11 image (Figure~\ref{fig:1989.11_image}), and
that their larger-than-normal fitting residuals were due to time
variation of a complex bright and/or dark emerging magnetic region at
the position of Feature E. Our Doppler image of 1989.11
(Figure~\ref{fig:1989.11_image}) shows that Feature E is indeed
changing and, what was only marginally detected as a subtle spot at
about phase 0.92 in the 1988.79 raw image (and which did not survive
our thresholding process) grew into a prominent dark spot at phase
0.89 by 1989.11. So we may have witnessed the birth of a spot (Feature
E) at latitude 25{\deg}. This spot first appeared as a weak bright
feature, but then quickly evolved into a prominent dark spot which, as
shown below, will then grow in area.

The polar spot also shows complex evolution. The large protuberance at
phase 0.54 in 1988.80 evolved into a detached spot at phase 0.53 and
latitude 50{\deg}, with a new protuberance at phase 0.70.  However,
since the 1989.11 image did not have many phases observed, and the
phases of some of these protuberances coincide with the phases
observed, we suspect that some of this structure may be an artifact of
the phase sampling which, when lacking sufficient image constraints,
tends to favor putting features at the sub-observer longitude. One
must resist over-interpreting detailed changes in the image structure
when phase sampling gets sparse. We believe that the 1989.11 image
does show that there is significant spot presence between latitudes
30\deg - 60\deg at phase 0.6, but some of the structural detail in
this area in the 1989.11 and 1989.72 (next section) images may be a
consequence of insufficient phase sampling. Finally, the large polar
spot protuberance near phase 0.09 in 1988.79 does appear to still be
present in 1989.11, though much broadened in longitude as though the
feature simply grew longitudinally by about 45\deg in the clockwise
direction.

\subsection{The 1989 Season Doppler Images}

We obtained three separate images in the 1989-90 observing season, at
epochs 1989.72, 1989.79, and 1989.83. Four light curves, for epochs
1989.73, 1989.79, 1989.86, and 1989.95 were presented by \cite{rod92},
and a light curve for epoch 1989.98 was presented by
\cite{moh93}. Multi-color light curves and a `3-spot' model for
1989.98 were also presented by \cite{zha94}.

\begin{figure}
\plotone{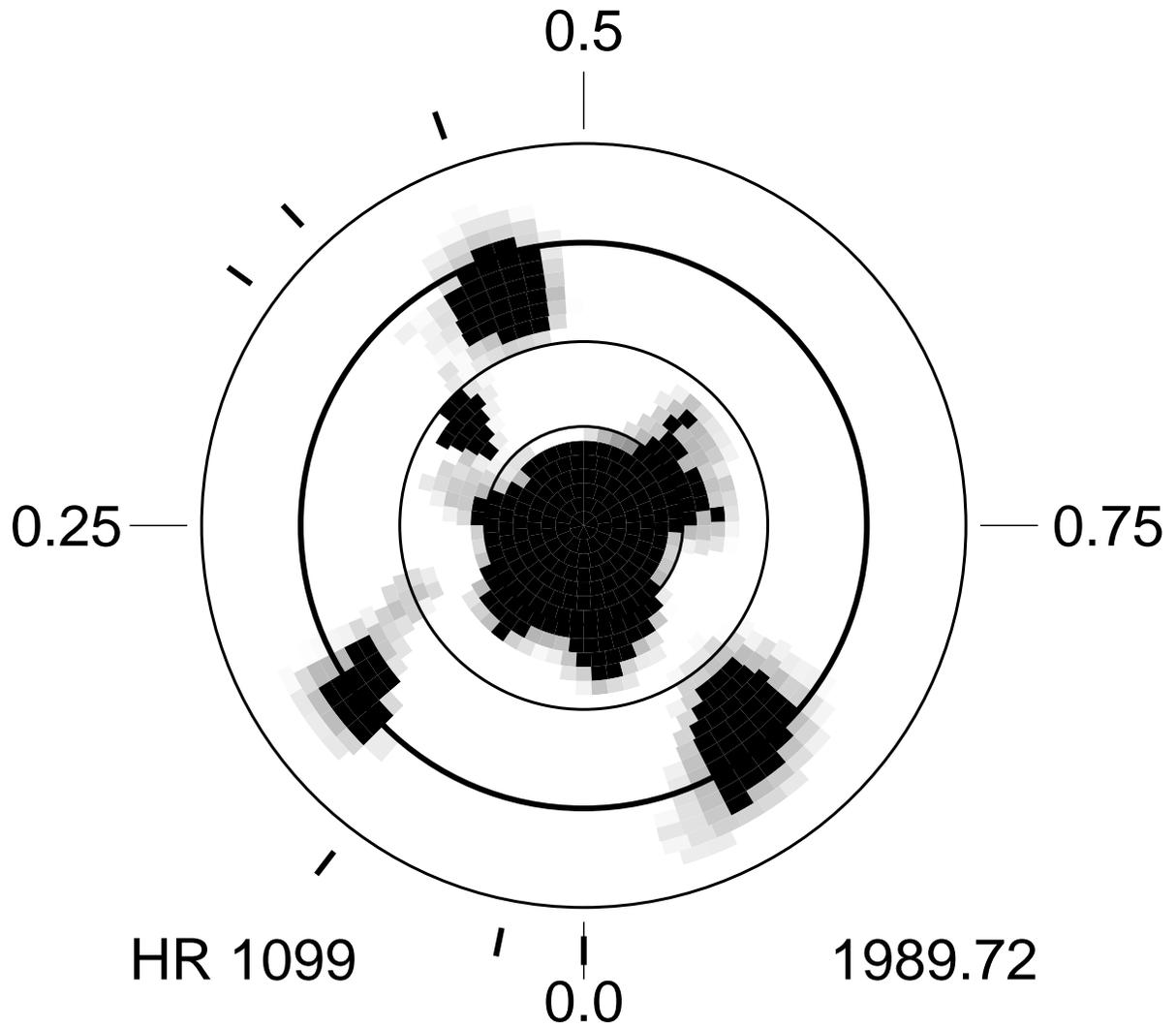}
\caption{HR 1099 raw (unthreshholded) Doppler image for 1989.72}
\label{fig:1989.72_raw}
\end{figure} 

\begin{figure}
\plotone{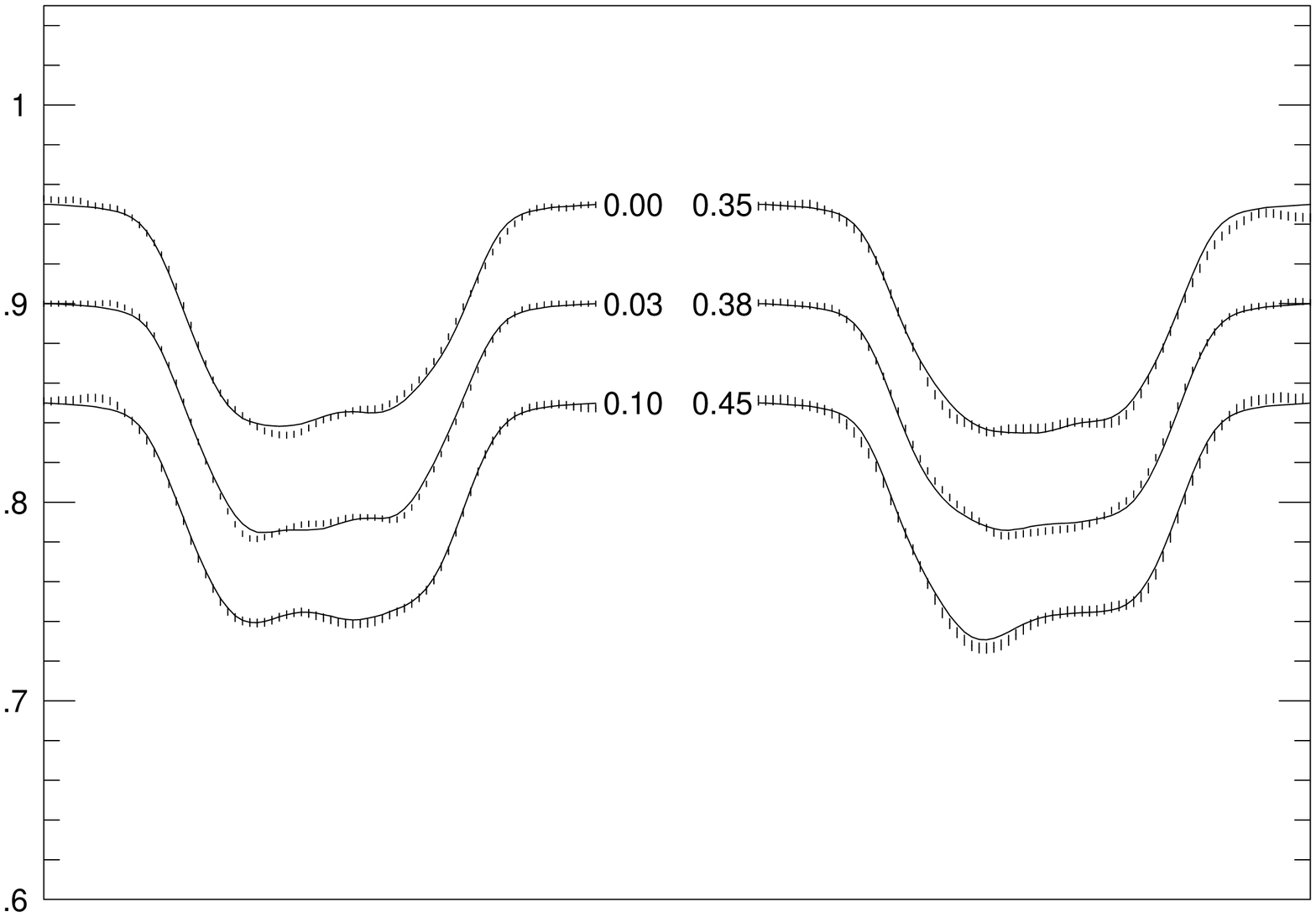}
\caption{Spectral line profiles and fits for 1989.72}
\label{fig:1989.72fits}
\end{figure} 

\begin{figure}
\plotone{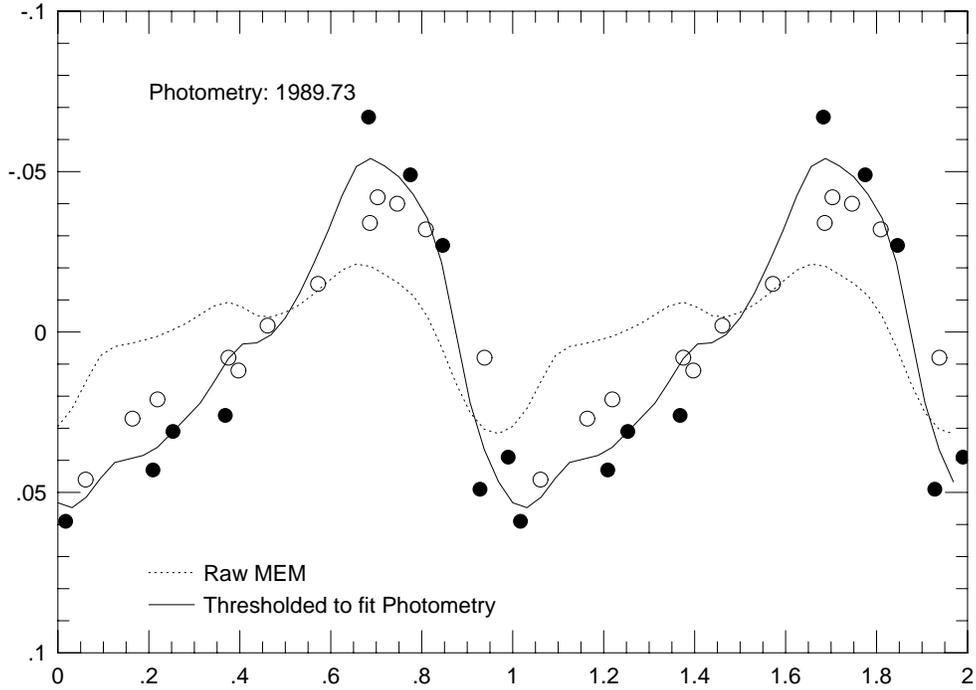}
\caption{HR 1099 light curve for the 1989.73 image. The solid points are from
epoch 1989.73. The open circles are from epoch 1989.79.}
\label{fig:1989.72_light}
\end{figure} 

\begin{figure}
\plotone{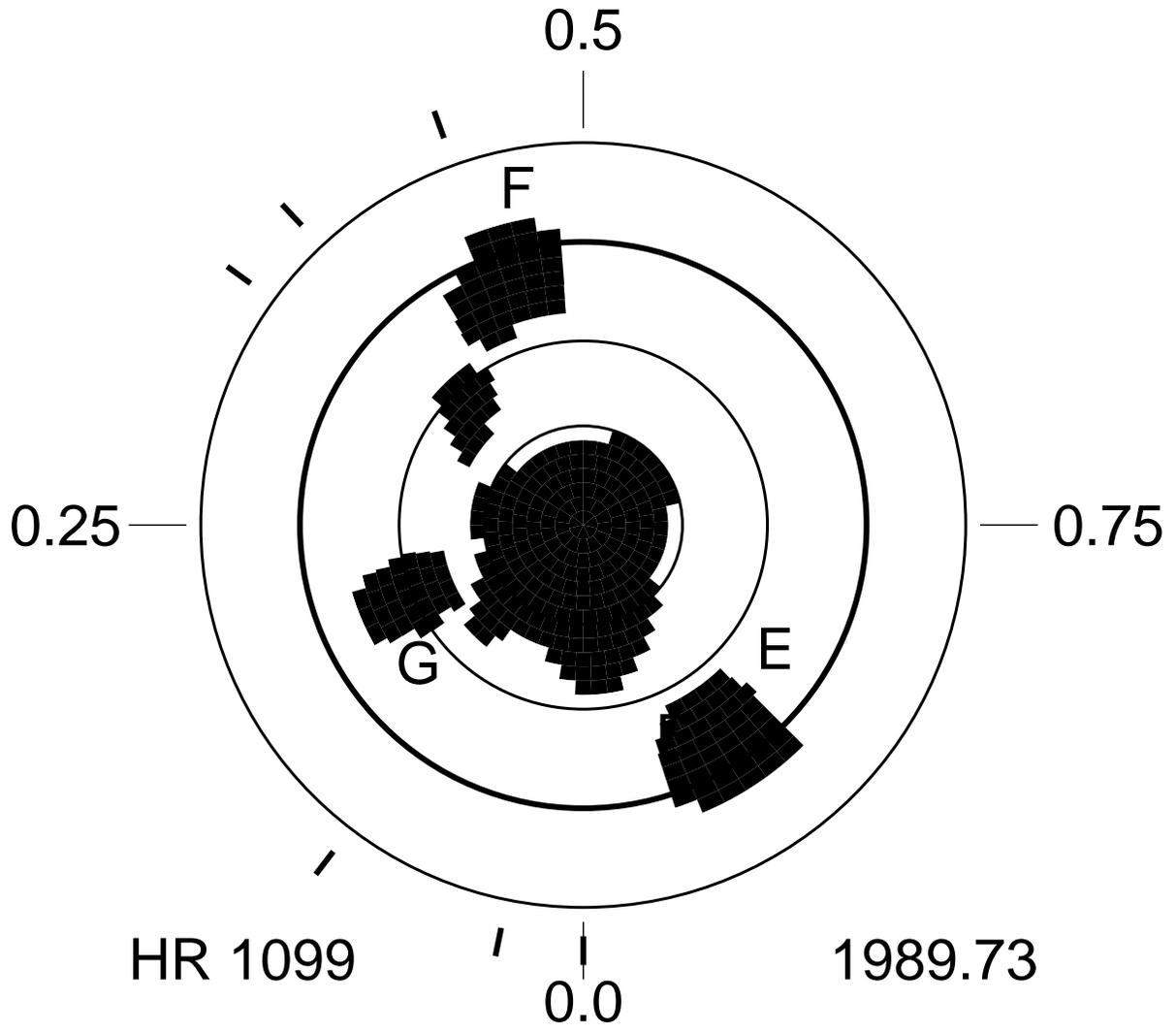}
\caption{HR 1099 thresholded Doppler image for 1989.73}
\label{fig:1989.72_image}
\end{figure} 

The raw MEM Doppler image for 1989.72 is shown in
Figure~\ref{fig:1989.72_raw} and the spectral line profiles and fits
in Figure~\ref{fig:1989.72fits}. The predicted light curve from our
raw image is shown as the dotted line along with the 1989.73 (solid
points) and 1989.79 (open points) photometry of \cite{rod92} in
Figure~\ref{fig:1989.72_light}. Threshholding to better fit the light
curve produced little change in the image. The area of the appendages
and the low-latitude spot had to be increased somewhat. The final
thresholded image, shown in Figure~\ref{fig:1989.72_image}, gives a
pretty good fit to the observed profiles. The area of the appendages
and of all the spots had to be increased somewhat and Feature G is
displaced to higher latitude in the final image.

\begin{figure}
\plotone{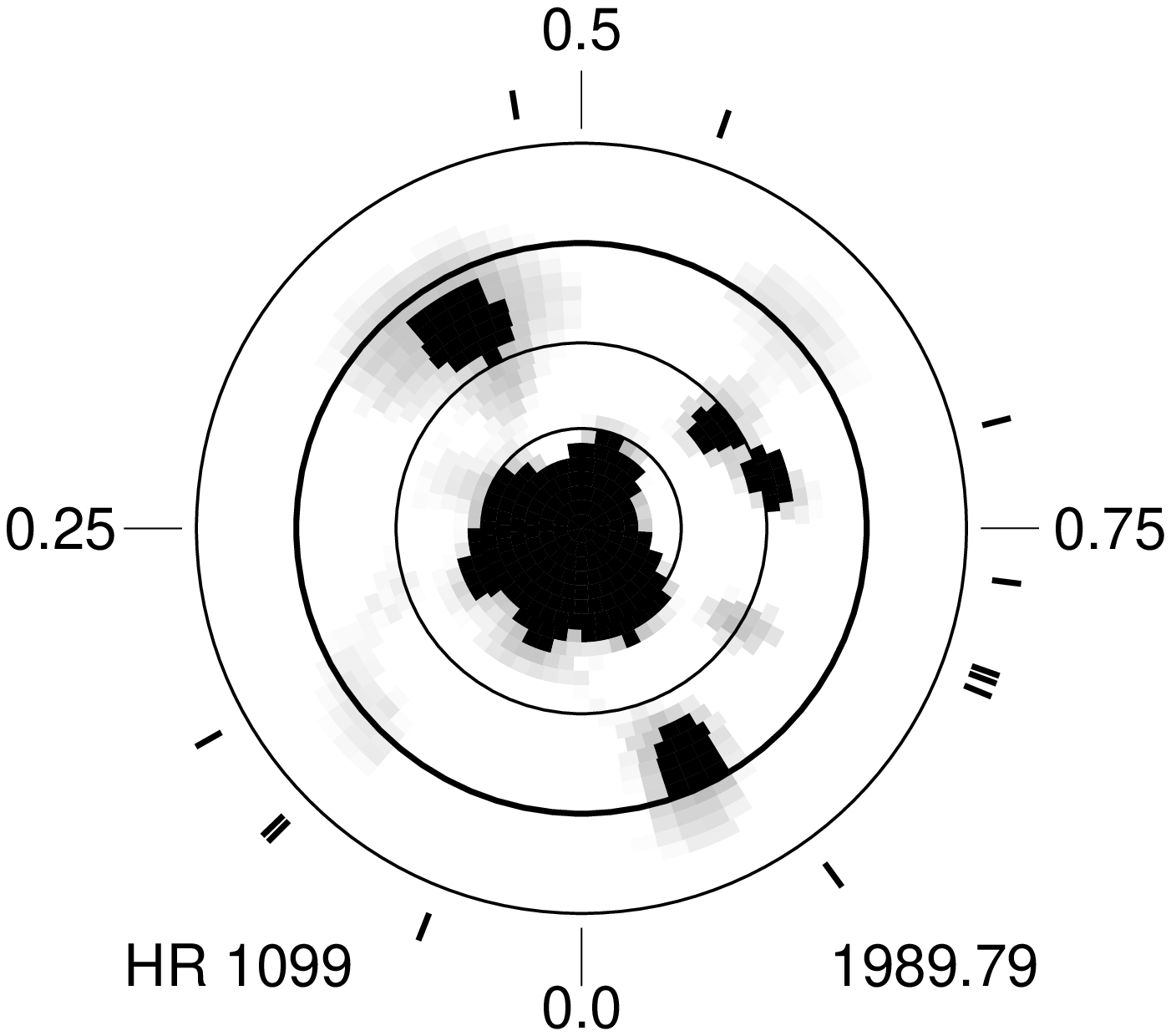}
\caption{HR 1099 raw (unthreshholded) Doppler image for 1989.79}
\label{fig:1989.79_raw}
\end{figure}

\begin{figure}
\plotone{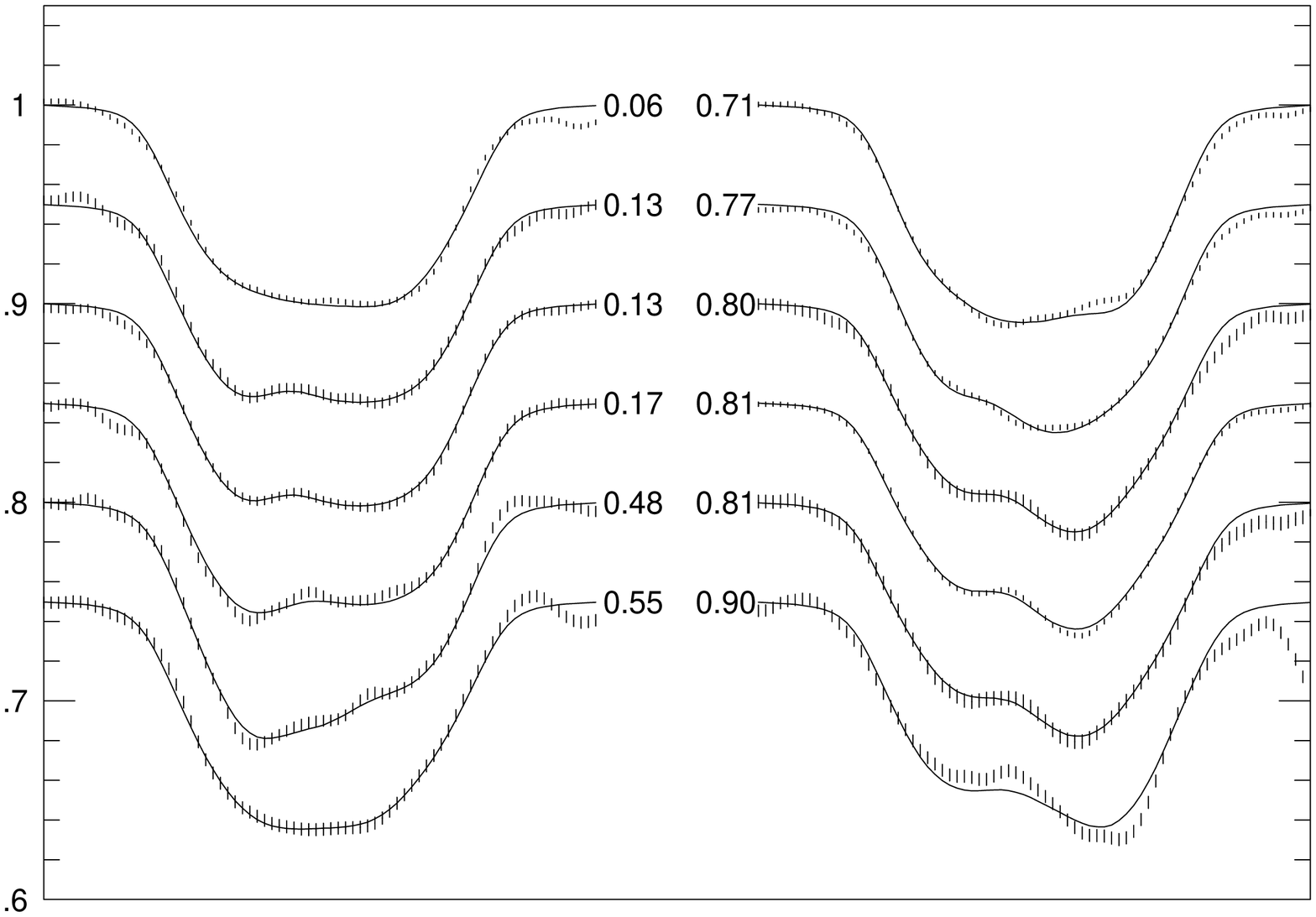}
\caption{Spectral line profiles and fits for 1989.79}
\label{fig:1989.79fits}
\end{figure}

\begin{figure}
\plotone{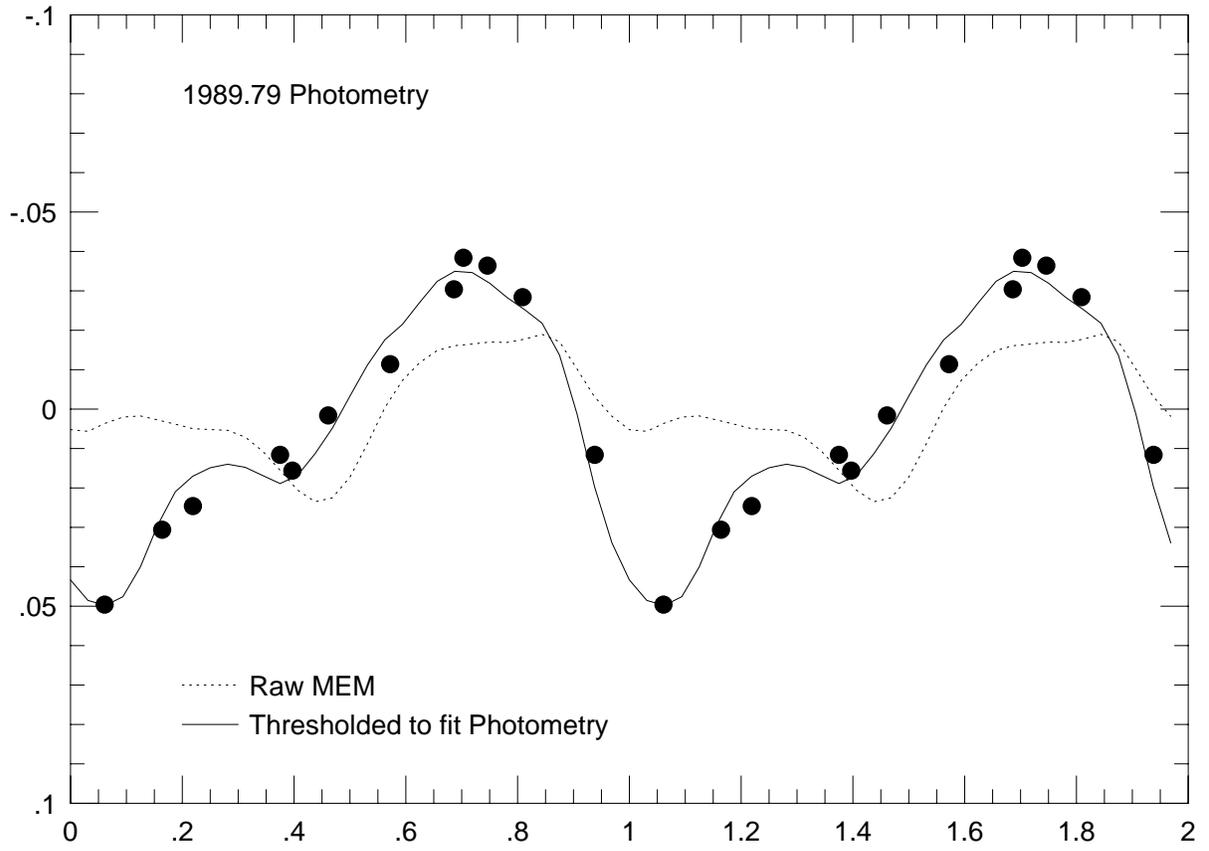}
\caption{HR 1099 light curve for 1989.79}
\label{fig:1989.79_light}
\end{figure}

\begin{figure}
\plotone{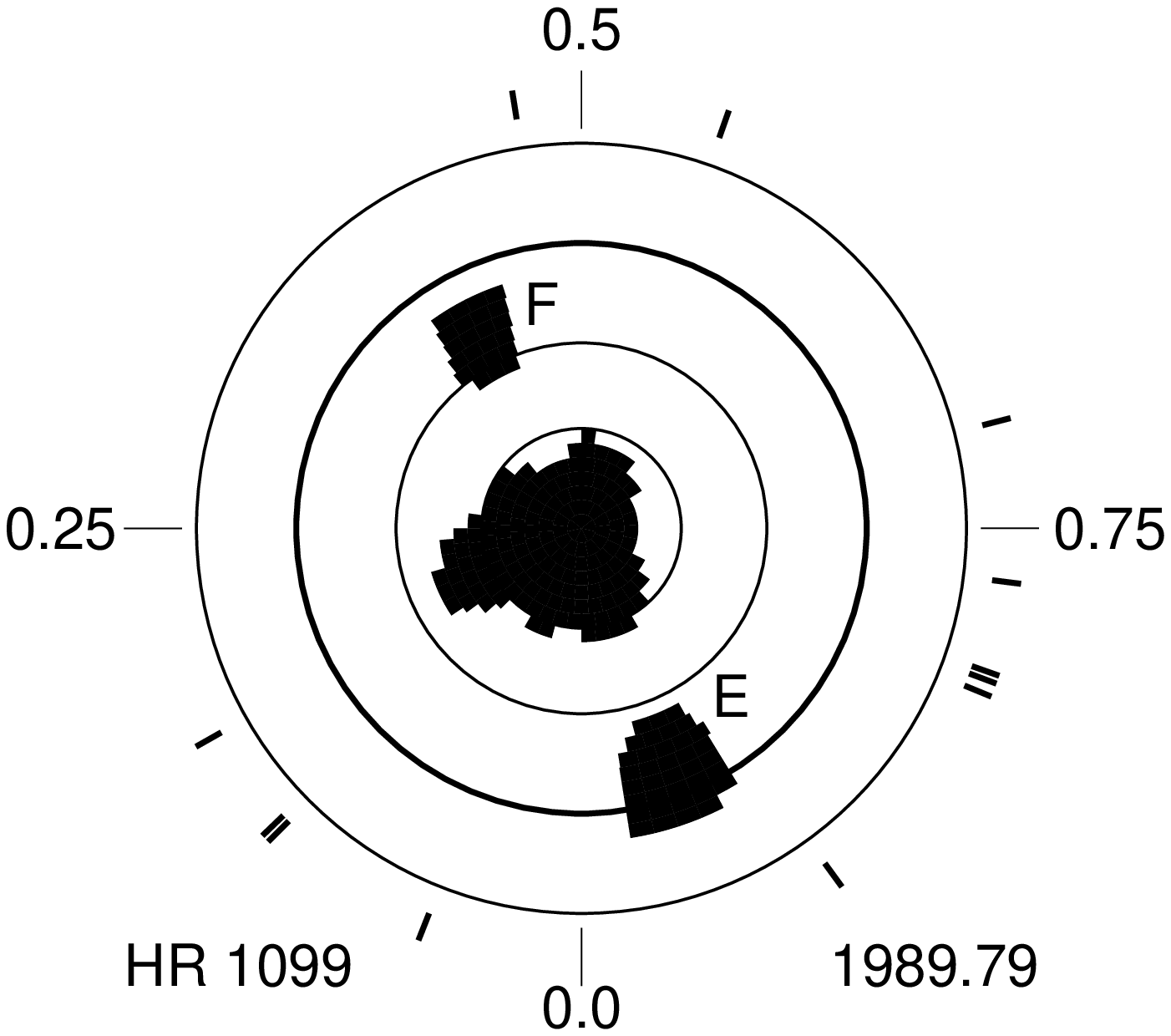}
\caption{HR 1099 thresholded Doppler image for 1989.79}
\label{fig:1989.79_image}
\end{figure}

The raw Doppler image for epoch 1989.79 is shown in
Figure~\ref{fig:1989.79_raw}; the spectral line profiles and fits are
shown in Figure~\ref{fig:1989.79fits}. The predicted light curve from
the raw image is shown as the dotted line in
Figure~\ref{fig:1989.79_light} along with the photometry (points) of
\cite{rod92}. The photometrically-constrained image is shown in
Figure~\ref{fig:1989.79_image} and the resulting light curve is shown
by the solid line in Figure~\ref{fig:1989.79_light}. A slightly better
fit was obtained by also removing the mid-latitude spot pair at phase
0.67 in the raw image.

\begin{figure}
\plotone{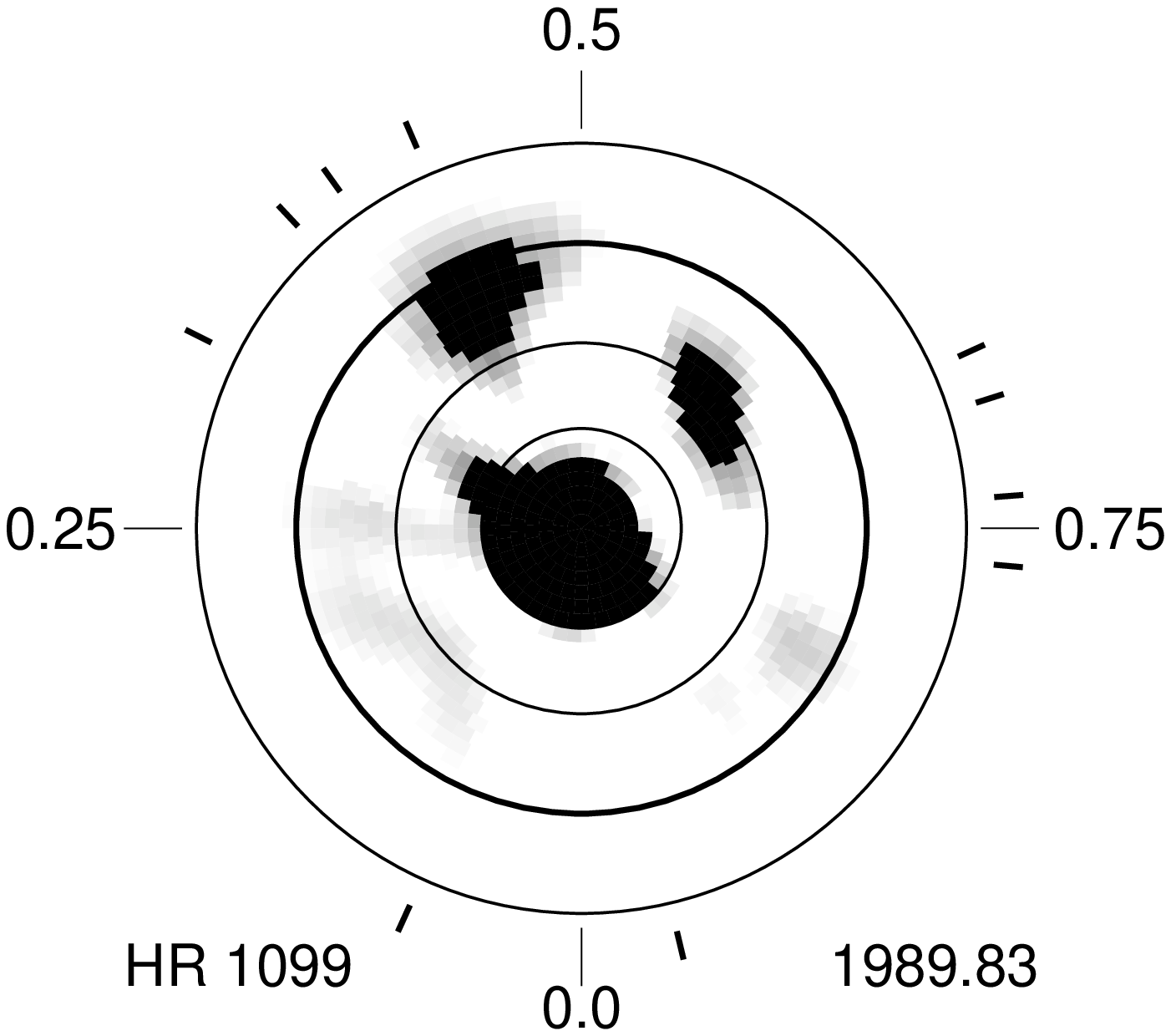}
\caption{HR 1099 raw (unthreshholded) Doppler image for 1989.83}
\label{fig:1989.83_raw}
\end{figure}

\begin{figure}
\plotone{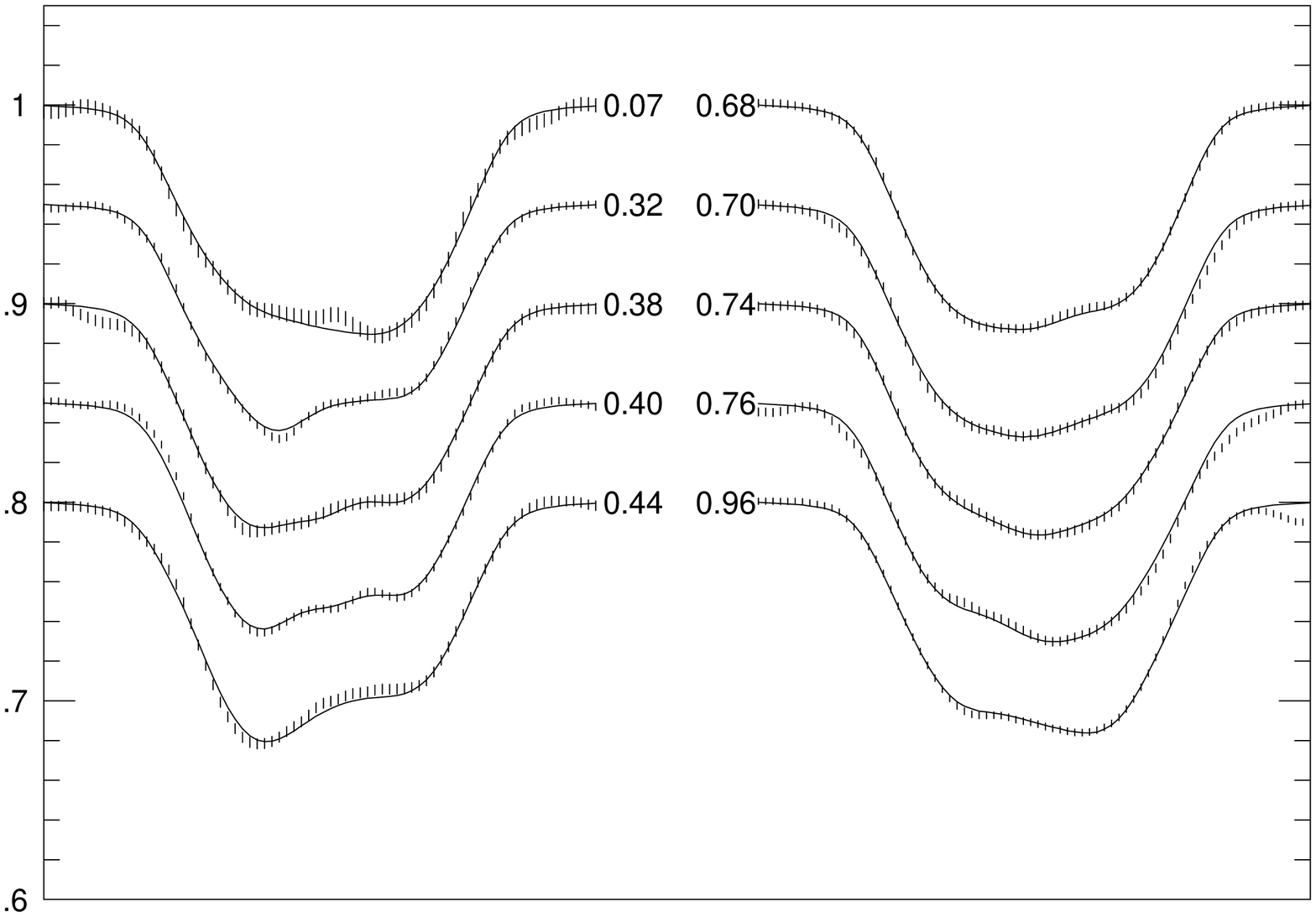}
\caption{Spectral line profiles and fits for 1989.83}
\label{fig:1989.83fits}
\end{figure}

\begin{figure}
\plotone{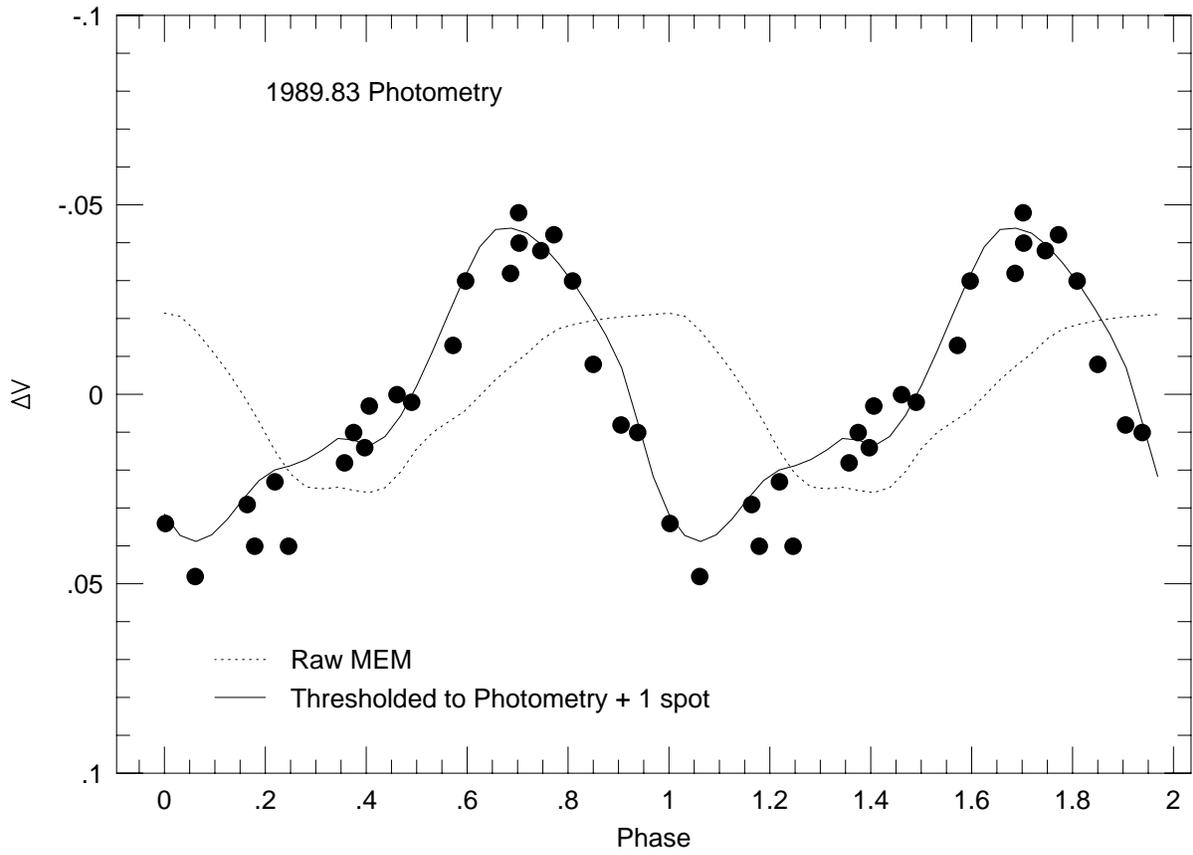}
\caption{HR 1099 light curve for 1989.83}
\label{fig:1989.83_light}
\end{figure}

Our raw image for epoch 1989.83 is shown in
Figure~\ref{fig:1989.83_raw}; the spectral line profiles and fits are
shown in Figure~\ref{fig:1989.83fits}. The predicted light curve is
shown as the dotted line in Figure~\ref{fig:1989.83_light} along with
the 1989.83 photometry (points) of \cite{rod92}. The strong minimum in
the light curve near phase 0.0 in 1989.83 strongly supports the
presence of a spot there, consistent with Feature E seen in previous
images, but this feature is clearly absent in the raw Doppler image of
1989.83 (Figure~\ref{fig:1989.83_raw}).

\begin{figure}
\plotone{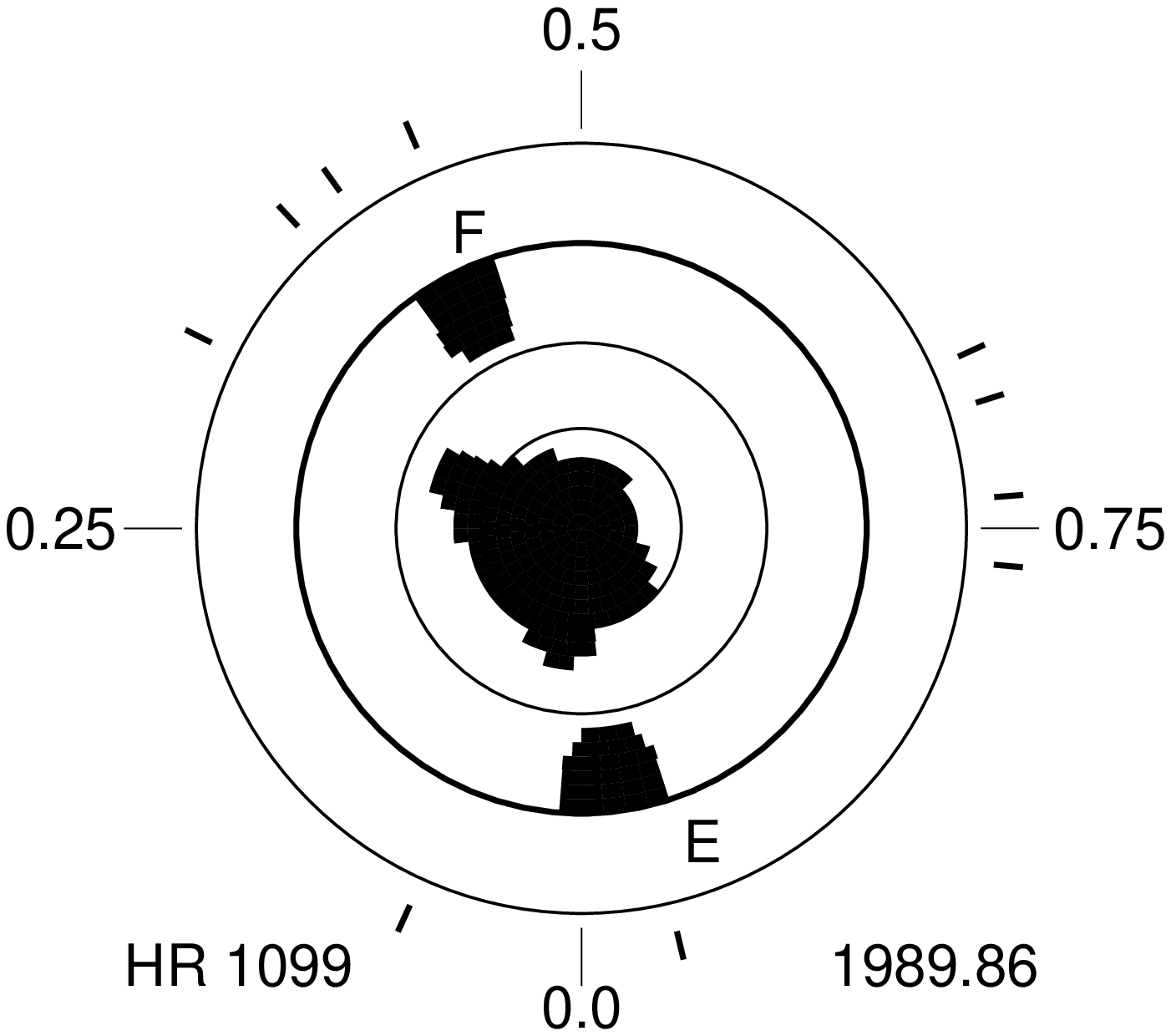}
\caption{HR 1099 thresholded Doppler image for 1989.83}
\label{fig:1989.83_image}
\end{figure} 

\begin{figure}
\plotone{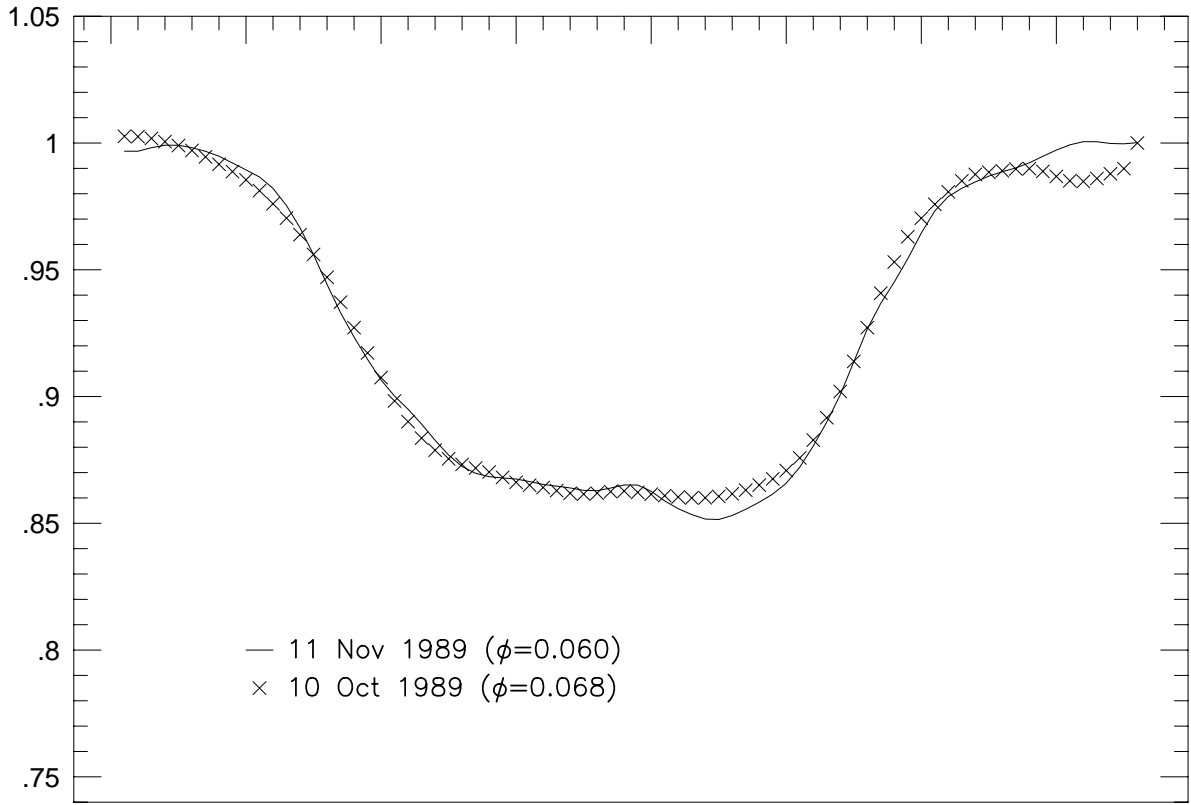}
\caption{HR 1099 Ca I 6439{\AA} line profiles taken one month apart 
near phase 0.064}
\label{fig:2phasesa}
\end{figure}

A clue as to this apparent discrepancy can be found by comparing two
line profiles taken near the same phase but about a month apart.
Figure~\ref{fig:2phasesa} shows two such profiles of the Ca I 6439
{\AA} line taken one month apart near phase 0.064. The two profiles
look significantly different. Note the excess absorption just to the
red of center for the profile taken in November (solid line) as
compared to the one taken in October (crosses). At this phase, the
low-latitude spot should have produced a `bump' just to the red of
line center and it is clearly not there. Possibly the spot
distribution has changed and the low-latitude spot no longer
exists. This, however, is not supported by the photometry. The other
alternative is that there is something else distorting the line
profile causing us to miss the low latitude spot. A likely possibility
is the presence of a hot spot, possibly even a flare visible in white
light. This would cause excess absorption instead of pseudo-emission
at that location in the line profile.

\begin{figure}
\plotone{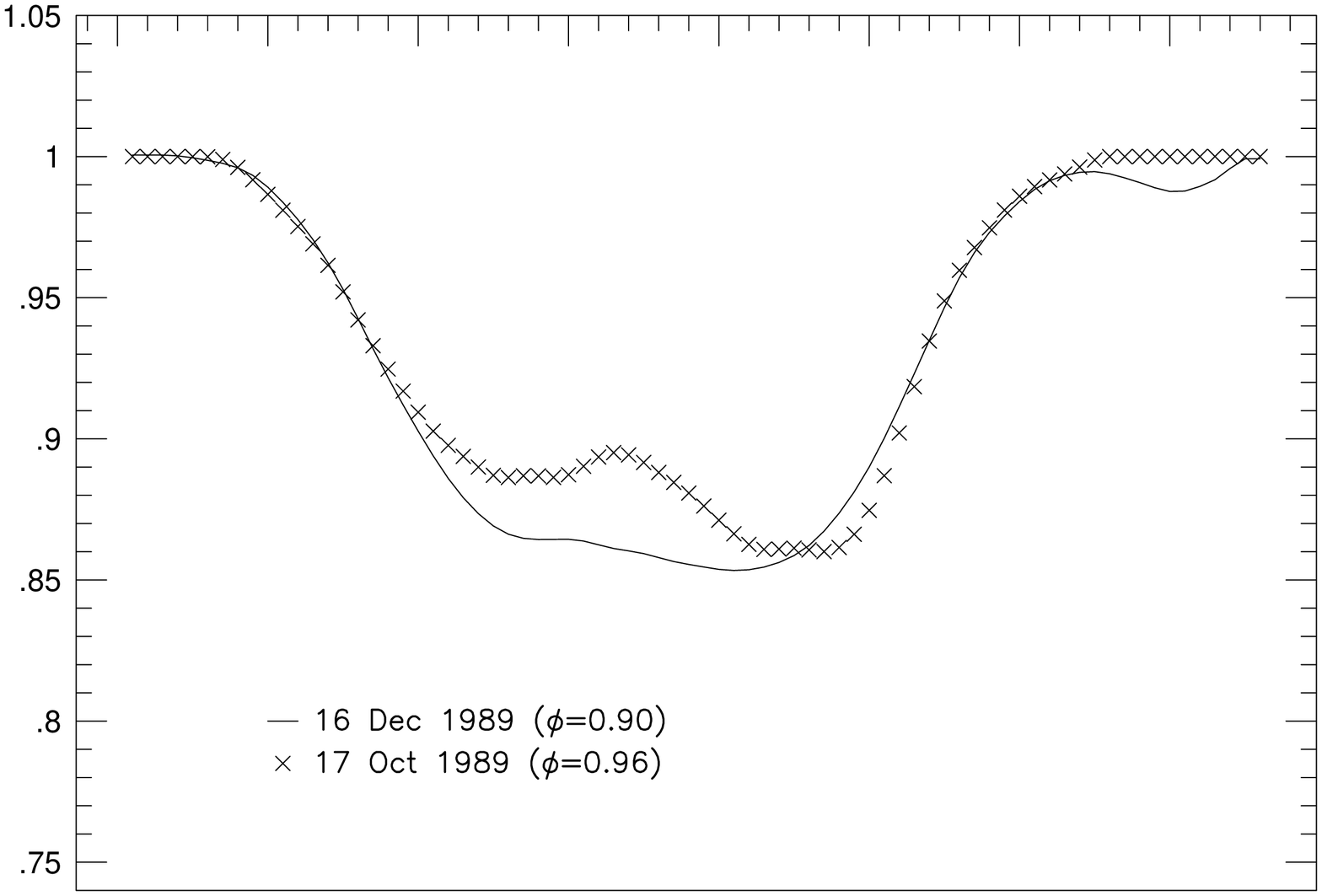}
\caption{HR 1099 Ca I 6439{\AA} line profiles taken two months apart 
near phase 0.93}
\label{fig:2phasesb}
\end{figure}

Figure~\ref{fig:2phasesb} shows two line profiles of (Ca I 6439 {\AA},
one (solid line) at $\phi$=0.90 taken on 16 Dec 1989 (this is the last
phase used for the 1989.83 image) and the other (crosses) at
$\phi$=0.96 taken on 17 Oct 1989. Here, the differences between the
two profiles are even more dramatic (the October profile has a lower
S/N ratio, but not enough to account for the differences in line
shape). Clearly either there are some rapid changes going on in the
spot distribution or a hot spot is greatly diminishing the effect of
the low-latitude spot on the shape of the spectral line.

\begin{figure}
\plotone{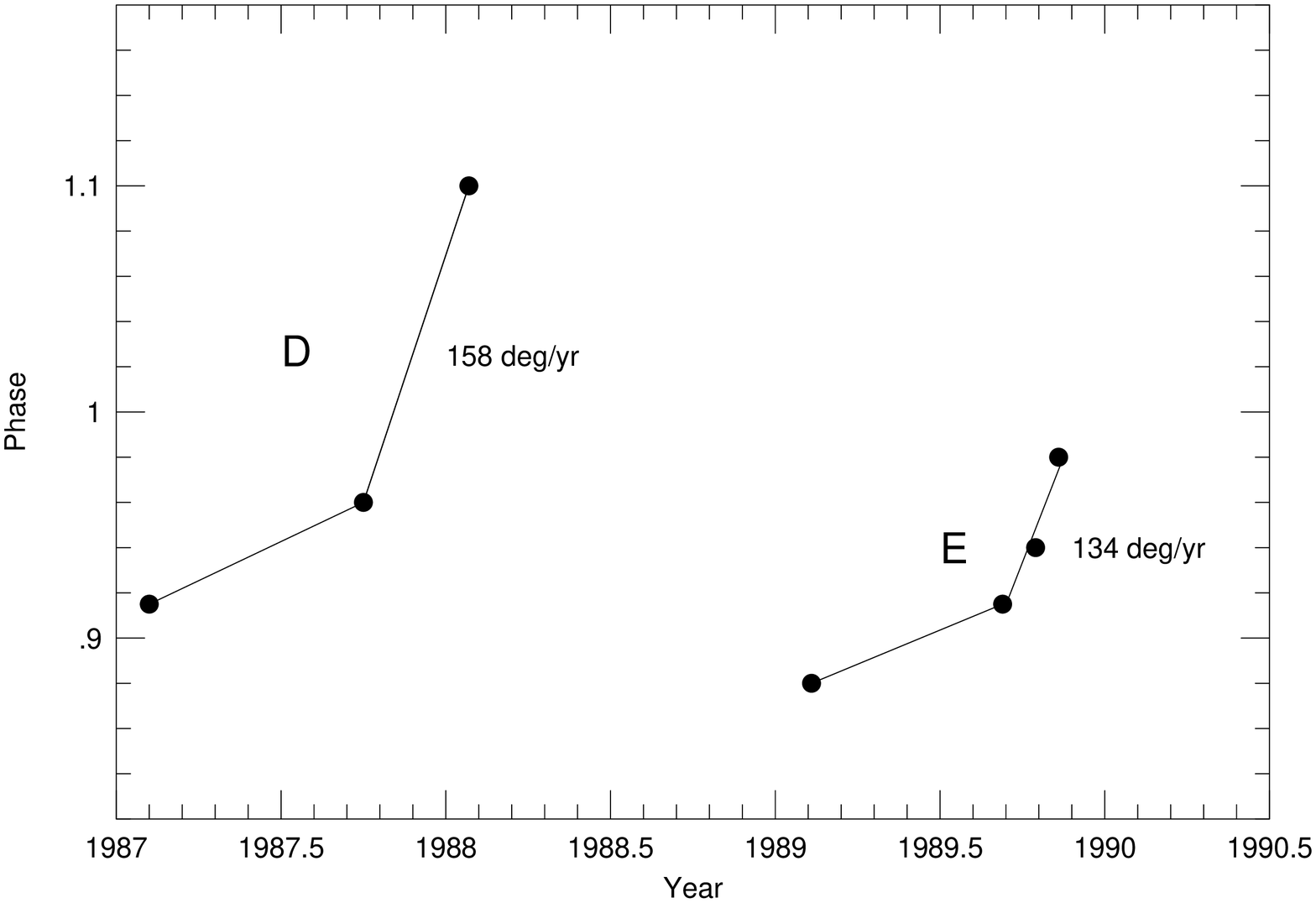}
\caption{Phases of Features D and E as a function of time.}
\label{fig:features_DE}
\end{figure}

Feature E continued to grow from 1989.11 to 1989.73 and there was a
pronounced drift in its longitude between the 1989.11, 1989.73,
1989.79, and 1989.86 images. It started at about phase 0.88 in the
1989.11 image, appeared at phase 0.915 in the 1989.73 image, at phase
0.940 in the 1989.79 image, and ended up at phase 0.980 in the 1989.86
image. Whether the present Feature E of this year is related to that
first seen in 1988.79 is not clear, but certainly this region of the
star has been an active spot-former. This region gave rise to two
major spots, Features D and E, which then migrated off to higher
phases at constant latitude. Figure~\ref{fig:features_DE} illustrates
this, showing the phases of Features D and E as a function of year.

This plot shows that the star formed two major spots near phase 0.9,
the first (Feature D) in early 1987.0, and the second (Feature E) in
early 1989.0. Both features showed a small and about equal drift rate
for the first 7 months before then moving off towards higher phases at
the same terminal rate of about 152{\deg} yr$^{-1}$. Of course, we
have no way of knowing if we are seeing the same spot from month to
month, but, at least for the 1989.7 to 1989.83 interval, where we have
high time resolution for the imaging, and good contemporaneous light
curves, the close agreement between Feature E's latitude and size, and
the smallness and uniformity of longitude advances makes this seem
very likely. Feature E did remain at a constant latitude of about
11{\deg}, and it also appeared to be steadily shrinking over this time
period, perhaps explaining why it was no longer visible by next year's
1990.69 image. So we can probably use the three closely-spaced images
from 1989.73 - 1989.86 to get an accurate estimate of the terminal
migration rate for a spot at 11\deg latitude. A formal least squares
fit gives 123 $\pm$ 17 degrees yr$^{-1}$ with respect to the orbital
frame at 11\deg latitude, and in the sense that the spot at this
latitude is rotating about the star's axis more slowly than the
orbit. While this may seem like an enormous migration rate, it is
still only about 1 part in 376 of the star's rotation period, and thus
any implied differential rotation is still quite small.

If Feature E is indeed migrating in longitude at a rate of 123{\deg}
yr$^{-1}$, then it might be expected to produce noticeable phase
smearing in our images. One simple consistency check is to verify that
this migration rate is not inconsistent with the longitudinal extent
of each spot over the time required to obtain each image. For the
1989.72 image, Feature E was 25{\deg} in longitudinal extent, and the
observations required 32 days to obtain. At a migration rate of
123{\deg} yr$^{-1}$, it would then have smeared no more than 11{\deg}
and is therefore consistent. The 1989.79 Feature E was 22{\deg} in
longitude extent, and required only 6.2 days to image, implying a
smearing of no more than 2{\deg} or less than a pixel width on the
image and again not inconsistent with the observed migration rate. The
1989.83 Feature E was also 22\deg in longitude extent, and required 33
days to image, implying a smearing of no more than 11{\deg}, and was
also consistent with the migration rate. While it is tempting to use
the slight elongation differences between spots, coupled with smearing
differences, to derive the true (unsmeared) size of Feature E, this is
probably over-interpreting the data as their sizes are also
threshhold-sensitive.

\cite{don92b} presented a Zeeman Doppler image of the radial component
of the surface magnetic field at epoch 1989.6 (their Fig. 10). They
observed a very strong longitudinal magnetic signature at phase 0.855.
This coincides very well in phase with our Feature E which would have
been at phase 0.87 at the time of the 1989.6 magnetic measurements of
\cite{don92b}. Their magnetic data was quite fragmentary, but their
subsequent analaysis indicated a large, bright, $>$1 kG monopolar
region at our phase 0.79 and latitude 5{\deg}. They cautioned against
the risk of interpreting their magnetic solution, and tried to make a
case for this magnetic region corresponding to their bright region or
plage at phase 0.78, but then wondered why their plage at phase 0.94
had no obvious counterpart in their magnetic image. Unfortunately,
their magnetic data and solution were not complete enough to
accurately constrain the longitude determination, so one can't say
with certainty whether their large monopolar region coincides exactly
with our Feature E, or is somewhat earlier in phase. Clearly though,
their large monopolar magnetic signature observed at phase 0.855 and
the presence of our recently-emerged Feature E near that phase and
latitude at that epoch (1989.6) argues that their strong longitudinal
magnetic signature was probably associated with our Feature E.

What then happened to Feature E? At a migration rate of 123 $\pm$ 17
degrees yr$^{-1}$, it would have migrated to phase 0.22 - 0.30 by the
next image at epoch 1990.69. It doesn't seem to be there. So Feature E
is an example of an isolated spot which emerged at low-latitude (about
11{\deg}), grew to maximum size in less than a year, migrated toward
increasing phase at fixed latitude, and then dissolved in less than a
year. At its peak size, it was probably associated with a strong
($\ge$ 1kG) longitudinal magnetic signature. Since Feature E grew to
full size in less than one year, and also completely disappeared in
less than a year, in images taken 1-year apart, such features as
Feature E could appear to come and go at random, appearing in one
image, but not in adjacent yearly images. It is precisely this type of
rapid spot evolution which make interpretation of many of our HR 1099
images with time gaps of a year or more so difficult.

The 1989-90 images also show a prominent spot at latitude 14\deg and
phase 0.45 which we hereafter refer to as Feature F. This is another
example of a spot which has emerged (or formed) at low-latitude. It
was probably not visible in 1989.11, but had fully emerged by
1989.72. It shows some longitude and latitude differences between the
1989.72, 1989.79, and 1989.83 images, but these are suspect for
several reasons. First, the time interval spanned by these three
images is only about 1.3 months, and one would expect the feature to
look very similar in size and shape over such a short time interval,
as was indeed the case for Feature E. Rather, Feature F was largest in
1989.72, shrunk significantly in 1989.79, and then grew slightly in
1989.83. Second, the phase coverage in that area of the star in
1989.79 was less than adequate, and can affect both the apparent
sizes and locations of features there. Third, Feature F does not move
at a uniform rate in longitude, as might be expected of a true
migratory motion. Rather, it jumps abruptly between the 1989.72 and
1989.79 images, but then remains fixed in longitude between the
1989.79 and 1989.83 images. Finally, there is another small feature
near latitude 39{\deg} and phase 0.38 which is complicating the image
geometry in this general area of the star. Whether this is a real
feature, or an artifact of our phase sampling coupled with ghosting
from the true image is unknown. But clearly, the overall spot
distribution in this area of the star is complex and variable, and
Feature E is not well-isolated from this evolving geometry. Deriving
accurate migration rates requires spots which are well-isolated, fixed
in shape and size, and moving smoothly at a constant rate in longitude
and/or latitude. Feature F fails on all accounts.

There is also a spot which appeared at latitude 28\deg and phase 0.15
in 1989.73. This spot, hereafter referred to as Feature G, was not
present 7 months earlier in the 1989.11 image. Tracing the evolution
of the spot geometry in this general area of the star from 1989.73 to
1989.86, it looks very much as if Feature G rapidly moved poleward and
merged with the polar spot by 1989.79, and then became the phase 0.31
protuberance on the polar spot by 1989.86. If so, it would be our
third case (see previous discussion of Features B, and C) of evidence
for poleward and clockwise spot migration, culminating in the merging
of the spot with the polar spot. However, this interpretation is
admittedly non-unique and verges on over-interpreting the image
set. For example, the small feature near latitude 39{\deg} and phase
0.38 in the 1989.73 image may instead have become the phase 0.31
protuberance on the polar spot in the 1989.86 image. Other scenarios
are also possible. There is also a protuberance at phase 0.98 on the
polar spot in the 1989.73 image which may have migrated to phase 0.01
in the 1989.86 image. Again, however, the outline of the polar spot is
changing too drastically to draw such a conclusion, or to derive
reliable migration rates.

\cite{zha90} reported observing a remarkable optical flare on HR 1099
on Dec.  14/15, 1989. This was followed up with photometry by
\cite{hen91} who observed an even larger flare 12 hours later and
concluded from the color of the flare that it covered about 8\% of the
surface of the K1 subgiant. The flare event increased the mean light
level of the star by 3\% in B, and required 3 months for the star to
return to its pre-flare mean light level.  The large flare reported by
\cite{hen91} occured at orbital phase 0.62. There is no obvious
feature at this phase in our 1989.86 image, which might likely
correspond to this flare, but, of course, such an energetic flare
could have happened almost anywhere on the observable disk of the
star.

\begin{figure}
\plotone{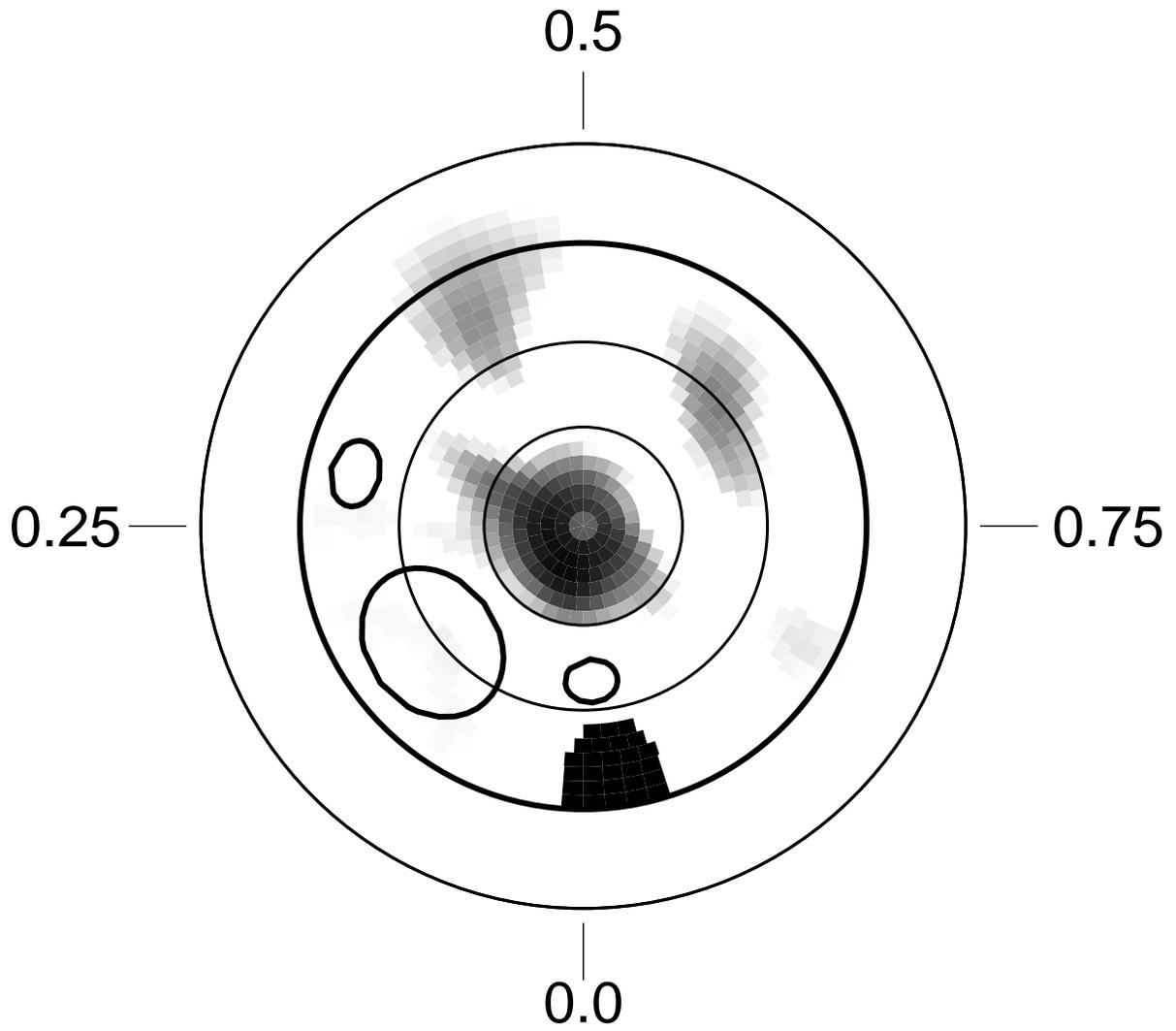}
\caption{Photometric spot model from Zhai et al. (1994) superimposed
on the MEM image.}
\label{fig:solutions4}
\end{figure}

HR 1099 was also the focus of an intense MUSICOS coordinated observing
campaign in late 1989 as reported by \cite{zha94} and by
\cite{foi94}. \cite{zha94} used a simple `3-spot' model to fit their
epoch 1989.96 multi-color light curve observations, and focused on
fitting an unusual phase shift of the light minimum among their
different photometric bands. They were able to model the principal
features of their 5-color light curves using two cool spots and one
hot spot, with two of their spots (one hot and one cool) adjacent to
one and other and creating a large temperature gradient in longitude
to account for the phase shift in different bands. A comparison of
their 3-spot solution with our nearest-in-time Doppler image (1989.86)
is instructive. Their 3-spot solution is shown superimposed on a
grayscale version of our 1989.83 raw Doppler image in
Figure~\ref{fig:solutions4}.

Their Spot 1 was a hot spot, 17\deg in diameter located at phase 0.99
and latitude 30{\deg}. It agrees quite well in phase and size with our
Feature E, and may suggest that Feature E was not simply a dark spot,
but rather had associated bright emission. This also makes sense with
regard to our discussion earlier in this section concerning
Figure~\ref{fig:2phasesa} where we suspected that emission was filling
in the line profile and making Feature E hard to detect from line
profile information alone. The longitudes of their Spot 1 and our
Feature E agree exactly. The latitude for their Spot 1 was slightly
north of our Feature E, but certainly within the latitude accuracy of
both methods.

Their Spot 2 was a cool spot 42\deg in diameter and situated at phase
0.15 and latitude 29{\deg}. There is no obvious counterpart to this
feature in our images. It may be that this feature is a result of
their method's attempt to parameterize the polar spot with its two
large protuberances at phases 0.01 and 0.31 as a single high-latitude
circular spot which splits this phase difference.

Their Spot 3 was a cool spot 16\deg in diameter and located at phase
0.29 and latitude 16{\deg}. Again, there is no obvious feature in our
images which corresponds precisely with this spot, though it is near
the phase 0.31 polar spot protuberance, and may be a reflection of
that feature. It might also correspond to our Feature F since it has
the same latitude though is fairly far away in longitude. Allowing
that distance for correspondence, it could as well correspond to
almost any feature on our image. As is usual for spot solutions from
light curve fitting, they did not detect the presence of the polar
spot. So except for the possible correspondence of their Spot 1 with
our Feature E, the overall agreement between the two solutions is not
particularly inspiring.

\cite{foi94} reported further on the extensive MUSICOS 1989 campaign
on HR 1099, including extensive Doppler imaging spectroscopic
observations. Unfortunately, they did not present any Doppler images
in that paper. Instead they simply used the spot solution of
\cite{zha94} discussed above to interpret their results, so any of
their conclusions based on the geometry from this `3-spot' solution
should be viewed with due caution.

\subsection{The 1990 Season Doppler Images}

We obtained 4 separate images in the 1990-91 observing season.
Unfortunately, we found only a single light curve for 1991.16,
presented by \cite{moh93}, with which to threshhold and further
constrain our imagery.

\begin{figure}
\plotone{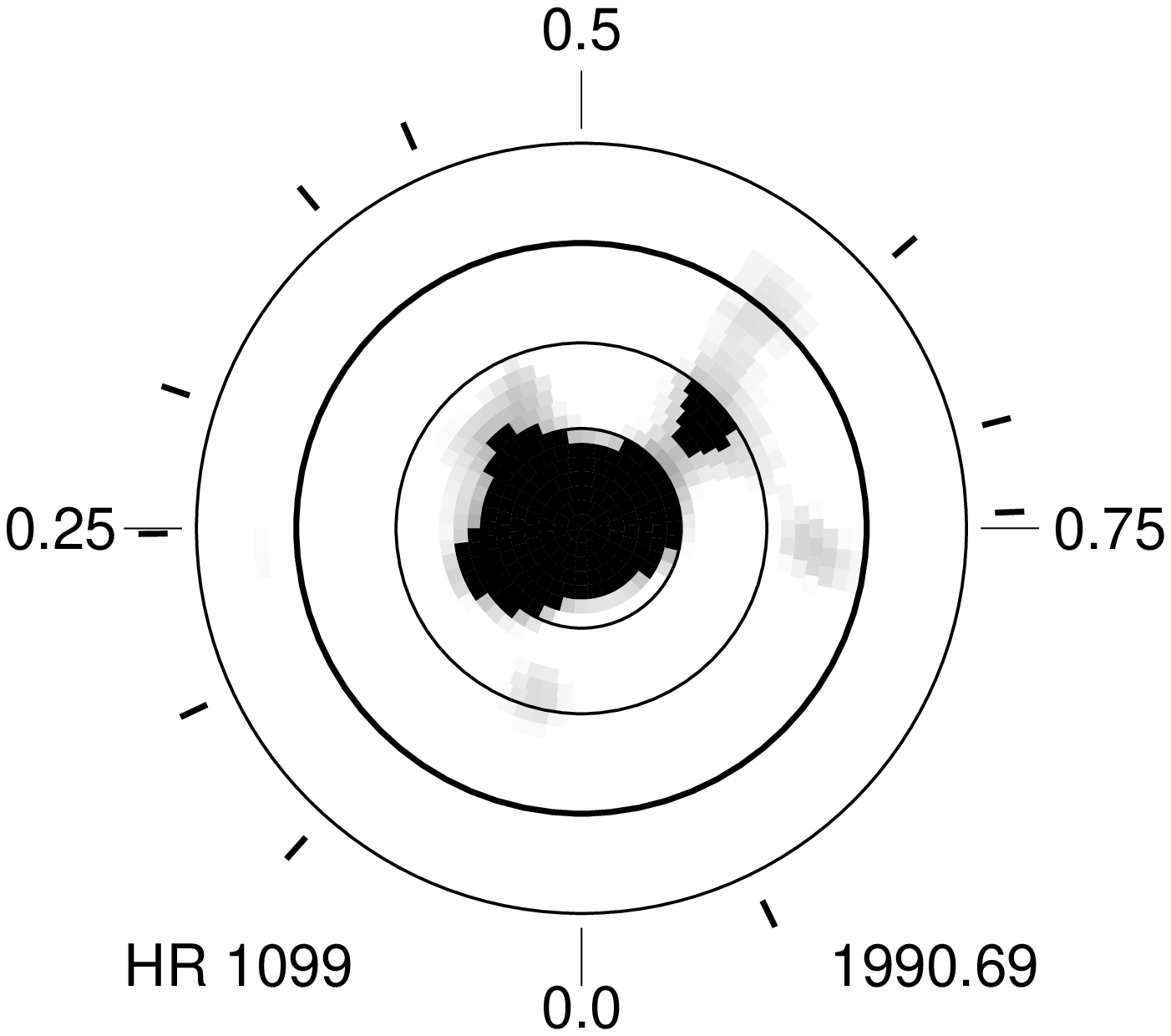}
\caption{HR 1099 raw (unthreshholded) Doppler image for 1990.69}
\label{fig:1990.69_raw}
\end{figure} 

\begin{figure}
\plotone{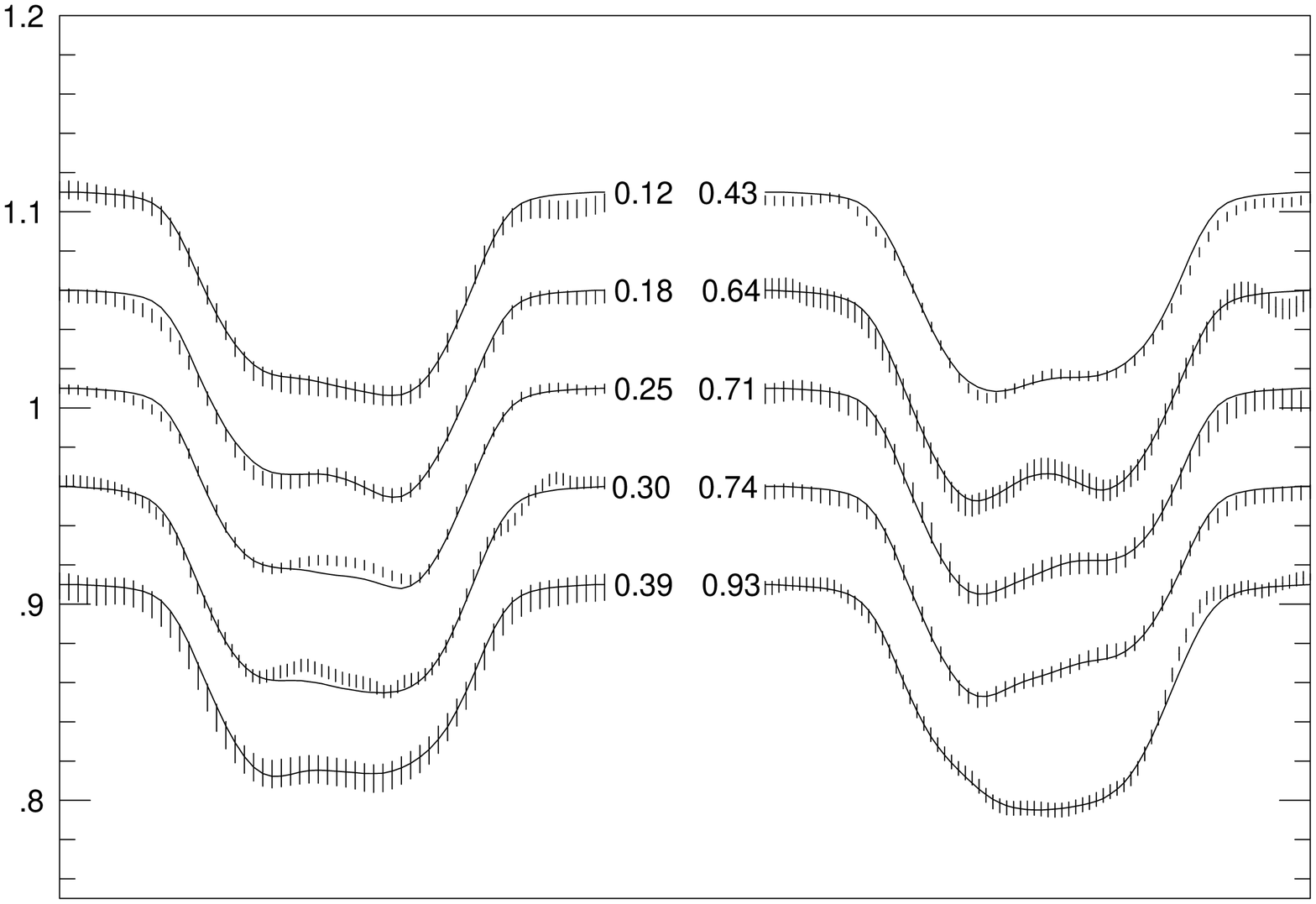}
\caption{Spectral line profiles and fits for 1990.69}
\label{fig:1990.69fits}
\end{figure} 

The Doppler image for epoch 1990.69 is shown in
Figure~\ref{fig:1990.69_raw} and the spectral line profiles and fits
in Figure~\ref{fig:1990.69fits}. Since we didn't have a light curve
for this epoch, we cannot present a photometrically-constrained
image. It shows a rather simple spot distribution this year, with a
fairly featureless polar spot and only a single detached lower
latitude feature near phase 0.64. As we have learned from previous
threshholded images, this feature is not necessarily detached from the
polar spot, and its exact size and latitude may be threshhold
dependent.

\begin{figure}
\plotone{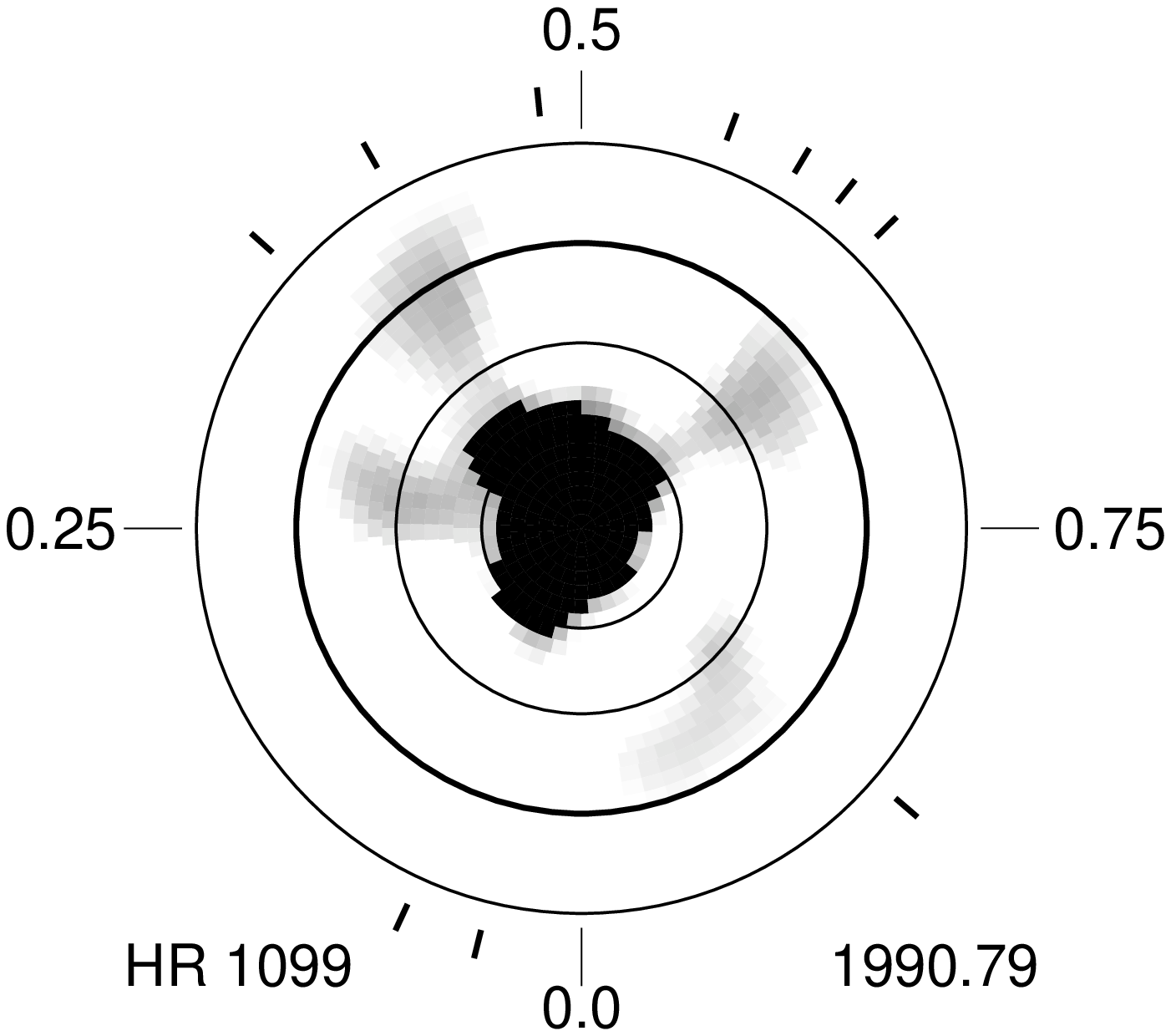}
\caption{HR 1099 raw (unthreshholded) Doppler image for 1990.79}
\label{fig:1990.79_raw}
\end{figure} 

\begin{figure}
\plotone{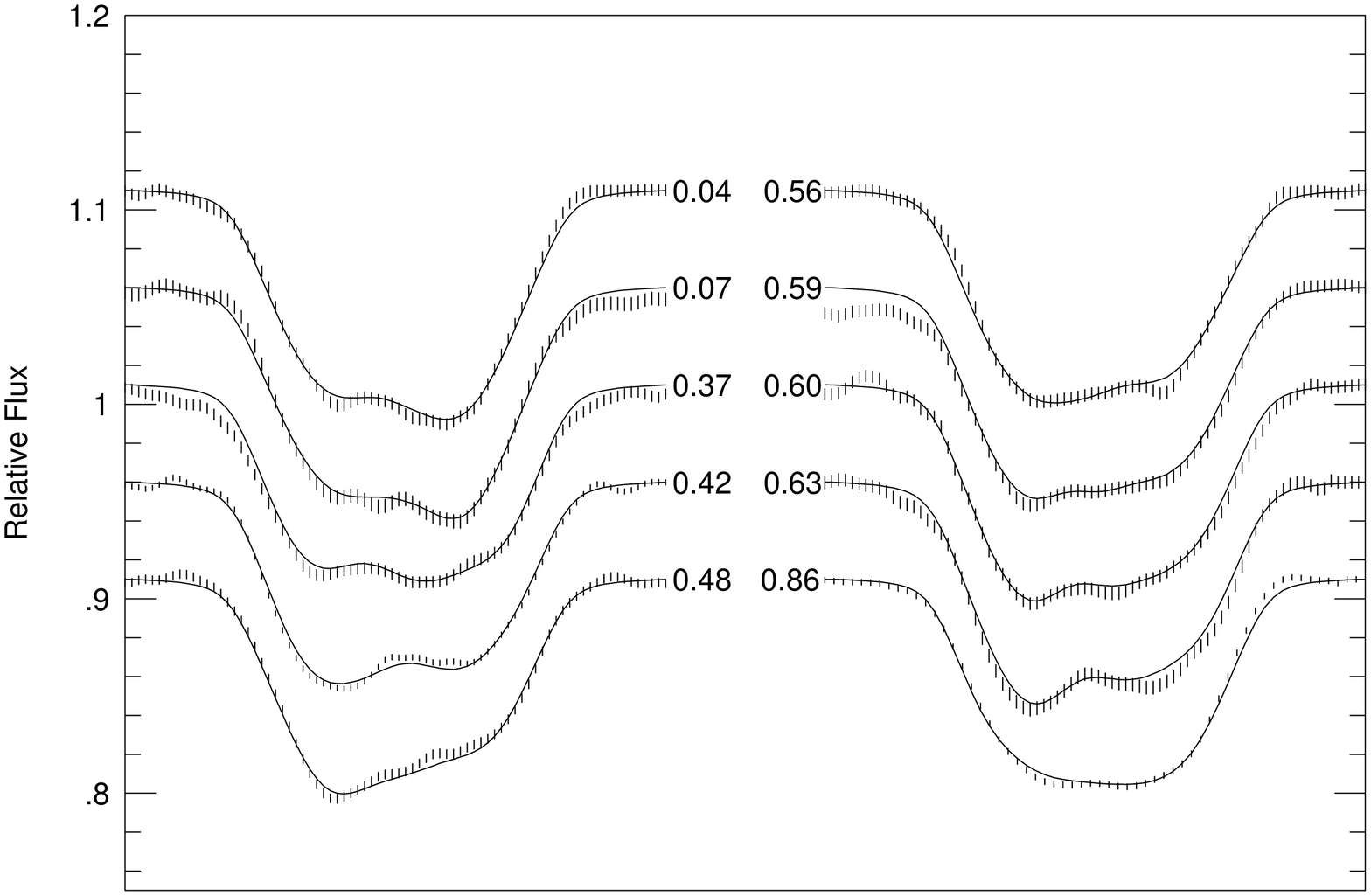}
\caption{Spectral line profiles and fits for 1990.79}
\label{fig:1990.79fits}
\end{figure} 

Likewise, the raw unthreshholded image for epoch 1990.79 is shown in
Figure~\ref{fig:1990.79_raw} and the spectral line profiles and fits
in Figure~\ref{fig:1990.79fits}. There is still some indication of a
feature near phase 0.64, but it appears at a rather low level in this
unthreshholded image. The polar spot outline has again changed to such
a degree that we cannot find any simple rotation which matches the
previous image. Also, the polar spot protuberance near phase 0.09 may
well be a `phase ghost' of the two observed phases at 0.04 and 0.07
since they are the only two observed phases on that side of the star
and are rather isolated. As we've seen, poor phase coverage leads to
such phase ghosting at isolated phases. Finally, there is also some
low-level indication of a feature near the equator near phase 0.4, the
location for Feature F from the 1989.83 image, but this cannot be
regarded as significant without further threshholding constraints.

\begin{figure}
\plotone{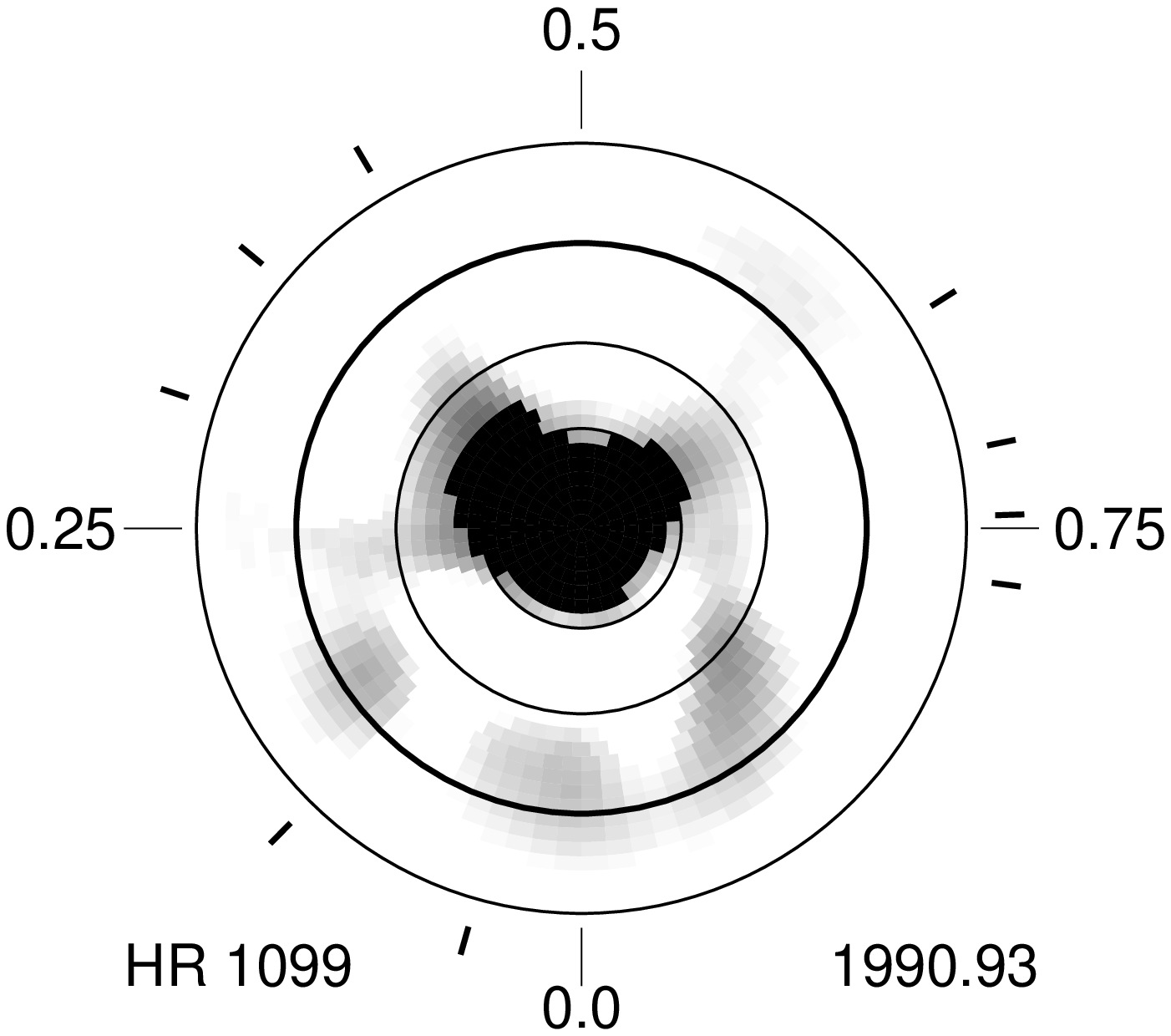}
\caption{HR 1099 raw (unthreshholded) Doppler image for 1990.93}
\label{fig:1990.93_raw}
\end{figure} 

\begin{figure}
\plotone{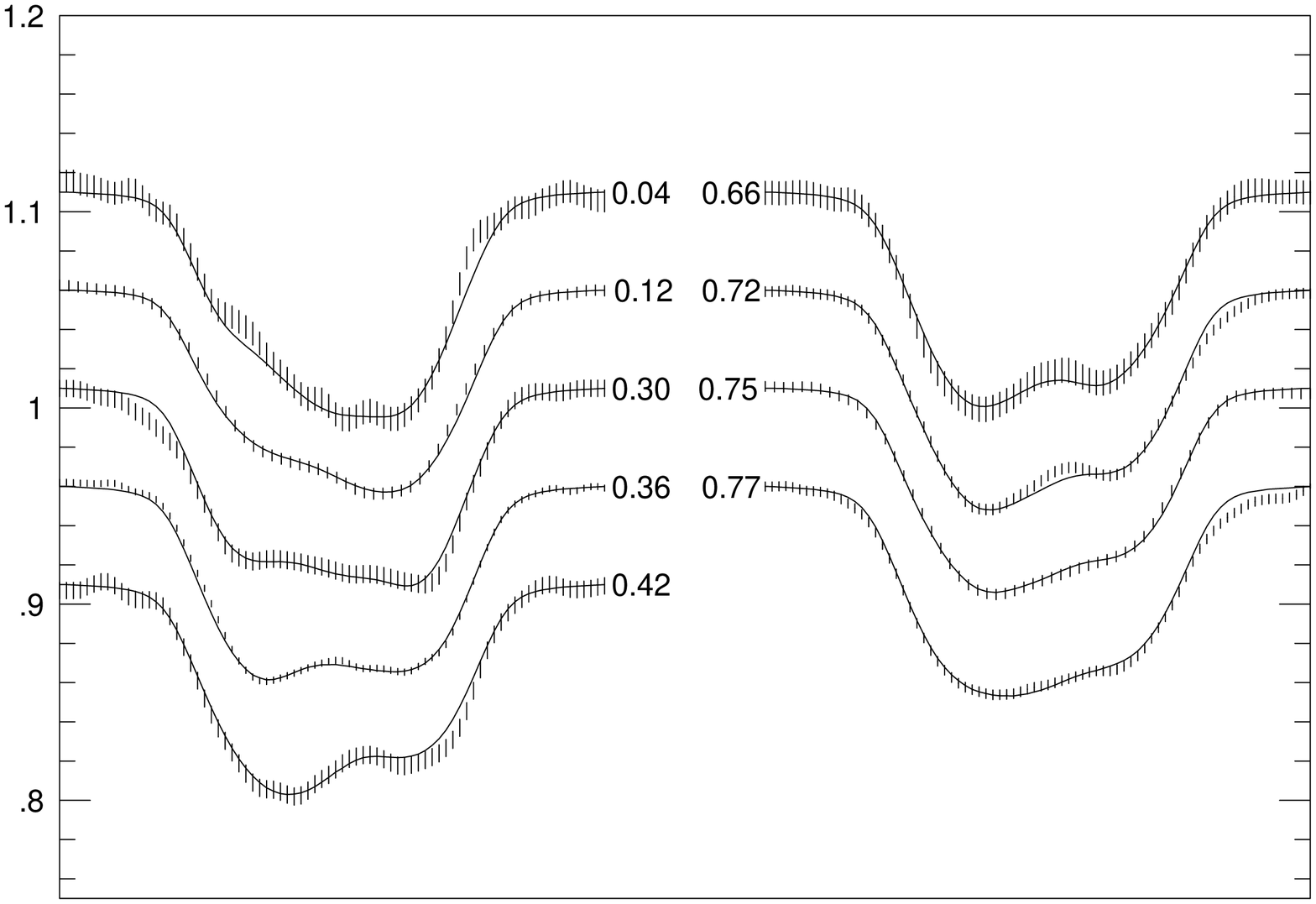}
\caption{Spectra line profiles and fits for 1990.93}
\label{fig:1990.93fits}
\end{figure} 

The raw unthreshholded image for epoch 1990.93 is shown in
Figure~\ref{fig:1990.93_raw} and the spectral line profiles and fits
in Figure~\ref{fig:1990.93fits}. Again, differences in the polar spot
outline preclude simple rotations to match the previous images and
thereby determine the polar spot rotation rate. There does, however,
seem to be a persistent suggestion of a stationary protuberance near
phase 0.3 - 0.4 in these successive images, and again there is a hint
of something near phase 0.64 at a low level. There is also now a hint
of significant low-level spot activity near the equator in the phase
0.9 to 0.2 region, almost as though a large annulus of dark spots,
some 45\deg in radius is emerging. But this is all at a quite
low-level and cannot be regarded as significant without further
constraints from light curves.

\begin{figure}
\plotone{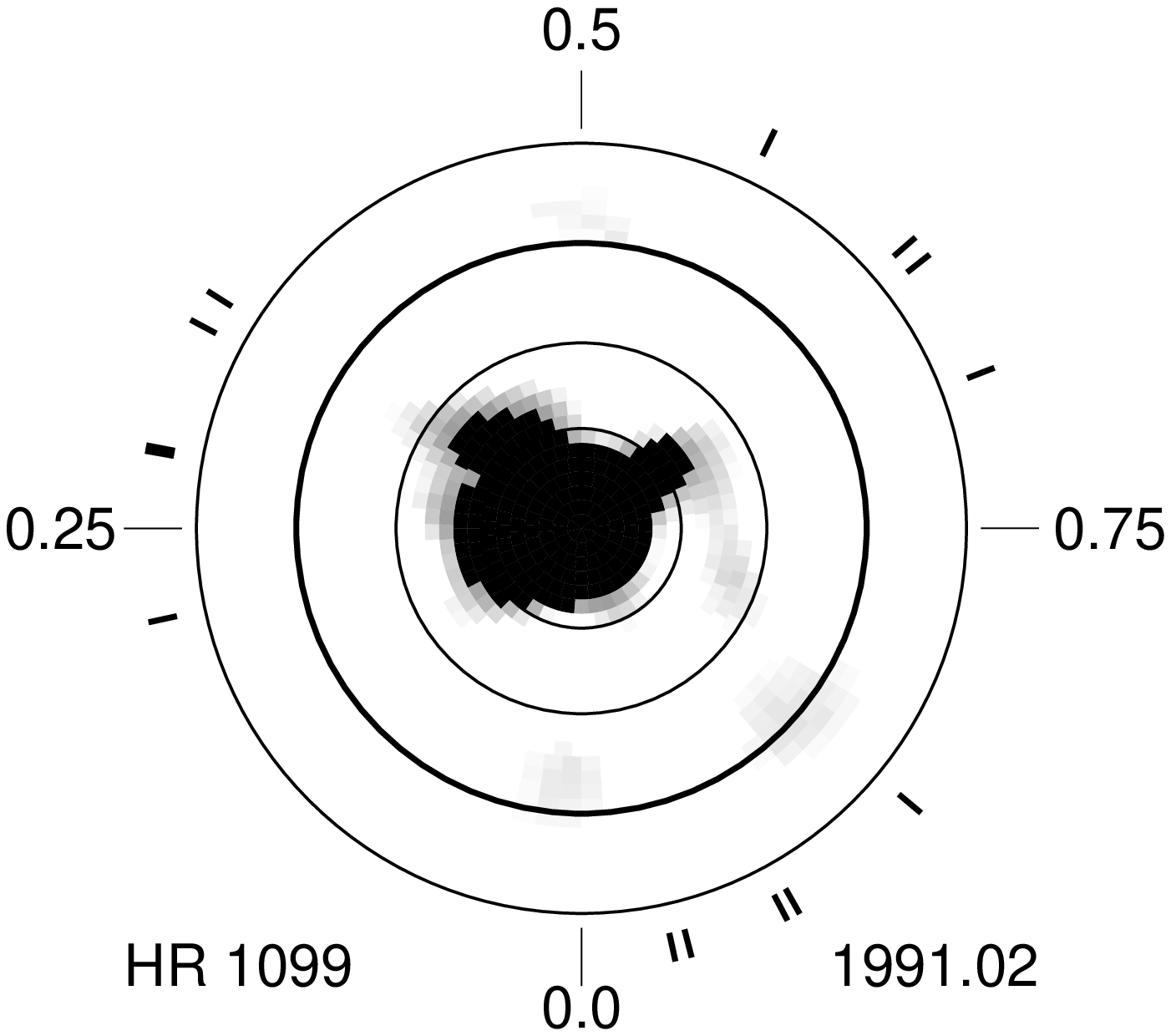}
\caption{HR 1099 raw (unthreshholded) Doppler image for 1991.02}
\label{fig:1991.02_raw}
\end{figure} 

\begin{figure}
\plotone{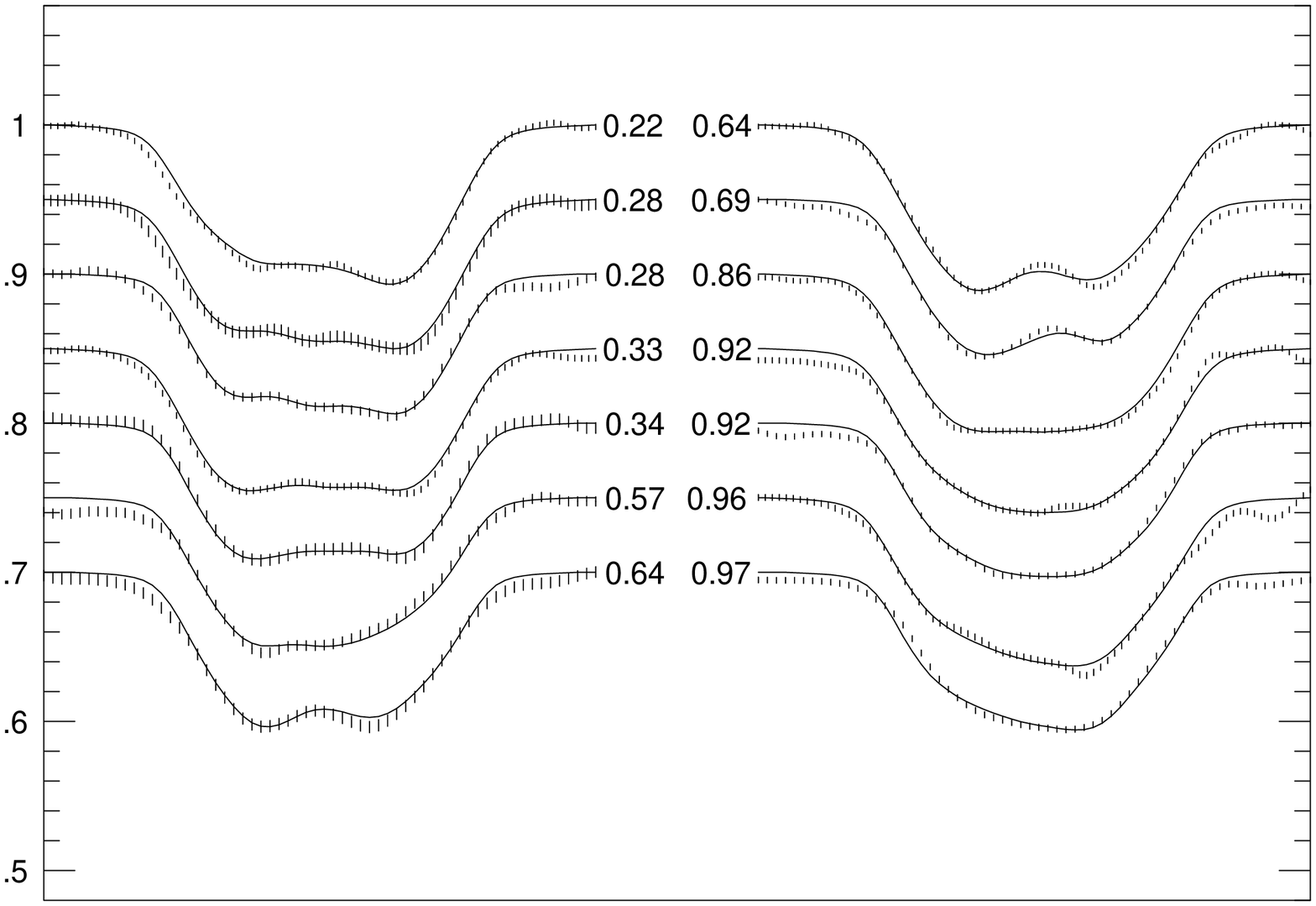}
\caption{Spectral line fits for 1991.02}
\label{fig:1991.02fits}
\end{figure} 

\begin{figure}
\plotone{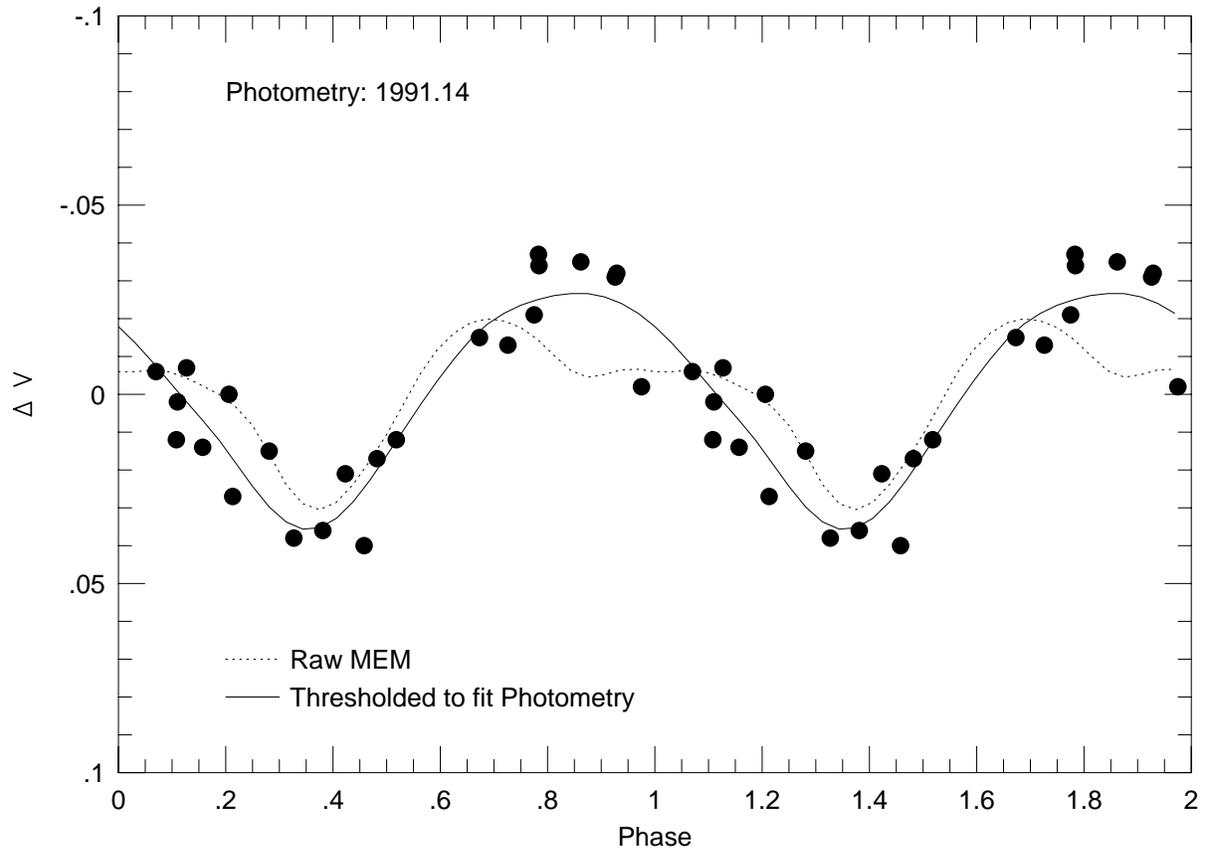}
\caption{1991.14 light curve used for the 1991.02 HR 1099 image}
\label{fig:1991.02_light}
\end{figure} 

\begin{figure}
\plotone{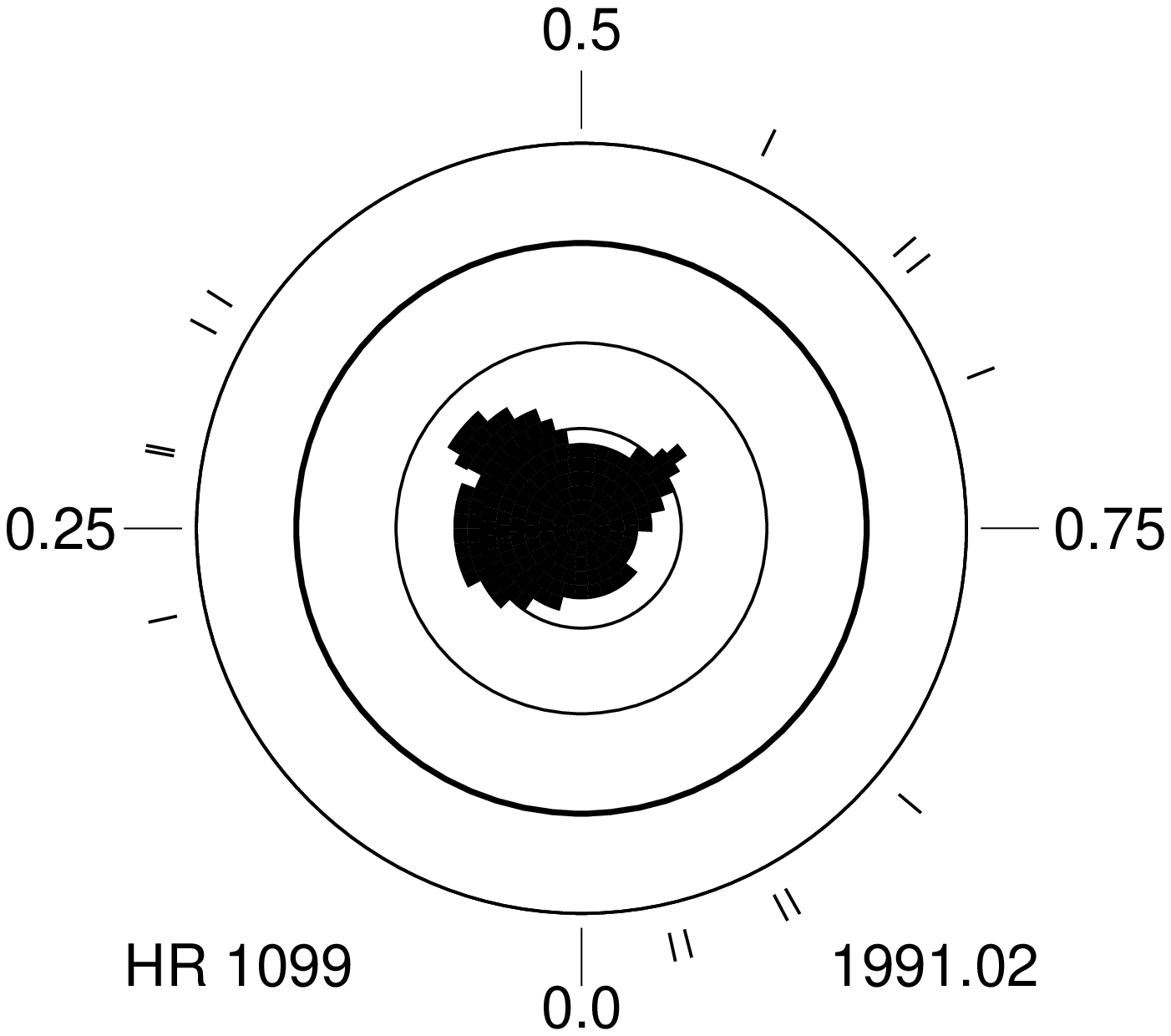}
\caption{HR 1099 thresholded Doppler image for 1991.02}
\label{fig:1991.02_image}
\end{figure} 

For the 1991.02 image, we did have a light curve with which to
threshhold and further constrain our image, though the threshholding
process did not affect the raw image greatly. The raw 1991.02 image is
shown in Figure~\ref{fig:1991.02_raw} and the spectral line profiles
and fits in Figure~\ref{fig:1991.02fits}. The predicted light curve
from this raw image is shown as the dotted line in
Figure~\ref{fig:1991.02_light} along with the 1991.14 photometry
(points) of \cite{moh93}. The predicted fit from the raw MEM image is
good except for a dip around phase 0.82. Applying the threshold cleans
this up nicely and an excellent fit is obtained. The photometric image
is shown in Figure~\ref{fig:1991.02_image} and its predicted
theoretical light curve is shown as the solid line in
Figure~\ref{fig:1991.02_light}. As can be seen, the addition of the
light curve data produced almost no change in the image, except to
sharpen up the protuberance at phase 0.64 somewhat, and provide strong
evidence that it is truly attached to the polar spot. Also, one can
get a good feeling here for the level of accuracy required in fitting
the light curve to derive accurate spot shapes.

Looking back at all four images for this observing season, one sees
that the polar spot was probably totally isolated all this season,
with no detached lower-latitude spots. Also, the phase of the 0.64
polar spot protuberance was quite well-fixed in longitude throughout
the 4-month interval spanned by these images. If this feature had been
migrating at the rate of the low-latitude Feauture E (123{\deg}
yr$^{-1}$), it would have moved some 40\deg or 0.11 in phase, and
would have moved from phase 0.64 in 1990.69 to phase 0.75 by
1991.02. This would have been readily apparent and clearly did not
occur. In fact, this feature, at a latitude of about 60\deg and
apparently attached to the polar spot, is consistent with little or no
relative longitude motion with respect to the orbital reference
frame. The obvious inference again is that spots near or at the pole,
or otherwise merged with the polar spot, are rotating essentially at
the orbital period, and that the polar region of HR 1099 is tightly
locked into synchronicity with the orbit. The broad protuberance at
phase 0.3-0.4 is also fairly well-reproduced in the 1991.02 image, as
though it has remained stationary in phase across these four
images. Again, both of these observations suggest that latitudes of
60\deg and above are quite tightly synchronized to the orbital frame,
whereas latitudes near the equator rotate more slowly.

This season, we have another independent, detailed Doppler image with
which to do a comparison check of our image. \cite{don92b} presented
both a `temperature' Doppler image and a corresponding Zeeman Doppler
image for epoch 1990.9. Both of their images were done using the Fe I
5497.520 {\AA} line. Our 1991.02 image is quite close in time and,
being constrained by a light curve, is also the best choice for
comparison with their imagery. The agreement between our Doppler image
and theirs is again excellent. We both see an isolated polar spot with
a narrow protuberance which descends to about latitude 50\deg near
phase 0.64 and has a characteristic triangular shape, coming down to a
sharp point at latitude 50{\deg}. We also both see a broad
protuberance on the polar spot near phase 0.31. Even the detailed
shape of this broad protuberance in our 1991.02 Doppler image is
well-reproduced in the \cite{don92b} image. They do however see a weak
(400 K cooler) spot near phase 0.3 at the equator, whereas this
feature does not show up strongly in any of our four images for this
season. There might just be a slight hint of it showing up as a
low-level extension from the polar spot near phase 0.34 which extends
down to about 15\deg in our raw 1991.02 image, and there was also hint
of low-latitude spots near phase 0.35 in our raw 1990.79 image, but
they do mention that their line profile at phase 0.34 is substantially
noisier and less well-fitted than their other profiles. They also
noted minor departures of their fit to the line profile at phase 0.293
where the core of the predicted profile was slightly too low, and cite
some difficulties with line blends in the blue wing of this line. So
our disagreement on this feature is probably not significant, and all
indications are that it is not a strong feature.

\cite{don92b} also presented a Zeeman Doppler image of the toroidal
component (i.e. along lines of constant latitude) of the surface
magnetic field. The reader is referred to their discussion for a more
detailed summary of their B field results. Basically, they found that
the polar spot is encircled by a ring of toroidal clockwise-directed
field of $-$300 G strength which lies at about 60{\deg} latitude. This
ring is not complete, but has a gap which extends about 60{\deg} in
longitude and is centered near phase 0.65. The ring coincides quite
precisely with the edge of the polar spot in both their image and
ours, and the gap in the ring sits right at the location of the polar
spot protuberance we both see at phase 0.64. They also found a second
area of +700 G toroidal field component situated at phase 0.27 and
latitude 20{\deg}, at the position of their low-latitude feature, for
which we detected no counterpart as described above.

\cite{don92b} explained the coincidence of the gap in the ring with
the polar spot protuberance at phase 0.65 as due to the fact that this
dark area of the ring, while probably magnetic, was not detected in
the circular polarization measurements because of its low
brightness. They further explained that the polar field distribution
was a ring (rather than a polar cap) because of the same effect: the
polar spot regions are too dark to contribute enough light for field
detection in the circular polarization measurements. Our Doppler
images agree very closely with their image and strongly support this
interpretation of the correspondence between the magnetic and
temperature images of the polar spot at this epoch. The polar spot is
thus seen as a permanent region of strong surface magnetic field, with
field strengths at least as strong - and probably much stronger than -
the 300 G fields they detected around the spot periphery. 

\subsection{The 1991 Season Doppler Images}

\begin{figure}
\plotone{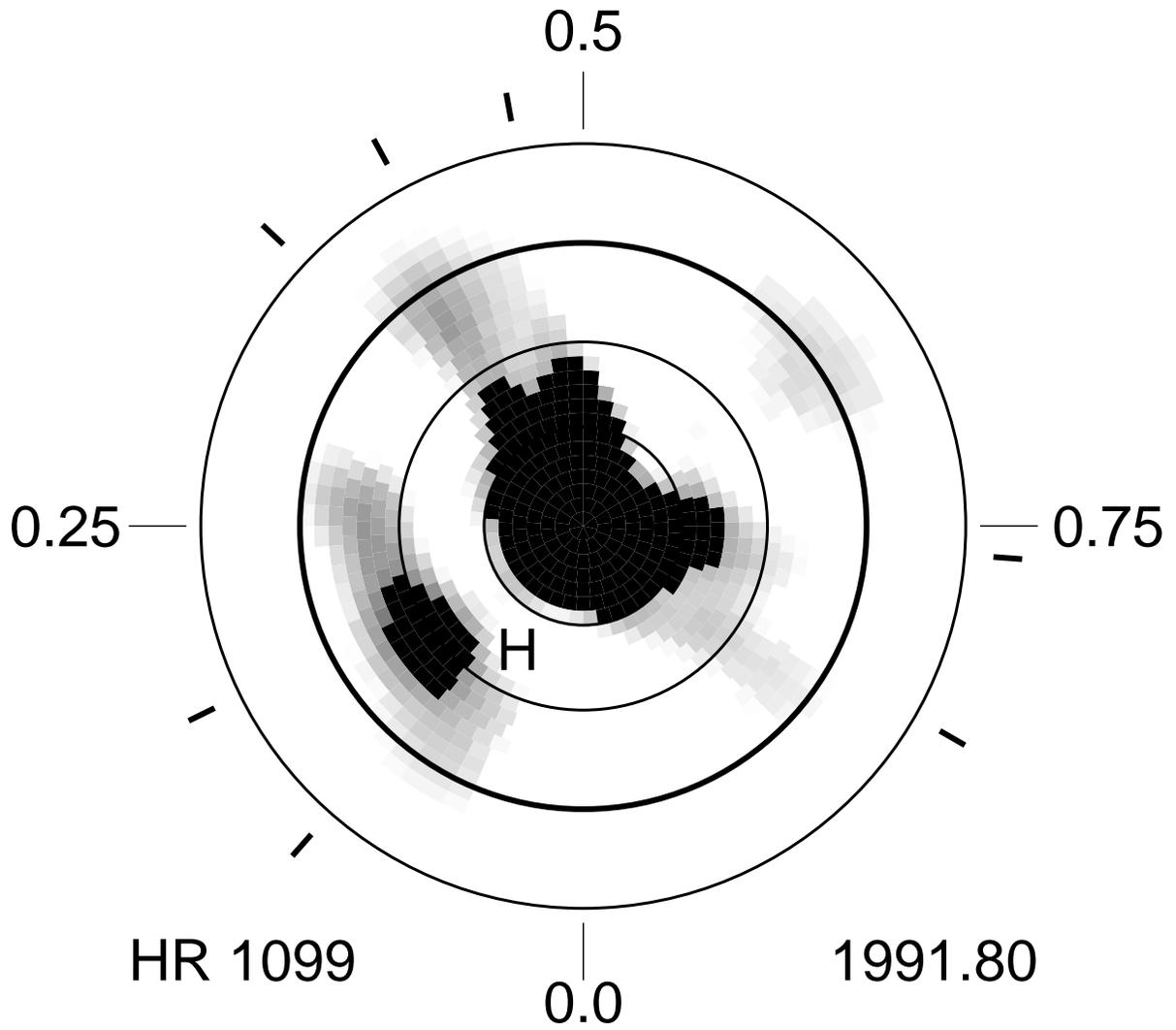}
\caption{HR 1099 raw (unthresholded) Doppler image for 1991.80}
\label{fig:1991.80_raw}
\end{figure}

\begin{figure}
\plotone{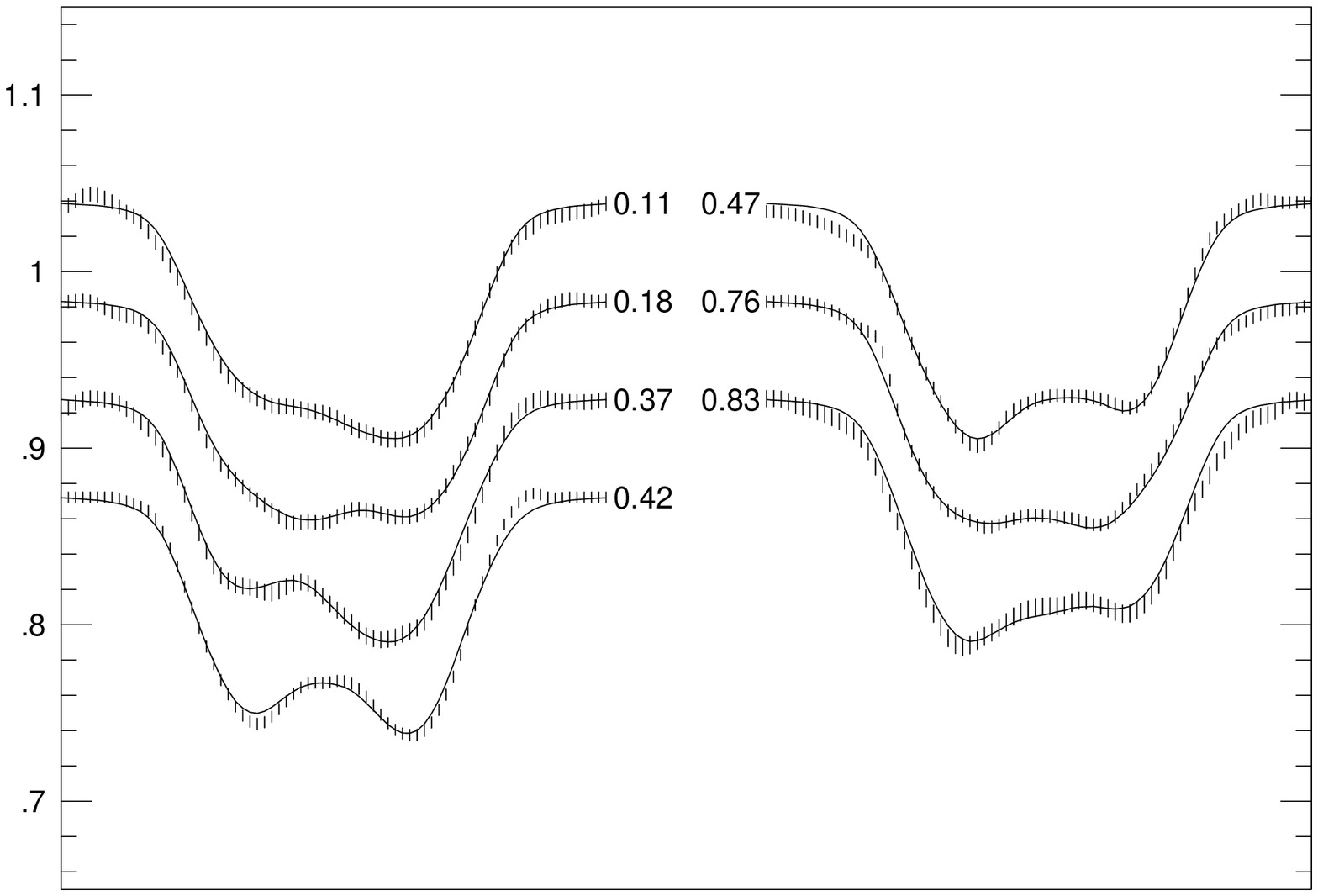}
\caption{Spectral line fits for 1991.80}
\label{fig:1991.80fits}
\end{figure}

\begin{figure}
\plotone{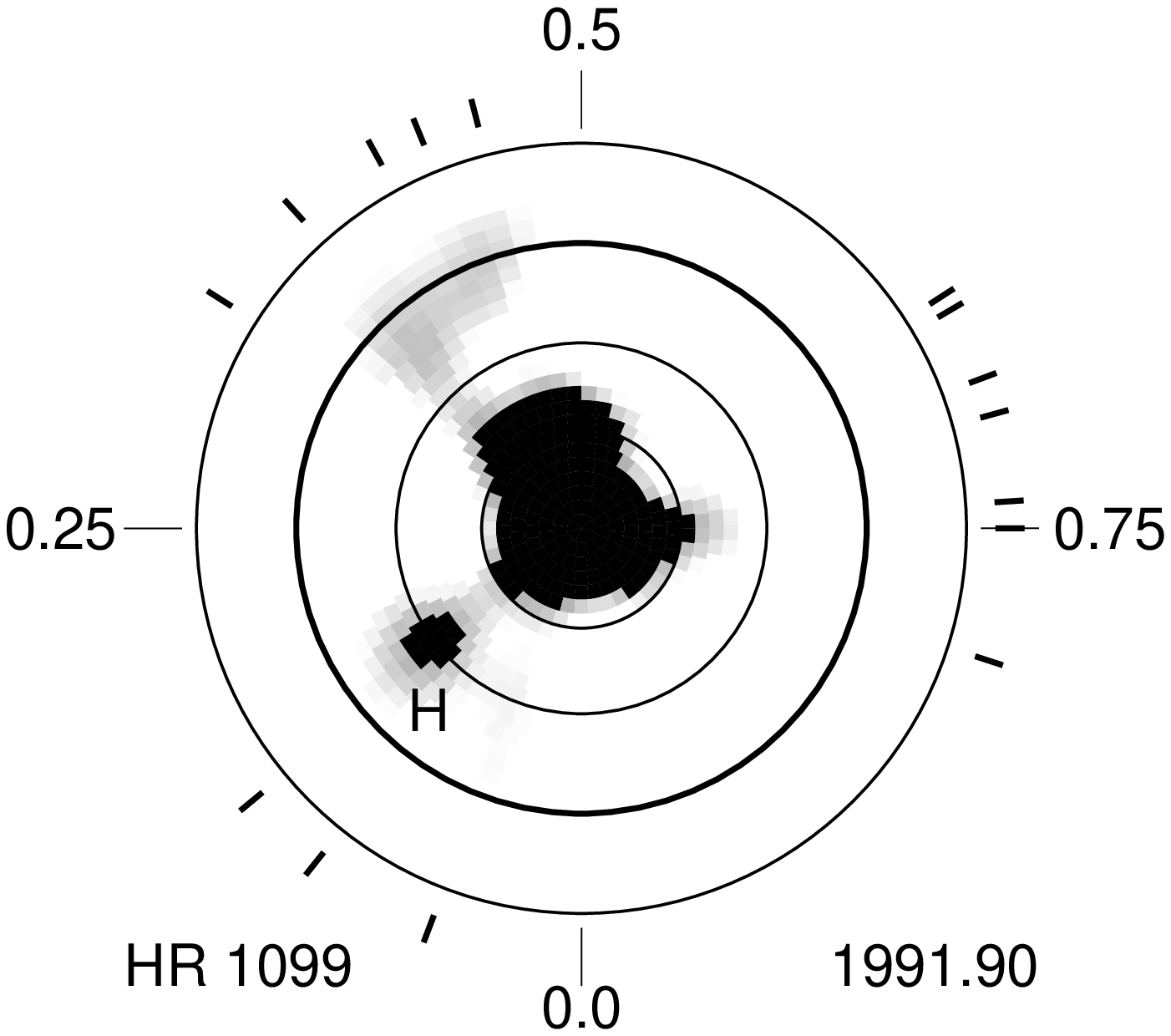}
\caption{HR 1099 raw (unthresholded) Doppler image for 1991.90}
\label{fig:1991.90_raw}
\end{figure}

\begin{figure}
\plotone{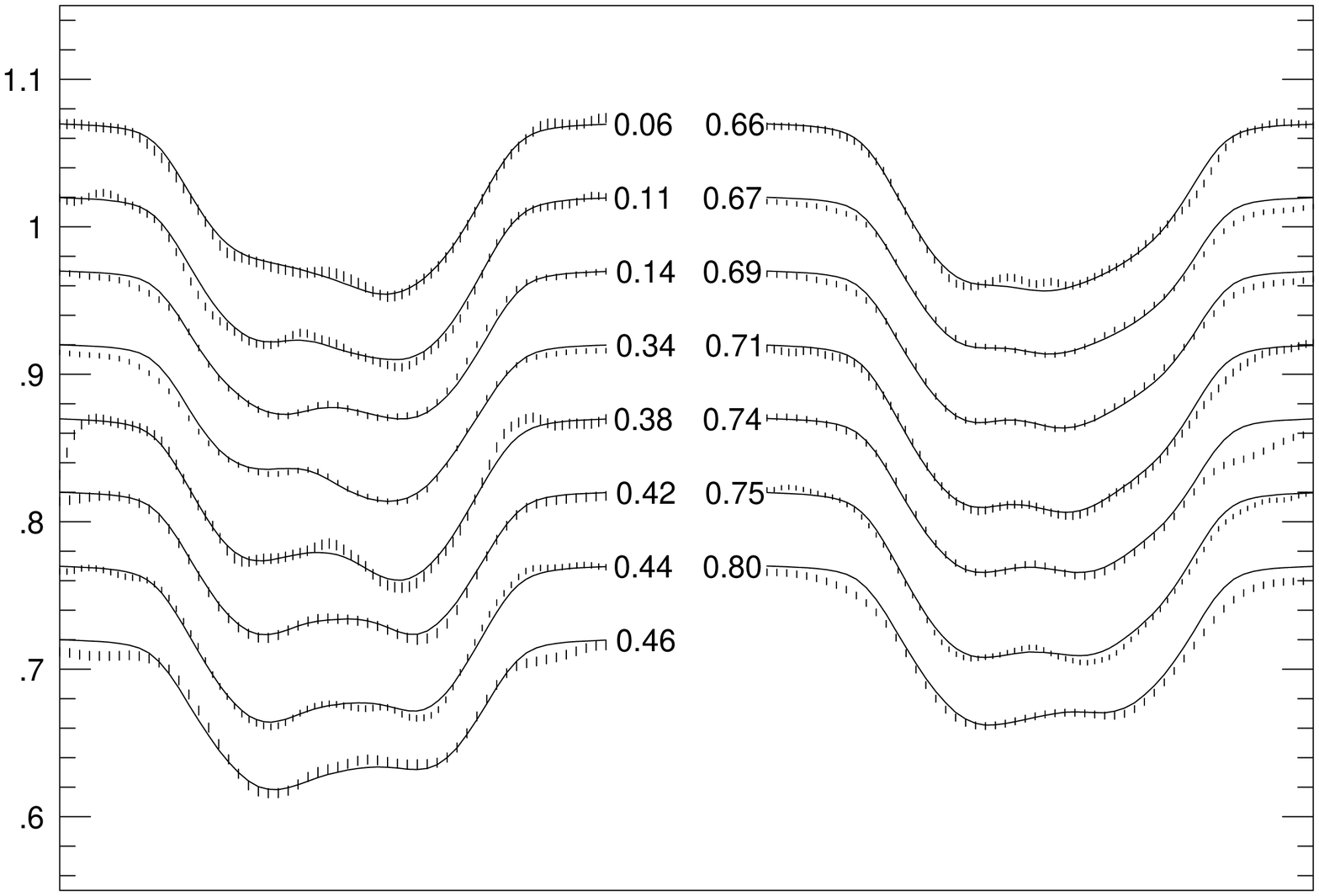}
\caption{Spectral line profiles and fits for 1991.90}
\label{fig:1991.90fits}
\end{figure}

\begin{figure}
\plotone{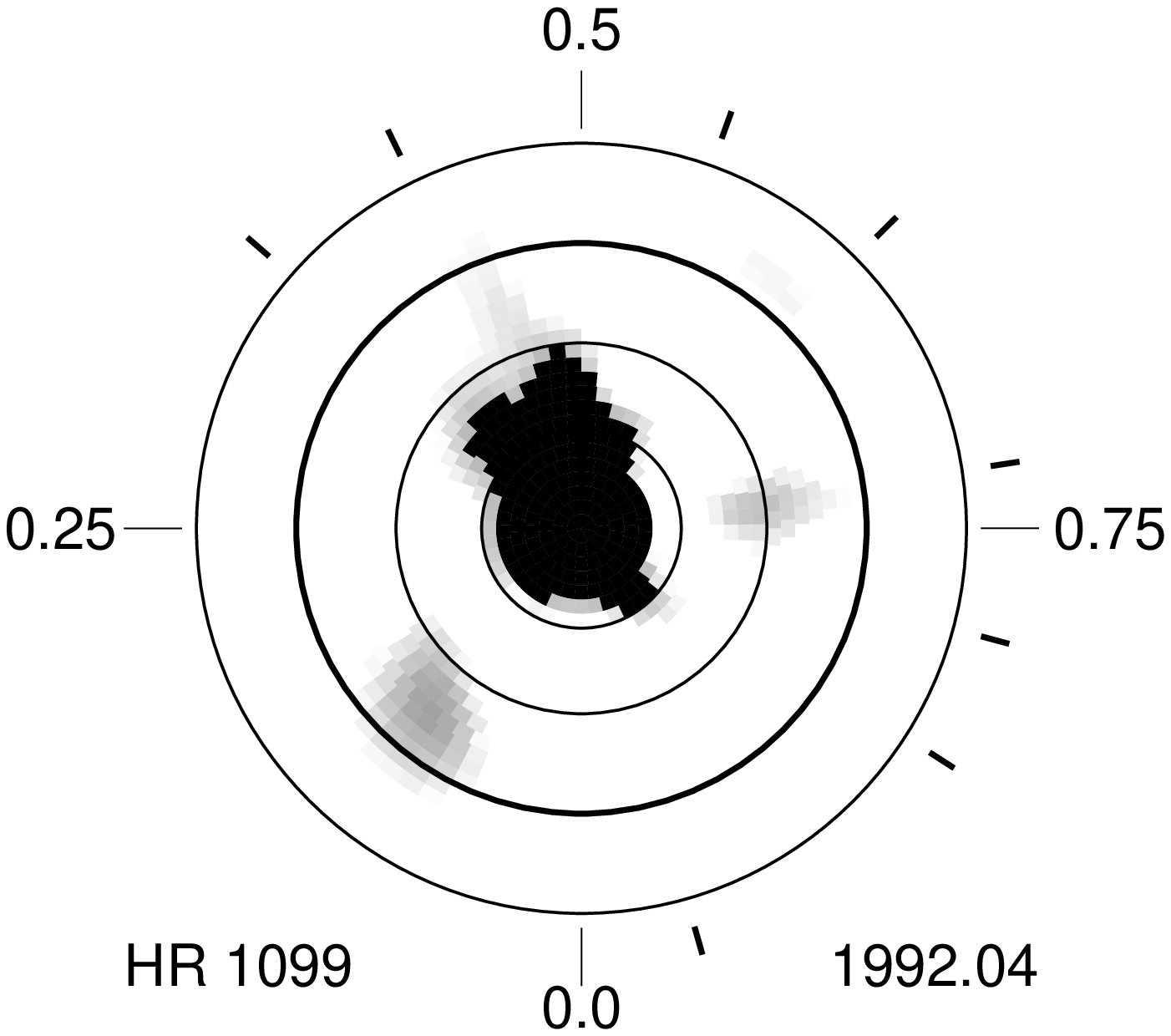}
\caption{HR 1099 raw (unthresholded) Doppler image for 1992.04}
\label{fig:1992.04_raw}
\end{figure}

\begin{figure}
\plotone{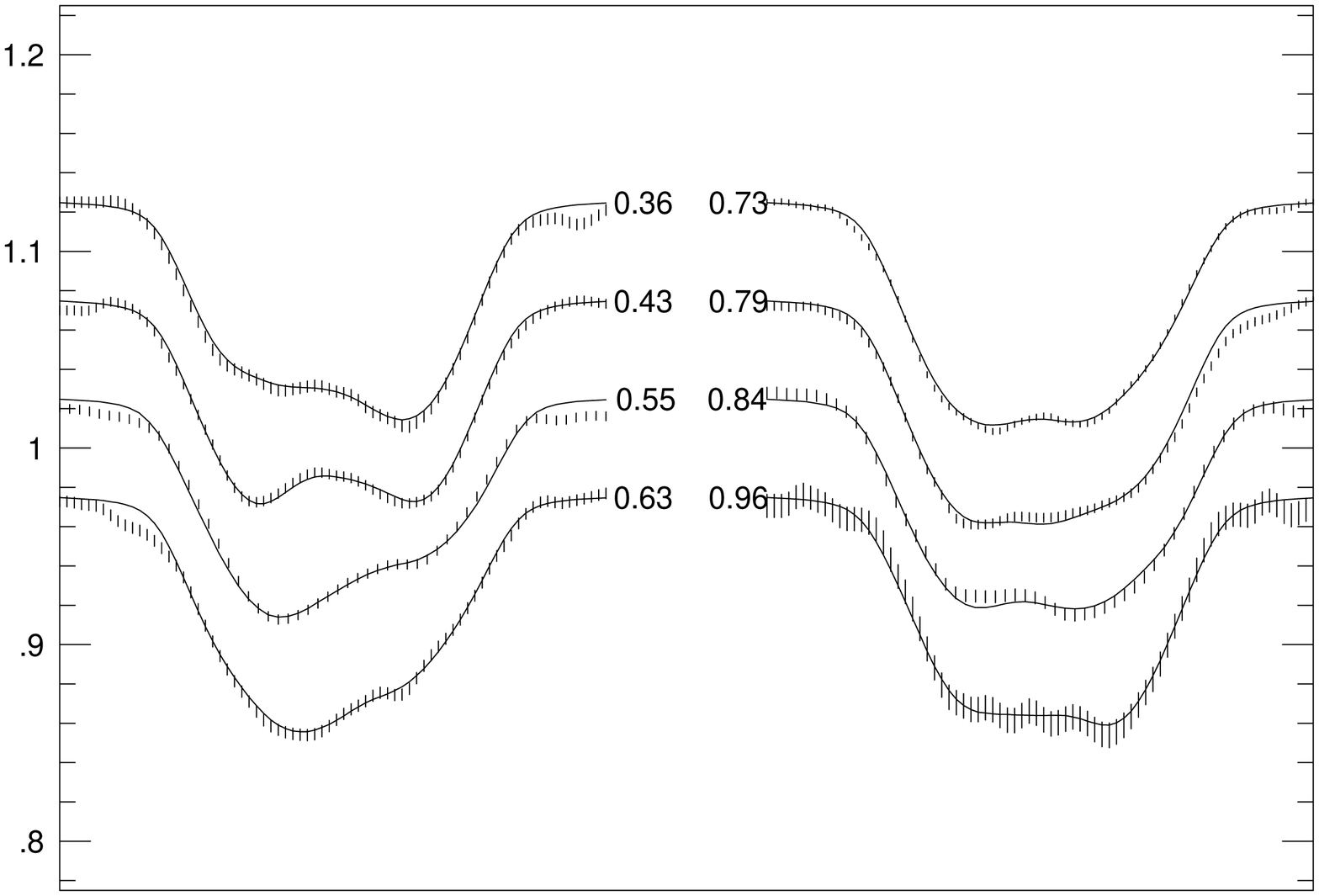}
\caption{Spectral line profiles and fits for 1992.04}
\label{fig:1992.04fits}
\end{figure}

We obtained three images in the 1991-2 observing season. The raw image
for 1991.80 is shown in Figure~\ref{fig:1991.80_raw}, and its spectral
line profiles and fits are shown in Figure~\ref{fig:1991.80fits}. The
raw image for 1991.90 is shown in Figure~\ref{fig:1991.90_raw}, and
its spectral line profiles and fits are shown in
Figure~\ref{fig:1991.90fits}. The raw image for 1992.04 is shown in
Figure~\ref{fig:1992.04_raw}, and its spectral line profile fits are
shown in Figure~\ref{fig:1992.04fits}. A light curve for 1991.16 was
presented by \cite{moh93}, but was probably too far away in time to be
useful.

The images from this season show the appearance and subsequent rapid
disapperance of a spot at phase 0.15 and latitude 30{\deg}. We
hereafter refer to this spot as Feature H. It was not evident in
1991.02, but was strong in 1991.80, and then faded rapidly, without
moving, to a low-level remnant by 1992.04. There was a broad and
fairly stable protuberance on the polar spot at about phase 0.45 which
remained quite well-fixed at this phase, again indicating that
features at these high latitudes are very closely-synchronized to the
orbital frame. Essentially no longitudinal motion of this feature is
seen (to within the limits of uncertainty in determining longitudes on
a spot whose shape is not absolutely constant) over the 2.9 month
interval spanned by the images.

The phase 0.75 polar spot protuberance in the 1991.80 image is
probably not the phase 0.64 protuberance of the previous year migrated
by 0.11 phases since it does not appear with consistent strength in
all of our images for the 1990-91 season. It may be a `phase ghost' in
the 1990.80 image, and merely a result of the fact that we observed
the star at relatively few phases on that side of the star in 1991.80,
one of which was at phase 0.76. Indeed, low-level radial `ghosts' can
be seen extending from the polar spot at both of the two observed
phases on that side of the star in the 1991.80 image.

\subsection{The 1992-93 Season Doppler Images}

We obtained four Doppler images at epochs 1992.68, 1992.80, 1992.95,
and 1993.15 this observing season. We could only find photometry for
1992.91, presented by \cite{zha93} and 1992.83 from \cite{dra94} so
only these two epochs are adequately threshholded to the
photometry. The others are presented only as spectral
images. \cite{nef93} reported a continuous IUE monitoring of HR 1099
in December, 1992 as part of the MUSICOS 92 campaign. HR 1099 was also
the subject of an EUVE study by \cite{dra94}. They obtained light
curves in the 60-180A passband in August and October, 1992 and
concluded that most of the EUV flux was coming from hot coronal plasma
associated with an especially bright coronal structure on the active K
subgiant.

\begin{figure}
\plotone{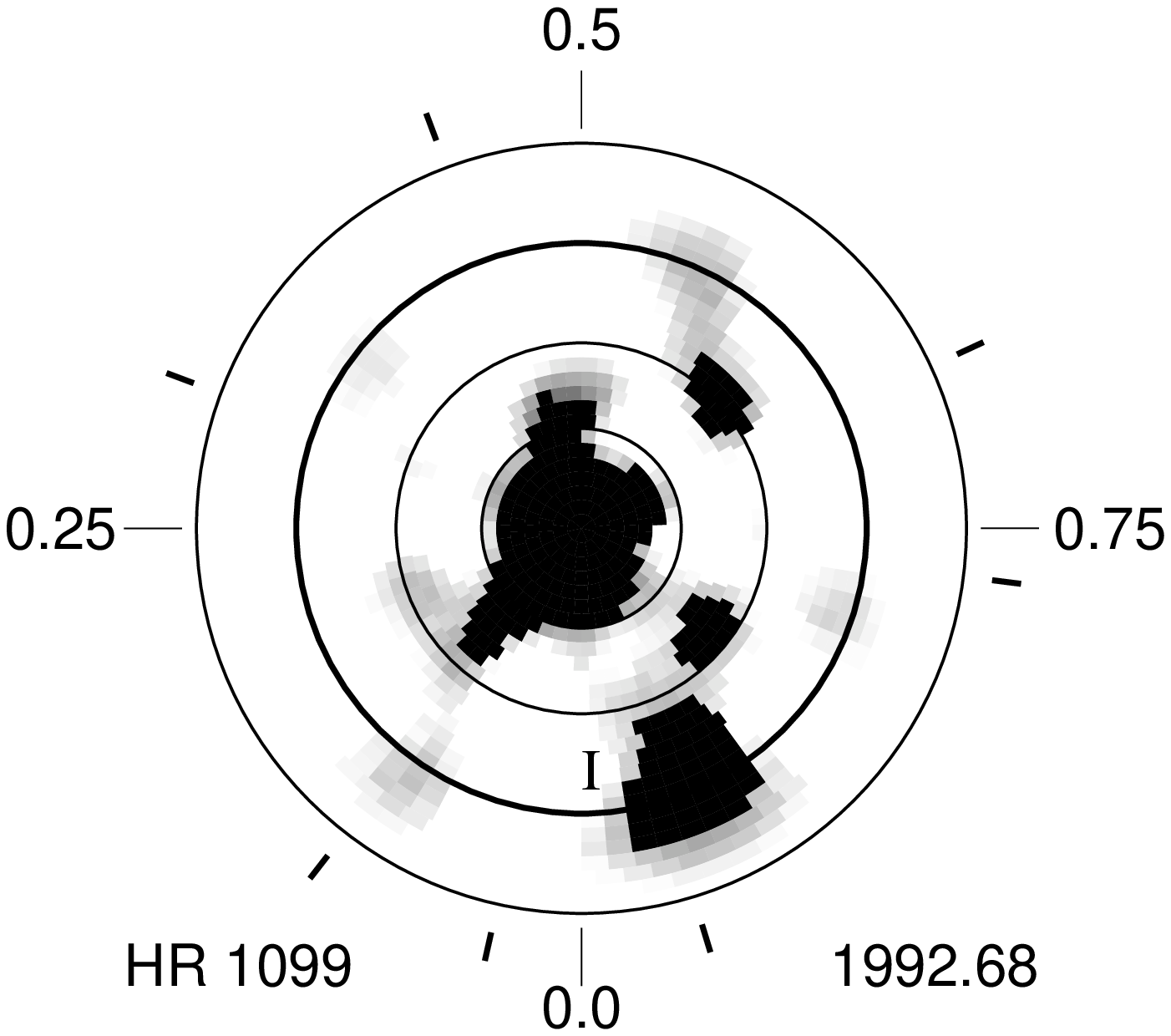}
\caption{HR 1099 raw (unthreshholded) Doppler image for 1992.68}
\label{fig:1992.68_raw}
\end{figure} 

\begin{figure}
\plotone{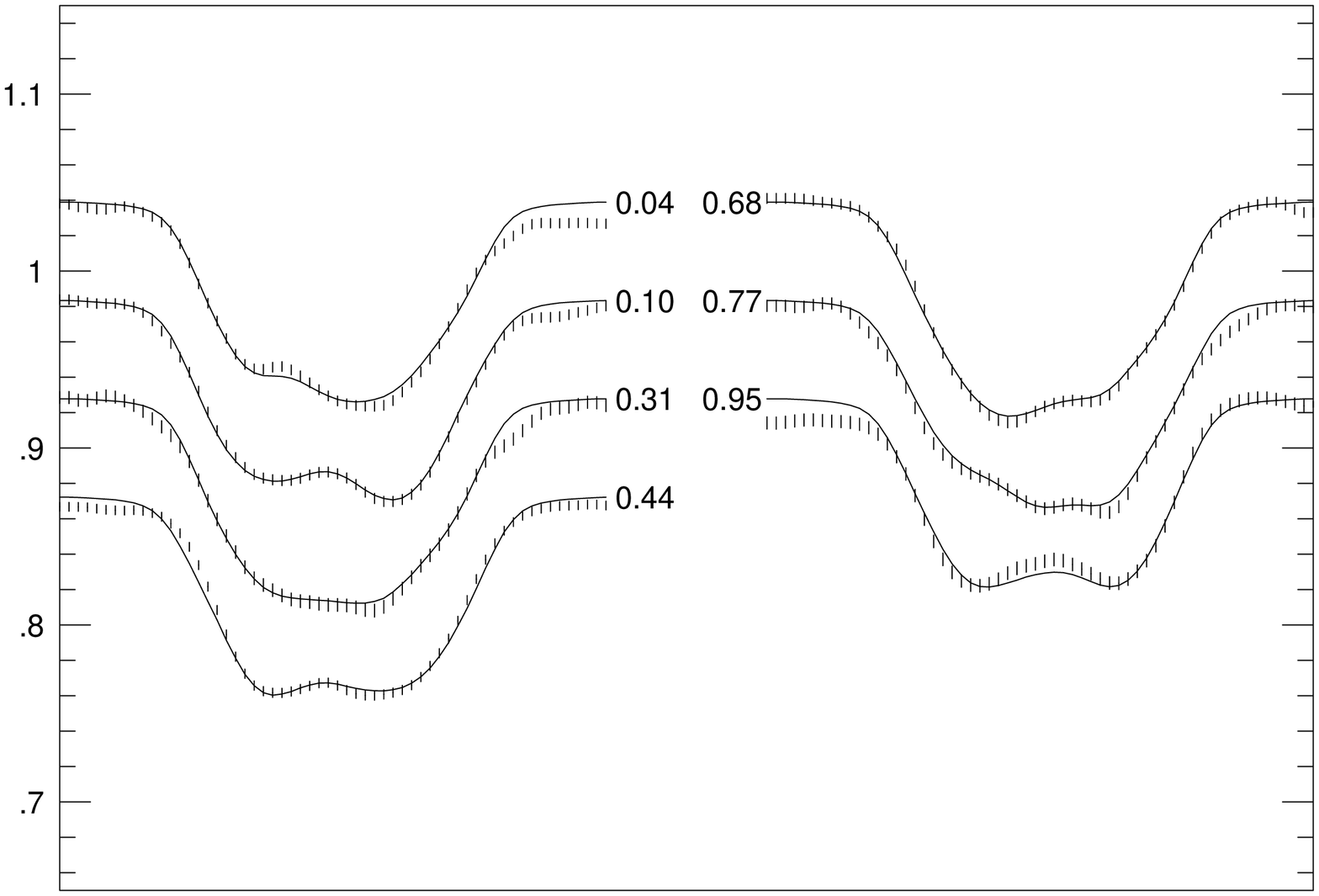}
\caption{Spectral line profiles and fits for 1992.68}
\label{fig:1992.68fits}
\end{figure} 

\begin{figure}
\plotone{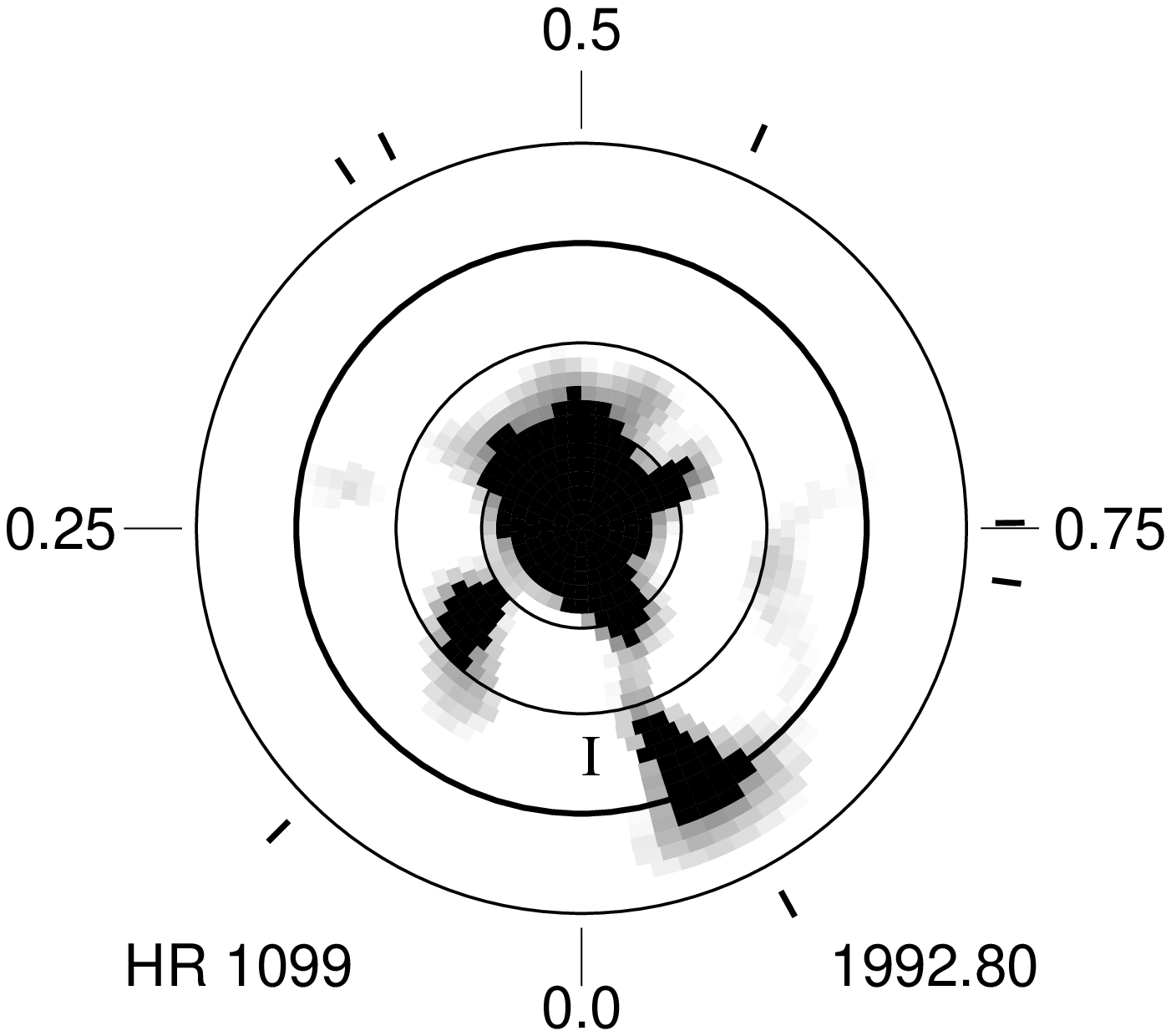}
\caption{HR 1099 raw (unthreshholded) Doppler image for 1992.80}
\label{fig:1992.80_raw}
\end{figure} 

\begin{figure}
\plotone{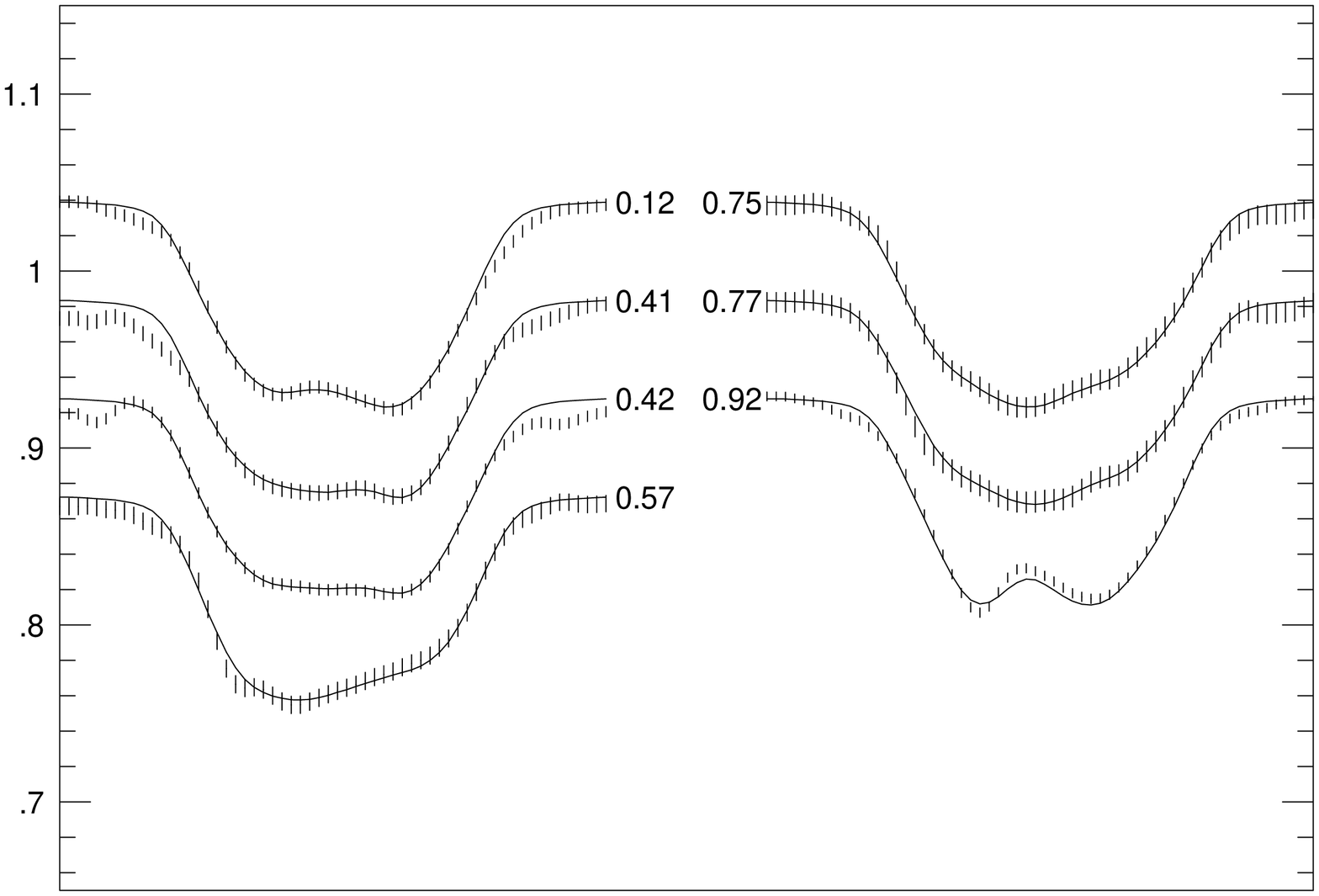}
\caption{Spectral line profiles and fits for 1992.80}
\label{fig:1992.80fits}
\end{figure} 

\begin{figure}
\plotone{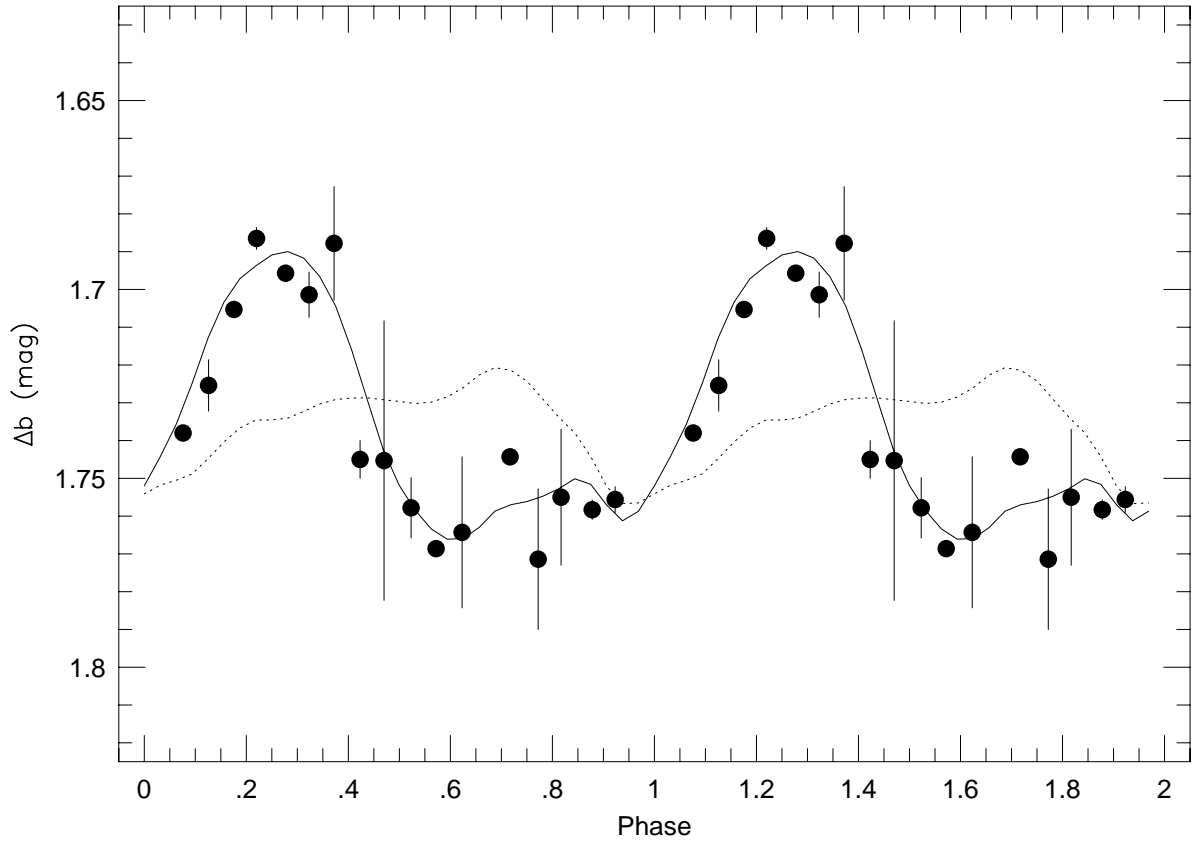}
\caption{$b$-band photometry for HR 1099 in 1992.83}
\label{fig:1992.83light}
\end{figure} 

\begin{figure}
\plotone{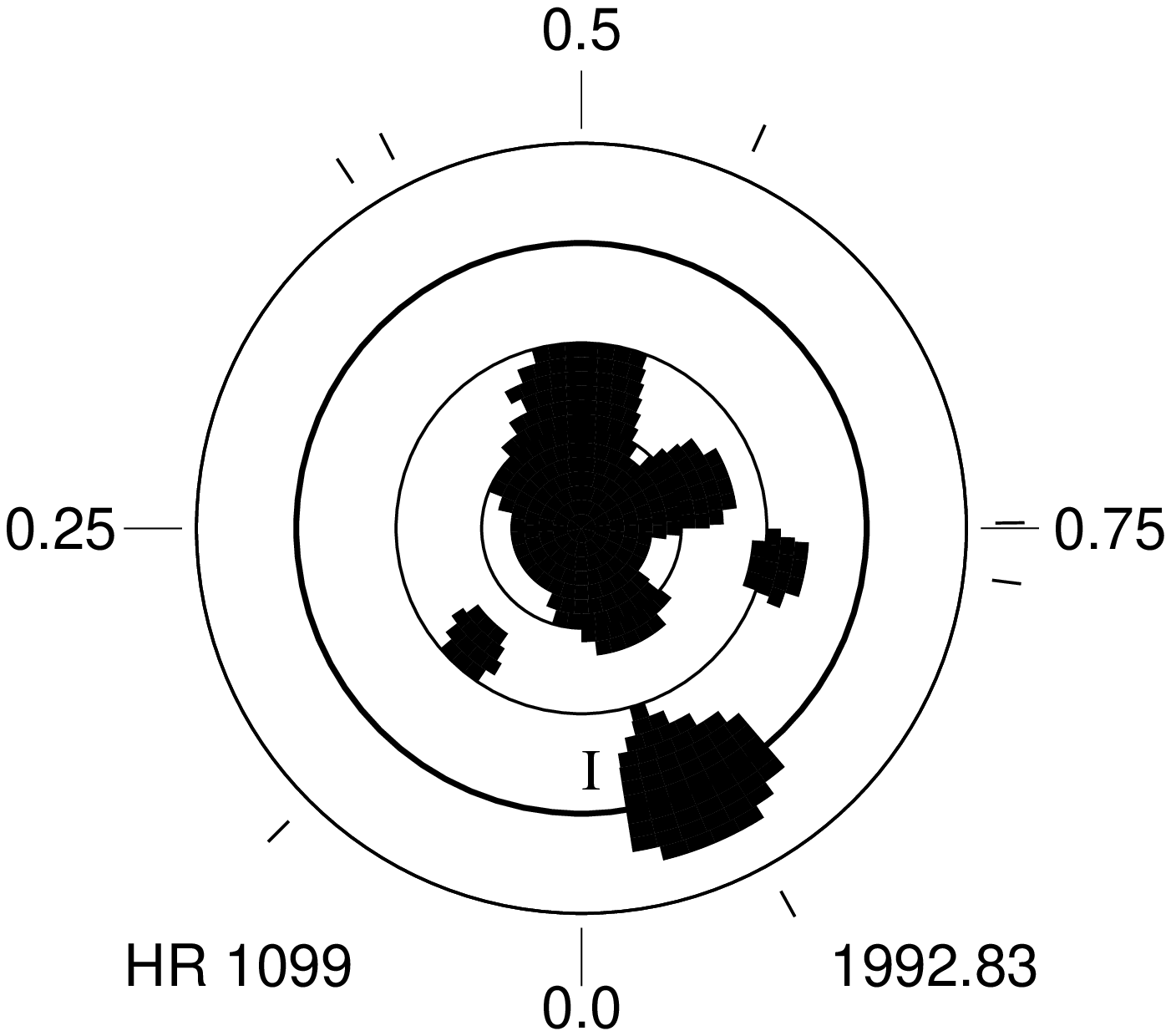}
\caption{HR 1099 thresholded Doppler image for 1992.83}
\label{fig:1992.80_image}
\end{figure} 

\begin{figure}
\plotone{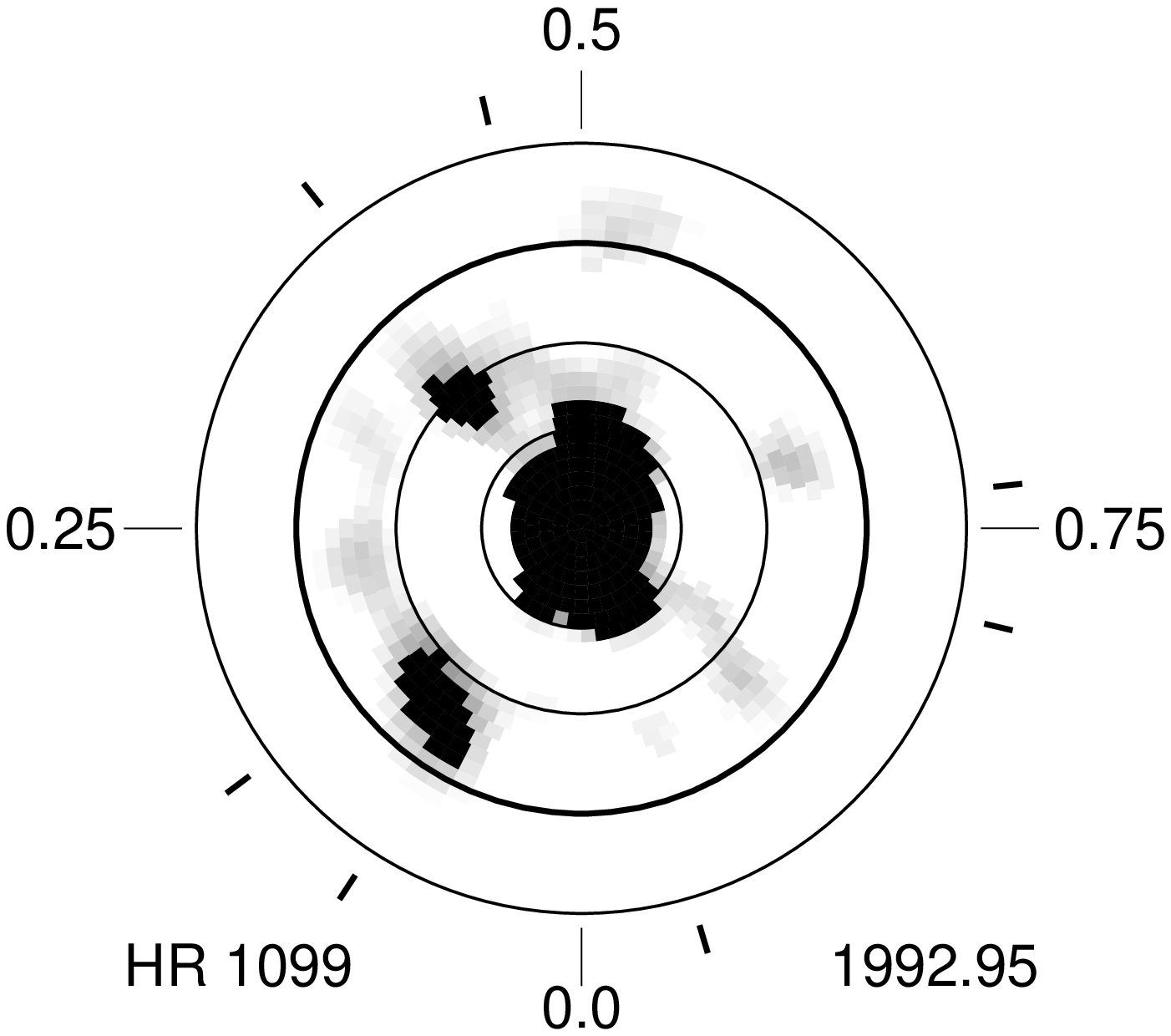}
\caption{HR 1099 raw (unthresholded) Doppler image for 1992.95}
\label{fig:1992.95_raw}
\end{figure} 

\begin{figure}
\plotone{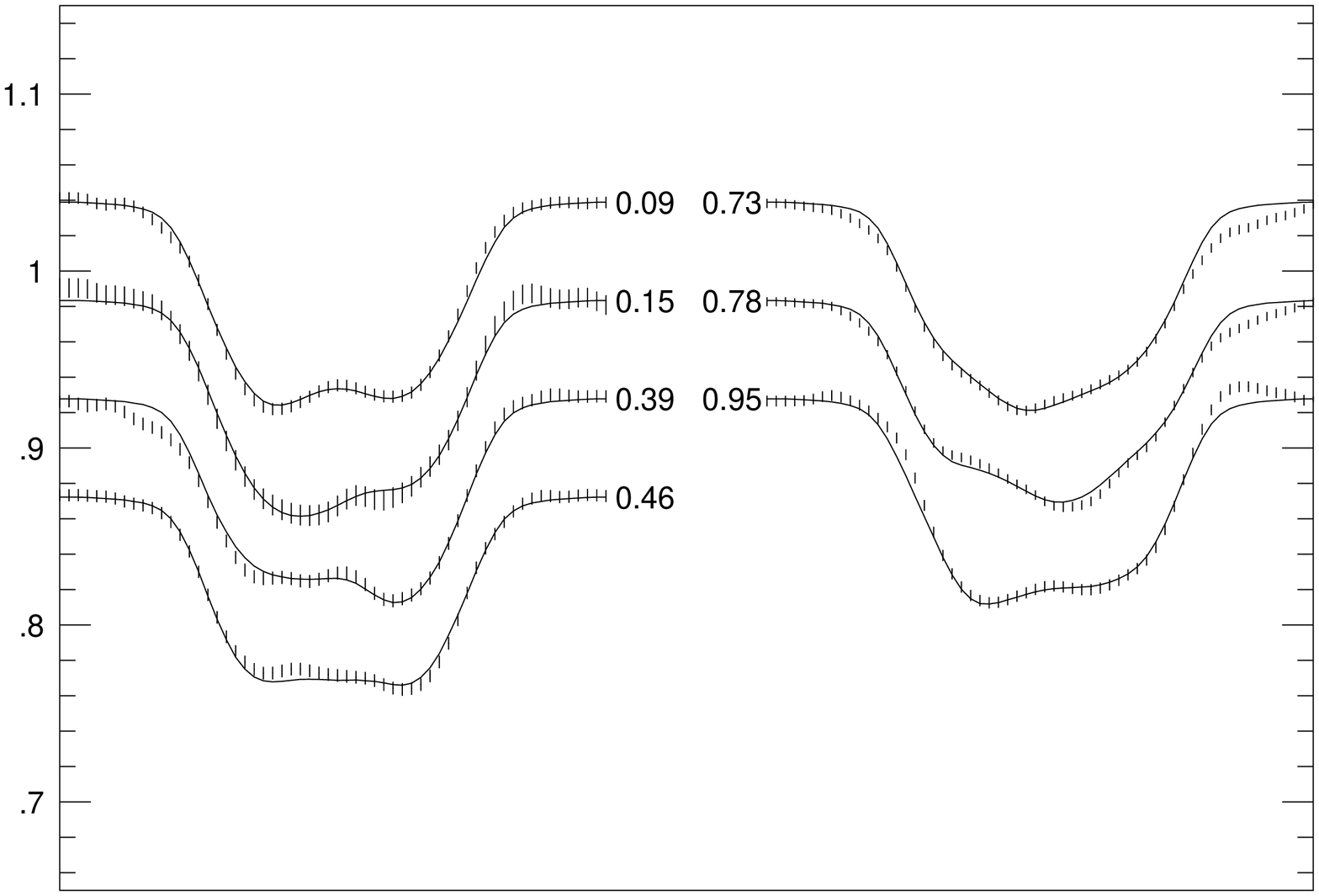}
\caption{Spectral line profiles and fits for 1992.95}
\label{fig:1992.95fits}
\end{figure} 

\begin{figure}
\plotone{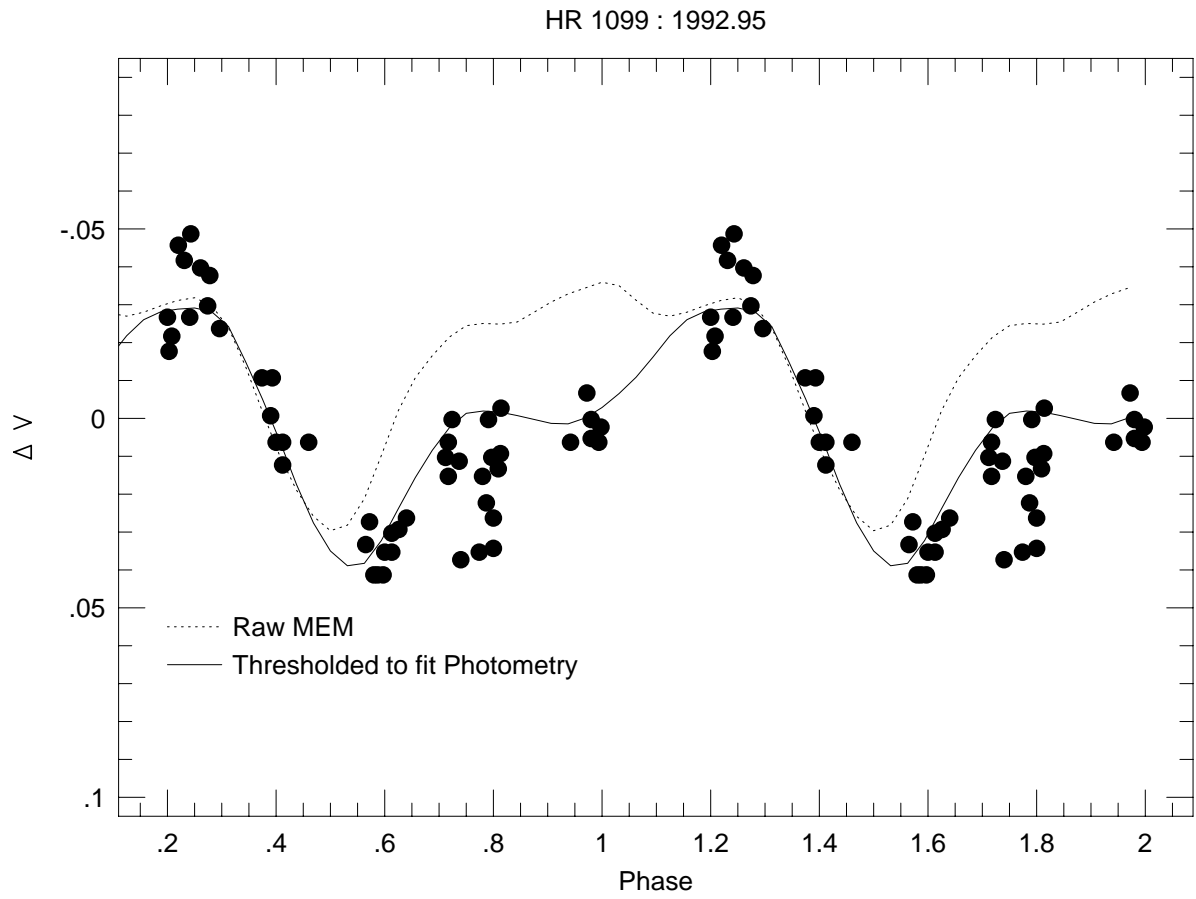}
\caption{HR 1099 light curve for 1992.95}
\label{fig:1992.95_light}
\end{figure} 

\begin{figure}
\plotone{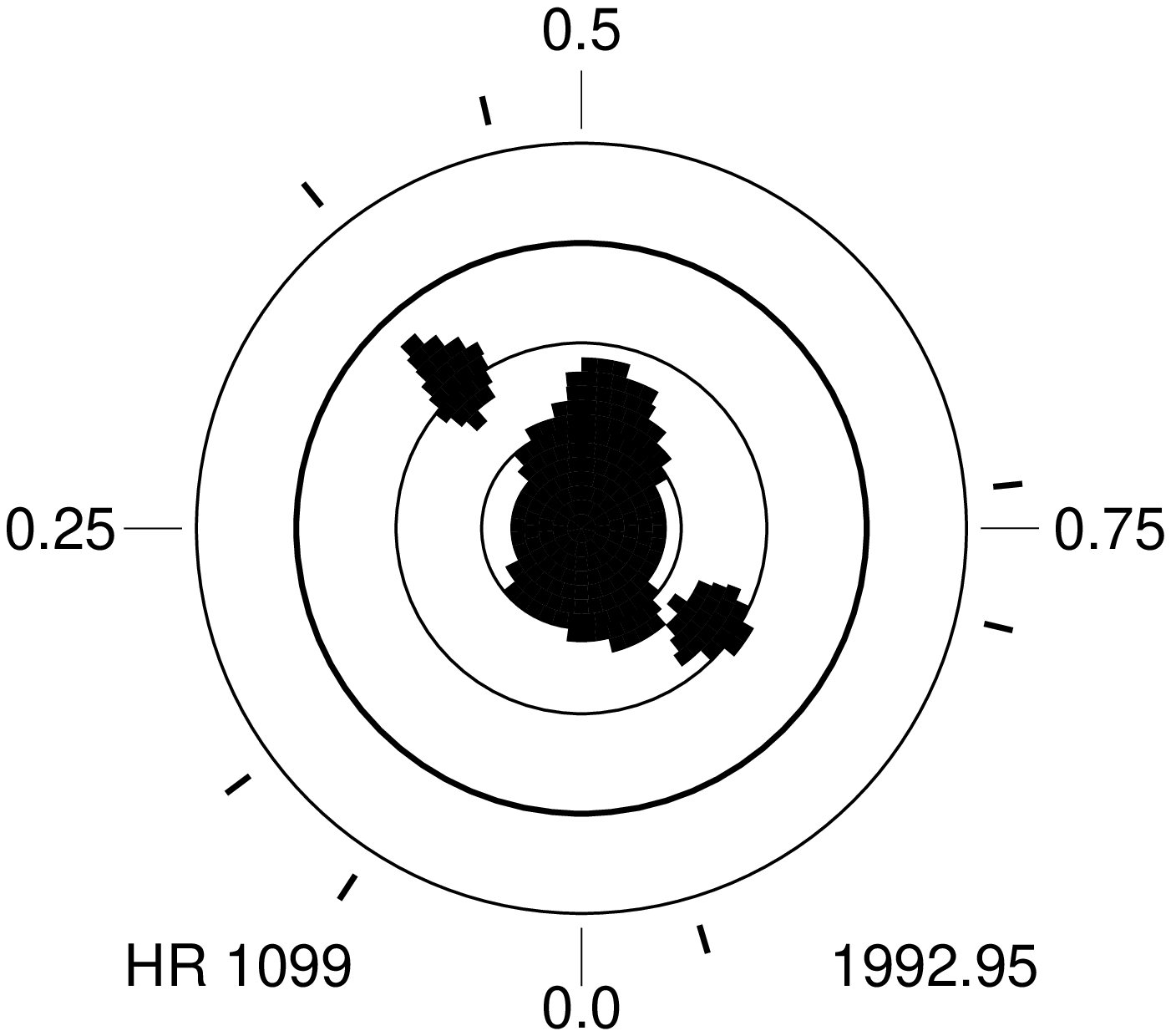}
\caption{HR 1099 thresholded Doppler image for 1992.95}
\label{fig:1992.95_image}
\end{figure} 

\begin{figure}
\plotone{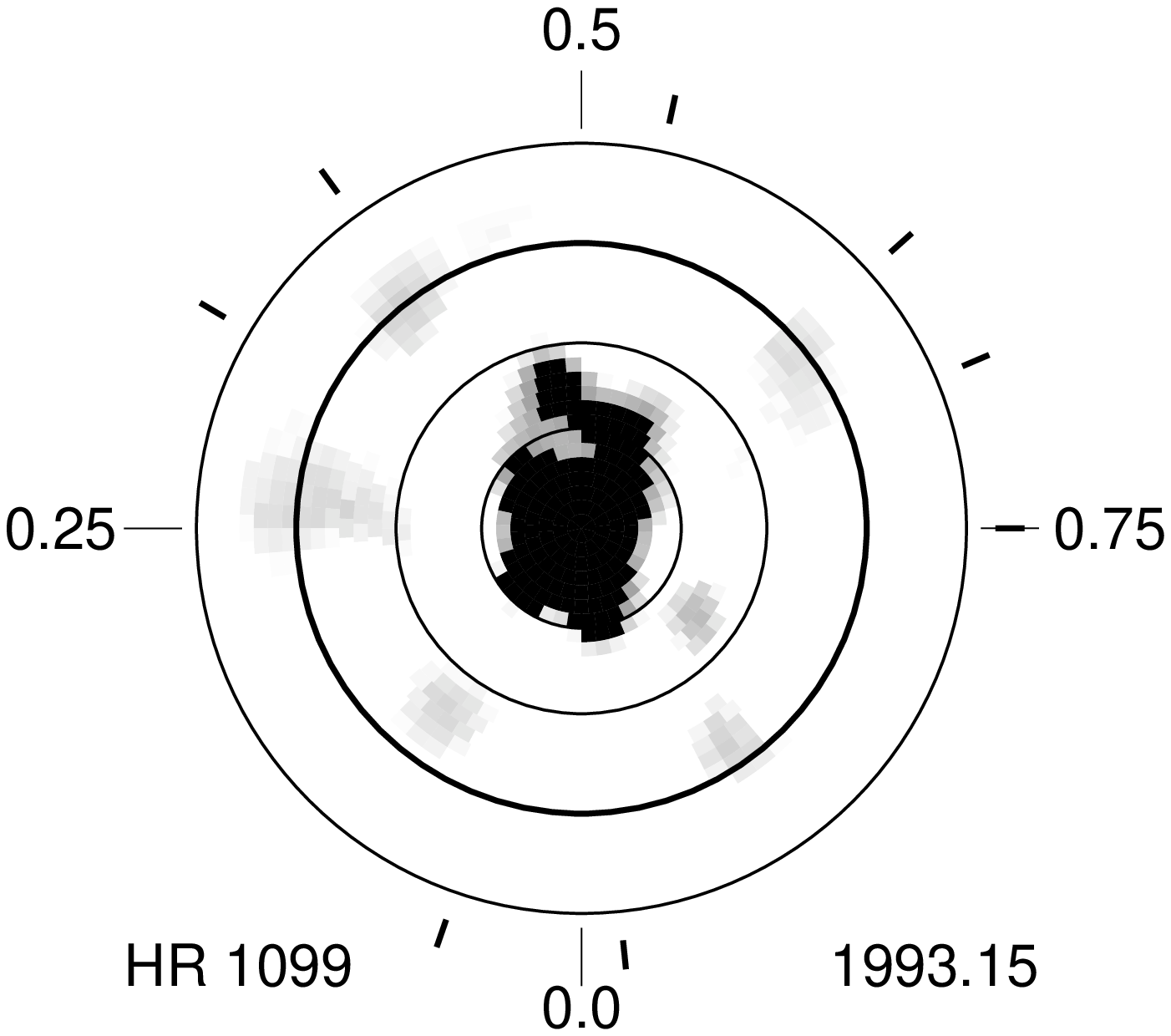}
\caption{HR 1099 raw (unthresholded) Doppler image for 1993.15}
\label{fig:1993.15_raw}
\end{figure} 

\begin{figure}
\plotone{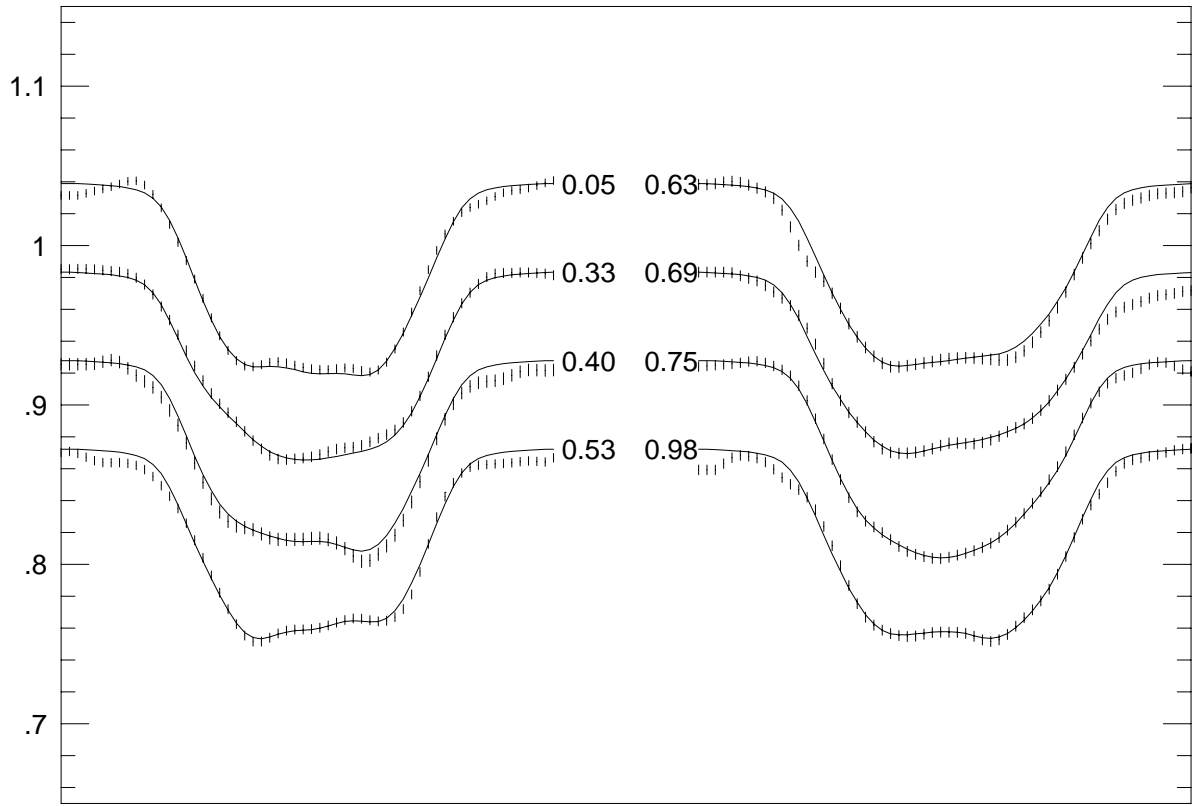}
\caption{Spectral line profiles and fits for 1993.15}
\label{fig:1993.15fits}
\end{figure} 

The unthresholded image for 1992.68 is shown in
Figure~\ref{fig:1992.68_raw}, with the corresponding spectral line
profiles and fits in Figure~\ref{fig:1992.68fits}. The unthresholded
image for 1992.80 is shown in Figure~\ref{fig:1992.80_raw}, with its
corresponding spectral line profiles and fits in
Figure~\ref{fig:1992.80fits}. The predicted photometric variation for
1992.80 is shown as the dotted line in Figure~\ref{fig:1992.83light}
along with the b-band photometry (points) of \cite{dra94}. The final
1992.80 image, as thresholded by the 1992.83 light curve is shown in
Figure~\ref{fig:1992.80_image}. The unthresholded image for 1992.95 is
shown in Figure~\ref{fig:1992.95_raw}, with its corresponding spectral
line profiles and fits in Figure~\ref{fig:1992.95fits}. The predicted
light curve for the raw 1992.95 image is shown as the dotted line in
Figure~\ref{fig:1992.95_light} along with the 1992.91 photometry
(points) of \cite{zha93}. The final 1992.95 image, as thresholded by
the 1992.91 light curve is shown in Figure~\ref{fig:1992.95_image},
and its corresponding light curve is the solid line in
Figure~\ref{fig:1992.95_light}. Finally, the unthresholded 1993.15
image is shown in Figure~\ref{fig:1993.15_raw}, with its corresponding
spectral line profiles and fits in Figure~\ref{fig:1993.15fits}.

The 1992.83 and 1992.95 thresholded images
(Figure~\ref{fig:1992.80_image} and Figure~\ref{fig:1992.95_image})
both provide much better fits to the photometry than do the raw
images. The thresholded images are generally consistent with the raw
images except for the low-latitude spot at phase 0.125 in 1992.95
which did not survive the added constraints imposed by the light curve
data. This serves as a warning and a reminder that one must be wary of
the reality of low-latitude features in our raw Doppler images in
cases where good light curves have not been incorporated into the
solution.

The spot distribution this season changed significantly from the
1992.04 image taken 8 months earlier, though there are some
similarities. The strong protuberance on the polar spot in the 1992.68
image at phase 0.12-0.14 and a similar but apparently detached spot in
the 1992.80 image looks at first to be associated with the persistent
Feature H from the previous year's images. However, this feature was
also present in the raw image of 1992.95 and unfortunately did not
survive the light curve threshholding process, and thus may be
spurious. There was also a large isolated spot (hereafter referred to
as Feature I) which appeared prominently near the equator at phase
0.93 in both the 1992.68 and 1992.83 images. There was no
corresponding feature in either the raw or threshholded 1992.95 image,
so we cannot say whether it would have threshholded away in similar
fashion. It was, however, a rather large feature and appeared at very
nearly the same latitude and longitude in the two entirely independent
data sets, suggesting that, unlike the previous feature whose
appearance and position changed randomly, it may well have been real.

Feature I appeared to shrink fast enough to have completely
disappeared by the 1992.95 image. If so, it would be an example of a
spot well-observed in the act of shrinking, and shows that a major
spot can disappear on this star in less than 5-6 months. Feature I
also showed no significant evidence of either longitude or latitude
movement while disappearing. Note that its fixed longitude is not
inconsistent with the 123{\deg} yr$^{-1}$ longitudinal migration rate
derived from Feature E. Over the 0.12 years between the 1992.68 and
1992.80 images, it would have drifted only 15\deg (3 pixels) in
longitude, and its longitudinal position probably cannot be determined
to that precision due to phase ghosting. It is worthwhile to mention
though that Feature I appeared at essentially the same location as
both Features D and E, as though this were some stable preferred
region of spot formation on HR 1099.

There were also two mid-latitude isolated spots near 30\deg - 36\deg
latitude and at phases 0.63 and 0.87 in the 1992.68 raw image. They
were absent in the 1992.80 raw image, though the phase 0.87 spot did
survive the threshholding process to re-appear in the final 1992.95
thresholded image. Unfortunately, without accurate light curves with
which to threshhold all the images, these ephemeral features are
simply too near the limits of the present dataset to further quantify.

Finally, there is actually a fairly high similarity between the shapes
of the polar spot among this season's raw images. In all cases, there
is a distinct tendency for a protuberance near phase 0.5. This
protuberance appears to be well-fixed in longitude. The polar spot
also looks very similar in shape and fixed in phase between the
1992.95 and 1993.15 images, once again implying that the polar spot,
or at least its high-latitude edge is closely synchronized to the
orbital frame.

The EUV light curve of \cite{dra94} showed a maximum at about orbital
phase 0.0, and minimum near phase 0.5. On our corresponding 1992.80
Doppler image, phase 0.0 coincides with the time of maximum visibility
of both low-latitude spots and phase 0.5 corresponds to the time when
both are completely hidden from view. If the EUV emission is at all
related to the starpot distribution, this phase correspondence may
indicate that the EUV emission is strongly spatially associated with
the two low and mid-latitude spots in the 1992.80 image. This is not
unexpected, since, following the solar coronal hole analogy,
low-latitude spots like Feature I are probably associated with newly
emerged mixed-polarity magnetic flux which has only recently become
organized enough to form a substantial UMR, and is probably still
surrounded with the litter of residual, hot, mixed-polarity emitting
regions. \cite{dra94} pointed out that the minimum of their Stromgren
{\it b} light curve appeared to lead the maximum in EUV emission by
about 90{\deg}, but they found no other obvious relationship between
the two light curves. Their light curve is well-fit by our Doppler
image solution and provided a good check on our Doppler image
solution, but yields no obvious further insight into the relationship
between spots and EUV emission.

\section{Discussion}

\subsection{The omnipresent polar spot}

Clearly the most prominent feature of our images of HR 1099, and a
recurring theme in other's images of rapidly-rotating spotted stars,
is the ever-present large cool polar spot. Since the reality of large
cool polar spots is still not universally accepted by the cool-star
community (for example see \cite{byrne96}), a few words in their
defense seem in order here. Perhaps the best argument for their
reality is that we do {\it not} see them on some more
moderately-rotating RS CVn stars, for example $\sigma$ Gem (Hatzes
1993), nor do we ever see them on other types of stars we have imaged
(Ap stars). In some other cool-star cases, like ZAMS (zero-age main
sequence) Pleiades stars (\cite{sto96}), they are at high latitudes,
but are {\it not} symmetrically placed on the pole as would be
expected if they were an artifact of our inability to model line
profiles properly. We also recently completed a study (\cite{hat96})
in which we showed that the inclination dependence of line core
flattening in a sample of stars could not be explained by gravity
darkening, by differential rotation, or by a bright equatorial band,
but was quite consistent with the inclination aspect-dependence
produced by a polar dark spot. Other researchers have also found these
polar spots both in intensity images and in Stokes parameter
imagery. Finally, phase-independent TiO features in the spectra of
some low-inclination spotted stars strongly support the presence of
large cool polar spots. So the reality of these polar spots now seems
reasonably well-established, and the large cool polar spot on HR 1099
has persisted for the 11-year span of this study.

\subsection{Comparison of the images with `few-circular-spot' solutions}

Most spotted stars rotate too slowly to be amenable to Doppler
imaging. In these cases, one must resort to fitting light curves with
simple models involving 2 or 3 circular spots. However, these simple
spot models are known to be quite non-unique since a light curve alone
offers little constraint for the large set of possible image
solutions. It was therefore of particular interest to check the
results of such `few-circular-spot' models against the more detailed
Doppler imagery of HR 1099.  This comparison would reveal how well
these simple models actually did, and thereby help us to better assess
their effectiveness for more slowly-rotating spotted stars.  We showed
such comparisons of few-circular-spot solutions against Doppler images
for epochs 1981.70, 1982.74, and 1989.83. In all cases, our Doppler
images and others' `few-circular-spot'solutions {\it both} adequately
fit the broadband light curves. The Doppler images, however, {\it
also} fit a set of line flux profiles and are thus much more highly
constrained. Unfortunately, the agreement between these
`few-circular-spot' solutions and the Doppler images was not very
encouraging, nor was the agreement between the circular-spot solutions
of different researchers for the same epoch. Clearly, the light curves
alone just do not contain enough information to adequately constrain
the spot solutions.  So, while some instances of correspondence can be
found, these circular-spot solutions do not seem to yield much useful
information on the spot distribution. In fact, for all the comparison
cases presented in this paper, these solutions were quite misleading.

Conversely, agreement of Doppler images of HR 1099 among different
researchers, for example our 1988.80 and 1991.02 images with the
Doppler images of Donati et al. (1992), is generally quite
excellent. Here, most of the subtle details of the shape of the polar
spot are faithfully reproduced between the two research groups, though
these images were derived from completely independent data sets, with
different imaging software.

\subsection{Where do the spots first emerge?}

It is probable that small spots can emerge almost anywhere on the
star, showing up as changes in the outlines and appendages of existing
spots, or as short-lived isolated spots. With only one, or at most
several images per year (100-200 rotations for HR 1099), it is not
possible to conclude much about the emergence of such small-scale
features from our image set. However, one fundamental result revealed
by the image set is that major spots do frequently `emerge'
(i.e. first appear) at quite low latitudes. Features D, E and I are
three examples of this, having first appeared on or near the
equator. Other equatorial and low-latitude spot emergences were also
seen.

This well-determined observational result contradicts the theoretical
explanation of polar spots put forth by Sch\"ussler and Solanki
(1992). In their model, coriolis forces cause buoyant magnetic flux
generated deep within the star to rise along paths parallel to the
star's rotation axis rather than along radial paths. Their model then
predicts that flux should erupt only at high latitudes, thereby
perhaps explaining the ubiquitous polar spots on these stars. However,
the common-place emergence of low-latitude spots on the very rapid
rotator HR 1099 is in direct contradiction to this prediction, and may
be a problem for the Sch\"ussler and Solanki model. Their idea was
further developed by \cite{schussler96} who found that there is
usually a range of latitudes over which magnetic flux can emerge, with
the mean latitude shifting toward the poles for increasing
rotation. This model is now in good basic agreement with the polar and
mid-latitude distribution of spots on young rapid rotators, but still
does not explain the frequent emergence of major flux at low-latitudes
in our HR 1099 Doppler images. While their theoretical model looks
quite promising, it must be further explored to see if it can also
explain this observed low-latitude spot emergence.

\subsection{Tracking the movements of individual spots}

It has often been assumed that the `migrating photometric waves' in
the light curves of RS CVn stars are attributable to longitudinal
migration of spots on a differentially-rotating star. The phase drift
of spotted star light curves is certainly an easily and accurately
measured quantity, and attempts have thus been made to derive the
differential rotation of spotted stars from such light curve phase
drifts. In the presence of latitudinal shear, spots at different
latitudes would have different rotation periods, migrating in
longitude with respect to the orbit, causing the light curve to
migrate in phase accordingly. In this simple picture, spots appearing
above or below some intermediate co-rotation latitude (i.e. like the
solar butterfly diagram for sunspots) could cause the photometric
light curve minimum to migrate in either direction.

Our imagery reveals that, at least for HR 1099, the situation is far
more complex. The phase of the photometric minimum is affected
strongly by the appearance and disappearance of a number of
relatively-fixed large spots, as well as by the movements of some of
these spots, and also by changes in their shapes and areas. It is
clearly not valid to invert these phase drifts through so simple a
parameterization as longitude migration of one or two spots. {\it The
phase of the migrating wave by itself reveals essentially nothing
about the details of the differential rotation pattern on HR 1099}.

\begin{figure}
\plotone{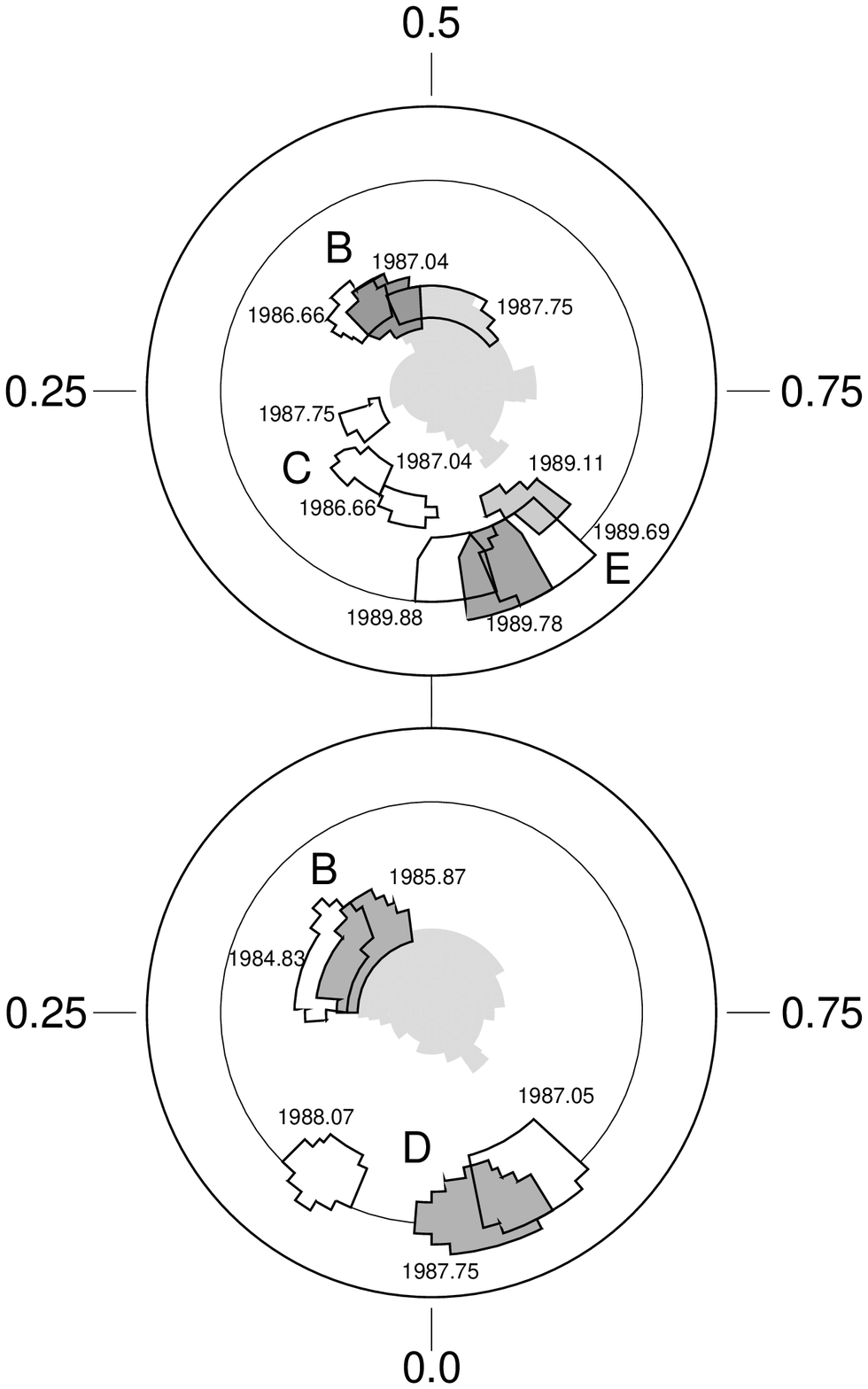}
\caption{Plots of the migration tracks of four long-lived prominent spots}
\label{fig:migration}
\end{figure}

Most spots come and go too quickly to track reliably with,
on-the-average, one image per year. However, there were a number of
isolated, long-lived spots which we were able to track from
year-to-year. Figure~\ref{fig:migration} shows the tracks of Features
B, C, D, and E in the frame of reference of the orbit. Thus all
migrations are shown here as occuring with respect to the orbital
frame, with the secondary (hotter) star fixed at phase 0.5. (The
position of spot D in 1988.07 was inferred from the photometry. A spot
was required at that position for the predicted photometry from our
1987.75 Doppler image to fit the 1988.07-epoch photometry of
\cite{moh93}).

Feature B, at high latitude, lasted for almost 3 years and was seen
migrating clockwise and poleward very slowly with respect to the
orbital frame. It apparently merged with the polar spot by 1987.75.
Feature C, at intermediate latitude, also migrated clockwise, its
angular velocity still slower than the orbit but faster relative to
the orbit than Feature B. It also followed a poleward-drifting
clockwise spiral path finally merging with the polar spot. Features D
and E also drifted clockwise, but even faster relative to the orbit
than B or C.

The tracks of both Features B and C suggest that some spots which
emerge at low or intermediate latitude do migrate poleward, and then
apparently merge with the polar spot. Since the dark spots presumably
trace magnetic flux, this must mean that some of the magnetic flux
which emerges at lower latitudes on HR 1099 then spirals pole-ward in
a clockwise manner (slower than the orbit), and eventually merges with
the polar spot. It is not yet clear whether this flux is of the same
or opposite polarity to the polar spot, and thus whether these
poleward-migrating low-latitude spots reinforce or cancel the polar
spot field.

Feature B also appeared to get somewhat {\it stretched} in longitude
as it approached the polar spot, and its overall track is quite
reminiscent of the annulus of toroidal field found by \cite{don92b}
encircling the polar spot of HR 1099 in 1990.9. Perhaps spots such as
Feature B become sheared in longitude as they approach the pole, and
thereby contribute to an annular ring of strong field around the polar
spot.  Feature B may thus be showing us directly how field lines get
wrapped around the pole into the toroidal structure seen by Donati et
al. (1992). Perhaps then, the `polar spot' is simply that region where
the field lines get wrapped sufficiently tightly around the pole, like
spaghetti around a fork, to create fields strong enough to suppress
convection in the photosphere, and thus create the cool polar spots.

\subsection{The differential rotation of HR 1099 and other RS CVn stars}
 
\begin{figure}
\plotone{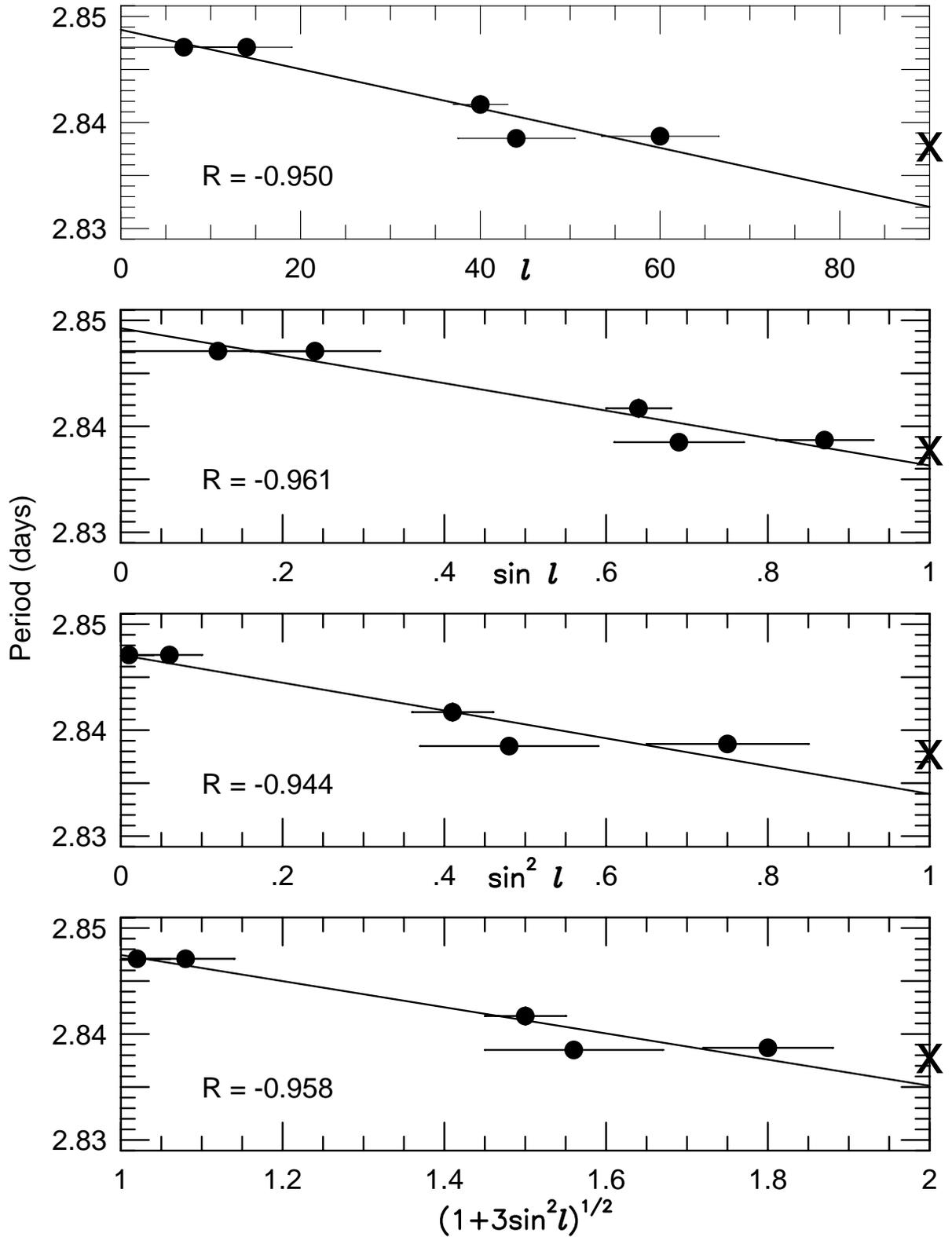}
\caption{The rotation period of individual spot features as a function
of various parameterizations of latitude. The location of a
synchronously-rotating pole is shown by the cross at the right edge of
each plot}
\label{fig:diffrot}
\end{figure}

The migration rates of these long-lived features were then used to
derive the `differential rotation' or rotation period of the star as a
function of latitude. Figure~\ref{fig:diffrot} shows the measured
period vs. several simple parameterizations of latitude ($\ell$).

The Sun's rotation period depends nearly linearly on $sin^2 \ell$, and
we have thus used that as one of the simple latitude parameterizations
chosen for Figure~\ref{fig:diffrot}. As can be seen, the rotation
period correlates well with all the chosen
parameterizations. Unfortunately, the rather small number of data
points and the scatter in the data precludes deriving a more precise
functional form from this data set.
 
Using the standard solar parameterization for differential rotation,
one expects angular velocity to vary linearly with $sin^2 \ell$ where
$\ell$ is the latitude. A formal linear least squares fit of this
functional form to our 5 points, including the differing error bars in
both dimensions, gives:

\begin{equation}
\Omega (^\circ/day) = 126.52 \ (\pm 0.003) + 0.47 \ (\pm 0.06) \
sin^2\ \ell \
\end{equation}

The amount of differential rotation is then generally parameterized by
$\alpha$, the ratio of the difference between polar and equatorial
velocities to the equatorial velocity:

\begin{equation}
{\alpha} = {{ {\Omega_{equator}-\Omega_{pole}}}\over { {\Omega_{equator} } }}
\end{equation}

Solid body rotation results in $\alpha$ = 0, whereas equatorial
acceleration has $\alpha > $ 0 and polar acceleration results in
$\alpha < 0$. Using the above $sin^2 \ell$ fit gives $\alpha_{HR
1099}$=--0.004($\pm$0.0004) whereas $\alpha_{Sun}$=+0.197. Thus the
differential rotaion implied by starspot motions on HR 1099 is of
opposite sign to the Sun: the poles rotate faster than the equator,
with about a 50 times smaller differential. The same result (poles
rotating faster than the equator) was also found for the spotted star
UX Ari by \cite{vog91}, though in that case, the amount of
differential rotation was only about 10 times less than solar, the
equator was synchronized to the orbit, and the pole actually rotated
faster than the orbit.

Interestingly, the rotation period of HR 1099 also correlates quite
well (bottom panel of Figure~\ref{fig:diffrot}) with $(1 + 3 sin^2
\ell)^{1/2}$, an abscissa proportional to the surface strength of a
centered, axisymmetric magnetic dipole field. Given also that an
extrapolation of the observed rotation periods suggests
near-synchronous rotation at the pole, these observations may well be
indicating that spot migration motions are dominated by a large,
aligned, magnetic dipole within HR 1099 which is co-rotating with the
orbital frame. The relatively high surface field strengths near the
pole would then enforce near-corotation of high-latitude spot
features, while the decreasing field strength with decreasing latitude
would allow progressively faster `slip' or clockwise longitude
migration of spots, just as observed.

Is the pole truly synchronized with the orbit, or actually going
slightly faster? The extrapolations of all the fits in
Figure~\ref{fig:diffrot} do hit slightly below the location of the
synchronously-rotating pole (marked by the cross at the right edge of
each plot). If, for argument's sake, the pole rotates faster than the
orbit, this would indicate that it is the intermediate-high latitudes
(50\deg to 60\deg) which are synchronized to the orbit, with the pole
then rotating faster than the orbit. This brings to mind the work of
\cite{sch82} who calculated the effects of tidal coupling in RS CVn
binaries. He predicted that, when the tidal coupling torque vanishes
(maximum degree of synchronism achieved), part of the star should be
rotating faster than the orbit, and part more slowly. Co-rotation with
the orbit should occur at some intermediate latitude (the co-rotation
latitude). If our simple linear least squares fits and extrapolations
do indicate that the pole rotates faster than the orbit, then the
implied differential rotation pattern may indeed fit the predictions
of \cite{sch82}. The corresponding implication also is that HR 1099 is
at or near a state of maximum degree of tidal coupling.

On the other hand, given the sizes of the error bars, and the scatter
of the few data points, the fits are quite consistent with synchronous
polar rotation. We certainly did not find any evidence of trackable
features at the pole which were rotating faster than the orbit. All
indications from the images are that features migrated progressively
more slowly with respect to the orbit as latitude increased, becoming
indistinguishable from stationary at the highest discernable
latitudes.

There is also no {\it a priori} reason to expect that differential
rotation should vary linearly with $sin^2 \ell$ as it does for the
Sun, and thus we should not artificially force such a fit upon the
data and then expect it to give a valid result upon extrapolation to
the pole. Indeed, the $sin \ell$ parameterization does extrapolate
quite closely to synchronous polar rotation. One problem is that the
latitude range of Features B1 and B2 are quite large, and difficult to
determine precisely, having extended at various times from at least
40\deg to 70\deg latitude. More likely, all we can probably say at the
moment is that, if Features B1 and B2 are high enough in latitude to
be representative of the polar zones, then the polar zones are nearly
synchronized with the orbit.

One simple explanation for polar synchronization is that the star
conserved angular momentum as it expanded upon leaving the main
sequence. The star may have been tidally locked to the orbit at all
latitudes before its ascent up the subgiant branch. As it expanded in
radius, conservation of angular momentum would have slowed its outer
layers and perhaps also preferentially slowed its equatorial
latitudes, causing them to rotate more slowly than the higher
latitudes. This would leave the polar regions still closely
synchronized to the orbital angular velocity and perhaps only slightly
slower, whereas the equatorial regions would be rotating more slowly.

A completely different explanation is that perhaps the large polar
spot is the footprint of a strong dipole magnetic field which is
firmly anchored to the highly-synchronized core of the main sequence
progenitor of this expanding subgiant, at a field strength of several
kilogauss. The magnetic energy density of this dipole could be
stronger than the kinetic energy density of the differential shearing
motions of the gas. This emergent dipole field would then dominate the
gas motions at high latitudes, producing apparent synchronization of
the very high-latitude spot features with the orbit.

Whatever the precise form of the rotation period's latitude
dependence, the basic picture which emerges is one of high latitude
spots on HR 1099 rotating in near or perfect synchronism with the
orbit, and of a rotation period increasing with decreasing
latitude. The equator is thus rotating more slowly than the poles,
opposite to the solar case. And while the Sun shows a 20\% difference
between polar and equatorial angular velocity in the photosphere as
determined from sunspots, starspots show only about a 0.4\% difference
on HR 1099. While this is much less than photospheric sunspot motions,
{\it it is quite similar to the small differential rotation values
observed in solar coronal holes}.

\cite{nav94} measured the solar rotation rate using solar coronal
holes. They found that solar coronal holes show an average rotation
rate which is almost constant with latitude, for latitudes above
30\deg and, at lower latitudes, increases to become quite similar to
that determined from sunspots. For latitudes above 20\deg the
differential rotation rate determined by the average rotation rate of
coronal holes is in the same sense (poles rotate slower than equator)
but about a factor of 10 less than that determined from sunspot
groups.  However, dividing the data set up into two distinct groups,
isolated holes and polar hole extensions, they see a more complicated
behavior. Polar hole extensions show slowest rotation at intermediate
latitudes, with rotation speeding up both toward the equator and
toward the pole. Isolated coronal holes, on the other hand, show
slowest rotation at the highest latitudes, a flat plateau of constant
rotation rate at intermediate latitudes, and then rotation rate
increasing steadily to values in excess of that from sunspots at the
lowest latitudes. So if the solar case is any lesson here, the true
situation is probably much more complex than simple monotonic
variation of rotation rate with latitude.

Given that starspots also bear a striking morphological resemblance to
solar coronal holes, this similarly low degree of differential
rotation of starspots and solar coronal holes is perhaps not
altogether surprising. It is consistent with the basic picture
suggested by \cite{vop83} that starspots are essentially just coronal
hole-like structures (footprints of the unipolar magnetic regions of a
current-free global coronal magnetic field), but where the global
field is multi-kgauss-strength rather than the few-gauss-strength
global fields of the Sun.

Of course one should not attempt to draw too many inferences from the
differential rotation of a single complex binary system such as HR
1099. But it can perhaps help to interpret the indirect differential
rotation signatures derived from light curve stability measurements
for a large number of RS CVn and other active late-type stars. Our
results of very small differential rotation for HR 1099 (present work)
and for UX Ari (\cite{vog91}) are in excellent accord with the results
of Hall (1991) who found that differential rotation signatures (from
light curve stability) of a large sample of spotted late-type stars
{\it decrease} rapidly as angular velocity increases. Rapidly-rotating
spotted stars such as HR 1099 thus appear to approach solid-body
rotation (at least as far as spot movements are concerned).

If there is a strong (multi-kgauss) axisymmetric magnetic dipole
within HR 1099 (which is causing the large cool polar spots), it may
well be expected to globally dominate the movements of these
starspots, just as the current-free coronal field of the Sun does in
the case of coronal hole structures (\cite{wan93}). A 5-10 kG dipole
field strength at the pole would still be 2.5-5 kG strength at the
equator, and thus strong enough to compete seriously with the shearing
forces of photospheric differential rotation. Newly emerging magnetic
flux would not emerge into a field-free environment, but rather would
be obliged to immediately connect with this powerful external global
dipole field, and any tendency for shearing due to differential
rotation would be strongly resisted by the current-free global
potential field. This picture might provide a simple explanation for
the observed trend toward solid-body rotation of starspots with
increasing stellar rotation velocity. More rapid rotation gives rise
to a stronger current-free global dipole field which, in turn,
enforces increasingly solid-body rotation of the spots (the unipolar
magnetic footprints of this field) as the rotational angular velocity
increases.

\subsection{Spot/Activity cycles}

There have been many claims of activity cycles on spotted RS CVn stars
over the years, as people search for the analog of the well-known
11-year sunspot cycle. On the Sun, the area of the polar coronal hole
varies dramatically and periodically, in anti-phase with the sunspot
cycle (\cite{bra89}, \cite{wal81}, \cite{sim79}). If starspots are
indeed stellar analogs of solar coronal hole structures, then probably
the best hope of detecting a dynamo cycle is to look for periodicity
in the {\it area} of the polar spot. Properly-thresholded Doppler
images do give a good unambiguous map of the polar spot, and
variations in its area are easy to measure from Doppler images
providing the thresholding process has been done very consistently
from one season to the next. We have taken great care to establish a
uniform threshold for all our images, and have cross-checked our
thresholding process with fits to both the line profiles and the
broadband light curves.

\begin{figure}
\plotone{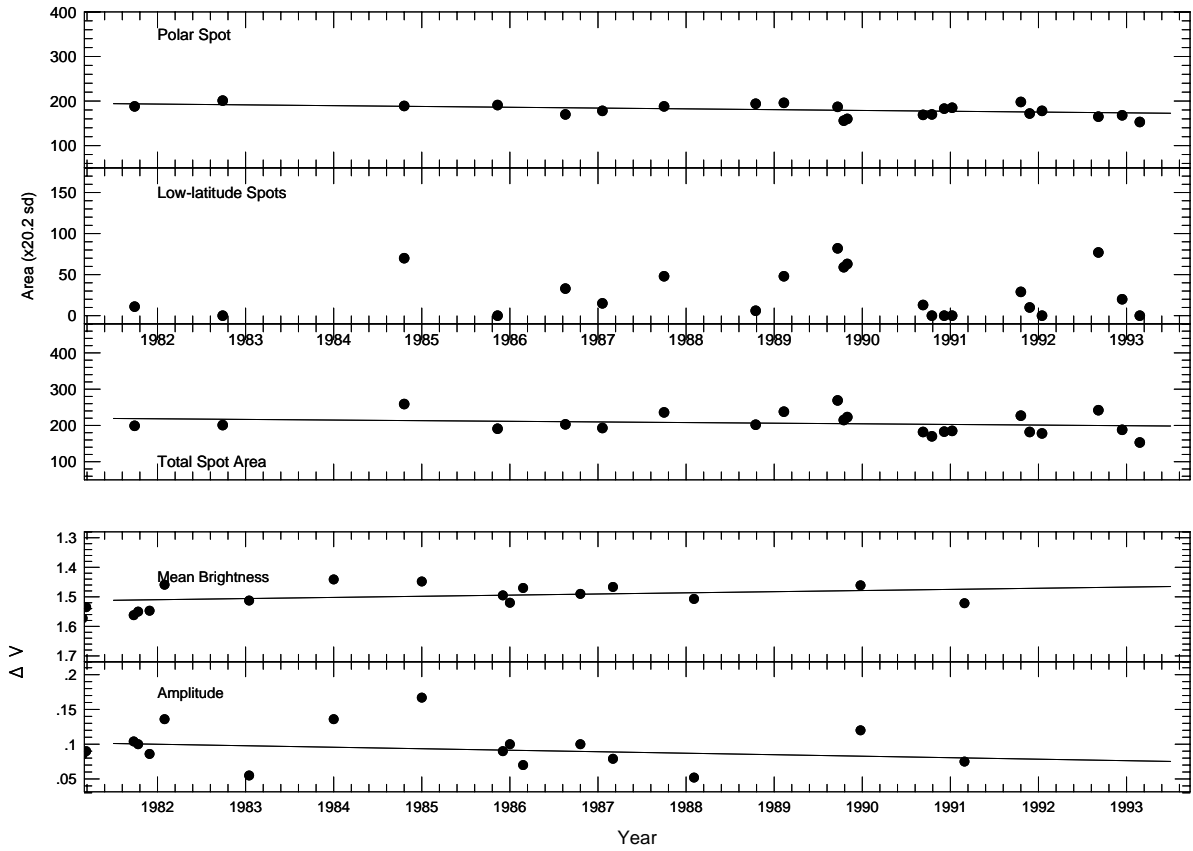}
\caption{Plots of the polar, low-latitude, and total spot area vs. time
(upper panel), and mean light and light amplitude vs. time (lower panel).}
\label{fig:meanlight}
\end{figure}

The top panel of Figure~\ref{fig:meanlight} shows polar, low-latitude,
and total spot areas (percent of visible hemisphere as a function of
time. There is an overall decrease in the polar spot area of about
11\% from 1982--1993 (the solid line represents a linear
regression). The lower panel shows the mean brightness level and light
amplitude for HR 1099 from \cite{moh93}. The line for the mean light
data represents a linear regression using the full span (1975-1991) of
those data and yields an overall brightness increase of 0.045 mag over
the 12-year span of the Doppler images. This observed increase agrees
well with a predicted increase of 0.042 magnitudes (calculated from
the Doppler images) and would seem to indicate that {\it the mean
light level here is, to first order, a good proxy of (polar) spot
area}.

\begin{figure}
\plotone{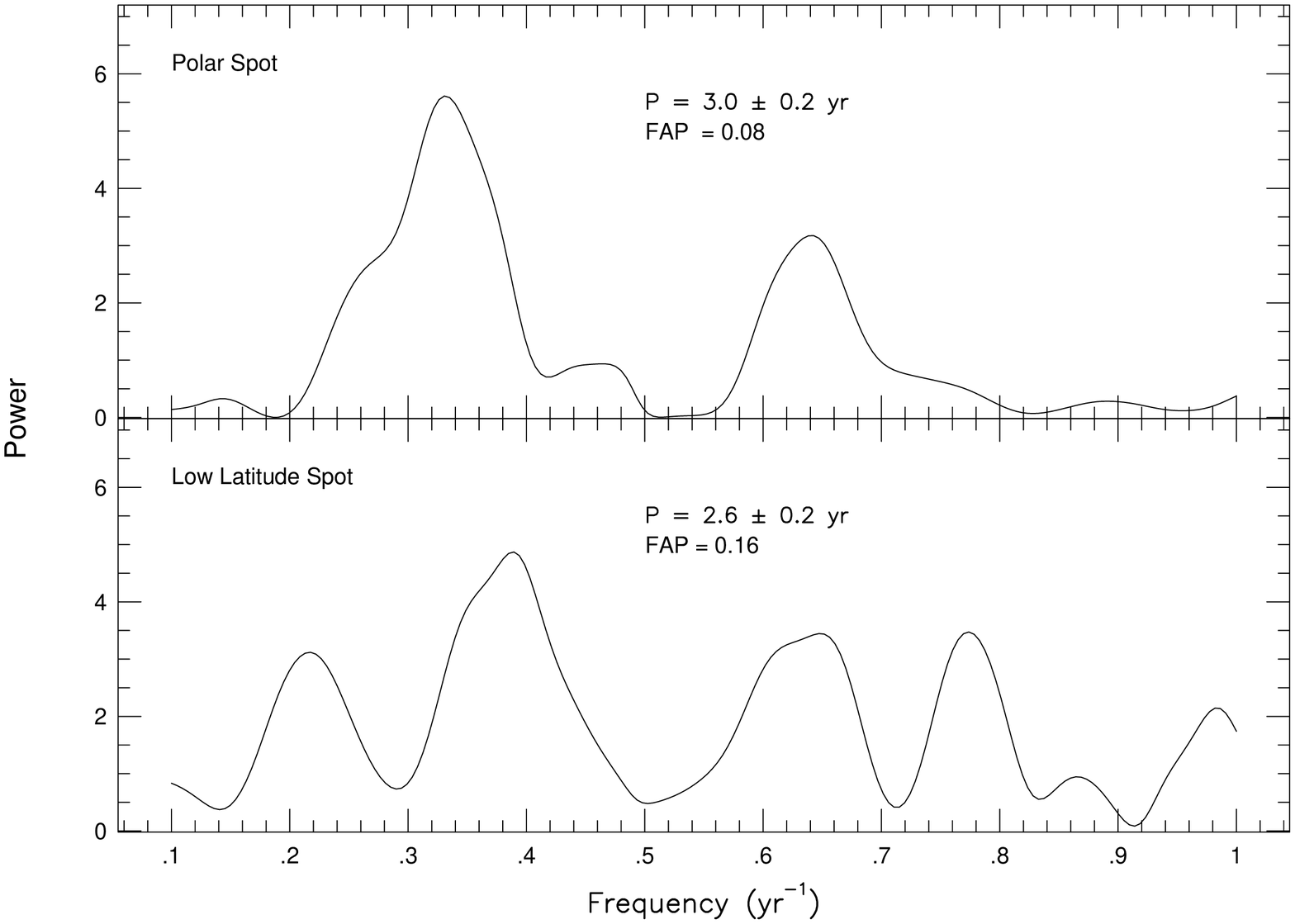}
\caption{Periodogram analysis of polar and low-latitude spot areas}
\label{fig:cycleft}
\end{figure}

A Scargle-type periodogram analysis of the de-trended polar spot area
is presented in Figure~\ref{fig:cycleft}. The upper panel is for the
detrended polar spot area, while the lower panel is for the total
low-latitude spot area. The analysis showed maximum power at a period
of 3.0 $\pm$ 0.2 yr for the polar spot area, with a false-alarm
probability (FAP) of the maximum peak of 0.08. For the low-latitude
total spot area, the analysis yielded a period of 2.6 $\pm$ 0.2 yr
with a FAP of 0.16.  Unfortunately, the quantity and quality of the
data are probably not sufficient to determine if the variations of the
polar spot and low-latitude spots indeed have different periods.

\begin{figure}
\plotone{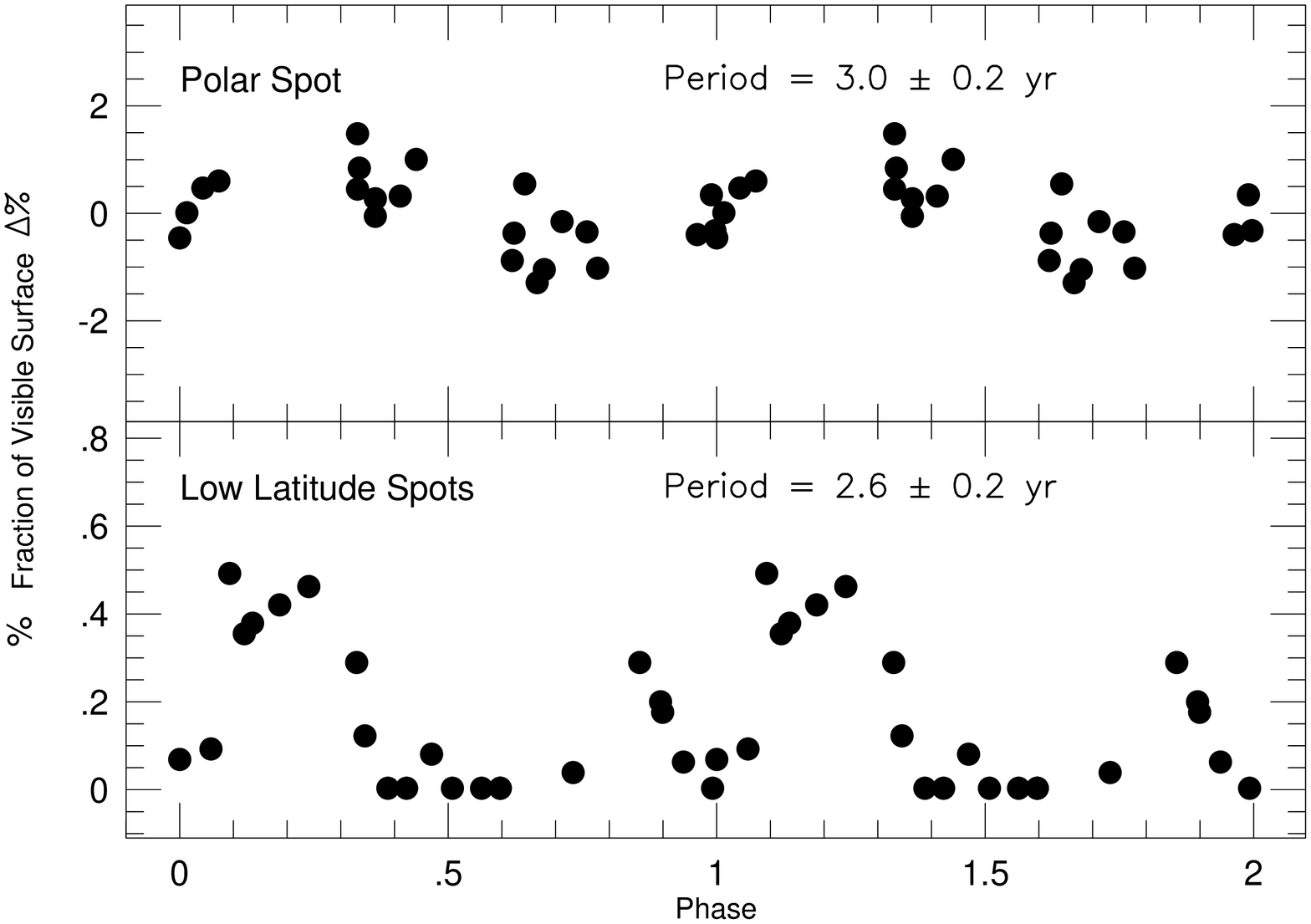}
\caption{Phase diagrams for polar and low-latitude spot areas}
\label{fig:cyclephase}
\end{figure} 

Figure~\ref{fig:cyclephase} shows the best phase diagrams of the polar
and low-latitude spot areas as provided by their respective best
periods from the periodograms. The upper panel is for the polar spot
area, phased at 3.0 year period, and the lower panel is for the total
low-latitude spot area, phased at 2.6 year period.

While there is a hint of periodicity near 3.0 years for the polar
spot, with a false-alarm probablity of 8\%, it cannot yet be regarded
as compelling evidence of a cycle. At this point, we can only conclude
that both the polar feature and low-latitude spots {\it may} show
marginal evidence for an approximately 3-year periodicity in the
variations of their total areas. Of course, we should not expect the
solar analogy to be perfect. It may be that the strong dipole fields
inside these RS CVn stars do not vary or periodically reverse in a
manner similar to the solar cycle, and that perhaps the polar spot
also does not show significant periodic area variation.

\cite{bal96} attempted to derive a dynamo interpretation of stellar
activity cycles based on 25 years of records of Ca II H and K
observations of single lower main sequence stars, 100 of which showed
cycles. They showed that ratio of magnetic cycle $P_{cyc}$ to rotation
period $P_{rot}$ is expected theoretically to be the observed
equivalent of the stellar dynamo number {\it D}. They argue that
$P_{cyc}/P_{rot}$ should scale as {\it D} to some power, where the
power is a positive constant greater than 1/3. Thus a plot of
log($P_{cyc}/P_{rot}$) vs. log({\it D}) should be a straight line with
slope which gives that exponent. HR 1099, with a 3-year cycle comes in
well-below the linear extrapolation of their relation, which predicts
a cycle period of about 5.2 years, but there is a lot of scatter in
their observed relation. Given their scatter, a cycle period for HR
1099 anywhere between about 4-8 years would still be consistent with
their relation.

\subsection{Starspots as Analogues of Solar Coronal Holes}

The detection by \cite{don92b} of strong ($\ge$ 300 G) largely
monopolar surface magnetic fields associated with the edges of the
polar spot in 1990.9, and 700 G monopolar fields associated with the
low-latitude region near phase 0.27, are further support for the idea
(\cite{vgb87}, \cite{vop83}) that the large, essentially permanent
polar spot on HR 1099 and other rapidly rotating RS CVn stars may be
closely analogous to coronal hole structures seen in x-ray images of
the Sun. Solar coronal holes are well-known to be the footprints of
large unipolar magnetic regions. The overall appearance, basic
morphology, approximately solid-body rotation, and temporal evolution
of solar coronal holes are indeed strikingly similar to the spots on
HR 1099 and other RS CVn stars. In particular, the Sun shows a large,
roughly circular polar coronal hole (typical diameter of 40\deg) with
frequent attached `protuberances' (these are called {\it polar hole
extensions} in the jargon of solar astronomers) which descend to lower
latitudes (for a recent reference see \cite{nav94}). Also present are
occasional `isolated' coronal holes at lower latitudes, strikingly
similar in size and evolution to the low and intermediate spots on HR
1099. The Sun's polar coronal hole area varies cyclically in
anti-phase with the 11-year sunspot cycle and is intimately connected
with the spot cycle as part of a more complex extended cycle
(\cite{bra89}, \cite{wal81}, \cite{sim79}).

\cite{wan93} showed that, on the Sun, polar hole extensions form when
the superposition of the axisymmetric polar field and non-axisymmetric
emergent flux at low latitudes distorts the global coronal field
configuration, and thus the polar hole boundaries. The global field
configuration is thus a complex, time-variable average over opposite
polarity flux which is both merging and decaying. The `corridors of
open field lines' formed in the merging of such flux patterns give
rise then to large unipolar magnetic regions (UMR's) within which form
the coronal holes. We suspect a similar complex process is at work
shaping the polar starspot and other spots on HR 1099.

We thus believe that solar coronal holes are a close analog to the
spots we see on the RS CVn stars, the main difference being only the
magnetic field strength. On the Sun, magnetic surface field strengths
inside coronal holes are only 10-20 G. This is not strong enough to
significantly affect convective energy transport in the photosphere,
but does affect plasma motions in the low-density corona, creating
regions of lower density and temperature in the plasma wind outflow
above the spot (hole), and giving rise to prominently visible
structures in x-ray images of the corona. However, on the RS CVn
stars, magnetic flux densities are several orders of magnitude larger,
leading to umbral-strength fields with high filling factor inside the
spot. These fields dominate the convective motions which transport
heat to the surface, thereby lowering the effective surface
temperature within the spot by 1000 K to 1500 K.

\subsection{Is HR 1099 hiding a multi-kG aligned global dipole field?}

Solar polar coronal holes are well-known to be the direct
manifestations of the axisymmetric component of the Sun's global
magnetic field, a 1-2 G periodically reversing dipolar field aligned
with the Sun's rotation axis. The large coronal hole at each pole is a
consequence of the unipolar fields at the rotation poles. The polar
fields are of opposite polarity, and periodically reverse polarity at
sunspot maximum. According to the basic Babcock solar dynamo model,
some fraction of this dipole field is transformed, by differential
rotation and convection, into a toroidal field which then gives rise
to the buoyant magnetic flux ropes whose emergence at low-intermediate
latitudes forms sunspot groups. If the solar analogy of coronal holes
holds true for starspots, we might then expect, as also suggested by
\cite{don92b}, that the large, permanent polar spot on HR 1099 is also
a direct manifestation of a global dipolar magnetic field
closely-aligned with the star's rotation axis. Since coronal holes,
and by analogy starpsots, are regions of predominantly unipolar or
`open' field, the field lines of the observable polar spot must
reconnect (out beyond the star's Alfven radius) with an opposite
polarity region somewhere else on the star, most probably a spot at
the other pole.

Since essentially every rapidly rotating RS CVn star shows a polar
spot, and since it is highly unlikely that we are seeing the same pole
in all these cases, the ubiquitous presence of polar spots indicates
that these stars probably have similar large polar spots at {\it both}
poles. Thus the prominent polar spots seen on HR 1099 and all other
rapidly rotating RS CVn stars are probably mirrored by a corresponding
opposite polarity spot at the other pole (as in the solar case). These
large, essentially permanent polar spots are thus probably the direct
manifestation of these stars' global dipole fields. But whereas the
Sun's poloidal field is only 1-2 G, HR 1099's dipole field, based on
the magnetic measurements of \cite{don92b}, would be at least 300 G
(and probably more like several kG in strength following our analogy).

Dipoles of several kG strength are of course well-known for the
magnetic Ap stars, but are generally highly tilted (oblique) with
respect to the star's rotation axis, whereas the dipoles we suspect
for the RS CVn stars, based on the symmetry of the polar spots, would
be highly aligned. Two obvious origins come to mind for these dipoles,
fossil fields left over from the frozen-in field of the proto-stellar
cloud which have somehow survived the collapse phase, or
dynamo-generated fields arising from the mechanical energy of
convection and rotation in these deeply convective and
rapidly-rotating stars. That more slowly rotating RS CVn stars
apparently do not show strong aligned magnetic dipoles leads us to
believe that they are not fossil fields, but rather are a product of
the deep convective zones and tidally-enforced rapid rotation of the
RS CVn stars.

\section{Summary discussion: dipoles and dynamo modes}

The cool polar spots on HR 1099 and many other rapidly rotating active
late-type stars must, if we've learned anything from the magnetic
fields of sunspots, be telling us that there are multi-kgauss-strength
fields filling these polar spots and suppressing convection in the
photosphere to a degree that the surface is cooled sufficiently to
produce visible spots, just as occurs in sunspot umbrae. However, from
the standpoint of locations, areas, shapes, and differential rotation,
starspots look much more like solar coronal holes than giant
sunspots. In fact, the resemblance between the polar spot of HR 1099
with its variable `protuberances' and the solar polar coronal hole
with its `polar hole extensions' is quite striking. Both straddle the
pole and have similar areas and shapes.

Starspots also show surprisingly little differential rotation, quite
unlike sunspots, but again very similar to solar coronal holes. On the
Sun, coronal holes are the footprints of the unipolar magnetic regions
(UMR's) of the global dipolar coronal field, and their shapes are
maintained rigidly (in the face of considerable photospheric shear) by
a few-gauss current-free global dipole coronal field as described by
\cite{wan93}, the same dipole field which produces the polar coronal
holes. The differential rotation of solar coronal holes is governed by
the stability of the globally-averaged current-free coronal field
rather than by the latitudinal shear of the solar photosphere. If the
solar coronal hole analogy holds true for starspots, then {\it the
differential rotation derived from starspots on RS CVn stars may
actually be revealing very little about the latitudinal shear of the
photosphere, but rather may be more a consequence of the stability of
a powerful, current-free global dipole field}. This global field is
probably of multi-kgauss-strength at the pole (and only half as strong
at the equator if dipolar), orders of magnitude stronger than the
few-gauss global fields of the Sun. Presumably, both the strength and
stability of this global field increases with stellar angular rotation
velocity (i.e. dynamo strength) thereby offering a simple explanation
for why the observed `differential rotation' signature from starspot
light curves decreases as angular velocity increases among active
stars.

That large polar spots are a quite common feature of rapidly rotating
spotted stars, suggests to us that there is a large polar spot at {\it
both} poles of the star. If so, it is reasonable to assume that these
UMR polar spots are of opposite polarity and that their field lines
connect out beyond the Alfven radius. These polar spots thus probably
bear witness to a strong dipole or low-order multi-pole magnetic field
within the star, co-axial with the star's rotation axis. Whether this
axisymmetric field is produced solely by an axisymmetric dynamo or is
the result of accumulation of flux from non-axisymmetric dynamos is
not clear. Probably both contribute. Indeed, the track of Feature B
may be showing us this winding up of field lines around the pole into
the toroidal magnetic structure seen by Donati et al. (1992).

We thus conclude from our Doppler imaging studies that the starspots
on HR 1099 and many other rapidly-rotating late-type stars are
essentially powerful stellar analogs of solar coronal hole structures
(UMR's), but formed by multi-kgauss strength global magnetic fields
rather than the few-gauss global coronal fields of the Sun. The
persistent cool polar spots and increasing stability of spot features
with increasing rotation velocity leads us also to conclude that {\it
rapidly rotating and highly convective stars with prominent polar
spots are probably all hiding multi-kgauss axisymmetric dipole
fields}.

But could such strong axisymmetric dipolar fields on spotted RS CVn
stars have escaped previous direct detection? Actually, they probably
have already been detected on the most active dMe flare stars (see
below), but are harder to detect on rapidly rotating RS CVn stars like
HR 1099 for several reasons. First, any purely axisymmetric field
component would show no rotational modulation. Second, these very
active stars are also quite rapid rotators, and their rotational line
broadening makes traditional Zeeman-splitting detections
difficult. Finally, most of the magnetic flux would be quite
effectively hidden from detection in the optical region of the
spectrum by being concentrated in the very cool (and hence optically
dark) polar spots. Some of this flux has now indeed been detected
directly (at least from areas adjacent to the dark regions) via Zeeman
Doppler images of these ubiquitous polar spots. But the bulk of the
flux probably remains hidden from optical observers in the very low
optical surface brightness spot regions.

Zeeman measurements in the infrared, where spot-to-photosphere
contrast is much less, and/or using spectral features such as TiO
which are strongly enhanced in the dark spots, may soon be able to
routinely measure these hypothesized axisymmetric dipole
fields. \cite{saar85} used the KPNO FTS at 2.2 microns and detected
3.8 kG fields covering 73\% of the surface of the dM3.5e flare star AD
Leo. Furthermore, these fields were from active regions outside of
dark spots. Fields in the dark spots are probably much
stronger. Further recent infrared Zeeman measurements at 2.2 microns
with CSHELL on the IRTF by \cite{saar96} indicate 3 kG fields covering
60\% of the visible surface of the heavily spotted RS CVn star II Peg
, and 2.8 kG fields covering 50\% of the visible surface of the nearly
pole-on K5Ve star Gl 171.2A. The bulk of this magnetic flux is
apparently from the dark spots themselves, which dominate the TiO
features in the infrared. Similarly, \cite{saar96} reports 4 kG fields
covering 60-70\% of the visible surfaces of the active flare stars AD
Leo and YZ CMi, and 3.5 kG fields covering 70\% of the visible surface
of the heavily spotted star LQ Hya. \cite{krull96} also report a
recent detection of 2.6-3.8 kG fields covering about 50\% of the
surface of the M4.5Ve flare stars Gliese 729 and EV Lac. Thanks to the
new generation of IR echelles, we now seem to be crossing the
threshold to routine direct detection of these hypothesiszed global
dipole fields in rapidly rotating RS CVn stars. Such multi-kgauss
global dipole fields are, of course, well-known and quite common among
the Ap stars, though in most of those cases, the field axis is highly
inclined to the rotation axis, as though the field were some fossil
remnant frozen into the star. Here, the precise alignment of the field
to the rotation axis probably indicates a dynamo-induced process for
its creation. Indeed the axisymmetric dipole is a strongly-preferred
dynamo mode.

Our 11-year HR 1099 Doppler image set also revealed that at least some
of the magnetic flux which emerges at lower latitudes migrates
poleward to join with the polar spot. A polar starspot is thus
probably the equivalent of the solar polar coronal hole, complete with
variable `protuberances' (polar hole extensions) but filled with
multi-kgauss-strength magnetic fields similar to those found in
sunspot umbrae. The stability (lack of differential rotation) of spots
is probably maintained by this powerful current-free global dipole
field and, in HR 1099, this dipole field is apparently also tightly
synchronized to the orbit. Given that this dipole probably penetrates
through to the core of the star, this is not altogether
surprising. The dipole is probably rooted to the rotation rate of a
core which was long ago synchronized to the orbit, and remains so
today, even after the outer layers of the star expanded (and slowed)
as the star left the main sequence.

We thus propose the following context and simple two-pronged analogy
by which to understand starspots in terms of solar activity. Starspots
are much like sunspot umbrae by virtue of their large (multi-kgauss)
field strengths, which are strong enough to inhibit convection and
cause a large enough temperature drop in the photosphere to be easily
visible in the optical. However, as regards their global properties
(sizes, shapes, locations, differential rotation, and migration), they
are a direct analog of solar coronal holes (UMR's), with their
structure and evolution heavily influenced by a strong axisymmetric
multi-kgauss global dipole field which has been built up within the
star by powerful dynamo action.

While HR 1099 reveals evidence for spots emerging at low latitude and
migrating poleward (at a latitude migration rate of 6-30 m/s) to
eventually merge with the polar spot, we don't yet know whether these
spots {\it reinforce} the polar spot with like-polarity flux, or
instead {\it destroy} it in some sort of feedback process which may
ultimately produce a periodic polarity-reversal of the polar spot.
Whichever the case, magnetic flux which emerges at lower latitudes
probably separates out into unipolar magnetic regions, and then
migrates poleward to merge, either constructively or destructively,
with the spots at each pole. The axisymmetric dipole field within HR
1099, if related to polar spot area, {\it may} show marginal evidence
for $\approx$ 3-year periodicity (i.e. a dynamo cycle), but probably
varies only weakly, if at all, in total strength.

As shown theoretically by \cite{mos95} from mean field dynamo models,
these stars probably excite both axisymmetric and non-axisymmetric
dynamo fields. The models show that, at moderate values of the Taylor
number, stable non-axisymmetric fields (perpendicular-dipole topology)
can be excited in deep convective shells, and this flux rises to the
surface, emerging at intermediate-latitudes to form spots.  These
theoretical studies apparently have some difficulty producing flux
emergence at low latitudes, but should be pursued to see if they can
be made to do so at these and even {\it lower} latitudes, since HR
1099 does indeed frequently form spots at or near the equator. At
large Taylor numbers (large differential rotation), only axisymmetric
dynamo solutions are stable, and one would therefore expect the
axisymmetric field (i.e. polar spots) to become more prominent with
increasing angular velocity. There is now some observational support
for this. The more slowly-rotating RS CVn stars (Hatzes 1993) exhibit
only intermediate-latitude spots, presumably created from such
non-axisymmetric dynamos, and do {\it not} exhibit the permanent polar
spot signature of an axisymmetric dynamo mode.

The large polar starspots and low-latitude spots on HR 1099 and other
rapidly rotating stars probably indicate that both types of dynamos
are excited within these stars, and furthermore that the flux
contribution from the axisymmetric component increases with angular
rotation velocity, resulting in more prominent and long-lived polar
spots. Some of the flux which emerges at low and intermediate
latitudes from the less stable non-axisymmetric dynamo modes clearly
makes its way to the poles on HR 1099 and may either simply accumulate
there to {\it reinforce} the axisymmetric dipole field, or may {\it
erode} it if of opposite polarity, perhaps as part of a long-term
polarity-reversing periodic process, as occurs on the Sun.  Further
Zeeman Doppler imagery, particularly in the infrared, of stars with
polar spots could be very helpful in sorting out this important
question.

Based on the Solar coronal field analogy, and on the efficiency of the
axisymmetric dynamo mode, we suspect that, close to the pole, the
globally-averaged field assumes the topology of an axisymmetric
dipole. But perhaps the field topology at lower latitudes within the
polar spot is simply a continuation of the toroidal pattern detected
around the polar spot periphery by Donati et al. (1992). If so, as
perhaps also suggested by the track of Feature B on HR 1099, magnetic
flux is literally `winding up' around the pole. The field strength
would thus increase with latitude, and one would be able to detect
such toroidal fields only up to a latitude where their strength was
sufficient to suppress convection (i.e. the edge of the spot). This
would produce a strong bias for field detection precisely at the
boundary of the polar spot, just as observed. The latitude of the
polar spot edge is exactly where the fields would be the strongest
before becoming {\it invisible} within the spot. At lower latitudes,
the fields would be too weak to detect. Above that latitude, the
photosphere would then turn dark, forming the polar spot, and the
tightly-coiled fields within the polar spot would be effectively
hidden from view.

Clearly the axisymmetric and non-axisymmetric dynamo modes need to be
further explored through realistic and self-consistent dynamo
modeling. The relative contributions of each should be predicted as a
function of rotation period, and can then be directly checked through
Doppler images of stars over a range of rotation periods. HR 1099
suggests that both modes are probably active in the rapidly rotating
RS CVn stars, that magnetic flux which emerges at low latitudes does
migrate poleward, and furthermore that there may be a pronounced
winding of low-latitude emergent flux into toroidally-wrapped flux
near or even in the polar spot. Does this winding of flux continue
right up to the polar singularity, or is this toroidal flux being
wound tightly around an axisymmetric dipole field? In either case,
such complex non-potential field configurations might well be expected
to give rise to persistent high temperature/energetic phenomena within
or above the polar spot, as detected by \cite{dupree96} in EUVE
coronal emission from stars with polar spots. These details of the
{\it interaction} between axisymmetric and non-axisymmetric dynamo
modes must be quantified, with the aim of understanding how these
modes interact over the long term, to see if they can indeed build up
sizeable polar fields, and if this field is stable in strength and/or
polarity, or if it varies periodically. It is an area of theoretical
inquiry now ripe for exploration, and should produce predictions that
can be sensibly tested on spotted stars through Doppler imagery.

Finally, the large cool polar spot on HR 1099 is presumably a direct
look at the `magnetic brake' which is ultimately responsible for
slowing the rotation of many rapidly-rotating single late-type
stars. The tidally-locked spotted subgiant star in the HR 1099 binary
system is thus probably ``standing on the brakes'', but alas in vain
in the face of tidal locking to the large angular momentum reservoir
of the orbit.

\acknowledgments

We are very grateful to the Lick and McDonald Observatory TAC's who
made this long-term project possible through major amounts of
observing time over the past decade. We are also very grateful for
support from the NSF under grants AST-9115376 to SSV and AST-9116478
and AST-9315115 to APH. We also gratefully acknowledge the efforts of
G. Donald Penrod in helping to develop the Doppler imaging method.

\appendix

\clearpage
\begin{deluxetable}{lcccc}
\tablewidth{30pc}
\tablecaption{HR 1099: Observations for 1981.70 Image}
\tablehead{
\colhead{Phase} &\colhead{JD 2440000+} &\colhead{Exp. (min)} & \colhead{$S/N$} &
\colhead{Station} }
\startdata
0.254 &  4863.892 & 48 & 150 &    Lick\\
0.297 &  4864.014 & 64 & 170 &    Lick\\
0.300 &  4866.855 & 96 & 220 &    Lick\\ 
0.349 &  4866.997 & 64 & 210 &    Lick \\
0.575 &  4896.016 & 48 & 230 &    Lick\\
0.643 &  4864.996 & 48 & 190 &    Lick\\
0.877 &  4896.874 & 32 & 150 &    Lick\nl
0.943 &  4897.060 & 32 & 180 &    Lick\nl
\enddata
\end{deluxetable}

\clearpage
\begin{deluxetable}{lcccc}
\tablewidth{30pc}
\tablecaption{HR 1099: Observations for 1982.74 Image}
\tablehead{
\colhead{Phase} &\colhead{JD 2440000+} &\colhead{Exp. (min)} & \colhead{$S/N$} &
\colhead{Station} }
\startdata
0.091 &  5240.847 & 96 & 360 &    Lick\\
0.149 &  5241.012 & 48 & 300 &    Lick\\
0.337 &  5250.847 & 48 & 300 &    Lick\\
0.460 &  5241.895 & 96 & 140 &    Lick\\ 
0.645 &  5216.880 & 96 & 280 &    Lick \\
0.677 &  5253.860 & 48 & 250 &    Lick\\
0.693 &  5217.017 & 48 & 150 &    Lick\nl
0.723 &  5253.993 & 48 & 275 &    Lick\nl
0.985 &  5251.899 & 48 & 200 &    Lick\nl
\enddata
\end{deluxetable}
\clearpage

\begin{deluxetable}{lcccc}
\tablewidth{30pc}
\tablecaption{HR 1099: Observations for 1983.74 Image}
\tablehead{
\colhead{Phase} &\colhead{JD 2440000+} &\colhead{Exp. (min)} & \colhead{$S/N$} &
\colhead{Station} }
\startdata                                 
0.050   & 6046.649& 30 &  310 &   Lick\nl
0.122   & 6046.853& 30 &  190 &   Lick\nl
0.204   & 5981.818& 30 &  600 &   Lick\nl
0.253   & 5981.956& 30 &  550 &   Lick\nl
0.282   & 5982.041 & 30 &  580 &   Lick\nl
0.546   & 5979.952 & 30 &  500 &   Lick\nl
0.566   & 5980.009 & 20 &  450 &   Lick\nl
0.581   & 5980.052 & 13 &  200 &   Lick\nl
0.638   & 6042.642& 20 &  410 &   Lick\nl
0.682   & 6042.768& 20 &  500 &   Lick\nl
0.856   & 5980.832 & 20 &  430 &   Lick\nl
0.928   & 5981.034 & 30 &  540 &   Lick\nl
\enddata
\end{deluxetable}
\clearpage

\begin{deluxetable}{lcccc}
\tablewidth{30pc}
\tablecaption{HR 1099: Observations for 1985.87 Image}
\tablehead{
\colhead{Phase} &\colhead{JD 2440000+} &\colhead{Exp. (min)} & \colhead{$S/N$} &
\colhead{Station} }
\startdata                                
0.050   & 6372.980 & 30  &  560  & Lick\nl
0.125   & 6395.903 & 20  &  420  & Lick\nl
0.266   & 6370.764 & 30  &  370  & Lick\nl
0.349   & 6371.000 & 30  &  325  & Lick\nl
0.374   & 6427.826 & 30  &  370  & Lick\nl
0.564   & 6368.774 & 20  &  400  & Lick\nl
0.643   & 6368.996 & 20  &  500   & Lick\nl
0.701   & 6372.000 & 30  &  350  & Lick\nl
0.843   & 6400.778 & 60  &  250  & Lick\nl
0.918   & 6369.776 & 20  &  330  & Lick\nl
0.973   & 6372.770 & 20  &  445  & Lick\nl
\enddata
\end{deluxetable}
\clearpage

\begin{deluxetable}{lcccc}
\tablewidth{30pc}
\tablecaption{Observations for 1986.66 Image}
\tablehead{
\colhead{Phase} &\colhead{JD 2440000+} &\colhead{Exp. (min)} & \colhead{$S/N$} &
\colhead{Station} }
\startdata
0.180 & 6659.969 & 30 & 450 &  Lick\nl
0.259 & 6663.003 & 25 & 350 &  Lick\nl
0.352 & 6720.051 & 25 & 300 &  Lick\nl
0.525 & 6660.950 & 20 & 430 &  Lick \nl
0.845 & 6659.020 & 30 & 450 &  Lick \nl
0.892 & 6661.991 & 25 & 470 &  Lick \nl 
\enddata
\end{deluxetable}
\clearpage

\begin{deluxetable}{lcccc}
\tablewidth{30pc}
\tablecaption{Observations for 1987.04 Image}
\tablehead{
\colhead{Phase} &\colhead{JD 2440000+} &\colhead{Exp. (min)} & \colhead{$S/N$} &
\colhead{Station} }
\startdata           
0.210 & 6807.615 & 30 & 450 & Lick 80''\nl
0.268 & 6807.781 & 30 & 460 & Lick 80''\nl
0.340 & 6810.830 & 60 & 160 & Lick 80''\nl
0.384 & 6813.788 & 30 & 280 & Lick 80''\nl
0.620 & 6811.623 & 60 & 330 & Lick 80''\nl
0.670 & 6814.619 & 20 & 375 & Lick 80''\nl
0.690 & 6811.819 & 30 & 230 & Lick 80''\nl
0.756 & 6814.842 & 40 & 310 & Lick 80''\nl
0.913 & 6809.612 & 40 & 325 & Lick 80''\nl
0.977 & 6809.794 & 60 & 350 & Lick 80''\nl
\enddata
\end{deluxetable}
\clearpage

\begin{deluxetable}{lcccc}
\tablewidth{30pc}
\tablecaption{ Observations for 1987.73 Image}
\tablehead{
\colhead{Phase} &\colhead{JD 2440000+} &\colhead{Exp. (min)} & \colhead{$S/N$} &
\colhead{Station} }
\startdata           
0.069 & 7076.802 & 30 & 450 & Lick 80'' \nl
0.151 & 7077.035 & 20 & 430 & Lick 80'' \nl
0.312 & 7051.952 & 30 & 460 & Lick 80'' \nl
0.420 & 7077.801 & 30 & 540 & Lick 80'' \nl
0.590 & 7049.912 & 30 & 450 & Lick 80'' \nl
0.627 & 7050.009 & 20 & 400 & Lick 80'' \nl
0.726 & 7075.831 & 30 & 570 & Lick 80'' \nl
0.796 & 7076.028 & 30 & 580 & Lick 80'' \nl
0.945 & 7050.911 & 25 &     & Lick 80'' \nl
0.974 & 7050.992 & 20 & 390 & Lick 80'' \nl
\enddata
\end{deluxetable}
\clearpage

\begin{deluxetable}{lcccc}
\tablewidth{30pc}
\tablecaption{Observations for 1988.80 Image}
\tablehead{
\colhead{Phase} &\colhead{JD 2440000+} &\colhead{Exp. (min)} & \colhead{$S/N$} &
\colhead{Station} }
\startdata           
0.083 & 7462.775 & 60 & 500 & Lick 80'' \nl
0.155 & 7465.817 & 36 & 400 & McDonald coud{\'e} \nl
0.190 & 7465.917 & 15 & 400 & McDonald coud{\'e} \nl
0.444 & 7463.801 & 50 & 230 & McDonald coud{\'e} \nl
0.555 & 7432.901 & 60 & 600 & Lick 80'' \nl
0.589 & 7432.997 & 60 & 500 & Lick 80'' \nl
0.781 & 7464.756 & 50 & 580 & Lick 80'' \nl
0.834 & 7464.907 & 30 & 200 & McDonald coud{\'e} \nl
0.854 & 7464.964 & 24 & 280 & McDonald coud{\'e} \nl
0.876 & 7465.025 & 45 & 350 & Lick 80'' \nl
0.925 & 7433.949 & 40 & 620 & Lick 80'' \nl
\enddata
\end{deluxetable}
\clearpage

\begin{deluxetable}{lcccc}
\tablewidth{30pc}
\tablecaption{ Observations for 1989.11 Image}
\tablehead{
\colhead{Phase} &\colhead{JD 2440000+} &\colhead{Exp. (min)} & \colhead{$S/N$} &
\colhead{Station} }
\startdata           
0.049 & 7570.513 & 14  & 260 & ESO \nl
0.182 & 7559.539 & 26  & 285 & ESO \nl
0.232 & 7562.519 & 14  & 350 & ESO \nl
0.342 & 7568.506 & 14  & 265 & ESO \nl
0.488 & 7571.758 & 50  & 500 & Lick coud{\'e}\nl
0.533 & 7574.724 & 60  & 250 & Lick coud{\'e}\nl
0.696 & 7569.510 & 14  & 370 & ESO \nl
0.886 & 7561.536 & 14  & 300 & ESO \nl 
\enddata
\end{deluxetable}
\clearpage

\begin{deluxetable}{lcccc}
\tablewidth{30pc}
\tablecaption{Observations for 1989.69 Image}
\tablehead{
\colhead{Phase} &\colhead{JD 2440000+} &\colhead{Exp. (min)} & \colhead{$S/N$} &
\colhead{Station} }
\startdata           
0.000 & 7788.870 & 35 & 500 & Lick  coud{\'e} \nl
0.032 & 7788.972 & 50 & 550 & Lick  coud{\'e} \nl
0.104 & 7757.961 & 50 & 380 & Lick coud{\'e} \nl
0.350 & 7789.862 & 30 & 350 & Lick coud{\'e} \nl
0.380 & 7789.966 & 35 & 449 & Lick coud{\'e} \nl
0.445 & 7758.935 & 50 & 300 & Lick  coud{\'e} \nl
\enddata
\end{deluxetable}
\clearpage

\begin{deluxetable}{lcccc}
\tablewidth{30pc}
\tablecaption{ Observations for 1989.78 Image}
\tablehead{
\colhead{Phase} &\colhead{JD 2440000+} &\colhead{Exp. (min)} & \colhead{$S/N$} &
\colhead{Station} }
\startdata           
0.060 & 7811.751 & 40 & 600 & Lick 80'' \nl
0.125 & 7814.774 & 60 & 300 & McDonald coud{\'e} \nl
0.128 & 7814.774 & 40 & 350 & Lick 80'' \nl
0.168 & 7814.898 & 45 & 300 & McDonald coud{\'e} \nl
0.475 & 7815.768 & 45 & 280 & McDonald coud{\'e} \nl
0.554 & 7815.991 & 45 & 250 & McDonald coud{\'e} \nl
0.710 & 7810.763 & 40 & 600 &  Lick 80'' \nl
0.770 & 7810.917 & 35 & 600 & Lick 80'' \nl    
0.804 & 7813.862 & 40 & 260 & McDonald coud{\'e} \nl
0.807 & 7811.035 & 35 & 650 & Lick 80'' \nl
0.812 & 7813.886 & 25 & 220 & McDonald coud{\'e} \nl
0.900 & 7816.974 & 45 & 250 & McDonald coud{\'e} \nl
\enddata
\end{deluxetable}
\clearpage

\begin{deluxetable}{lcccc}
\tablewidth{30pc}
\tablecaption{Observations for 1989.88 Image}
\tablehead{
\colhead{Phase} &\colhead{JD 2440000+} &\colhead{Exp. (min)} & \colhead{$S/N$} &
\colhead{Station} }
\startdata           
0.068 & 7845.827 & 45 & 210 & McDonald coud{\'e} \nl
0.320 & 7843.717 & 40 & 400 &  Lick 80'' \nl \nl
0.380 & 7846.712 & 90 & 270 & McDonald coud{\'e} \nl
0.400 & 7843.935 & 40 & 420 &  Lick 80'' \nl
0.435 & 7846.870 & 60 & 300 & McDonald coud{\'e} \nl
0.680 & 7844.733 & 40 & 370 &  Lick 80'' \nl
0.701 & 7844.788 & 50 & 330 & McDonald coud{\'e} \nl
0.738 & 7844.890 & 35 & 330 &  Lick 80'' \nl
0.764 & 7844.965 & 45 & 360 & McDonald coud{\'e} \nl
0.963 & 7876.744 & 40 & 450 &  Lick 80'' \nl
\enddata
\end{deluxetable}
\clearpage

\begin{deluxetable}{lcccc}
\tablewidth{30pc}
\tablecaption{HR 1099: Observations for 1990.69 Image}
\tablehead{
\colhead{Phase} &\colhead{JD 2440000+} &\colhead{Exp. (min)} & \colhead{$S/N$} &
\colhead{Station} }
\startdata           
0.116 & 8143.927 & 21 & 220 & ESO \nl
0.180 & 8149.786 & 28 & 300 & ESO \nl
0.248 & 8152.813 & 28 & 430 & ESO \nl
0.302 & 8135.942 & 35 & 170 & McDonald coud{\'e} \nl
0.390 & 8141.869 & 21 & 150 & ESO \nl
0.434 & 8144.828 & 21 & 440 & ESO \nl
0.636 & 8136.890 & 40 & 200 & McDonald coud{\'e} \nl
0.710 & 8142.774 & 28 & 320 & ESO \nl
0.744 & 8142.870 & 28 & 280 & ESO \nl
0.928 & 8134.879 & 45 & 300 & McDonald coud{\'e} \nl
\enddata
\end{deluxetable}
\clearpage

\begin{deluxetable}{lcccc}
\tablewidth{30pc}
\tablecaption{Observations for 1990.80 Image}
\tablehead{
\colhead{Phase} &\colhead{JD 2440000+} &\colhead{Exp. (min)} & \colhead{$S/N$} &
\colhead{Station} }
\startdata           
0.040 & 8174.927 & 60 & 260 & McDonald coud{\'e} \nl
0.070 & 8175.010 & 60 & 220 & McDonald coud{\'e} \nl
0.367 & 8175.849 & 60 & 270 & McDonald coud{\'e} \nl
0.418 & 8176.000 & 40 & 270 & McDonald coud{\'e} \nl
0.484 & 8201.726 & 40 & 330 & McDonald coud{\'e} \nl
0.557 & 8201.932 & 45 & 250 & McDonald coud{\'e}\nl
0.586 & 8167.963 & 40 & 250 & McDonald coud{\'e}\nl
0.606 & 8204.910 & 45 & 260 & McDonald coud{\'e} \nl
0.626 & 8204.966 & 45 & 230 & McDonald coud{\'e} \nl
0.863 & 8165.912 & 20 & 500 & Lick HS  \nl
\enddata
\end{deluxetable}
\clearpage

\begin{deluxetable}{lcccc}
\tablewidth{30pc}
\tablecaption{Observations for 1990.93 Image}
\tablehead{
\colhead{Phase} &\colhead{JD 2440000+} &\colhead{Exp. (min)} & \colhead{$S/N$} &
\colhead{Station} }
\startdata           
0.044 & 8231.691 & 60 & 180 & McDonald coud{\'e} \nl
0.124 & 8231.919 & 30 & 380 & Lick \nl
0.301 & 8229.583 & 60 & 240 & McDonald coud{\'e} \nl
0.360 & 8229.756 & 60 & 330 & McDonald coud{\'e} \nl
0.415 & 8229.906 & 50 & 240 & McDonald coud{\'e} \nl
0.660 & 8230.612 & 60 & 170 & McDonald coud{\'e} \nl
0.718 & 8230.768 & 55 & 400 & McDonald coud{\'e} \nl
0.745 & 8230.845 & 33 & 320 & Lick HS \nl
0.771 & 8230.918 & 60 & 430 & McDonald coud{\'e} \nl
\enddata
\end{deluxetable}
\clearpage

\begin{deluxetable}{lcccc}
\tablewidth{30pc}
\tablecaption{HR 1099: Observations for 1991.02 Image}
\tablehead{
\colhead{Phase} &\colhead{JD 2440000+} &\colhead{Exp. (min)} & \colhead{$S/N$} &
\colhead{Station} }
\startdata           
0.216 & 8260.558 & 60 & 400 & McDonald coud{\'e} \nl
0.278 & 8260.735 & 60 & 200 & McDonald coud{\'e} \nl
0.280 & 8263.580 & 28 & 275 & ESO \nl
0.328 & 8263.714 & 45 & 370 & McDonald coud{\'e} \nl
0.340 & 8266.585 & 21 & 200 & ESO \nl
0.572 & 8261.590 & 21 & 230 & ESO \nl
0.636 & 8264.587 & 21 & 200 & ESO \nl
0.644 & 8264.772 & 50 & 450 & McDonald coud{\'e} \nl 
0.691 & 8267.581 & 21 & 340 & ESO \nl
0.861 & 8259.551 & 40 & 360 & McDonald coud{\'e} \nl 
0.918 & 8259.712 & 60 & 400 & McDonald coud{\'e} \nl 
0.923 & 8262.559 & 21 & 400 & ESO \nl
0.960 & 8259.832 & 60 & 360 & McDonald coud{\'e} \nl 
\enddata
\end{deluxetable}
\clearpage

\begin{deluxetable}{lcccc}
\tablewidth{30pc}
\tablecaption{Observations for 1991.80 Image}
\tablehead{
\colhead{Phase} &\colhead{JD 2440000+} &\colhead{Exp. (min)} & \colhead{$S/N$} &
\colhead{Station} }
\startdata           
0.115 & 8549.721 & 50 & 280 & McDonald coud{\'e} \nl
0.177 & 8549.895 & 50 & 300 & McDonald coud{\'e} \nl
0.370 & 8558.952 & 30 & 260 & McDonald coud{\'e} \nl
0.421 & 8547.703 & 45 & 370 & McDonald coud{\'e} \nl
0.472 & 8547.895 & 30 & 330 & McDonald coud{\'e} \nl
0.762 & 8548.720 & 75 & 350 & McDonald coud{\'e} \nl
0.833 & 8548.926 & 70 & 240 & McDonald coud{\'e} \nl
\enddata
\end{deluxetable}
\clearpage

\begin{deluxetable}{lcccc}
\tablewidth{30pc}
\tablecaption{Observations for 1991.88 Image}
\tablehead{
\colhead{Phase} &\colhead{JD 2440000+} &\colhead{Exp. (min)} & \colhead{$S/N$} &
\colhead{Station} }
\startdata           
0.058 & 8580.775  & 45 & 230 & McDonald Coud{\'e} \nl
0.107 & 8580.913  & 50 & 300 & McDonald Coud{\'e} \nl
0.140 & 8586.800  & 28 & 380 & ESO \nl
0.342 & 8581.580  & 28 & 450 & ESO \nl
0.383 & 8581.695  & 45 & 230 & McDonald Coud{\'e} \nl
0.420 & 8584.638  & 35 & 240 & ESO \nl
0.438 & 8581.852  & 50 & 380 & McDonald Coud{\'e} \nl
0.460 & 8587.590  & 49 & 240 & ESO \nl
0.659 & 8579.641  & 45 & 320 & McDonald Coud{\'e} \nl
0.665 & 8579.657  & 28 & 450 & ESO \nl
0.693 & 8582.575  & 28 & 400 & ESO \nl
0.707 & 8579.776  & 45 & 300 & McDonald Coud{\'e} \nl
0.740 & 8585.553  & 42 & 340 & ESO \nl
0.750 & 8579.899  & 50 & 480 & McDonald Coud{\'e} \nl
0.800 & 8588.550  & 28 & 320 & ESO \nl
\enddata
\end{deluxetable}
\clearpage

\begin{deluxetable}{lcccc}
\tablewidth{30pc}
\tablecaption{ Observations for 1992.06 Image}
\tablehead{
\colhead{Phase} &\colhead{JD 2440000+} &\colhead{Exp. (min)} & \colhead{$S/N$} &
\colhead{Station} }
\startdata           
0.364 & 8635.558  & 40 & 260 & McDonald Coud{\'e} \nl
0.428 & 8635.742  & 45 & 315 & McDonald Coud{\'e} \nl
0.555 & 8641.777  & 10 & 320 & Lick HS \nl
0.626 & 8613.601  & 40 & 260 & McDonald Coud{\'e} \nl
0.726 & 8636.588  & 40 & 500 & McDonald Coud{\'e} \nl
0.792 & 8636.774  & 50 & 340 & McDonald Coud{\'e} \nl
0.841 & 8676.642  &  8 & 250 & Lick HS \nl
0.956 & 8631.564  & 47 & 120 & McDonald Coud{\'e} \nl
\enddata
\end{deluxetable}
\clearpage

\begin{deluxetable}{lcccc}
\tablewidth{30pc}
\tablecaption{Observations for 1992.69 Image}
\tablehead{
\colhead{Phase} &\colhead{JD 2440000+} &\colhead{Exp. (min)} & \colhead{$S/N$} &
\colhead{Station} }
\startdata           
0.035 & 8872.994  & 13 & 300 & McDonald SE \nl  
0.105 & 8878.870  & 10 & 290 & Lick HS \nl
0.307 & 8870.931  & 13 & 230 & McDonald SE \nl
0.443 & 8876.990  & 20 & 320 & Lick HS \nl
0.681 & 8871.992  &  9 & 280 & McDonald SE \nl
0.770 & 8877.920  & 15 & 250 & Lick HS \nl
0.953 & 8869.926  & 11 & 230 & McDonald SE \nl
\enddata
\end{deluxetable}
\clearpage

\begin{deluxetable}{lcccc}
\tablewidth{30pc}
\tablecaption{Observations for 1992.82 Image}
\tablehead{
\colhead{Phase} &\colhead{JD 2440000+} &\colhead{Exp. (min)} & \colhead{$S/N$} &
\colhead{Station} }
\startdata           
0.125 & 8912.980  & 15 & 240 & Lick HS \nl
0.407 & 8913.780  & 10 & 200 & Lick HS \nl
0.425 & 8913.830  & 10 & 270 & Lick HS \nl
0.568 & 8936.940  & 20 & 170 & Lick HS \nl
0.748 & 8911.91   & 20 & 150 & Lick HS \nl
0.770 & 8914.810  & 10 & 140 & Lick HS \nl
0.920 & 8937.945  & 15 & 320 & Lick HS \nl
\enddata
\end{deluxetable}
\clearpage

\begin{deluxetable}{lcccc}
\tablewidth{30pc}
\tablecaption{Observations for 1992.95 Image}
\tablehead{
\colhead{Phase} &\colhead{JD 2440000+} &\colhead{Exp. (min)} & \colhead{$S/N$} &
\colhead{Station} }
\startdata           
0.092 & 8969.641  & 25 & 260 & McDonald SE \nl
0.148 & 8969.798  & 15 & 150 & McDonald SE \nl
0.392 & 8967.653  & 30 & 240 & McDonald SE \nl
0.464 & 8967.858  & 40 & 250 & McDonald SE \nl  
0.734 & 8968.624  & 15 & 330 & McDonald SE \nl
0.782 & 8968.775  & 18 & 340 & McDonald SE \nl
0.954 & 8957.900  & 15 & 280 & McDonald SE \nl
\enddata
\end{deluxetable}
\clearpage

\begin{deluxetable}{lcccc}
\tablewidth{30pc}
\tablecaption{ Observations for 1993.14 Image}
\tablehead{
\colhead{Phase} &\colhead{JD 2440000+} &\colhead{Exp. (min)} & \colhead{$S/N$} &
\colhead{Station} }
\startdata           
0.053 & 9020.611  & 15 & 320 & McDonald SE \nl
0.335 & 9052.625  & 20 & 270 & Lick HS \nl
0.400 & 9021.594  & 30 & 210 & McDonald SE \nl
0.534 & 9044.676  & 14 & 290 &  McDonald SE \nl
0.634 & 9050.636  & 20 & 240 & Lick HS \nl
0.686 & 9053.619  & 15 & 250 & Lick HS \nl
0.750 & 9022.585  & 30 & 260 & McDonald SE \nl
0.984 & 9051.629  & 20 & 260 & Lick SE \nl
\enddata
\end{deluxetable}

%

\clearpage

%

\end{document}